\documentclass[11pt, a4paper, margin=0.25in]{article}

\usepackage[multidot]{grffile}

\usepackage{geometry}
\usepackage[pagewise, displaymath, mathlines]{lineno}
\usepackage{enumitem}
\usepackage{setspace}

\usepackage{amsmath}
\usepackage{amsthm}

\usepackage[single]{accents}
\usepackage[nodayofweek]{datetime}
\usepackage{natbib, url}
\usepackage{graphicx}
\usepackage{threeparttable}
\usepackage{rotating}
\usepackage{xr-hyper}
\usepackage{booktabs} 
\usepackage[bookmarks,bookmarksnumbered=true,colorlinks=true,
     linkcolor=blue,urlcolor=blue,citecolor=blue,breaklinks = true,backref=page]{hyperref}

\usepackage{doi}
\newif\ifempty
\providecommand{\noopsort}[1]{} 

\input{./commands.sty}

\newcommand{\footremember}[2]{%
  \footnote{#2}
  \newcounter{#1}
  \setcounter{#1}{\value{footnote}}%
}

\newdateformat{mydate}{\twodigit{\THEDAY}{  }\monthname[\THEMONTH] \THEYEAR}
\usepackage{scalerel,amssymb}
\usepackage[multidot]{grffile}

\usepackage{array}
\newcolumntype{L}{>{\centering\arraybackslash}m{5cm}}

\newcommand{\Lowx}{\textrm{Low}}

\usepackage{amsthm}
\title{Modern Statistical Models and Methods for Estimating Fatigue-Life and Fatigue-Strength Distributions from Experimental Data}
\author{%
  William Q. Meeker\footremember{isu}{Department of Statistics, Iowa State University}%
  \and Luis A. Escobar\footremember{lsu}{Department of Experimental Statistics, Louisiana State University}%
  \and Francis G. Pascual\footremember{wsu}{Department of
    Mathematics and Statistics, Washington State University}
  \and Yili Hong\footremember{vt}{Department of Statistics, Virginia Tech}
  \and Peng Liu \footremember{jmp}{JMP Statistical Discovery LLC}
  \and Wayne M. Falk\footremember{med}{U.S. Food and Drug Administration}
  \and Balajee Ananthasayanam \footremember{hon}{Honda Aero}
}
\mydate

\begin{document}

\maketitle

\begin{abstract}
Engineers and scientists have been collecting and analyzing fatigue
data since the 1800s to ensure the reliability of life-critical
structures. Applications include (but are not
limited to) bridges, building structures, aircraft and spacecraft
components, ships, ground-based vehicles, and medical devices.
Engineers need to estimate \SN{}
relationships (Stress versus Number
of cycles to failure), typically with
a focus on estimating small quantiles of the \textit{fatigue-life}
distribution.
Estimates from this
kind of model are used as input to models (e.g., cumulative damage
models) that predict failure-time distributions under varying stress
patterns. Also, design engineers need to estimate lower-tail
quantiles of the closely related
\textit{fatigue-strength} distribution.
The
history of applying incorrect statistical methods is nearly as long
and such practices continue to the present. Examples include
treating the applied stress (or strain) as the response and the
number of cycles to failure as the explanatory variable in
regression analyses (because of the need to estimate fatigue-strength
distributions) and ignoring or otherwise mishandling censored
observations (known as runouts in the fatigue literature).
The first part of the paper reviews the traditional
modeling approach where a fatigue-life
model is specified. Then we show how this specification induces a
corresponding fatigue-strength model. The second part of the
paper presents a novel alternative modeling approach where a
fatigue-strength model is specified and a corresponding
fatigue-life model is induced. We explain and illustrate the
important advantages of this new modeling approach.
\end{abstract}


\newpage

\tableofcontents

\newpage

\section{Introduction}
\subsection{Motivation}
\label{section:motivation}
Engineers and scientists have been collecting and analyzing fatigue
data since the 1800s to ensure the reliability of life-critical structures.
Applications include (but are not limited to)
bridges, building structures, aircraft and spacecraft components,
ships, ground-based vehicles, and medical devices.  Because of its
importance, fatigue has been and continues to be the most widely
studied failure mechanism.  Hundreds of technical papers
describing fatigue data are published each year.  Even today, many
of these papers are not using appropriate statistical methods.
\textit{Current} standards and handbooks such as
\citet{ISO12107:2012}, \citet{E739-10:2015}, and \citet{MMPDS2021}
describe and recommend \textit{archaic statistical methods}
developed from the late 1940s to the late 1960s.
Modern statistical methods and the availability of computational
power allow engineers to fit needed nonlinear regression models and
properly handle runouts (right-censored observations).  The modern
 statistical methods that provide improved statistical inference, however, are not
presented in engineering standards.  While this paper will not
immediately remedy
the omission of these topics, the information presented in this
paper can guide inclusion in future revisions of engineering
standards and handbooks.

\subsection{Laboratory Experiments to Obtain \SN{} Data}
\label{section:laboratory.experiments.obtain.sn.data}
In laboratory testing, under cyclic stress or strain loading
(often a sine wave where the amount of stress or strain
is usually given in terms of amplitude), test units accumulate
damage---crack initiation and subsequent growth.
Depending on the material and range of
stress to be applied in the test it is sometimes appropriate to control stress
(or displacement that causes stress) amplitude  and in other applications
it is more appropriate to control strain amplitude.
Textbooks such as \citet{Dowling2013} provide more details about how
fatigue tests are conducted.
For
consistency, we will generally use the word ``stress'' except in the
numerical examples where strain-control was used.

The damage accumulation rate depends on levels of stress.
A representative sample of
specimens from some production process (e.g., selected from
multiple heats or batches) will be tested.
These specimens should be randomly assigned to test conditions and
order of testing.  The number and location of stress levels, the
number of test specimens, and the allocation of the test specimens
needs to be specified in a
purposeful way (described further in
Section~\ref{section:concluding.remarks}).  Fatigue tests can be conducted
to control  stress amplitude, displacement (which is proportional to
stress), or strain amplitude.
Some fatigue testing machines can test only one
specimen at a time. Other machines are available to do stressing
simultaneously on multiple specimens.
Typically units are tested at a fixed stress amplitude until failure
or a prespecified censoring point, whichever comes first. Unfailed
units are known as runouts or right-censored observations and are an
important part of the data.

\subsection{Motivating Examples}
\label{section:motivating.examples}
This section presents motivating examples based on fatigue
tests with three different materials: a composite material and two
metals with different characteristics. As will be shown in subsequent
sections, the features of the different data sets will suggest
regression models with different characteristics.

\begin{example}
\label{example:laminate.panel.data}
\Exampletitle{Fatigue-Life Data from a Test on a Laminate Panel.}
Four-point out-of-plane bending tests were run on 25 specimens of
carbon eight-harness satin/epoxy laminate panels at each of five
different stress levels. There were ten runouts at the two lowest
stress levels. Fatigue life was considered to be the number of
cycles until a specimen fractured. These data were previously
analyzed (using models different from those presented in this paper)
in \citet{ShimokawaHamaguchi1987},
\citet{PascualMeeker1999}, and \citet[][Chapter~17]{MeekerEscobarPascual2021}.

Figure~\ref{figure:LaminatePanelScatter.plots}a is a log-log
scatterplot of the laminate panel \SN{} data with the
response (thousands of cycles) on the vertical axis.
Figure~\ref{figure:LaminatePanelScatter.plots}b plots the same
data with the response on the horizontal axis. It is
common practice to plot a regression response on the vertical axis. In the
fatigue literature, however, the response is plotted on the
horizontal axis and we will follow that convention here.
Although it is partially obscured by the runouts, there is strong
statistical evidence of curvature in the data on log-log
scales. Such curvature is ubiquitous in \SN{} data,
especially when there are tests at low stress levels resulting
in long failure times (i.e., high-cycle fatigue or HCF).
\begin{figure}[ht]
\begin{tabular}{cc}
(a) & (b) \\[-3.2ex]
\rsplidapdffiguresize{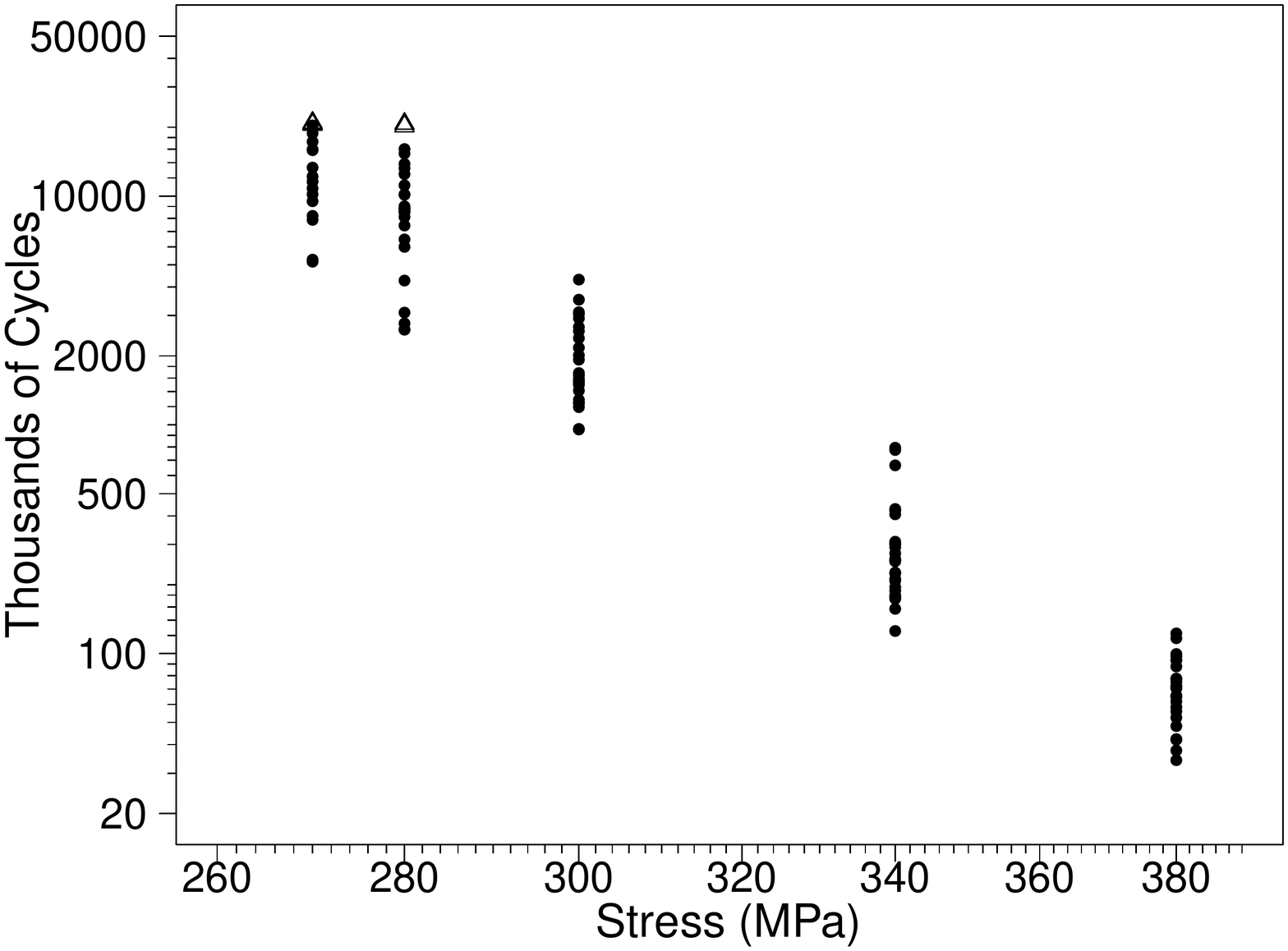}{3.25in} &
\rsplidapdffiguresize{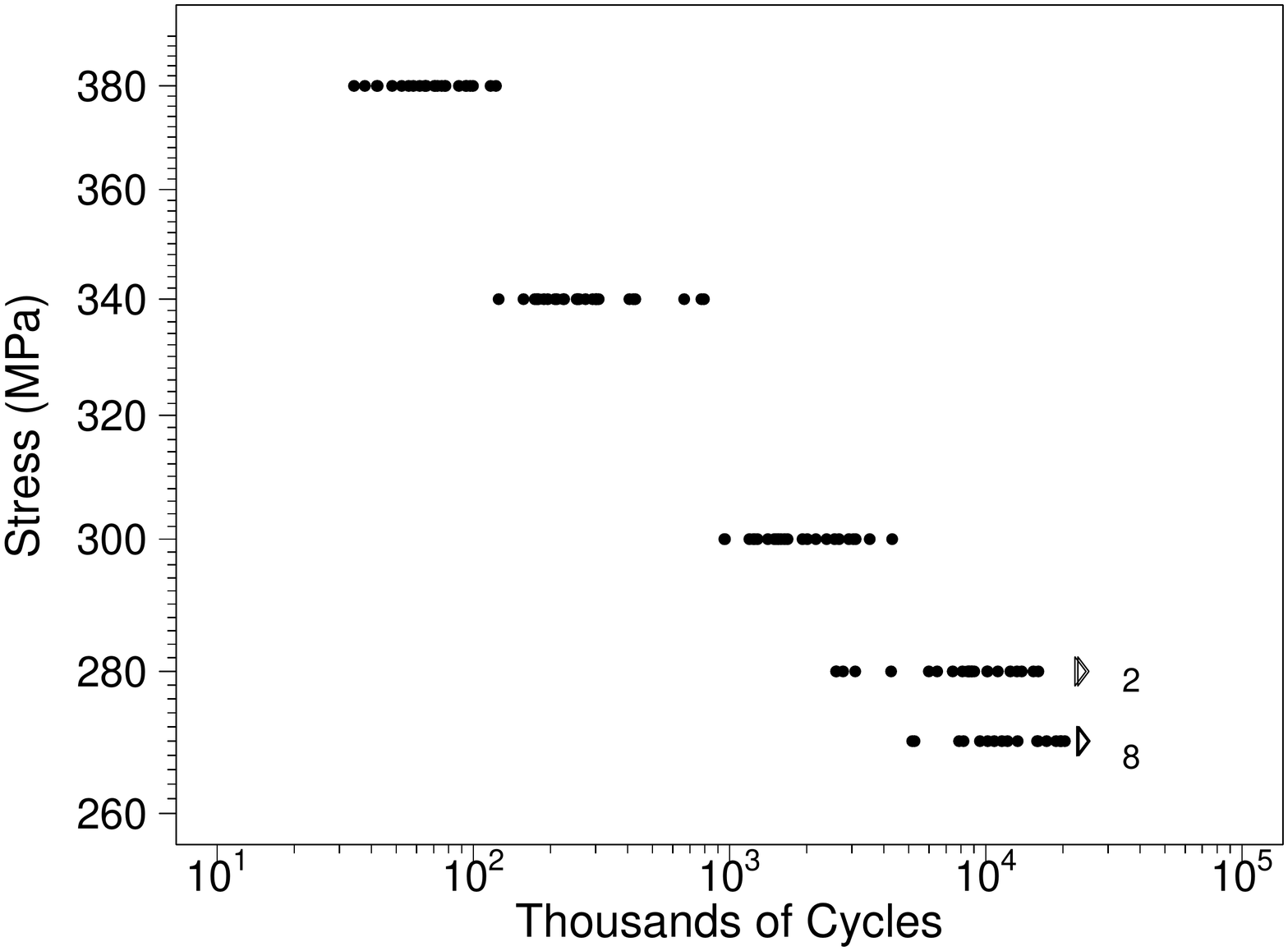}{3.25in}
\end{tabular}
\caption{Laminate panel \SN{} data with number of cycles on
  the vertical axis~(a) and the horizontal axis~(b).}
\label{figure:LaminatePanelScatter.plots}
\end{figure}

Figure~\ref{figure:LaminatePanelProbability.plots} shows Weibull and
lognormal probability plots for the laminate panel fatigue-life data
with estimates of the cdfs (the lines going through the nonparametric
estimate points) at
each level of stress.
\begin{figure}[ht]
\begin{tabular}{cc}
(a) & (b) \\[-3.2ex]
\rsplidapdffiguresize{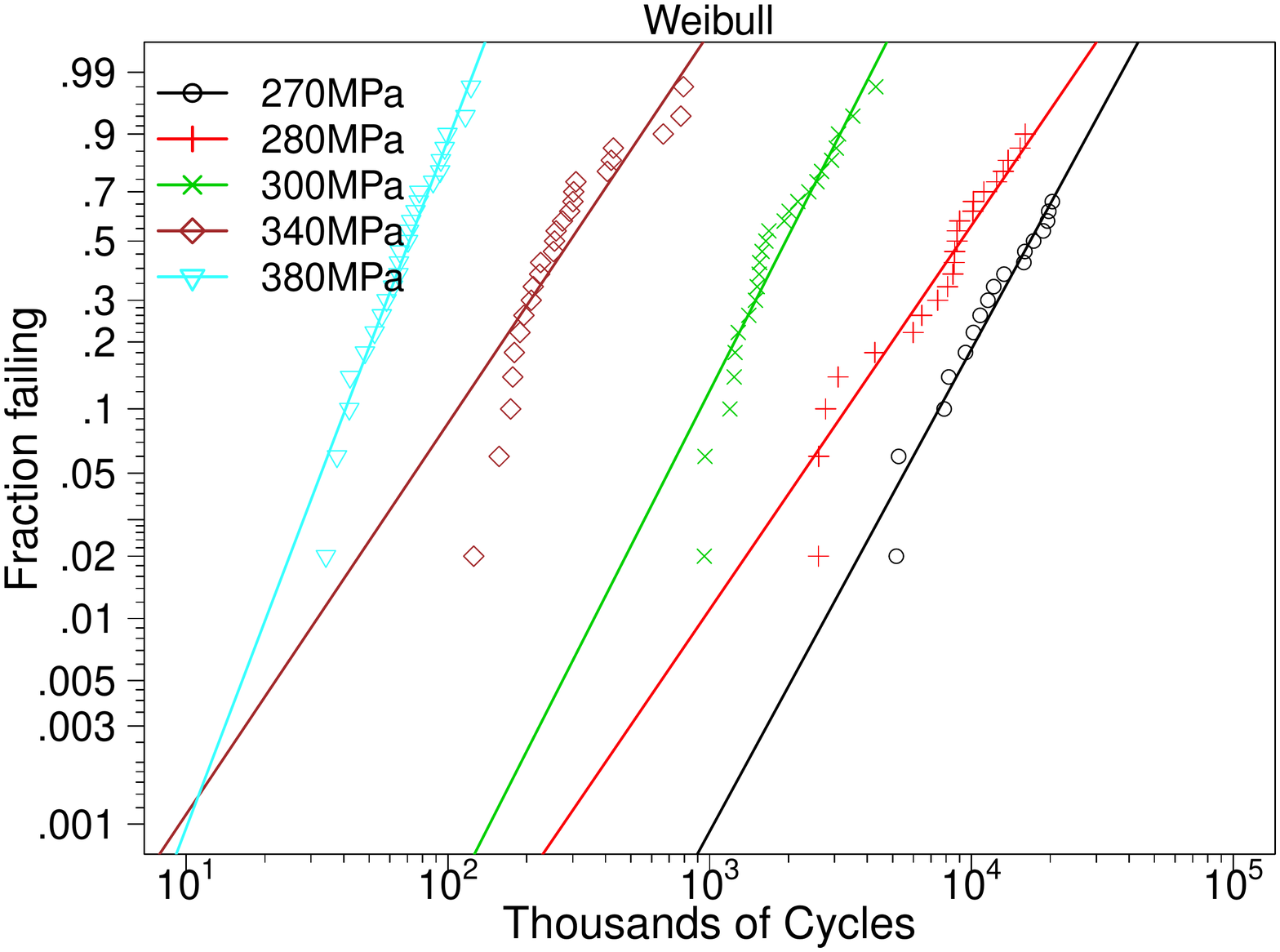}{3.25in} &
\rsplidapdffiguresize{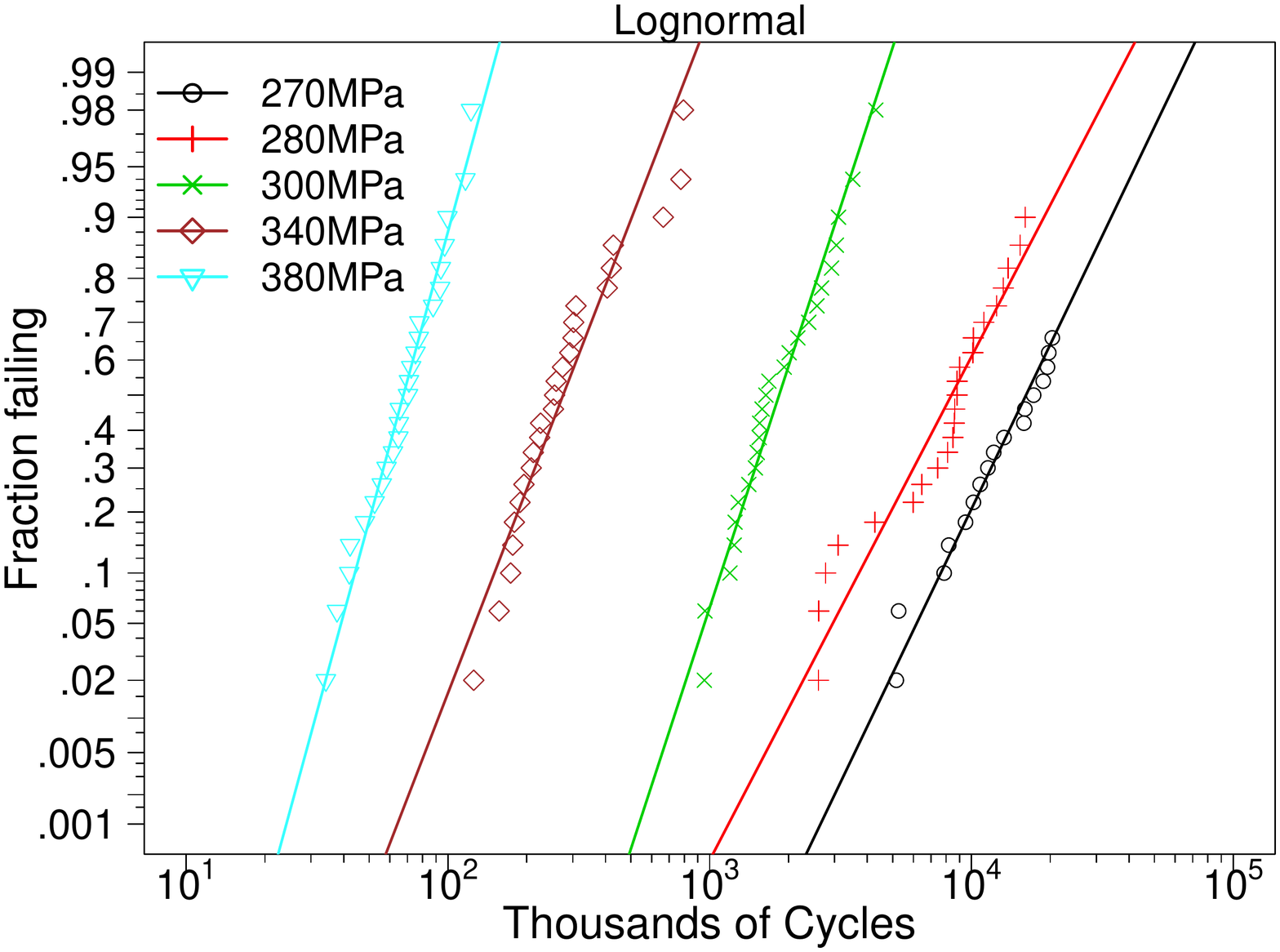}{3.25in}
\end{tabular}
\caption{Laminate panel \SN{} data Weibull~(a) and lognormal~(b)
  probability plots with separate distributions fit to each
  level of stress.}
\label{figure:LaminatePanelProbability.plots}
\end{figure}
These plots suggest that the lognormal distribution provides a
better description of the data. The slopes of the cdf-estimate lines
in Figure~\ref{figure:LaminatePanelProbability.plots}b
tend to decrease from left to right. This implies more spread in the data
at lower stress levels, another common characteristic of \SN{}
data. Section~\ref{section:fitting.fatigue.life.model} will present
an appropriate fatigue-life regression model to describe these data.
\end{example}

\begin{example}
\label{example:Ti64.data}
\Exampletitle{Fatigue-Life Data from a Test on Ti64 Specimens.}
Ti-6Al-4V (Ti64) is an alloy of titanium, aluminum, and vanadium
that has a high strength-to-weight ratio and corrosion
resistance. Because of these properties, Ti64 is used widely in
aerospace applications. Data from a fatigue test are shown in
Figure~\ref{figure:Ti64350FRm1.plots}a. Units were subjected to
cyclic loading at a temperature of $350\degreesf$ with a stress
ratio $R=-1$ (fully reversed loading with a zero-mean stress) with
stress amplitudes of 60 (37 specimens), 70 (12 specimens), 80 (11
specimens), and 90 (12 specimens) ksi (thousands of pounds per square inch).
More units were tested
at 60 ksi because it was expected that a smaller proportion of
tested units would fail there. Of the 37 specimens tested at
60 ksi, 28 were runouts that survived between 30,000 and 46,505
thousand cycles (indicated by the triangles pointing to the right in
the figure). Figure~\ref{figure:Ti64350FRm1.plots}a
indicates strong curvature in the \SN{}
relationship and increasing spread at lower stress levels.
\begin{figure}[ht]
\begin{tabular}{cc}
(a) & (b) \\[-3.2ex]
\rsplidapdffiguresize{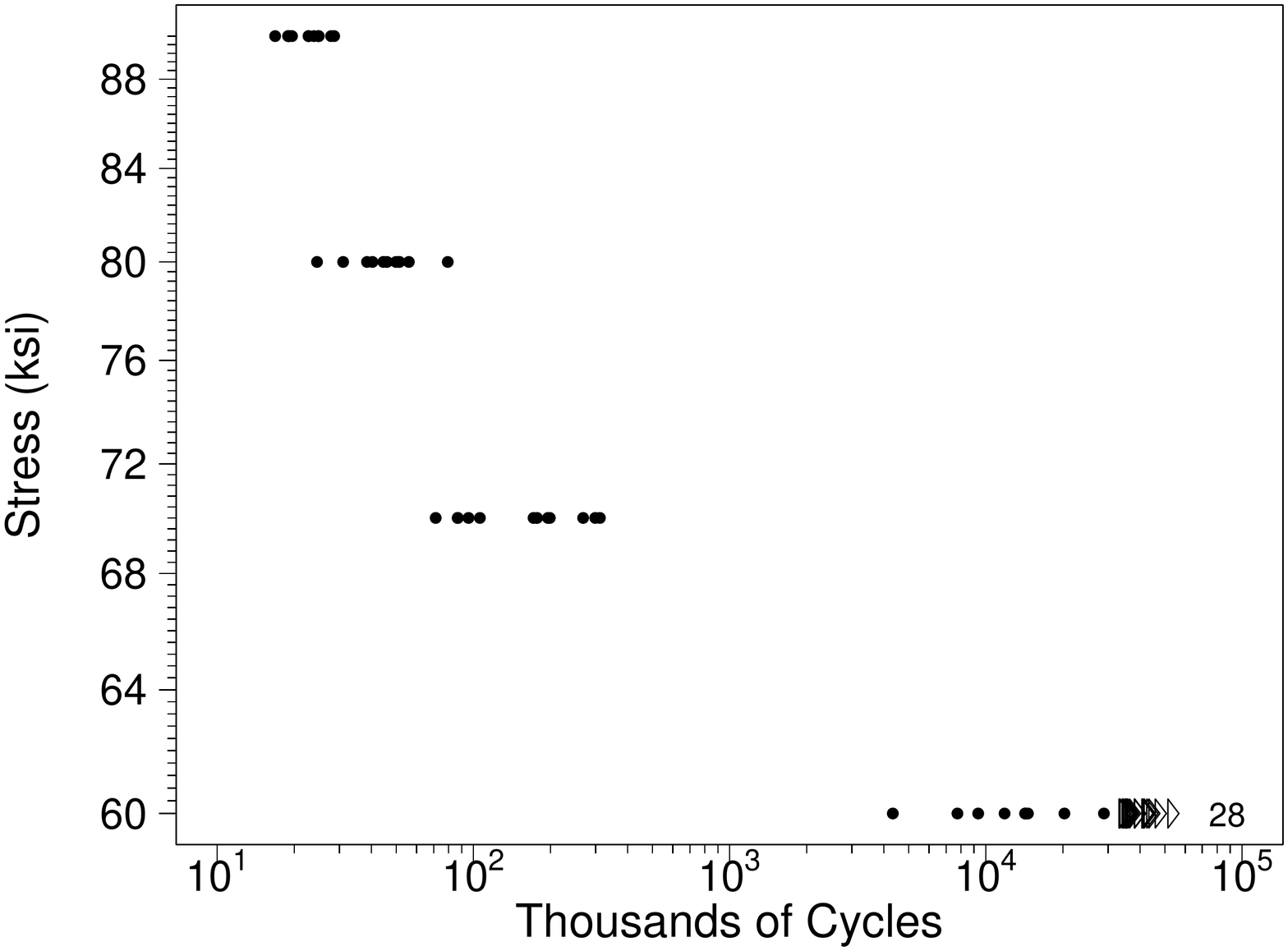}{3.25in} &
\rsplidapdffiguresize{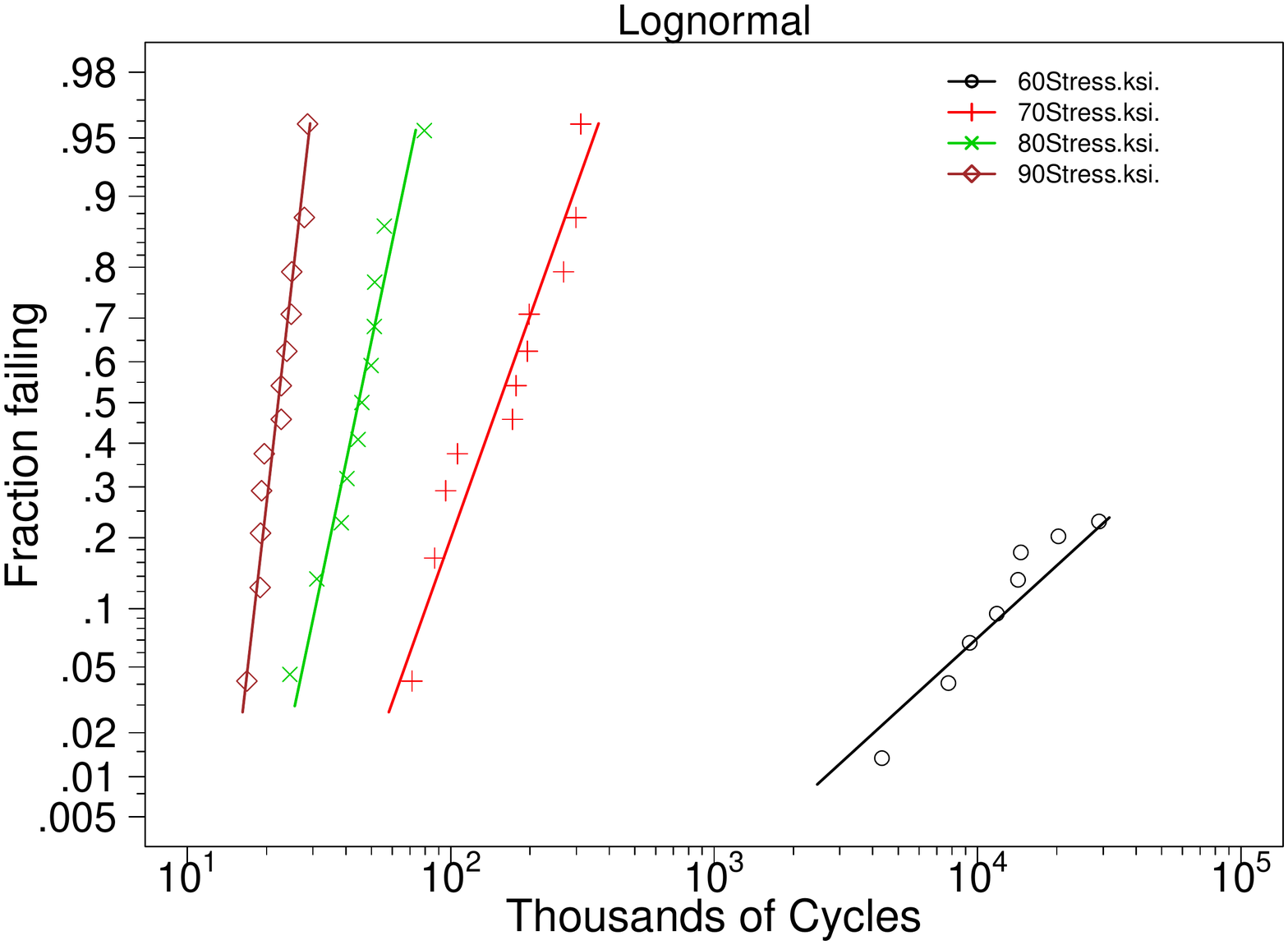}{3.25in}
\end{tabular}
\caption{Ti64 \SN{} Data scatter plot~(a) and lognormal
  probability plot~(b) with separate lognormal distributions fit to each
  level of stress.}
\label{figure:Ti64350FRm1.plots}
\end{figure}
Figure~\ref{figure:Ti64350FRm1.plots}b is a lognormal probability
plot for the Ti64 data. For these data, the lognormal distribution
provides a better fit than the Weibull distribution (see
Figure~\ref{S.figure:Ti64.lognormal.Weibull.probability.plots} in
the Appendix for a side-by-side comparison). The changes in slopes
of the fitted lines (corresponding to estimates of the lognormal shape
parameters) indicate the increase in spread at lower
stress levels.
\end{example}

\begin{example}
\label{example:Nitinol02.data}
\Exampletitle{Fatigue-Life Data from a Test on Superelastic Nitinol
  Specimens.}  Nitinol is an alloy of nickel and titanium able to
accommodate large recoverable strains via martensitic phase
transition, an effect sometimes referred to as super-elasticity.
Nitinol has found numerous successful applications in
implantable medical devices which are designed to remain durable
beyond 100 million cycles.  Rotary bend fatigue tests with nitinol
straight wire specimens were conducted with target alternating
strain ranging from 0.28 to 2.66\%.  The material specification,
sample preparation and test procedure, and interpretation of
the results can be found in \citet{Weaver_etal2022}.  Rotating bend
produces inherently fully reversed loading with a zero-mean stress.
All tests were conducted in phosphate buffered saline maintained at
$37 \pm 2\degreesc$ to approximate \textit{in vivo}
conditions. Tests were run until fracture or until completion of 1
billion cycles. Surviving units are runouts. The data from tests
conducted at one of two laboratories is
plotted in Figure~\ref{figure:Nitinol02Model.plots}a,
resulting in 46 fractures and 20 runouts.
The nitinol data set used in \citet{Weaver_etal2022}
also contains a variable ``Exact Strain'' that
results after applying a small
correction to five nominal strain
levels. Because the size of the correction varies from unit to unit,
this results in a substantial increase in the number of strain
levels making it impossible to use some of the diagnostics we
want to illustrate.

\begin{figure}[ht]
\begin{tabular}{cc}
(a) & (b) \\[-3.2ex]
\rsplidapdffiguresize{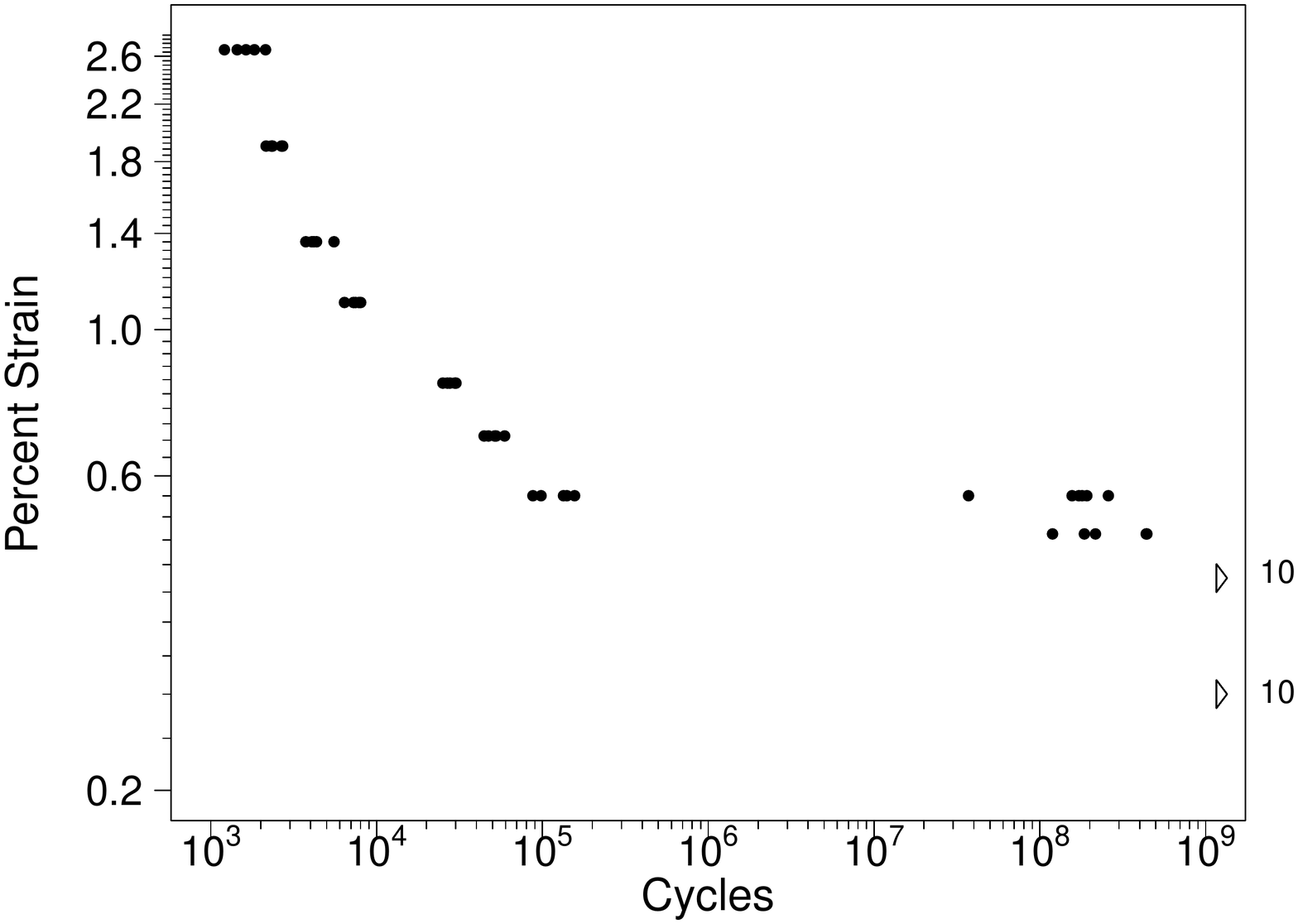}{3.25in} &
\rsplidapdffiguresize{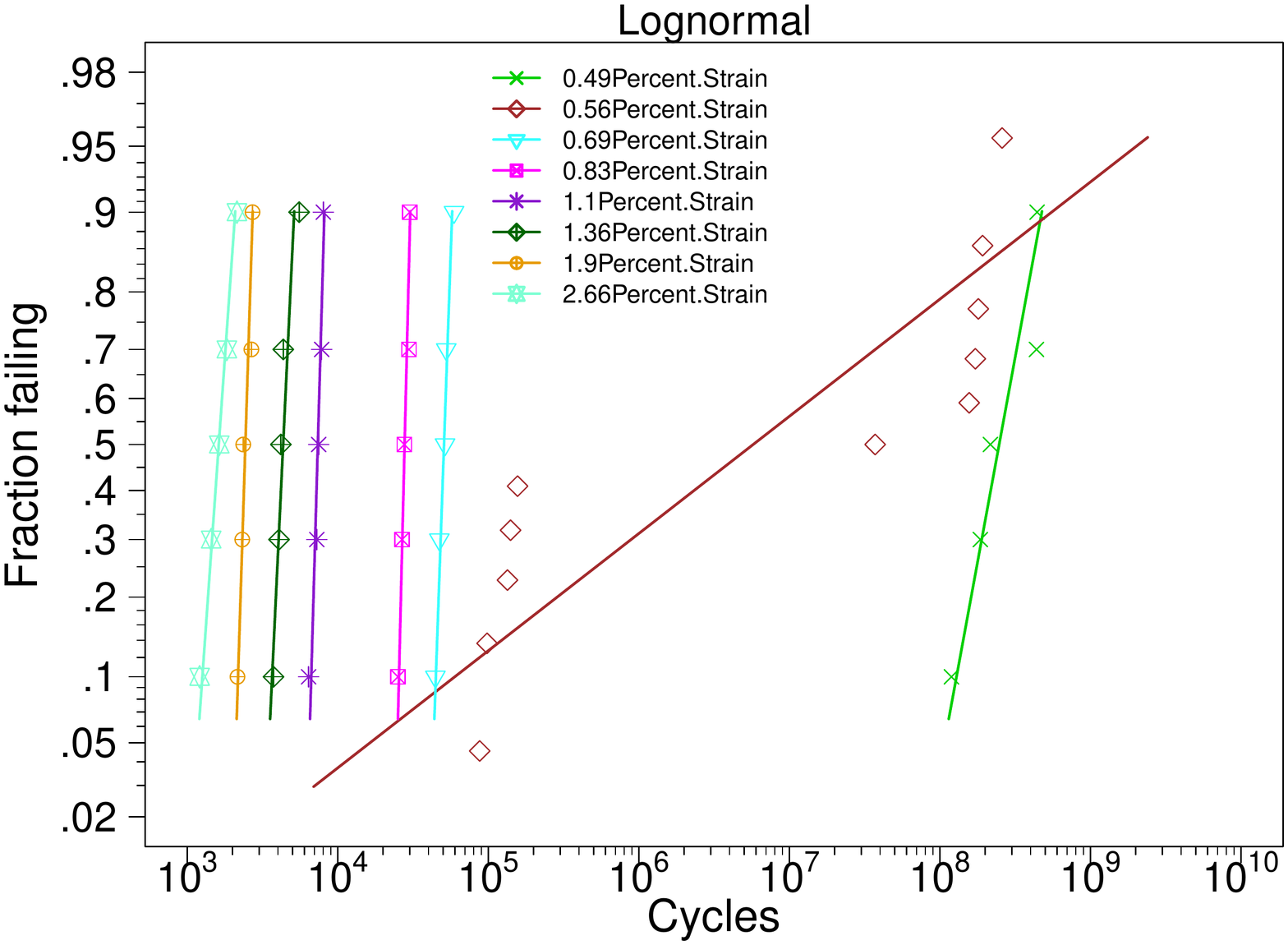}{3.25in}
\end{tabular}
\caption{Nitinol \SN{} Data scatter plot~(a) and lognormal
  probability plots~(b) with separate distributions fit to each
  level of strain.}
\label{figure:Nitinol02Model.plots}
\end{figure}
Similar to the Ti64 data in Example~\ref{example:Ti64.data}, the
scatter plot in Figure~\ref{figure:Nitinol02Model.plots}a shows an
\SN{} relationship with strong curvature and increased spread at low
levels of strain. Figure~\ref{figure:Nitinol02Model.plots}b is a
lognormal probability plot for the nitinol data. The most striking
feature in the plot is at 0.56\% strain where there were five early
failures and, after a gap, five later failures. This kind of bimodal
behavior is often seen in \SN{} data of standard metallic materials,
especially at intermediate levels of stress or strain where cycling
might be either elastic or plastic. In nitinol, however, early fractures
predominately initiate at small inclusions.  The inclusions initiate
propagation sooner under conditions of cyclic martensitic
transformation than under conditions of purely elastic cycling.
For the other levels of strain, either a
lognormal or a Weibull distribution provides an excellent
description of the data (see
Figure~\ref{S.figure:Nitinol.lognormal.Weibull.probability.plots} in
the Appendix for a side-by-side comparison).
\end{example}

\subsection{Fatigue Life, Fatigue Strength, and System Reliability}
\label{section:fatigue.life.and.fatigue.strength.system.reliability}
This section describes, at a high level, the relationship between
fatigue life and fatigue strength---two closely-related random
quantities that are of interest when studying the reliability of a
system or component that is subject to failure from fatigue caused
by cyclic loading. Technical details about this relationship are
given in
Section~\ref{section:relationship.fatigue.life.fatigue.strength}
and~\ref{section:alternative.approach.modeling.sn.data}.
This description is followed by brief
explanations of how laboratory-test results are used to quantify
system reliability in two important application areas.

\subsubsection{The Relationship Between Fatigue Life and Fatigue Strength}
\label{section:relationship.fatigue.life.and.fatigue.strength}

\begin{figure}
\begin{tabular}{cc}
\phantom{xxxxx}Lognormal fatigue-strength  densities (a) &\phantom{xxxxx}Weibull fatigue-strength  densities (b) \\[-3.2ex]
\rsplidapdffiguresize{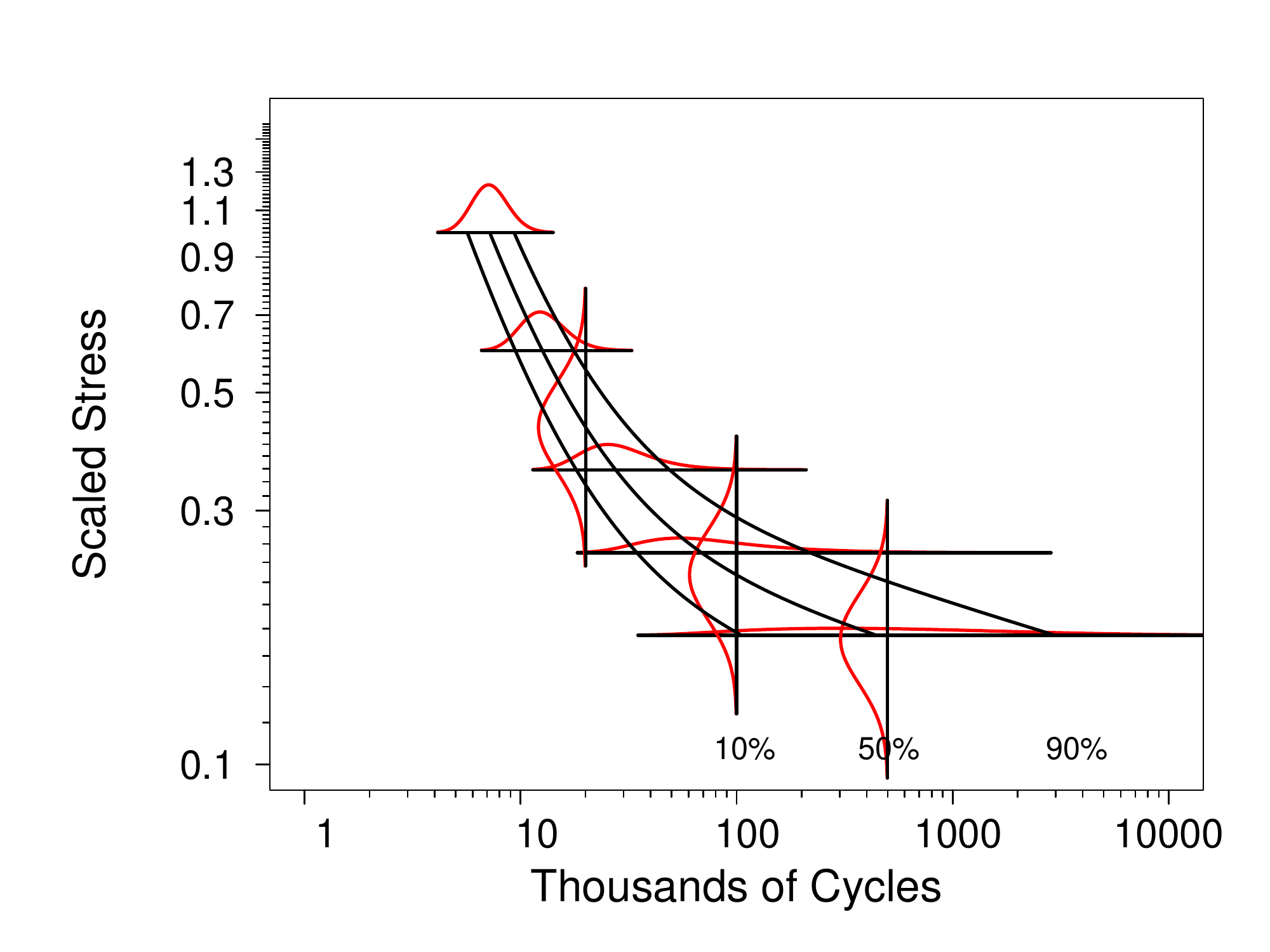}{3.25in}&
\rsplidapdffiguresize{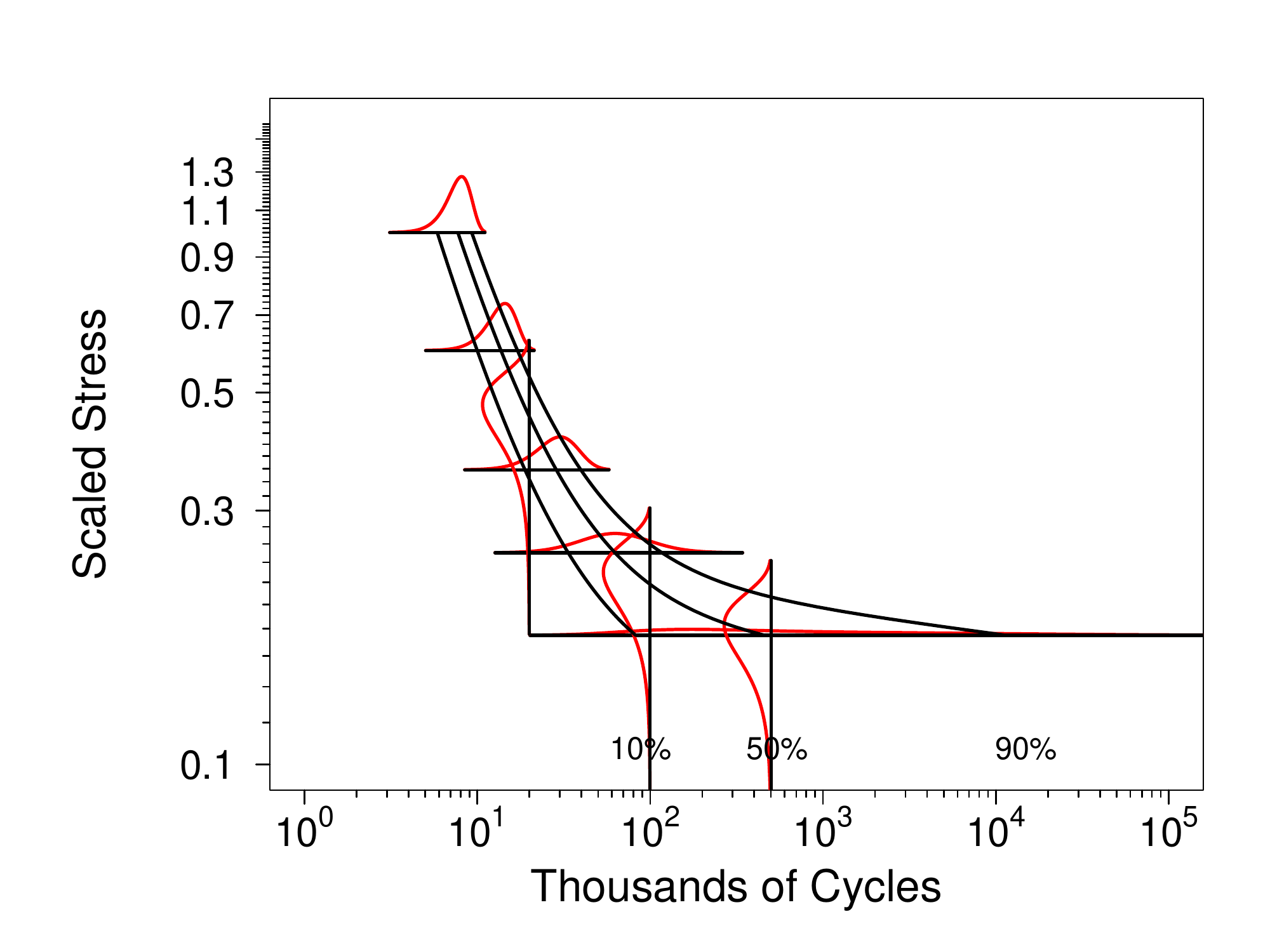}{3.25in}\\
\end{tabular}
\caption{Plot showing  lognormal fatigue-strength  densities
  (vertical, constant spread/shape)
  and corresponding \textit{induced} fatigue-life densities (horizontal)~(a); Plot showing Weibull fatigue-strength  densities
  (vertical, constant spread/shape)
  and corresponding \textit{induced} fatigue-life densities (horizontal)~(b).}
\label{figure:fatigue.life.fatigue.strength.plots}
\end{figure}

Fatigue life $N$ is defined as the time (number of cycles) when a
unit fails from repeated cyclic loading. Failure can be defined in
different ways, depending on the application. Examples include time
to fracture of a specimen, crack initiation, crack reaching a
critical size, or a specimen experiencing irreversible deformation;
in a composite-material structure, failure might be defined as time
of the beginning of a delamination.  A fatigue-life probability
model describes the distribution of $N$ and is generally given as a
function of stress amplitude $S_{e}$ (although other variables
such as stress ratio, temperature, and dwell time, are sometimes
used as factors in a fatigue experiment and included in a larger
regression model).
The horizontal densities in
Figure~\ref{figure:fatigue.life.fatigue.strength.plots} are
fatigue-life densities.

The fatigue-strength random variable $X$ is defined as the
\textit{level of stress} at which a unit would fail at a given
number of cycles, $N_{e}$.  Fatigue strength is \textit{not
  observable} because test-stress levels, $S$, are specified
experimental factors and the number of cycles at failure, $N$, is
random. It is possible, however, to estimate the distribution of $X$
by using fatigue-life data (if failures are observed at more than one
level of stress).  The vertical densities in
Figure~\ref{figure:fatigue.life.fatigue.strength.plots}a are
lognormal fatigue-strength densities and
those in Figure~\ref{figure:fatigue.life.fatigue.strength.plots}b are Weibull
fatigue-strength densities (lognormal and Weibull are
the most commonly used probability distributions for modeling
fatigue-life data). For the examples in
Figure~\ref{figure:fatigue.life.fatigue.strength.plots}, the scale
parameters of the fatigue-strength distributions depend on the
given number of cycles, $N_{e}$, and the shape parameters are
constant.

There are two ways to  view the relationship between the fatigue-life model and
the fatigue-strength model. Specification of a fatigue-life
(fatigue-strength) implies a corresponding fatigue-strength
(fatigue-life) model. Relatedly, the models generally have the
same quantile lines, as illustrated in
Figure~\ref{figure:fatigue.life.fatigue.strength.plots}. The
relationship between the models can also be expressed in a
mathematically precise manner, as it is in
Sections~\ref{section:relationship.fatigue.life.fatigue.strength}
(specified fatigue-life model and induced fatigue-strength model)
and~\ref{section:alternative.approach.modeling.sn.data}
(specified fatigue-strength model and induced fatigue-life model).

Correspondingly, there are two different approaches for modeling
fatigue-life data. The traditional approach is to specify a model
for the observable fatigue-life random variable $N$. Then there is an induced
model for the fatigue-strength random variable $X$. Alternatively
(as illustrated in
Figure~\ref{figure:fatigue.life.fatigue.strength.plots}), one can
specify a fatigue-strength model which then induces the fatigue-life
model that is used as a basis for defining a likelihood in terms of
the observable $N$. These approaches are equivalent (and the
distributions of $N$ and $X$ have the same form) \textit{if and only
  if} the \SN{} relationship is linear on log-log scales (unlike
Figure~\ref{figure:fatigue.life.fatigue.strength.plots}).  The
important advantages of using the new
specify-the-fatigue-strength-model approach when the \SN{}
relationship is \textit{nonlinear} are described in
Section~\ref{section:important.advantages.of.specifying.fatigue.strength.distribution}.

\subsubsection{Using experimental fatigue-life data to determine
  the safe life of an aircraft turbine engine disk} The primary
threat for an aircraft engine turbine disk failure is the initiation
and growth of a fatigue crack. Such failures could endanger the
continued safe operation of the aircraft and, in the worst-case
scenario, could lead to loss of the aircraft. To avoid catastrophic
disk failures, the design life (also known as safe life
and  approved life) of the
disks in commercially operated jet engines are required to be
computed by the engine manufacturer and disclosed to the FAA
\citep{FAA33.70-1}. Similar policies are used in Europe \citep{EASA2018} and
in other places around the world.
When a part reaches its safe life, it must be retired from service.

The safe life is specified in terms of the
number of flight take-off-landing cycles. Turbine disk lifetimes are
affected by many factors including engine design, flight mission,
and the materials used.  The FAA \citep{FAA33.70-1} dictates that
the acceptable part risk level (i.e., safe life) be the lower 95\%
confidence bound for the 0.001 quantile (also known as B0.1 life) of
the turbine disk failure-time distribution.

Fatigue tests at various levels of stress, temperature, and dwell
time, are conducted using simple material coupons (titanium alloys
are typically used in the cool parts of an engine, and nickel alloys
are used in the hot parts of the engine). Statistical methods are
used with the experimental data to estimate lower-tail \SN{} curves
(also known as quantile curves) and to compute corresponding lower
confidence bounds. Based on a given flight mission, the operating
conditions of the engine are specified in terms of variables such as
engine rotating speed, metal temperature and their gradients across
the part, and duration of the maximum load. Using these operating
conditions, detailed finite-element stress analyses are conducted to
quantify the stress at various critical locations in the complex
geometry of the part
\citep[e.g.][Appendix~N]{MattinglyHeiserPratt2002}. Using the stress
values and the lower confidence bounds of the lower tail \SN{} quantile
curves, the safe-life limit for the disk is determined. This is
known as the safe-life method and a general outline of the procedure
is given in~\citet[][Figure~1]{FAA33.70-1} and~\citet[][Figure
  N.1]{MattinglyHeiserPratt2002}. Additional safety margins and
methods are added to make the safe-life limit more conservative and
reduce the risk of part failure. One such method, known as the
Damage Tolerance approach, is applied by adding required periodic
inspections of critical parts using nondestructive evaluation (NDE)
methods. For this method, fatigue-test results are also an important
input for determining the inspection schedule, as described
in~\citet{FAA33.70-2}.

\subsubsection{Using experimental fatigue-life data to
  assure the reliability of medical devices}
Implantable medical devices must safely survive fatigue loading,
often for more than a decade.  In the case of cardiovascular
devices, such as stents and heart valves, the design life is 10 or
15 years, corresponding to 400 or 600 million cardiac cycles
\citep{FDA2010, ISO2018}.
Medical device manufacturers establish safety by comparing the
stress analysis of the in-vivo cycle to the component's fatigue
strength.  Typically, the regulatory requirement is that the
component must be shown to have greater than 0.90 reliability with
95\% confidence at the design life $N_{e}$.

The component's fatigue-strength distribution at the design life is
estimated by collecting experimental fatigue data on representative
test units over a range of test amplitudes. For each unit, cycles to
fracture is recorded, or the unit is right censored if it survives
until the design life. Appropriate statistical methods are used to compute
lower confidence bounds for quantiles of the
fatigue-strength distribution at the design life $N_{e}$.

In its simplest form, the estimation of a fatigue-strength
distribution is used to calculate a factor of safety.  In this case,
a lower 95\% confidence bound for the 0.1 quantile of the
fatigue-strength distribution at the design life $N_{e}$
(corresponding to 0.9 reliability) is computed.  The factor of
safety is calculated by dividing this bound by the highest cyclic
stress determined in the stress analysis of the in-vivo cycle.  If
this factor of safety is greater than one, then the component is
considered acceptable \citep{ASTM2017}.

Alternatively, estimation of the fatigue-strength distribution can
be used as input to the classical stress-strength interference model
to predict a probability of failure.
\citet{Shanmugam_et_al2019} illustrate this approach
\citep[see Section 23.2 of][for
  technical details of the stress-strength interference
  model]{MeekerEscobarPascual2021}. In
addition, \citet{HaddadHimesCampbell2014} developed a Bayesian
approach which uses \SN{} regression analysis to estimate the
probability of failure of a medical device, in the case of a complicated
in-vivo cycle.

\subsection{History and Literature Review}
\label{section:literature.review}
Scientific and engineering focus on understanding and managing
fatigue as a failure mechanism accelerated in the 1950s after there
were several fatigue-related failures in the rapidly growing area of
aerospace applications. These and some other earlier fatigue-related
reliability disasters are described in \citet[][Chapter~2]{Woo2020}.
Engineers collecting fatigue data in the 1950s through the 1970s
generally did not know how to properly handle censored data and
other complications that arise in \SN{} data. Stress was sometimes
treated as the response variable (due to the desire to estimate
fatigue-strength distributions) and runouts were either ignored or treated
as failures (practices that unfortunately still,
in some places, continue today).

Examples of early development of statistical theory for the
analysis of fatigue data include
\citet{FreudenthalGumbel1953,FreudenthalGumbel1954,Weibull1956,FreudenthalGumbel1956,BastenaireBastienPomey1961}.
Wayne Nelson (while working at GE
Corporate Research and Development) did pioneering work in
developing appropriate statistical methods for modeling and making
inferences from complicated \textit{censored} \SN{}
data. \citet{Nelson1984,Nelson1990a} illustrates these methods which
include maximum likelihood estimation, model-checking diagnostics,
and confidence intervals for fatigue-life distribution quantiles. He
illustrated the methods on data from a strain-controlled experiment
on a nickel-base super alloy used in high-temperature components of an aircraft
engine. Log-life was modeled as a quadratic function of
log pseudo-stress (strain multiplied by Young's modulus, a covariate) to
describe the characteristic curvature of most
HCF \SN{} data when
plotted on log-log scales. He also employed a log-linear
relationship for the lognormal shape parameter (also used in
Section~\ref{section:simple.sn.relationships}) to describe the
often-seen increase in spread at lower stress levels.

As mentioned in Section~\ref{section:motivation}, thousands of
papers have been published in the engineering literature describing
methods for modeling fatigue data, often for specific
applications. Dozens of statistical models for fatigue data have been suggested
in these papers.
\citet{CastilloFernandez-Canteli2009} review and describe many of these models.
We review several of the most
commonly used \SN{} models in
Section~\ref{section:statistical.models.estimate.fatigue.life}.  More
complicated \SN{} models are described and applied to the Ti64 and
nitinol data in
Section~\ref{section:other.sn.regression.relationships}.

The publication of early papers using maximum likelihood methods,
such as \citet{SpindelHaibach1979} and \citet{Nelson1984}, began a
trend that continues today, of more analysts using appropriate
methods for handling \SN{} data with runouts. More recently, papers
like \citet{Babuska_etal2016} and \citet{Castillo_et_al2019}
illustrate the use of Bayesian methods to fit and compare models
fit to fatigue data.  Bayesian methods can also properly handle
runouts and correctly quantify statistical  uncertainty.

\subsection{Contributions}
\label{section:contributions}
In addition to some review, this paper contains many new and important technical
results.
\begin{enumerate}
\item
A general, flexible, modular framework for statistical modeling
\SN{} data for which most of the models that have been used in the
thousands of published papers in the engineering literature, can be
viewed as special cases.
\item
Engineers (and some statisticians) have, conceptually, known about
the closely connected distributions of fatigue life and fatigue
strength \citep[e.g.,][]{FreudenthalGumbel1956,Weibull1956,BastenaireBastienPomey1961,CastilloGalambos1987}.
Because fatigue life is observable and fatigue strength is
not, there has not been a unified and flexible estimation method for these
distributions. We present a unified flexible model that
connects these two important distributions and allows for the
efficient use of \SN{} data to estimate both fatigue-life and fatigue-strength
distributions. These connected distributions are at the heart of the
framework in Contribution~1.
\item
Perhaps most importantly, and building on Contribution~2,
insights from \citet{Weibull1956} and \citet{BastenaireBastienPomey1961},
and an important contribution by \citet{Falk2019}, we show
how one can usefully specify a relatively simple fatigue-strength model
that will then induce an appropriate fatigue-life model.
We demonstrate and illustrate the important advantages of specifying
the fatigue model for \SN{} data in this manner.
\item
We describe the physical explanation for the curvature in the \SN{}
relationship and show how this curvature induces
fatigue-life distributions with an increased spread at lower levels of
stress.
\end{enumerate}

\subsection{Overview}
The remainder of this paper is organized as follows.
Section~\ref{section:statistical.models.estimate.fatigue.life}
outlines our general approach for modeling \SN{} data and
illustrates it on an example where a fatigue-life model is specified
and the corresponding \textit{fatigue-strength model is induced}.
Section~\ref{section:statistical.models.estimate.fatigue.strength.distributions}
describes fatigue-strength models and shows how a specified
fatigue-strength model \textit{induces a corresponding fatigue-life model}.
Section~\ref{section:estimating.parameters.diagnostics.inferences.quantiles}
briefly reviews likelihood and Bayesian methods for statistical
inference, including residual analysis for censored data and general methods for
estimating lower-tail quantiles of fatigue-life and fatigue-strength
distributions that engineers need.
Section~\ref{section:other.sn.regression.relationships} describes
and compares additional widely-used
nonlinear regression relationships for \SN{} data, describes
the important advantages of specifying a fatigue-strength
model that induces a fatigue-life model, and
illustrates the approach with two additional HCF
applications.  Section~\ref{section:concluding.remarks} provides
concluding remarks and outlines areas for future research.
To save space and improve readability, proofs,
various technical details, additional plots, and more detailed numerical results
have been relegated to appendices.

\section{Statistical Models for Fatigue \SN{} Data}
\label{section:statistical.models.estimate.fatigue.life}
As mentioned in Section~\ref{section:literature.review},
dozens of different statistical models have been suggested to describe \SN{}
data. This section introduces a modular framework that includes
most of these models as special cases. For a given \SN{} data set, appropriate
model components are chosen to comprise a specific model or
to describe the data. We encourage fitting and
comparing alternative statistical models.

\subsection{A Modular Framework for Modeling Fatigue \SN{} Data}
\label{section:modular.framework.modeling.fatigue.sn.data}

As described in Section~\ref{section:relationship.fatigue.life.and.fatigue.strength}, a statistical model for fatigue \SN{} experimental data has two
closely related random variables---fatigue life $N$ and fatigue
strength $X$.

There are two different ways to specify a statistical model for \SN{} data:
\begin{enumerate}
\item
Specify a model $F_{N}(t; S_{e}, \thetavec)$ for fatigue life $N$ as
a function of a given stress amplitude $S_{e}$ which will induce
(imply) a corresponding fatigue-strength model
(Section~\ref{section:relationship.fatigue.life.fatigue.strength}).
\item
Specify a model $F_{X}(x; N_{e}, \thetavec)$ for the fatigue-strength
random variable $X$ as a function of a given number
of cycles $N_{e}$. This model
will induce a corresponding fatigue-life model
(Section~\ref{section:alternative.approach.modeling.sn.data}).
\end{enumerate}
Here $\thetavec$ is a vector of unknown parameters (the nature of
which depends on the particular model components) to be estimated
from the \SN{} data. To simplify notation, we will usually suppress
the dependency of $F_{N}(t; S_{e})$, $F_{X}(x; N_{e})$, or their
corresponding density and quantile functions
on $\thetavec$.

Because $N$ is \textit{observable} (and $X$ is not), the first
approach for \SN{} model specification has been used most commonly
in practice and will be described and illustrated in the rest of
this section. The second approach is new, has important advantages,
and will be described in detail in
Section~\ref{section:alternative.approach.modeling.sn.data} and
illustrated in Examples~\ref{example:modeling.Ti64.sn.data}
and~\ref{example:modeling.superelastic.nitinol.sn.data}.  After
deciding whether to specify the model for $N$ or $X$, two model
components need to be specified:
\begin{itemize}
\item
A functional \SN{} regression relationship (decreasing, usually
nonlinear, continuous, and differentiable) describing how the
distribution of the observable fatigue life $N$ depends on the experimental factor stress amplitude
$S_{e}$ (or how  the
distribution of the not-observable fatigue strength $X$ depends on
the
given number of cycles $N_{e}$).
Although it is possible to have explanatory/experimental
variables other than Stress in a fatigue-life test, and such extensions
are straightforward, we will focus on a model with just this one
experimental factor and
describe the extensions in our concluding remarks.
\item
A probability model to
describe spread in the \SN{} data. In this paper we use the
log-location-scale family of distributions because they include the
lognormal and Weibull distributions that are used almost exclusively
for fatigue modeling. Extensions to other, more general, families of
distributions are possible, as described in
Section~\ref{S.section:generalization.specifiedfsm.induced.flm}
of the Appendix.
\end{itemize}

\subsection{Log-Location-Scale Probability Distributions}
\label{section:log.location.scale.distributions}
If $Y$ has a location-scale distribution, then $T=\exp(Y)$ has a
log-location-scale distribution with cdf
\begin{align}
\label{equation:lls.cdf}
 F(t;\mu, \sigma)&=\Phi\left [\frac{\log(t)-\mu}{\sigma}
\right ]=\Phi\left [ \log \left( \left[\frac{t}{\exp(\mu)} \right]^ {(1/\sigma)}
\right) \right ], \quad t > 0
 \end{align}
where $\Phi(z)$ is the cdf for the
particular standard
location-scale distribution, $\exp(\mu)$ is a scale parameter and
$\sigma$ is the shape parameter.
The most well-known log-location-scale distributions are the
lognormal ($\Phi(z)=\Phi_{\norm}(z)$ is the standard normal cdf),
and Weibull ($\Phi(z)=\Phi_{\sev}(z)=1-\exp[-\exp(z)]$ is the
standard smallest extreme value cdf) distributions.
See Chapter~4 of \citet{MeekerEscobarPascual2021} for more
information about these and other log-location-scale distributions.

\subsection{Basic \SN{} Relationships and Statistical Models for
  Fatigue Life}
\label{section:simple.sn.relationships}
This section describes simple \SN{} relationships that are
useful for specifying a fatigue-life model for \SN{} data.
More complicated \SN{} relationships that are better suited
when a fatigue-strength model is specified
are given in
Section~\ref{section:other.sn.regression.relationships}.

\subsubsection{A statistical model for fatigue-life}
Suppose that the logarithm of the fatigue-life random variable $N$
at a \textit{given} stress level $S_{e}$ is
\begin{align}
\label{equation:general.model.for.N}
  \log(N)&=\log[\gfun(S_{e};\betavec)]+\sigma_{N} \epsilon,
\end{align}
where $N=\gfun(S;\betavec)$ is a positive monotonically
decreasing \SN{} regression relationship of known form,
$\betavec$ is a vector of regression parameters,
$\sigma_{N}\epsilon$
is a random-error term and $\epsilon$ has a
location-scale distribution with $\mu=0$ and $\sigma=1.$
Then for any given stress level
$S_{e}$, $N$ has a log-location-scale distribution with cdf
\begin{align}
\label{equation:fatigue.life.failure.time.model.cdf}
F_{N}(t; S_{e})&=\Pr \left(N \le t; S_{e}\right)=\Phi
         \left (\frac{\log(t)-\log[\gfun(S_{e};\betavec)]}{\sigma_{N}}   \right), \quad \quad
  t>0, \,\, S_{e}>0,
\end{align}
where $\gfun(S_{e};\betavec)$ is a scale parameter and $\sigma_{N}$
is the shape parameter of the distribution of $N$.  The fatigue-life $p$ quantile is obtained by solving $p=F_{N}(t_{p}(S_{e});
S_{e})$ for $t_{p}(S_{e})$, giving
\begin{align}
\label{equation:fatigue.life.failure.time.model.quantile}
t_{p}(S_{e}) &= \exp(\log[\gfun(S_{e};\betavec)] + \Phi^{-1}(p)\sigma_{N}), \quad \quad
  0<p<1, \,\, S_{e}>0.
\end{align}

\subsubsection{The Basquin model}
\label{section:Basquin.model}
The \citet{Basquin1910} \SN{} relationship  (sometimes referred to as
  the \textit{inverse-power rule}) is
\begin{align*}
N&=A \times S^{-B},
\end{align*}
were $A$ and $B$ are
parameters. As usually presented in the engineering literature,
there is no random-error term in the relationship, which is
generally taken to
represent the relationship between a particular failure-time
distribution quantile (e.g., the median) and stress $S$.
  Taking logs, changing parameter
names and a sign, and adding a random-error term gives the statistical
model for fatigue life $N$
\begin{align}
\label{equation:basquin.failure.time.model}
\log(N) &= \beta_{0} + \beta_{1}\log(S) + \sigma_{N} \epsilon.
\end{align}
For any given level of stress $S_{e}$, $N$ has a
log-location-scale distribution with constant $\sigma_{N}$ with
cdf and quantile functions given by
(\ref{equation:fatigue.life.failure.time.model.cdf}) and
(\ref{equation:fatigue.life.failure.time.model.quantile}), respectively,
where $\log[\gfun(S_{e};\betavec)]=\beta_{0} + \beta_{1}\log(S_{e})$.
Basquin is the simplest and most widely used model for fatigue life.

\subsubsection{The Stromeyer relationship}
\label{section:stromeyer.model}
\citet{Stromeyer1914} introduced the \textit{fatigue-limit} \SN{} model
\begin{align}
\label{equation:stromeyer}
\log(N) &= \beta_{0} + \beta_{1}\log(S-\gamma) + \sigma_{N} \epsilon,
\quad S>\gamma,
\end{align}
which is a generalization to the Basquin relationship in
(\ref{equation:basquin.failure.time.model}).
The Stromeyer relationship describes (for $\gamma>0$) the
concave-up curvature
commonly seen in \SN{}~data when plotted on log-log scales.
For any given level of stress
$S_{e}$, $N$ has a log-location-scale distribution with cdf and
quantile functions given by
(\ref{equation:fatigue.life.failure.time.model.cdf}) and
(\ref{equation:fatigue.life.failure.time.model.quantile}), respectively,
where $\log[\gfun(S_{e};\betavec)]=\beta_{0}+ \beta_{1}\log(S_{e}-\gamma)$.
In this model, if $S_{e}$ is less than the fatigue-limit $\gamma$
(also known as an endurance-limit), lifetime is infinite---stress is
low enough that cycling does not cause permanent damage. Although
there are dissenters, \citep[e.g.,][]{Bathias1999}, it is widely
believed that fatigue-limits exist in hard metals like steel and
some titanium alloys but not in soft metals like aluminum or copper. Even
if fatigue-limits exist, it is unreasonable to assume that $\gamma$
would be constant in a process/population
because there are many additional sources of
variability that would affect such fatigue-limits
(e.g., surface finish, residual stresses, other
manufacturing variabilities, and environmental variables). This
motivates the random fatigue-limit (RFL) model described in
Section~\ref{section:RFL.model}.

\subsubsection{Box--Cox (power) transformation model}
\label{section:box.cox.relationship}

\citet[][page 96]{Nelson1990a} mentions the use of a power
transformation of stress instead of a log transformation.
Although the Box--Cox transformation is widely used to transform the response
in statistical modeling, it can also be used to transform
explanatory variables \citep[e.g.,][]{CarrollRuppert1988}.
\citet{MeekerEscobarZayac2003}
and \citet[][Section 18.5.5]{MeekerEscobarPascual2021} use a Box--Cox
transformation \SN{} model
\begin{align}
\label{inverse.power.generalization.statistical}
\log(N)&=\beta_{0} + \beta_{1}\nu(S, \lambda)+ \sigma_{N}\epsilon =\begin{cases}
\beta_{0} + \beta_{1}\left(\dfrac{S^{\lambda} -1}{\lambda}\right) +\sigma_{N}\epsilon & \text{if $\lambda \ne  0$}\\[2ex]
\beta_{0} + \beta_{1}\log(S) + \sigma_{N}\epsilon & \text{if $\lambda=0$}
\end{cases}
\end{align}
instead. Here $\nu(S_{e}, \lambda)$ is the Box--Cox power transformation of
stress. This transformation is  preferred because $\nu(S_{e},
\lambda)$ is continuous in the power
parameter $\lambda$ and the special case $\lambda=0$
corresponds to the Basquin relationship.
For any given level of stress $S_{e}$, $N$ has a
log-location-scale distribution with cdf and quantile functions
given by (\ref{equation:fatigue.life.failure.time.model.cdf}) and
(\ref{equation:fatigue.life.failure.time.model.quantile}), respectively,
where $\log[g(S_{e}; \betavec)]=\beta_{0}+ \beta_{1}\nu(S_{e}, \lambda)$.

\begin{figure}[ht]
\begin{tabular}{cc}
\phantom{XX}Box--Cox (a) & \phantom{XX}Modified
  Bastenaire (b) \\[-3.2ex]
\rsplidapdffiguresize{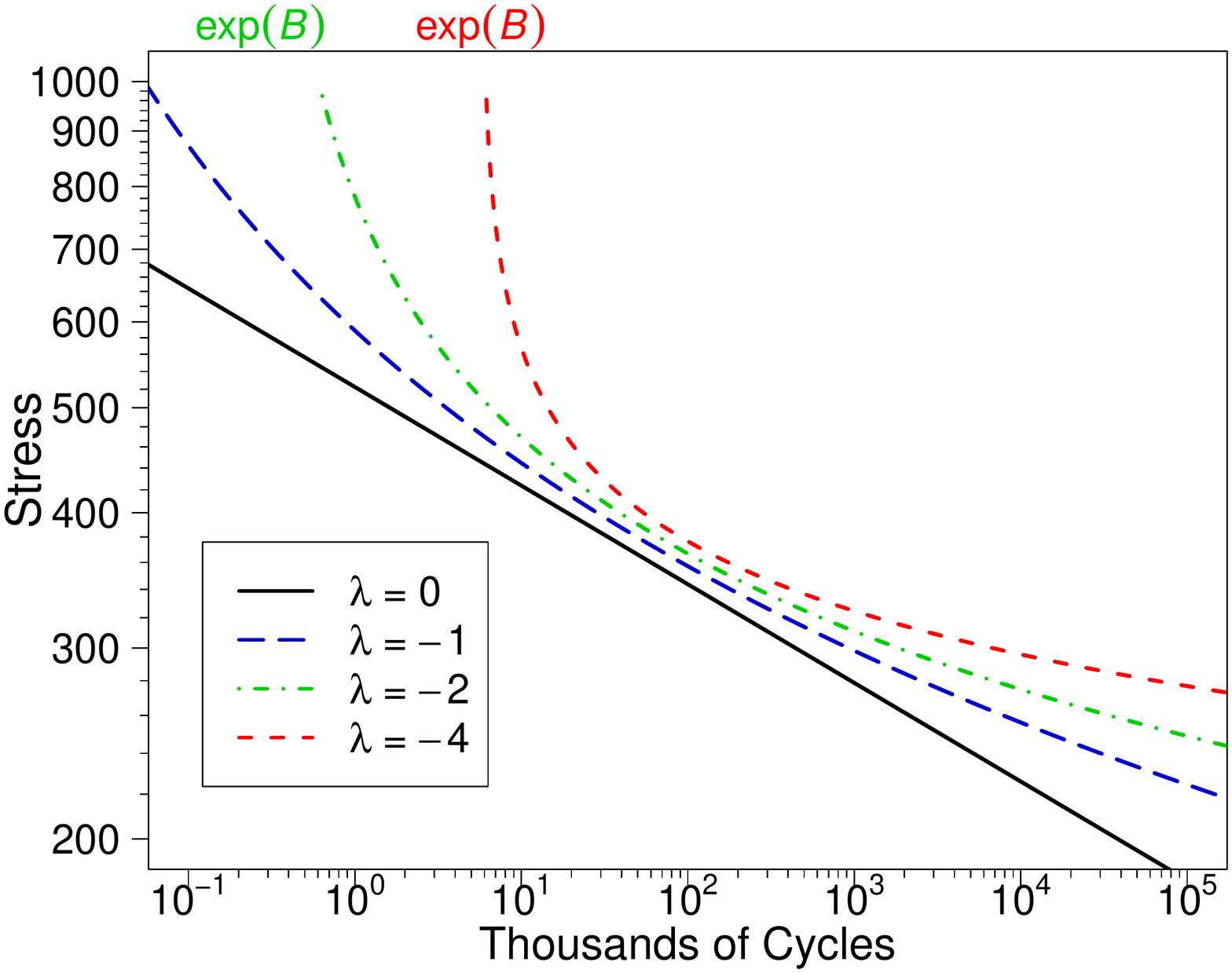}{3.25in}&
\rsplidapdffiguresize{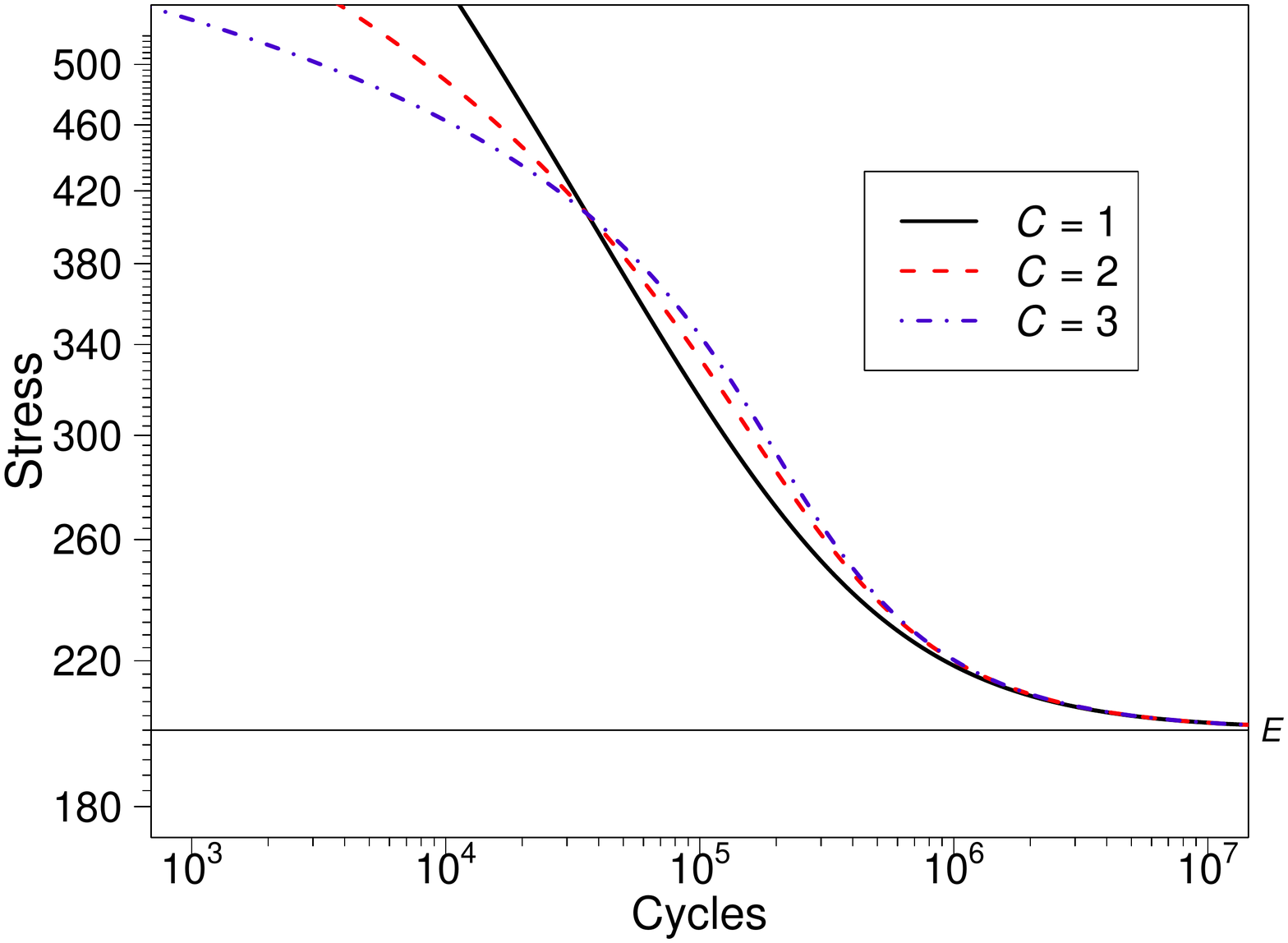}{3.25in}\\
\phantom{XX}Nishijima (c) & \phantom{XX}Coffin--Manson (d) \\[-3.2ex]
\rsplidapdffiguresize{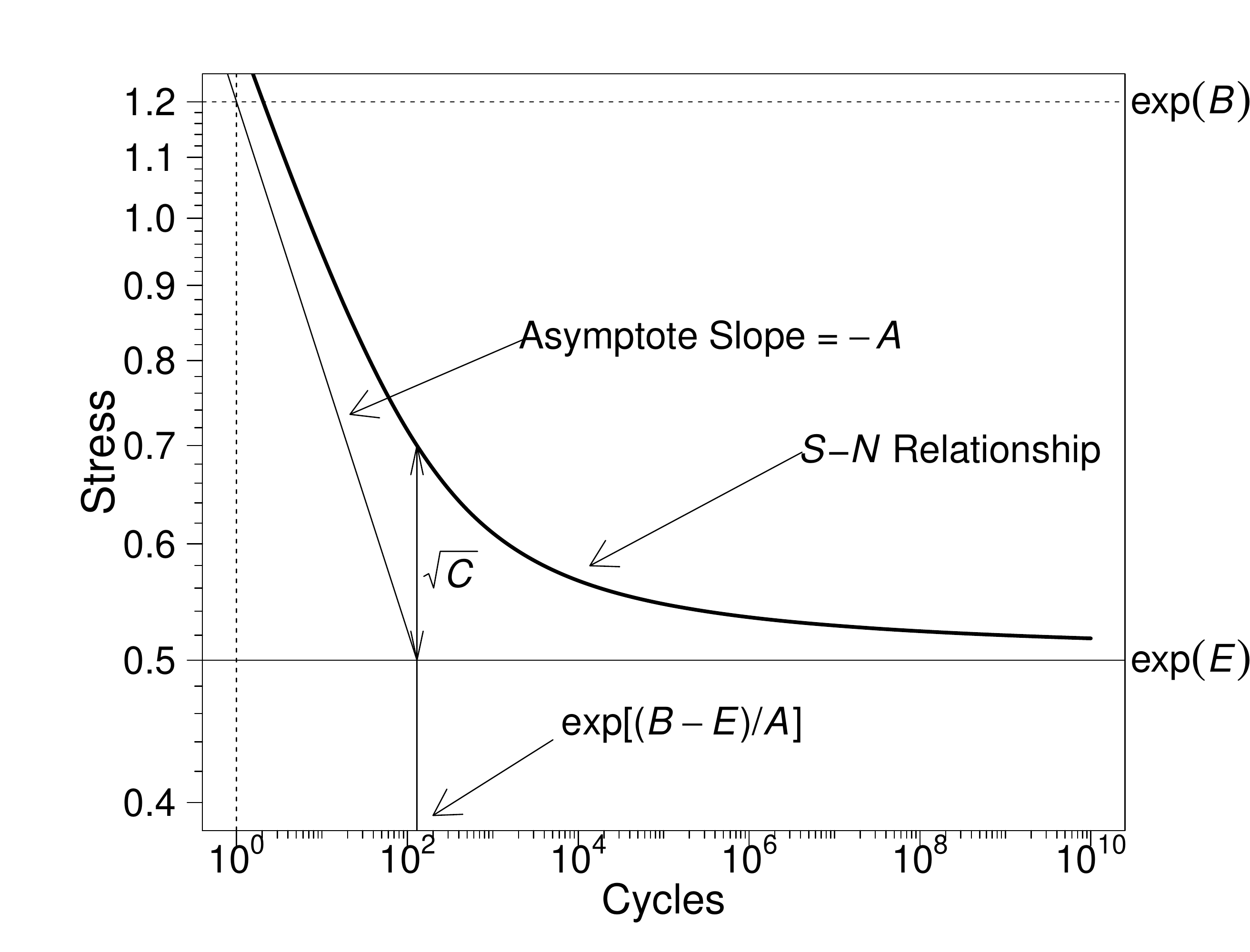}{3.25in}&
\rsplidapdffiguresize{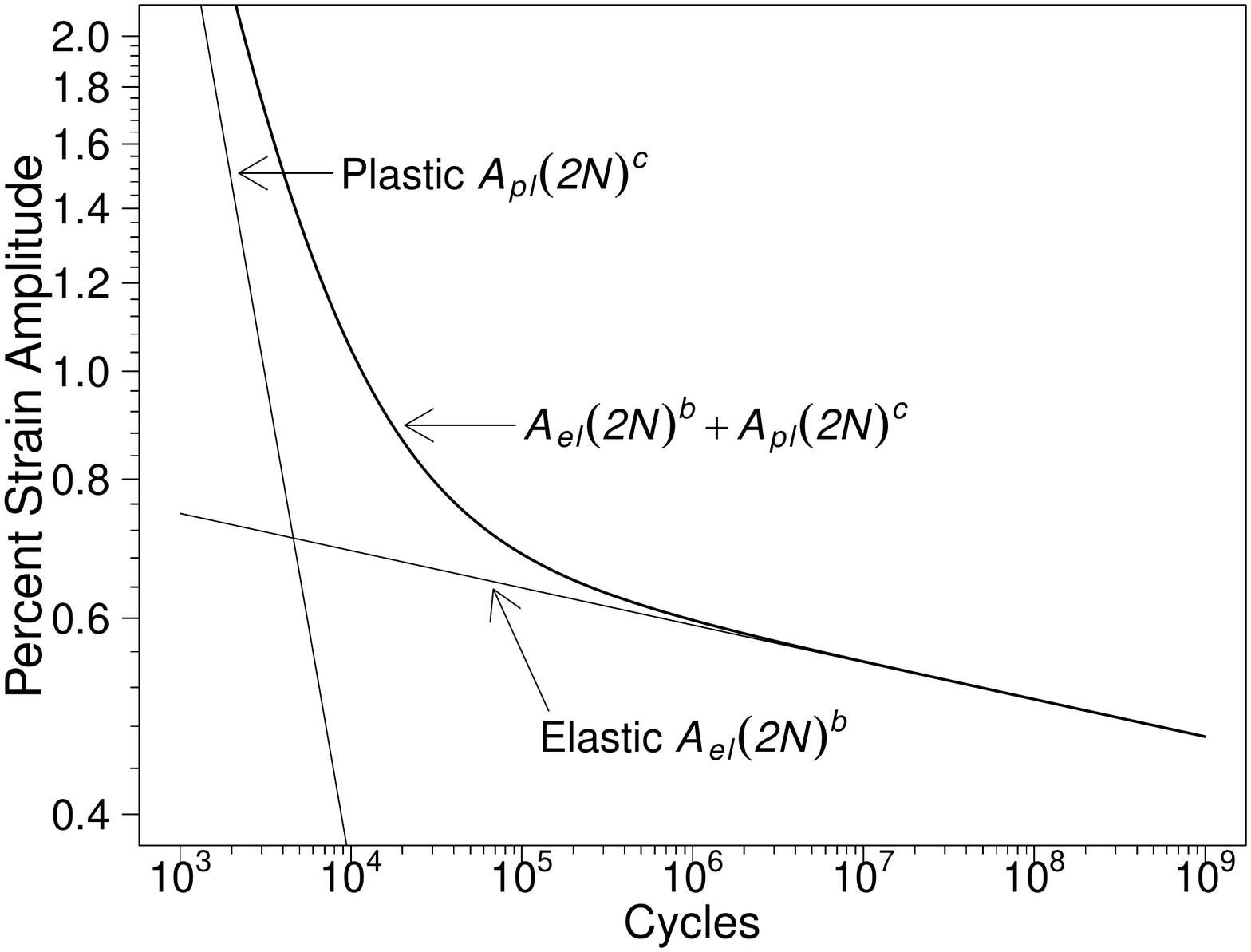}{3.25in}
\end{tabular}
\caption{\SN{} relationships:  Box--Cox~(Section~\ref{section:box.cox.relationship})~(a), modified
  Bastenaire~(Section~\ref{section:modified.bastenaire})~(b), Nishijima~(Section~\ref{section:nishijima.sn.relationships})~(c), and Coffin--Manson~(Section~\ref{section:coffin-manson.relationship})~(d).}
\label{figure:sn.relationships}
\end{figure}

For $\lambda<0$ and $\beta_{1}<0$ (values expected in \SN{}
applications) the Box--Cox relationship has a concave-up shape. In
contrast to the Stromeyer relationship, as shown in
Figure~\ref{figure:sn.relationships}a, there is vertical asymptote
in the \SN{} relationship at $B=\beta_{0} +\beta_{1}(-1/\lambda)$.
As described in
Sections~\ref{section:induced.fatigue.strength.distribution.vertical.asymptote}
and~\ref{section:induced.fatigue-life.distribution.vertical.asymptote},
this asymptotic behavior can lead to physically unreasonable model
features. When, however, the steep asymptotic behavior is outside
the range where the model would be used (e.g.,
Example~\ref{example:box-cox.loglin.laminate.panel.data}), there are
no practical problems. For example, the Box--Cox model nicely
describes the laminate panel data in
Example~\ref{example:laminate.panel.data},
as will be shown in Example~\ref{example:box-cox.loglin.laminate.panel.data}.

\subsubsection{A model component to describe nonconstant
  \texorpdfstring{$\sigma_{N}$}{sigmaN}}
\label{section:loglinear.sigma.component}
The models described earlier in this section do not account for nonconstant
$\sigma_{N}$ that is often seen in \SN{} data (e.g., the data
introduced in Examples~\ref{example:Ti64.data}
and~\ref{example:Nitinol02.data}).
Thus, for some data sets,
it is necessary to add an additional model component such as
\begin{align}
\label{equation:loglinest.sigma}
\sigma_{N} = \exp\left[\beta_{0}^{[\sigma_{N}]} +
\beta_{1}^{[\sigma_{N}]} \log(S_{e}) \right].
\end{align}
\citet{Nelson1984} used a quadratic \SN{} relationship and (\ref{equation:loglinest.sigma})
to estimate
\SN{} curves for a nickel-based superalloy. \citet{PascualMeeker1997}
used a Stromeyer \SN{} relationship  (Section~\ref{section:stromeyer.model}) and
(\ref{equation:loglinest.sigma}) to
describe the same data. Section~\ref{section:fitting.fatigue.life.model},
uses a Box--Cox relationship with~(\ref{equation:loglinest.sigma})
to describe the increase in spread
at the lower stress levels seen
in Figure~\ref{figure:LaminatePanelProbability.plots}.
Although this loglinear-$\sigma_{N}$ model component has been useful
in applications we want to note that its use will cause violations
of certain compatibility conditions,
described further in Section~\ref{section:characteristics.of.sn.models}.

\subsection{Linking Fatigue-Life and
  Fatigue-Strength Models}
\label{section:relationship.fatigue.life.fatigue.strength}
This section describes the relationship between a specified
fatigue-life model and the corresponding induced fatigue-strength model.

\subsubsection{The induced fatigue-strength model when
\texorpdfstring{$\gSN$}{g(S)} has neither a vertical nor a horizontal asymptote}
\label{section:induced.fatigue.strength.distribution.no.asymptotes}
Section~\ref{section:modular.framework.modeling.fatigue.sn.data}
defined the unobservable \textit{fatigue-strength}
random variable $X$ as the level of applied stress that would
result in a failure at a given number of cycles $N_{e}$. This
definition describes the close relationship to the observable
fatigue-life random variable $N$. In this model,
the distributions of $X$ and $N$ share the same random-error
term.  For the moment, suppose that $\log[\gfun(S_{e};\betavec)]$
has neither a horizontal nor a vertical asymptote.
To derive the distribution of $X$,
replace $N$ with $N_{e}$
and $S_{e}$ with $X$ in (\ref{equation:general.model.for.N}) giving
\begin{align}
\label{equation:reexpressed.fatigue.life.strength}
\log(N_{e})&= \log[\gfun(X;\betavec)] + \sigma_{N} \epsilon.
\end{align}
This shows that the common random-error term $\sigma_{N} \epsilon$ drives
the random variable $X$ at fixed $N_{e}$ as well as the random
variable $N$ at fixed $S_{e}$.  Also,
(\ref{equation:reexpressed.fatigue.life.strength}) implies that
$(\log(N_{e})-\log[\gfun(X; \betavec)])/\sigma_{N}= \epsilon$ has a
location-scale distribution with $\mu=0$ and $\sigma=1$.  Then
using~(\ref{equation:reexpressed.fatigue.life.strength}) and the
fact that $\gfun(S;\betavec)$ is monotonically decreasing in $S$, the cdf
of $X$ is
\begin{align}
\nonumber
F_{X}(x; N_{e})&=\Pr(X \le x; N_{e})=\Pr\left [\gfun(X;\betavec)>\gfun(x;\betavec) \right]\\
\nonumber
&=\Pr\left(\log [\gfun(X;\betavec)] > \log [\gfun(x;\betavec)]\right)=\Pr\left(-\log [\gfun(X;\betavec)] < -\log [\gfun(x;\betavec)]\right)\\
\nonumber
  &=\Pr\left(\log(N_{e})-\log [\gfun(X;\betavec)] < \log(N_{e})-\log [\gfun(x;\betavec)]\right).\\
  &=
\label{equation:general.fatigue.strength.cdf.ls}
  \Phi\left[\frac{\log(N_{e})-\log[\gfun(x;\betavec)]}{\sigma_{N}}
    \right],
\quad \quad   x>0, \,\, N_{e}>0.
\end{align}
Note that (\ref{equation:general.fatigue.strength.cdf.ls}) is a
log-location-scale distribution if and only if
the \SN{} relationship
$\log[\gfun(x;\betavec)]$ is a linear function of $\log(x)$ (i.e.,
the Basquin relationship in (\ref{equation:basquin.failure.time.model})).
For nonlinear \SN{} relationships,
the induced distribution for $X$ provides a theoretically
justified method for making inferences about fatigue-strength
distributions as a function of the given number of cycles $N_{e}$
and the fatigue-life model parameters $(\betavec, \sigma_{N})$.

Expressions for the corresponding pdf of $X$  are given in
Section~\ref{S.section:pdfs.for.induces.distributions}.  The $p$
quantile of the fatigue-strength distribution is obtained by solving
$F_{X}(x_{p}; N_{e})=p$ in
(\ref{equation:general.fatigue.strength.cdf.ls}) for $x_{p}$ giving
\begin{align}
\label{equation:fatigue.strength.quantile}
x_{p}(N_{e})&=\gfun^{-1}\left(\exp \left[\log(N_{e})-\Phi^{-1}(p)\sigma_{N} \right];\betavec \right), \quad \quad
  0<p<1, \,\, N_{e}>0.
\end{align}

\begin{example}
\label{example:basquin.induced.fatigue.strength}
\Exampletitle{The Induced Fatigue-Strength Model for the
  Basquin Relationship.}
This example provides details for the special-case induced fatigue-strength
model for the Basquin relationship, illustrated in
Figure~\ref{figure:BasquinFatigueLifeStrength} for the lognormal
and Weibull distributions.
\begin{figure}[ht]
\begin{tabular}{cc}
\phantom{XX}Lognormal (a) & \phantom{XX} Weibull (b) \\[-3.2ex]
\rsplidapdffiguresize{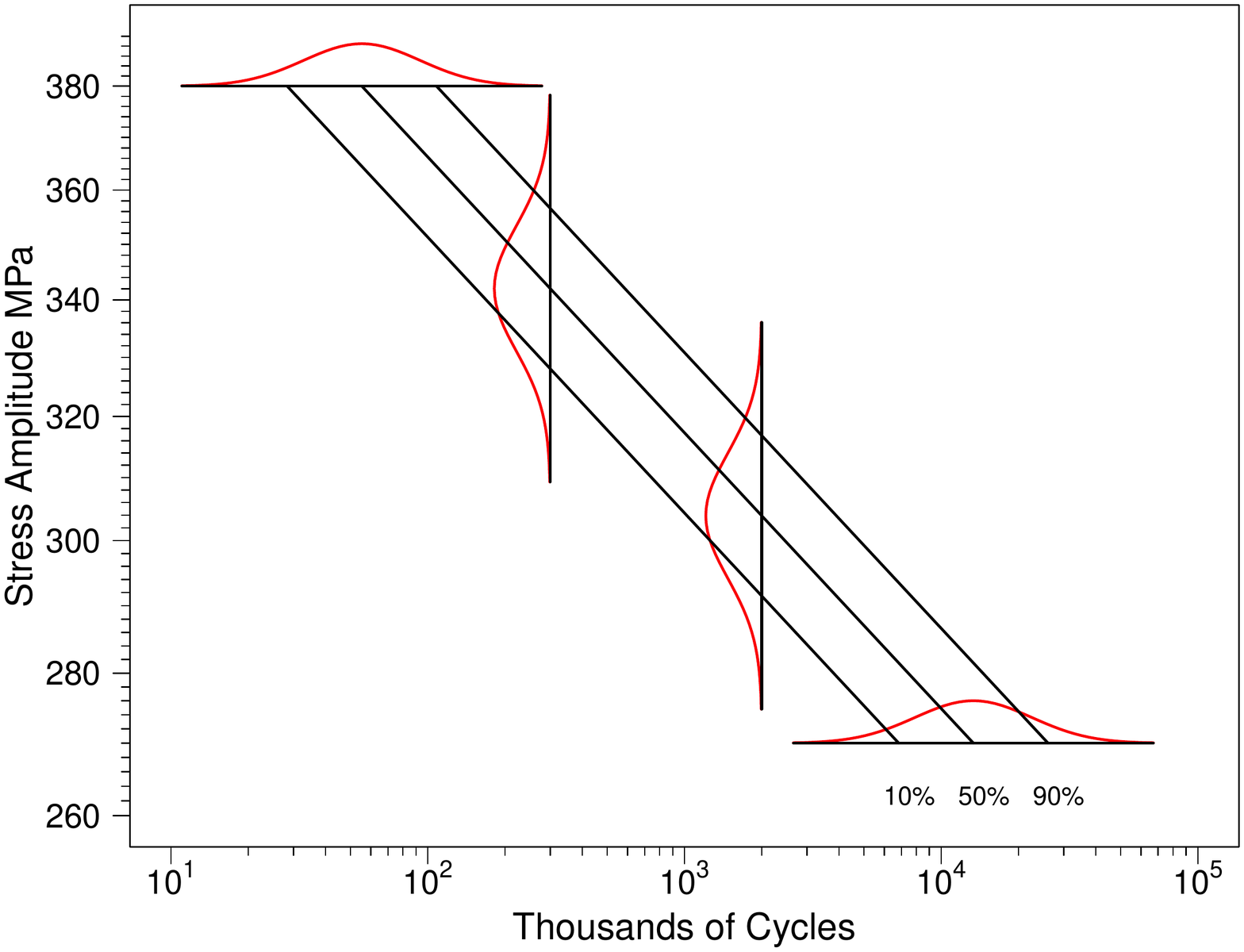}{3.25in} &
\rsplidapdffiguresize{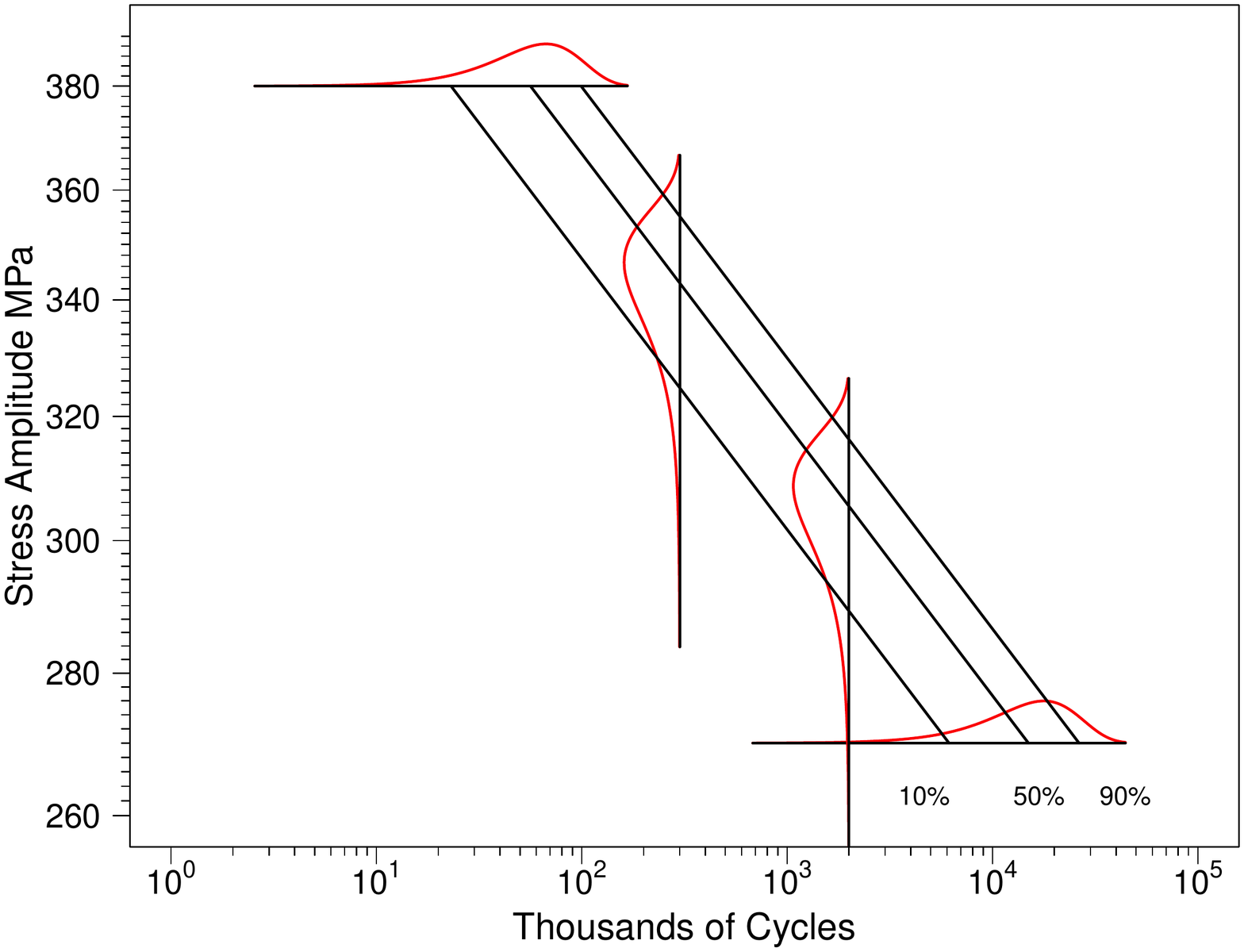}{3.25in}
\end{tabular}
\caption{Lognormal~(a) and Weibull~(b) fatigue-life (horizontal) and
  fatigue-strength (vertical) distributions for the Basquin \SN{} relationship.}
\label{figure:BasquinFatigueLifeStrength}
\end{figure}
Substituting $\log[\gfun(x;\betavec)]=\beta_{0} + \beta_{1}\log(x)$ into
(\ref{equation:general.fatigue.strength.cdf.ls}), with $\beta_{1}<0$
(because the \SN{} relationship is strictly decreasing), gives
\begin{align*}
  F_{X}(x;N_{e})
  &=\Phi \left [
      \frac{
      \log(N_{e})-[\beta_{0}+\beta_{1} \log(x)]}{\sigma_{N}}
      \right ], \quad \quad x>0, \,\, N_{e}>0\\
  &=\Phi \left [
      \frac{\log(x)- [\beta_{0}- \log(N_{e})]/|\beta_{1}|}{\sigma_{N}/|\beta_{1}|}
      \right ]
  =\Phi \left [
      \frac{\log(x)- [\beta^{\dagger}_{0}+\beta^{\dagger}_{1} \log(N_{e})]}{\sigma_{X}}
      \right ],
\end{align*}
where $\beta^{\dagger}_{0}=\beta_{0}/|\beta_{1}|$,
$\beta^{\dagger}_{1}=-1/|\beta_{1}|,$ with $\beta^{\dagger}_{1}<0,$   and
$\sigma_{X}=\sigma_{N}/|\beta_{1}|.$
This shows that the induced fatigue-strength
model has the same log-location-scale form with different parameters.
\citet{GroveCampean2008} give a similar result.
\end{example}

\subsubsection{The induced fatigue-strength model when
  \texorpdfstring{$\gSN$}{g(S)}
  has  a horizontal asymptote}
\label{section:induced.fatigue.strength.distribution.horizontal.asymptote}

When $\log[\gfun(x;\betavec)]$ has a horizontal asymptote at
$\log(S)=E$, \, $\lim_{x \downarrow
  \exp(E)} \log[\gfun(x;\betavec)]=\infty$ as illustrated in
Figure~\ref{figure:sn.relationships}b and~c.
Because $\left(\log(N_{e})-\log[\gfun(x;\betavec)]\right)/\sigma_{N}$ is
unbounded, the derivation of the cdf for
$X$ is the same as (\ref{equation:general.fatigue.strength.cdf.ls})
but because of the asymptote, for given $N_{e},$
\begin{align*}
  \lim_{x \downarrow \exp(E)} F_{X}(x; N_{e})&=
\Phi\left(\frac{\log(N_{e})-\infty}{\sigma_{N}}   \right)=0,
\end{align*}
and thus
\begin{align*}
  F_{X}(x; N_{e})&=
  \Phi\left(\dfrac{\log(N_{e})-\log[\gfun(x;\betavec)]}{\sigma_{N}}
  \right), \quad\quad  x> \exp(E), \,\, N_{e}>0.
\end{align*}
The range of $X$ depends on the unknown threshold parameter
$\exp(E)$, so the model
is not regular \citep[e.g., ][]{Smith1985}. The horizontal
asymptote also implies that
fatigue strength $X$ will never be less than $\exp(E)$.
The pdf of $X$ is the same as
(\ref{S.equation:induced.fatigue.life.pdf.ls}) in
Section~\ref{S.section:pdfs.for.induces.distributions},
except that it is positive only when~$x>\exp(E).$
The quantiles of $X$ are the same as
(\ref{equation:fatigue.strength.quantile}) but as $p\to 0,$
$x_{p}(N_{e}) \to \exp(E)$.

\subsubsection{The induced fatigue-strength model when
  \texorpdfstring{$\gSN$}{g(S)}
  has a vertical asymptote}
\label{section:induced.fatigue.strength.distribution.vertical.asymptote}

If $\log[\gfun(x;\betavec)]$ has a vertical asymptote (e.g., the
Box--Cox relationship in Section~\ref{section:box.cox.relationship}
and see Figure~\ref{figure:sn.relationships}a
with $\lambda<0$), then $\lim_{x \to
  \infty}\log[\gfun(x;\betavec)]=B$ and thus $B$ is
a lower bound for $\log[\gfun(x;\betavec)].$ Consequently,
$\left[\log(N_{e}) - B\right]/\sigma_{N}$ is an upper bound
for $\epsilon,$ but the range of $\epsilon$ is $(-\infty, \infty).$
To resolve this inconsistency, we modify
(\ref{equation:reexpressed.fatigue.life.strength}) and use
\begin{align*}
  \log(N_{e}) &= \log[\gfun(X;\betavec)] +  \sigma_{N}  \epsilon
  \, I[-\infty < \epsilon < \left(\log(N_{e}) - B\right)/\sigma_{N}],
\end{align*}
where $I[\cdot]$ is the indicator function. The derivation of the cdf for
$X$ is the same as (\ref{equation:general.fatigue.strength.cdf.ls}) but
because of the asymptote,
for given $N_{e},$
\begin{align*}
\lim_{x \to \infty}F_{X}(x; N_{e})&=
\Phi\left(\frac{\log(N_{e})-B}{\sigma_{N}}   \right) < 1.
\end{align*}
This implies that the distribution of $X$ has a discrete atom of probability
at $\infty.$ The size of the discrete atom is
\begin{align}
\label{equation:atom.of.prob.due.ver.asym}
1-\Phi\left(\frac{\log(N_{e})-B}{\sigma_{N}}   \right).
\end{align}
This discrete atom corresponds to the
limiting proportion of units for which $N>N_{e}$ as $x \rightarrow \infty$.
This can be interpreted as the (physically questionable)
proportion of units that
would survive $N_{e}$ cycles, even as stress approaches $\infty$.
The pdf of $X$ is the same as
(\ref{S.equation:induced.fatigue.life.pdf.ls}) in
Section~\ref{S.section:pdfs.for.induces.distributions}
but it does not integrate to 1 because of the discrete atom of
probability at $\infty$.  The quantiles of $X$ are the same as
(\ref{equation:fatigue.strength.quantile}) but, because of the discrete atom
at $\infty$, $x_{p}(N_{e})$ is finite only for $0 \le p <
\Phi\left[(\log(N_{e})-B)/\sigma_{N}\right].$
As an example,
Section~\ref{S.section:technical.details.distributions.box.cox} in
the Appendix gives details for the Box--Cox \SN{} model.

\subsubsection{Equivalence of fatigue-life and fatigue-strength
  quantile curves}
\label{section:equivalence.fatigue.life.fatigue.strength.quantile.curves}
For \SN{} relationships that have neither a horizontal nor a vertical asymptote,
the fatigue-life and fatigue-strength models have the same
quantile curves and thus when estimating a fatigue-life
model, one is simultaneously estimating the fatigue-strength
model.
For \SN{} relationships that have either a horizontal or a vertical
asymptote the fatigue-life and fatigue-strength quantile curves are
still equivalent except that quantile lines for certain values of
$p$ do not exist because of the
discrete atom of probability at $\infty$, as described in
Sections~\ref{section:induced.fatigue.strength.distribution.vertical.asymptote},~\ref{section:induced.fatigue-life.distribution.horizontal.asymptote},
and~\ref{section:quantile.curve.visualization}.
Section~\ref{S.section:quantile.curve.equivalence}
provides a proof of this result and more explanation about the exceptions.

\subsection{Choosing a Fatigue-Life Probability Distribution}
\label{section:choosing.fatigue.life.distribution}
The lognormal and Weibull distributions are the most commonly used
distributions in fatigue data analysis. This is because one or the
other often fits well and because there is physical motivation for
using them. \citet[][Section~4.6.2]{MeekerEscobarPascual2021} give
physics-of-failure arguments (based on cumulative damage mechanisms
like fatigue)
for using the lognormal distribution to describe time to
fracture from fatigue in ductile materials like metals, when there
is a single crack growing toward fracture. Mathematical
justification for this physical/chemical motivation is
given in \citet[][pages 36--37]{GnedenkoBelyayevSolovyev1969}
and \citet[][pages 133--134]{MannSchaferSingpurwalla1974}.
\citet[Section~4.6][]{CrowderKimberSmithEtAl1994},
\citet{CastilloFernandez-Canteli2009}, and
\citet[][Section~4.8.4]{MeekerEscobarPascual2021} describe
extreme-value-theory arguments for using the Weibull
distribution to describe time to fracture from fatigue in brittle
materials like ceramics or metals if there are potentially many
cracks competing to be the first to cause fracture (e.g., in a wire,
chain, gear, or bearing).

Although this kind of physical guidance is useful in deciding which
distribution to use, it is important to use probability plots like
those in Figures~\ref{figure:LaminatePanelProbability.plots},
\ref{figure:Ti64350FRm1.plots}b, and
\ref{figure:Nitinol02Model.plots}b to help make a decision. It is
also important to use sensitivity analysis to assess the effect of
alternative choices in the distribution (especially when the data do
not result in a definitive conclusion or when extrapolating into the
lower tail of a fatigue-life or fatigue-strength
distribution).  For additional illustrations of this,
Section~\ref{S.section:examples.comparing.lognormal.and.weibull.distributions.fit.sn.data}
provides a side-by-side comparison of lognormal and
Weibull distributions fit to nine different \SN{} data sets. For the four
data sets where tests were on wire specimens, the Weibull
distribution fits well (as predicted by extreme-value theory). For
the others (e.g., notched or hour-glass shaped metal specimens), the
data show that the lognormal distribution is a more
appropriate distribution (as predicted by the cumulative damage
theory).

\subsection{An Example of Fitting a Fatigue-Life Model}
\label{section:fitting.fatigue.life.model}
This section provides an example to illustrate the key ideas presented
earlier in this section and to set the stage for the remainder of
the paper.
\begin{example}
\label{example:box-cox.loglin.laminate.panel.data}
\Exampletitle{Fitting the Box--Cox/Loglinear-$\sigma_{N}$ \SN{} Model
  to the Laminate Panel Data.}  This example is a continuation of
Example~\ref{example:laminate.panel.data}. A description of the
noninformative joint prior distribution that was used,
and other details are in 
Section~\ref{S.section:more.details.laminate.panel.boxcox.loglinear}.
Figure~\ref{figure:LaminatePanelModel.plots}a is a lognormal
probability plot showing Bayesian cdf estimates. The estimates were
computed by taking the median of the empirical distribution of the
draws from the marginal posterior distributions of $F_{N}(t; S_{e})$
for a large number of values of $t$ for each of the five levels of
$S_{e}$ used in the experiment. The Bayesian estimates agree well
with the nonparametric estimates at all levels of $S_{e}$. For
estimation at $S_{e}=270$ MPa, corresponding 95\% credible intervals
are also plotted. These were obtained from the 0.025 and 0.975
quantiles of the empirical distribution of the draws from the
marginal posterior distribution $F_{N}(t; 270)$ for the same values
of $t$ used to compute the cdf estimates. The credible intervals are
narrow because of the large number of tested specimens with few
runouts.
\begin{figure}[ht]
\begin{tabular}{cc}
(a) & (b) \\[-3.2ex]
\rsplidapdffiguresize{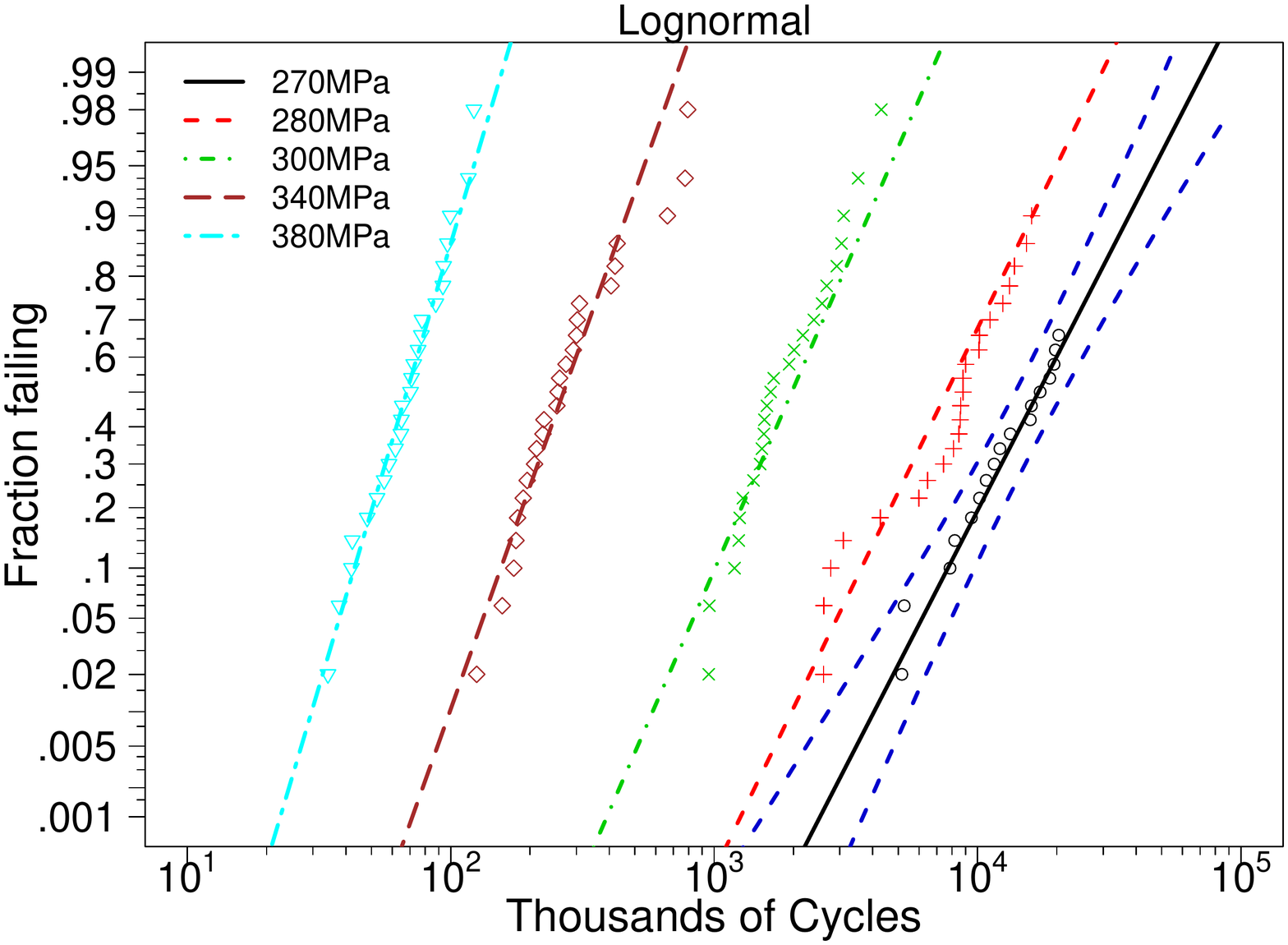}{3.25in} &
\rsplidapdffiguresize{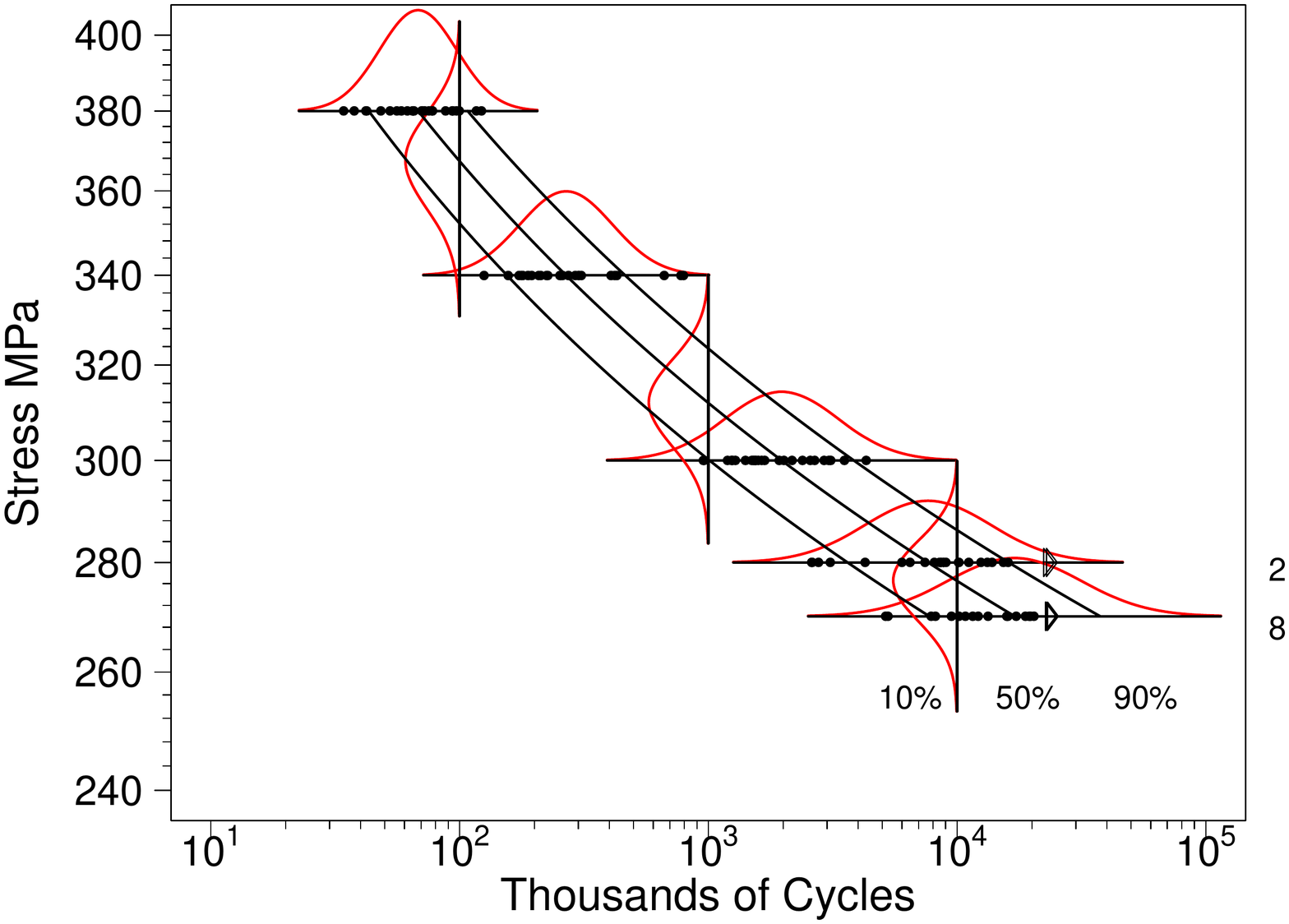}{3.25in}
\end{tabular}
\caption{Lognormal probability plot showing the cdf estimates from
  the Box--Cox/loglinear-$\sigma_{N}$ model fit to the laminate
  panel \SN{} Data~(a) and the corresponding model plot showing the (shared)
  quantile curves and density estimates for fatigue life (horizontal) and
  fatigue strength (vertical)~(b).}
\label{figure:LaminatePanelModel.plots}
\end{figure}

Figure~\ref{figure:LaminatePanelModel.plots}b is a fitted
\textit{model plot} showing the estimates of the 0.10, 0.50, and 0.9
quantile curves, along with estimates of the fatigue-life and fatigue-strength
densities superimposed on top of the \SN{} data that we first saw in
Figure~\ref{figure:LaminatePanelScatter.plots}b. The increase in
spread in the fatigue-life (horizontal) densities (due to the
loglinear-$\sigma_{N}$ component in the model) is
evident. Interestingly, the spread in the \textit{induced} fatigue-strength
(vertical) densities appears to be approximately constant.
\end{example}

\subsection{Compatibility Conditions and Characteristics of \SN{} Models}
\label{section:characteristics.of.sn.models}
\citet{Bastenaire1972} and \citet[][Chapter
  2]{CastilloFernandez-Canteli2009} describe various characteristics
or ``compatibility conditions'' that are required for
statistical models for fatigue to be sensible both
physically and probabilistically. This section reviews some of these
characteristics.

Perhaps the most important model characteristic is that the \SN{}
relationship (e.g., $N=g(S)$, corresponding to a particular quantile
curve) should be positive and monotonically decreasing---higher stress implies
shorter life.  Additionally, the fatigue-life model cdf $F_{N}(t; S_{e},
\thetavec)$ should be
\begin{itemize}
\item
Monotonically increasing in $t$ for fixed $S_{e}$ and
\item
Monotonically increasing in $S_{e}$ for fixed $t$.
\end{itemize}
The first condition is generally met for any of the continuous
cdfs typically used in fatigue-life
models and used in this paper. Whether the second condition
holds or not will depend on the nature of the regression model. Generally,
the condition will hold if cdfs for different stress levels like
those in Figure~\ref{figure:LaminatePanelModel.plots}a do not
cross. Equivalently, the condition will hold if quantile curves like those in
Figure~\ref{figure:LaminatePanelModel.plots}b do not cross.
Although desirable, it is not essential that these not-cross conditions hold
over the entire range of $t$ and $S_{e}$. It is, however, essential
that the conditions hold over the range of $t$ and $S_{e}$ where the
model is used.  \citet[][page 10]{Bastenaire1972} makes a similar
point.

For example, with the fitted quadratic model used in
\citet{Nelson1984} (mentioned in
Section~\ref{section:literature.review}), the quantile curves are
decreasing in pseudo-stress over the range of the data but begin to increase for
larger values of the pseudo-stress
used there---but this happens
only outside of the range of pseudo-stress where the model would be
used.
The model used for the laminate panel data in
Section~\ref{section:fitting.fatigue.life.model} allows $\sigma_{N}$
to be a loglinear function of stress. This causes the slopes of the
estimates of the lognormal cdfs in a lognormal probability plot (see
Figure~\ref{figure:LaminatePanelModel.plots}a) to depend on
stress, implying that the cdfs will cross. The crossing behavior,
however, is far away from the region where the model would be used.

Similar conditions can be stated for the fatigue-strength
model $F_{X}(x; N_{e})$. Recognizing that the
fatigue-life and fatigue-strength models have the same
quantile curves
(Section~\ref{section:equivalence.fatigue.life.fatigue.strength.quantile.curves})
shows that if the conditions hold for the fatigue-life model, they
also hold for the fatigue-strength model and vice versa.
This crossing behavior can be avoided by specifying a
fatigue-strength model with constant $\sigma_{X}$ which, if
there is curvature in the \SN{} relationship, will result in a change
in spread for the induced fatigue-life model for $N$. This
illuminates an important advantage in specifying the \SN{} model in
terms of the fatigue-strength distribution. This modeling approach
is described in detail in
Section~\ref{section:alternative.approach.modeling.sn.data}.

\section{Statistical Models for Fatigue-Strength}
\label{section:statistical.models.estimate.fatigue.strength.distributions}

\subsection{Estimating a Fatigue-Strength Distribution Using Binary Data}
Because fatigue strength is not directly observable, the traditional
way to estimate a fatigue-strength distribution at a given level $N_{e}$
\citep[e.g.,][]{LittleJebe1975, Nelson1990a,
  AwadDeJackKrivtsov2004,GroveCampean2008} has been to
\begin{itemize}
\item
Test a sample of $n$ units at different fixed stress levels $S_{i}$,
$i=1,\dots, n$.
The units are tested until
failure or the given value of $N_{e}$ cycles (whichever comes
first).
\item
Dichotomize the data to consist of only the runouts
(right-censored at $S_{i}$ because $X>S_{i}$) and failures (left-censored
at  $S_{i}$ because $X < S_{i}$). The actual failure times are ignored.
\item
Use binary regression methods (e.g., logit or probit regression,
possibly on $\log(S)$)
to estimate the fatigue-strength distribution at
$N_{e}$ cycles.
\end{itemize}
Data from the well-known and commonly used staircase method
\citep[e.g.,][]{PollakPalazottoNicholas2006, Muller_et_al2017}
provide useful estimates of the \textit{median} of the fatigue-strength
distribution, but not small quantiles that are needed in
high-reliability applications. This is because the method
concentrates observations near the center of the
fatigue-strength distribution. \citet{WuTian2014} review and suggest
an alternative sequential method when the goal is to estimate a
particular quantile of a distribution based on
binary data.

Dichotomizing \SN{} data to estimate fatigue-strength distributions
has serious disadvantages. Such methods are statistically
inefficient and limit the range of $N_{e}$ for which
fatigue-strength distributions can be estimated.
Sections~\ref{section:relationship.fatigue.life.fatigue.strength}
and ~\ref{section:fitting.fatigue.life.model} showed how fitting a
specified fatigue-life model can be used to estimate characteristics
of fatigue-strength distributions, using all of the available \SN{}
data (i.e., not ignoring the failure times).
Section~\ref{section:alternative.approach.modeling.sn.data} shows
how to use a \textit{specified fatigue-strength model} that also
uses all of the available \SN{} data to make inferences
for either fatigue-strength or fatigue-life distributions.

\subsection{Modeling \SN{} Data by Specifying a Fatigue-Strength Model}
\label{section:alternative.approach.modeling.sn.data}
The relationship between $F_{N}(t; S_{e})$ and $F_{X}(x; N_{e})$ described in
Sections~\ref{section:relationship.fatigue.life.and.fatigue.strength}
and~\ref{section:relationship.fatigue.life.fatigue.strength}
suggests an alternative path for specifying a statistical model for \SN{} data.
 Similar to \citet{Falk2019}, one can specify the form of
 the fatigue-strength distribution and use a specified \SN{}
 relationship to \textit{induce} a fatigue-life model that can be fit
 to the \SN{} data using statistical methods
 (e.g., maximum likelihood or Bayesian estimation) that can
 accommodate censored data.

\subsubsection{The advantages of specifying the fatigue-strength
 model to describe \SN{} data}
\label{section:important.advantages.of.specifying.fatigue.strength.distribution}
Specifying a fatigue-strength
model and having it induce the corresponding fatigue-life
model has important advantages.
\citet{Weibull1956} recognized these advantages but his ideas were,
unfortunately, lost over time, perhaps because fatigue-strength
cannot be observed directly. The most important advantage is that
fatigue-strength distributions generally have a simpler form than
fatigue life distributions. In particular, $F_{X}(x; N_{e})$ tends to have
constant shape/spread for different values of $N_{e}$ whereas
the fatigue-life distributions $F_{N}(t; S_{e})$ often have
increased spread and a different shape at lower levels of
stress. This was also noted by \citet{Hanaki_etal2003,Hanaki_etal2010}.
Empirically, Figure~\ref{S.figure:datasets.contrasting.spread.fl.fs} in
 Section~\ref{S.section:additional.motivation.specify.fatigue.strength}
provides six HCF examples where the vertical spread in
the data is relatively constant but the spread in the fatigue-life
distributions is larger at lower levels of stress.
Thus a model component to describe increasing spread in
fatigue life $N$ will usually \textit{not} be needed (further
physical explanation is given in
Section~\ref{section:physical.explanation.curvature.nonconstant.variance}). As
described in Section~\ref{section:characteristics.of.sn.models},
this implies that the compatibility conditions will hold
(e.g., quantile lines will not cross). More generally,
when there is curvature in the \SN{} relationship, the
induced fatigue-life distributions have features that
agree better with the physical nature of fatigue data---increased
spread at lower levels of stress.

\subsubsection{A statistical model for fatigue-strength}
\label{section:specifying.fatigue.strength.distribution}
Suppose that the logarithm of the fatigue-strength random variable
$X$ at a \textit{given} number of cycles $N_{e}$ is
\begin{align}
\label{equation:logx.general.strength.model}
\log(X) = \log[\hfun(N_{e};\betavec)] +  \sigma_{X} \epsilon,
\end{align}
where $S=\hfun(N;\betavec)$ is a positive monotonically
decreasing \SN{} regression relationship of known form,
$\betavec$ is a vector of regression parameters,
$\sigma_{X} \epsilon$ is a random-error term, and $\epsilon$
has a location-scale distribution with
$\mu=0$ and $\sigma=1.$
Then for any given number of cycles
$N_{e}$, $X$ has a log-location-scale distribution with cdf
\begin{align}
\label{equation:logx.general.strength.cdf}
  F_{X}(x; N_{e})&= \Pr(X \le x; N_{e}) = \Phi\left[ \frac{\log(x) -
      \log[\hfun(N_{e};\betavec)]}{\sigma_{X}}\right], \quad \quad
  x>0, \,\, N_{e}>0,
\end{align}
where $\hfun(N_{e};\betavec)$ is a scale parameter and $\sigma_{X}$
is the shape parameter of the distribution of $X$.
The fatigue-strength $p$ quantile
is obtained by solving $p=F_{X}(x_{p}(N_{e});N_{e})$ for $x_{p}(N_{e})$, giving
\begin{align}
\label{equation:quantiles.specified.fatigue.strength}
x_{p}(N_{e}) = \exp(\log[\hfun(N_{e};\betavec)] + \Phi^{-1}(p) \sigma_{X}), \quad \quad
  0<p<1, \,\, N_{e}>0.
\end{align}

\subsubsection{The induced fatigue-life model when
  \texorpdfstring{$\hSN$}{log[h(N)]}
has neither a vertical nor a horizontal
  asymptote}
\label{section:induced.fatigue.life.distribution.neither.vertical.nor.horizontal}
For the moment, suppose that the positive monotonically decreasing \SN{}
relationship $S=\hfun(N;\betavec)$ has neither a vertical nor a
horizontal asymptote.  Replacing $N_{e}$
with $N$ and $X$ with $S_{e}$ in
(\ref{equation:logx.general.strength.model}) gives
\begin{align}
\label{equation:distribution.general.h.sn.relationship}
\log(S_{e}) -\log[\hfun(N;\betavec)]&=  \sigma_{X} \epsilon.
\end{align}
In this role switching, $N$ at given $S_{e}$ replaces $X$ at
fixed $N_{e}$, but the random variables $X$ and $N$
have the same $\sigma_{X} \epsilon$ random-error term.  Equation
(\ref{equation:distribution.general.h.sn.relationship}) implies
that $(\log(S_{e})-\log[\hfun(N;\betavec)])/\sigma_{X} = \epsilon$
has a location-scale distribution with $\mu=0$ and $\sigma=1.$
Thus the induced cdf of $N$ is
\begin{align}
\nonumber
F_{N}(t; S_{e})&=\Pr(N \le t; S_{e})=\Pr\left [\hfun(N;\betavec)>\hfun(t;\betavec) \right]\\
\nonumber
&=\Pr\left(\log [\hfun(N;\betavec)] > \log [\hfun(t;\betavec)]\right)=\Pr\left(-\log [\hfun(N;\betavec)] < -\log [\hfun(t;\betavec)]\right)\\
\nonumber
  &=\Pr\left(\log(S_{e})-\log [\hfun(N;\betavec)] < \log(S_{e})-\log [\hfun(t;\betavec)]\right)\\
  &=
\label{equation:cdf.for.induced.fatigue.life}
  \Phi\left[\frac{\log(S_{e})-\log[\hfun(t;\betavec)]}{\sigma_{X}}   \right], \quad \quad
  t>0, \,\, S_{e}>0.
\end{align}
Expressions for the corresponding pdf of $N$ (needed to compute a
likelihood function) are given in
(\ref{S.equation:induced.fatigue.life.pdf.ls}) in
Section~\ref{S.section:pdfs.for.induced.distributions}.
Note that~(\ref{equation:cdf.for.induced.fatigue.life}) is a
log-location-scale distribution if and only if
$\log[\hfun(t;\betavec)]$ is a linear function of $\log(t)$ (i.e.,
the Basquin relationship in (\ref{equation:basquin.failure.time.model})).
For linear or nonlinear \SN{} relationships,
the induced distribution for $N$ provides a theoretically
justified method for making inferences about fatigue-life
distributions as a function of the fatigue-strength model parameters
$(\betavec, \sigma_{X})$.
The $p$
quantile of $N$ is obtained by solving $F_{N}(t_{p};S_{e})=p$ for
$t_{p}$ giving
\begin{align}
\label{equation:quantiles.induced.fatigue.life}
t_{p}(S_{e})&=\hfun^{-1}\left(\exp \left[\log(S_{e})-\Phi^{-1}(p)\sigma_{X} \right];\betavec \right), \quad \quad
  0<p<1, \,\, S_{e}>0.
\end{align}

\subsubsection{The induced fatigue-life model when
  \texorpdfstring{$\hSN$}{log[h(N)]}
has  a horizontal asymptote}
\label{section:induced.fatigue-life.distribution.horizontal.asymptote}
When $\lim_{t \to \infty} \log[\hfun(t;\betavec)]=E > -\infty$,
$\log[\hfun(N;\betavec)]$ has a horizontal asymptote, as illustrated
in Figure~\ref{figure:sn.relationships}b and~c. Note that both
axes in these plots are logarithmic so $\exp(E)=0.5$ in
Figure~\ref{figure:sn.relationships}c.
Because $E$ is a lower bound for $\log[\hfun(N;\beta)],$
$\left[\log(S_{e}) - E\right]/\sigma_{X}$ is an upper bound for
$\epsilon,$ but the range of $\epsilon$ in
(\ref{equation:distribution.general.h.sn.relationship}) is
$(-\infty, \infty).$ To resolve this inconsistency, we modify
(\ref{equation:distribution.general.h.sn.relationship}) and use
\begin{align*}
  \log(S_{e}) &= \log[\hfun(N;\betavec)] +  \sigma_{X}  \epsilon \,
    I\left [-\infty < \epsilon < \left(\log(S_{e}) - E\right)/\sigma_{X} \right],
\end{align*}
where $I[\cdot]$ is the indicator function.  The cdf for $N$ is
still (\ref{equation:cdf.for.induced.fatigue.life}) but
because $\lim_{t \to \infty} \log[\hfun(t;\betavec)]=E,$
\begin{align}
  \lim_{t \to \infty} F_{N}(t; S_{e})&=\Phi\left(\frac{\log(S_{e})-E}{\sigma_{X}}   \right)< 1,
\end{align}
which implies that the cdf $F_{N}(t;S_{e})$ has a discrete atom of
probability at $\infty.$ The size of the discrete atom is
\begin{align}
\label{equation:atom.of.prob.due.hor.asym}
1-\Phi\left(\frac{\log(S_{e})-E}{\sigma_{X}}   \right).
\end{align}
This discrete atom corresponds to the limiting proportion of units for which
 fatigue strength $X>S_{e}$, as $t \rightarrow \infty$.
This can be interpreted as
the proportion of units that, if tested at stress
 $S_{e}$, would not fail because $X>\exp(E)$, where $\exp(E)$ can be interpreted
as a fatigue limit.
The quantiles $t_{p}(S_{e})$ are the
same as in (\ref{equation:quantiles.induced.fatigue.life}), but
because of the discrete atom of probability at $\infty,$ $t_{p}(S_{e})$ is
only finite
for $0 < p < \Phi\left[(\log(S_{e})-E)/\sigma_{X}\right].$

\subsubsection{The induced fatigue-life model  when
  \texorpdfstring{$\hSN$}{log[h(N)]}
  has  a vertical asymptote}
\label{section:induced.fatigue-life.distribution.vertical.asymptote}
Consider, for example, the Box--Cox $\SN$ curves in
Figure~\ref{figure:sn.relationships}a which, for $\lambda<0$, have
vertical asymptotes at $B=\log(N).$ When $t$ decreases to
$\exp(B)$, $\log[\hfun(t;\betavec)]$ is unbounded. That is, $\lim_{t
  \downarrow\exp(B)}\log[\hfun(t;\betavec)]=\infty$.  The
cdf is obtained as in~(\ref{equation:cdf.for.induced.fatigue.life})
except that $\exp(B)$ is a threshold parameter (i.e.,
$\Pr[N\leq\exp(B)]=0$) and thus
\begin{align}
\label{equation:induced.fatigue.life.cdf.vertical.asymptote}
  F_{N}(t;
  S_{e})=\Phi\left(\dfrac{\log(S_{e})-\log[\hfun(t;\betavec)]}{\sigma_{X}}
  \right), \quad \quad
  t> \exp(B), \,\, S_{e}>0.
\end{align}
The quantiles $t_{p}(S_{e})$ of $F_{N}(t;S_{e})$ are the same as
(\ref{equation:quantiles.induced.fatigue.life}) but as $p\to 0,$
$t_{p}(S_{e}) \to \exp(B).$

The interpretation of this threshold parameter is similar to that
described in Section~\ref{section:box.cox.relationship}. For
models with a vertical asymptote, even as stress increases to high
levels, there is a positive value of $N$ that a unit could
survive. Of course that positive value could be a small fraction of
a cycle. Even so, it could be argued that \SN{} relationships with a
vertical asymptote are inconsistent with what happens
physically. As mentioned in
Section~\ref{section:box.cox.relationship}, however,
when the asymptotic behavior occurs far outside the range where the model
would be used, any inconsistency is not of practical concern.

\subsubsection{Visualization of the effect that \SN{} relationship
  coordinate asymptotes have on quantile curves and the
  induced fatigue-life distributions}
\label{section:quantile.curve.visualization}

\begin{figure}
\centering
\rsplidapdffiguresize{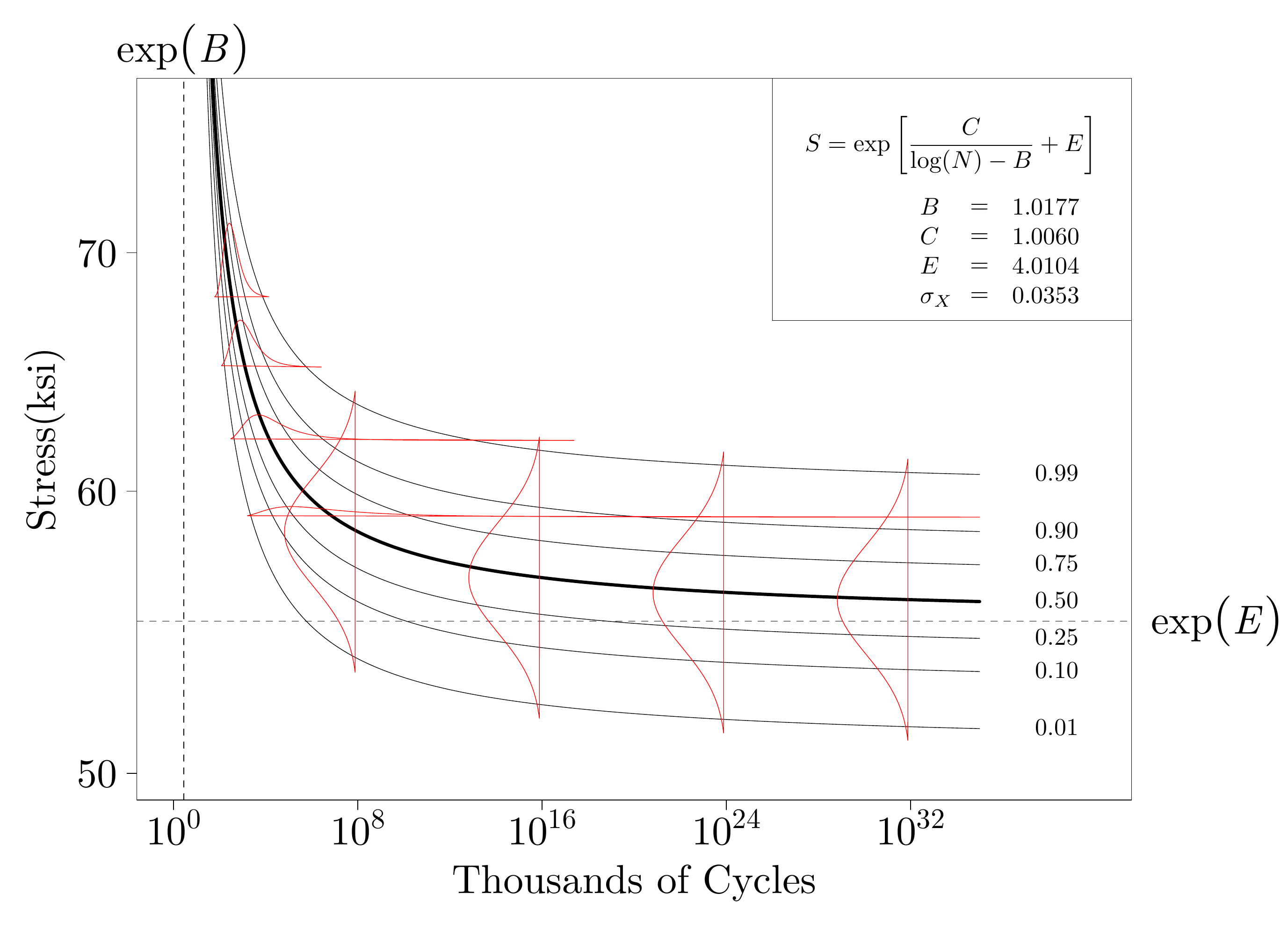}{5.0in}
\caption{A fitted rectangular-hyperbola \SN{} regression model.}
\label{figure:rectangular.hyperbola.regression.model}
\end{figure}

Figure~\ref{figure:rectangular.hyperbola.regression.model} is a
plot of a fitted rectangular hyperbola model that has a vertical
asymptote at $\exp(B)$ and a horizontal asymptote at $\exp(E)$. The
plot provides a visualization of the effects described in
Sections~\ref{section:induced.fatigue-life.distribution.horizontal.asymptote}
and~\ref{section:induced.fatigue-life.distribution.vertical.asymptote}.
The plot shows quantiles curves for
$p=0.01, 0.1, 0.25, 0.5, 0.75, 0.9,$ and 0.99.
The vertical densities at $N_{e}=10^{8}$, $10^{16}$,
$10^{24}$, and $10^{32}$ thousand cycles correspond
to lognormal fatigue-strength distributions with a constant $\sigma_{X}$
shape parameter. These densities look like normal densities because
the vertical axis is a log axis. Also shown are the horizontal
densities corresponding to the \textit{induced} fatigue-life distributions at
$S_{e} = 59, 62, 65$, and 68 ksi. The curves are quantile
curves for both the fatigue-strength and the fatigue-life
models.

The effect of the vertical asymptote at $\exp(B)$ is that the
induced fatigue-life distributions have a threshold parameter at
$\exp(B)$ and thus $\Pr[N < \exp(B)]=0$ for any value of $S_{e}$. It
is also interesting to see the relatively small amount of spread in
the induced fatigue-life distributions at high levels of stress.

The effect of the horizontal asymptote at $\exp(E)$ is more subtle.
As suggested by the exaggerated time range in
Figure~\ref{figure:rectangular.hyperbola.regression.model}, each
quantile curve has its own horizontal asymptote. Imagine the
fatigue-life distribution at a level of stress $S^{*}_{e}$ that is
infinitesimally below the asymptote for the 0.99 (top) quantile
curve in
Figure~\ref{figure:rectangular.hyperbola.regression.model}. That
fatigue-life distribution does not have a finite 0.99 quantile and
has an atom of probability at infinity that is infinitesimally
larger than 0.01. Another way to explain the atom at infinity is
that even as $N_{e} \rightarrow \infty$, $\Pr(X > S^{*}_{e}) \approx
0.01$ (the probability that fatigue strength is greater than
$S^{*}_{e}$---and failure will never occur---has a positive limit).

\section{Estimating \SN{} Model Parameters, Model-Fitting Diagnostics, and Making Inferences
  about Fatigue Distributions}
\label{section:estimating.parameters.diagnostics.inferences.quantiles}
This section briefly describes maximum likelihood and Bayesian
methods for fitting \SN{} models, estimating tail probabilities and
quantiles, and computing confidence or credible intervals for
quantifying statistical uncertainty (i.e., uncertainty due to limited
data). Both methods are well suited to handle runouts that
appear in many fatigue tests. Ordinary least squares should not be
used for estimation when there are runouts.

\subsection{Likelihood-Based Methods}
\label{section:likelihood.inference.methods}
Likelihood is the primary tool for making non-Bayesian inferences
when using advanced statistical models. The method is general,
versatile, and has been widely implemented in readily available software
for many different kinds of statistical models. Under mild
conditions (met in the applications in this paper), likelihood
methods have desirable statistical properties in large samples and
are generally difficult or impossible to beat even with small
samples. Among others, \citet{Severini2000} and \citet{Pawitan2013}
provide likelihood theory and methods.

\subsubsection{Log-likelihood for an \SN{} regression model with runouts}

Typical \SN{} data are $(S_{i}, N_{i}, \delta_{i}), i=1,\dots, n$
giving stress level $S_{i}$, number of cycles $N_{i}$, and a
censoring indicator $\delta_{i}$ for each of $n$ observations. The
log-likelihood for these data is
\begin{align}
\label{equation:location.scale.likelihood}
\loglike(\thetavec) &= \sum_{i=1}^{n}\left\{
\delta_{i} \log  \left[f_{N}(N_{i} ; S_{i}, \thetavec)
\right ] +
(1-\delta_{i}) \log \left [1- F_{N}(N_{i} ;S_{i}, \thetavec) \right]
\right \},
\end{align}
where
\begin{align*}
\delta_{i}&=
\begin{cases}
1 & \text{if $N_{i}$ is a failure time} \\
0 & \text{if $N_{i}$ is a runout time.} \\
\end{cases}
\end{align*}
Here, $F_{N}(N_{i} ;S_{i}, \thetavec)$ is the fatigue-life cdf
in~(\ref{equation:fatigue.life.failure.time.model.cdf}) when the
fatigue-life model is specified
and~(\ref{equation:cdf.for.induced.fatigue.life}) when the
fatigue-strength model is specified (and the fatigue-life
model is induced). Then $f_{N}(t; S_{i},
\thetavec)=d\,F_{N}(N_{i} ; S_{i}, \thetavec)/dt$ is the
corresponding pdf.
Sections~\ref{S.section:specified.fatigue.life.distribution}
and~\ref{S.section:pdfs.for.induced.distributions}
give expressions for the pdfs.
Standard optimization algorithms can be used to maximize
$\loglike(\thetavec)$.
\citet{LiuMeeker2024} provide implementation suggestions for
the nonlinear models used in this paper.

\subsubsection{Methods for computing confidence intervals when using
  likelihood-based inference}
In engineering applications, inferences are generally needed for
distribution tail probabilities and distribution quantiles.  For
non-Bayesian inference, basing confidence intervals for these
quantities on the distribution of of the likelihood-ratio statistic is
perhaps the most natural method to use. Coverage probabilities tend
to be close to the specified nominal confidence level.  The method is
computationally complicated but not hard to implement with modern
computing capabilities.
\citet{PascualMeeker1999} illustrate this approach for the RFL model.
\citet{LiuHongEscobarMeeker2024} outline algorithms to compute
likelihood-based confidence intervals for the fatigue-life
and fatigue-strength models described in this paper.

Wald confidence intervals are based on a quadratic approximation to
the profile log-likelihood function \citep{MeekerEscobar1995} and are
generally much easier to compute.  Wald intervals, however, tend to
have actual coverage probabilities that are smaller than the
specified nominal confidence level.  Bootstrap methods
\citep[e.g.,][]{EfronTibshirani1993} provide another method of
potentially improving on the Wald approximation.

\subsubsection{Equivalence of likelihood
pointwise confidence interval bands for cdfs and
quantiles} \citet{HongMeekerEscobar2008b} showed that a band of
pointwise confidence intervals for a cdf (e.g., the 270 MPa cdf
estimate in Figure~\ref{figure:LaminatePanelModel.plots}a) are
exactly the same as the band of pointwise confidence intervals for
quantiles if the confidence intervals are computed using the
likelihood ratio method (and approximately the same for Wald
intervals).  There are similar results relating bands of confidence
intervals for cdfs and quantiles for both fatigue-life and
fatigue-strength distributions for the models used in this paper.
Technical details are given in~\citet{LiuHongEscobarMeeker2024}.

Similarly, one can use confidence
intervals for fatigue-life model quantiles to obtain
confidence intervals for quantiles of the corresponding
fatigue-strength model. For
example, the value of stress $S_{e}$ for which the
likelihood-based lower confidence bound
$\tplower(S_{e})=N_{e}$ is then equivalent
to $\xplower(N_{e})$, the likelihood-based lower confidence bound
for the fatigue-strength distribution at $N_{e}$.
Again, technical details are given in~\citet{LiuHongEscobarMeeker2024}.
The importance of these results is that one can
use existing software that computes confidence intervals for fatigue-life
quantiles (or probabilities) to obtain confidence intervals for \textit{fatigue-strength}
quantiles (or probabilities).

\subsection{Bayesian Inference Methods}
\label{section:bayesian.inference.methods}
Over the past thirty years, there has been an increasing trend in
the proportion of applications where Bayesian methods are
used. We use Bayesian inference methods for fitting \SN{}
regression models because
engineers may have informative prior information for some of
the model parameters and because we have found that extensions to
models using random parameters (e.g., random batch effects) are
easier to implement using Bayesian methods.

\subsubsection{Specifying the joint prior distribution}
Bayesian inference requires the specification of a joint prior
distribution for the model parameters. For most applications and
certainly those considered here, there is usually a desire to use
noninformative or other minimally informative priors.
\citet{JohnsonFitzgeraldMartz1999} describe how they
developed a partially informative prior (while trying to be
minimally informative) to fit the RFL model.
\citet{GelmanSimpsonBetancourt2017} outline a general strategy for
specifying weakly informative priors.
\citet{TianLewis-BeckNiemiMeeker2022} describe methods for
specifying prior distributions in reliability applications for a
single distribution. The ideas  can be extended to regression models
and approximately to nonlinear regression models like those used in
our paper. \citet{LiuMeeker2024} outline a general strategy
for nonlinear regression based on a stable parameterization
and describe how noninformative or minimally informative prior
distributions can be specified for models like those used
in this paper.

\subsubsection{Generating and using draws from the joint
  posterior distribution} For the examples in this paper (with
additional details in the
Appendix), we use Bayesian methods to fit the \SN{} model and to
compute credible intervals for quantities of interest like
lower-tail quantiles of the fatigue-life and fatigue-strength
distributions. For each of the \SN{} models that we used in our examples,
a stable parameterization was specified
\citep[details are provided in][]{LiuMeeker2024} and a Stan
\citep{Stan-software} model was written and run using the RStan
\citep{RStan} interface to R \citep{Rsoftware}. For each model fit,
20,000 draws from the joint posterior distribution were computed and
saved. Then R functions were used to post-process these draws to
compute estimates and credible intervals for quantities interest
(e.g., the results in Figure~\ref{figure:LaminatePanelModel.plots})
and residuals used for diagnostic checking (e.g., the results in
Figure~\ref{figure:LaminatePanelResidual.plots}).

\subsubsection{Numerical methods  to obtain
  starting values and default joint prior distributions}
\label{section:numerical.methods.start.values.default.prior}
Robust algorithms for estimating the parameters
of nonlinear regression models (using either maximum likelihood or
Bayesian estimation) require careful attention to parameterization
and methods for finding starting values. Satisfactory starting
values can often be obtained by using simple moment estimates (e.g.,
sample means, variances, and linear regression). Our approach is to
define a parameterization where all parameters are unrestricted
without any ordering relationships. Optimizers tend to perform best
with such a parameterization and flat priors provide a natural
default joint prior distribution. In some applications, it is
necessary to replace the flat prior with an approximately flat normal (Gaussian)
distribution with an extremely large standard deviation (e.g., 10
times the standard error obtained from maximum likelihood
estimation). Exactly how these ideas are implemented depends on the
particular model. \citet{LiuMeeker2024} give details
for the models used in our examples.

\subsection{Using Residuals as Model-Checking Diagnostics}
\label{section:using.residuals.as.model.checking.diagnostics}
Although probability plots like those in
Figures~\ref{figure:LaminatePanelProbability.plots},~\ref{figure:Ti64350FRm1.plots}b,~\ref{figure:Nitinol02Model.plots}b,
and~\ref{figure:LaminatePanelModel.plots}b are useful
for detecting departures from the assumed model, such plots are
available only when experiments result in data with many
observations at each of some number of stress levels.  Frequently
\SN{} data have many stress levels with few repeats. In such
cases, residuals can be computed and these can be displayed in
various ways to see if they depart from what is expected under the
assumed model. \citet{Nelson1973}
describes regression analysis methods for censored data. The key
idea is that the residual for a censored observation is
correspondingly censored.

Here, residuals are defined as estimates of the $\epsilon$ error
variable in models such as (\ref{equation:general.model.for.N}) and
(\ref{equation:logx.general.strength.model}).  Such residuals are generally
known as \textit{standardized residuals} and should behave
approximately like
independent identically distributed (iid) observations with constant
spread from the assumed distribution.

Scatter plots of the residuals versus other variables, such as the
fitted values, stress, or other explanatory variables,
and potential explanatory variables (e.g., test
order and heat or batch) are useful. Systematic dependence of the
residuals on any such variable or systematic change in spread versus
such variables indicates a departure from the
assumed model. Special symbols (e.g., an upward-pointing triangle)
should be used to plot censored residuals. Heavy censoring can make
residual scatter plots difficult to interpret
\citep{Nelson1973}. For \SN{} data, most censoring occurs at the
lowest stress levels so scatter plots can
detect model departures. Suppose no serious departures are detected in such
scatter plots. In that case, probability plots of the residuals can be used
to check the adequacy of the assumed distribution of the $\epsilon$
error variable. In addition to checking the model fit for individual
fitted models, we have found that comparing residual plots for
across different fitted models for the same data to be particularly
useful. Section~\ref{S.section:comparison.sn.model.shapes}
gives a particular example.

Based on the \textit{specified fatigue-life} regression model
in~(\ref{equation:general.model.for.N}), the
standardized residuals (estimates of the $\epsilon_{i}$ error for
observation $i$)
are
\begin{align*}
 \epsilonhat_{i} &=
 \frac{\log(N_{i})-\log[\gfun(S_{i};\betavechat)]}{\sigmahat_{N}},
 \quad i=1,\dots, n.
\end{align*}
When there is a loglinear model for $\sigma_{N}$, as described in
Section~\ref{section:loglinear.sigma.component}, the
standardized residuals are
\begin{align}
\label{equation:standardized.residuals.fatigue.strength}
 \epsilonhat_{i} &=
 \frac{\log(N_{i})-\log[\gfun(S_{i};\betavechat)]}{\exp\left[\betahat_{0}^{[\sigma_{N}]} +
\betahat_{1}^{[\sigma_{N}]} \log(S_{i}) \right]},
 \quad i=1,\dots, n.
\end{align}
Based on the \textit{specified fatigue-strength} regression model
in~(\ref{equation:logx.general.strength.model}), the
standardized residuals
are computed from
\begin{align*}
 \epsilonhat_{i} &=
 \frac{\log(S_{i})-\log[\hfun(N_{i};\betavechat)]}{\sigmahat_{X}},
 \quad i=1,\dots, n.
\end{align*}
Then the $\exp(\epsilonhat_{i})$ values should, if the
assumed model is adequate, behave approximately like an
iid sample
from the assumed log-location-scale distribution.

Fatigue-life fitted values, a function of stress $S_{e}$, are
defined as estimates of the median lifetime $t_{0.50}(S_{e})$.
Fatigue-strength fitted values, a function of the number of
cycles $N_{e}$, are defined as estimates of the median strength
$x_{0.50}(N_{e})$. Because of the equivalence of fatigue-life
and fatigue-strength quantile curves
(as shown in Sections~\ref{S.section:quantile.curve.equivalence}
and~\ref{S.section:history.alternative.view}
 of the Appendix), the functions $t_{0.50}(S_{e})$ and $x_{0.50}(N_{e})$ map out
the same curve. For reasons described in 
Section~\ref{S.section:Ti64.additional.residual.analyses}, it
is better to plot residuals versus fatigue-life fitted values
rather than fatigue-strength fitted values.

\begin{example}
\label{example:box-cox.loglin.laminate.panel.residuals}
\Exampletitle{Residual Analysis for the Laminate Panel
  Box--Cox/Loglinear-$\sigma_{N}$ \SN{} Model.}
This is a continuation of
Example~\ref{example:box-cox.loglin.laminate.panel.data}
where the Box--Cox/loglinear-$\sigma_{N}$ \SN{} model
  was fit to the laminate panel data.
\begin{figure}[ht]
\begin{tabular}{cc}
(a) & (b) \\[-3.2ex]
\rsplidapdffiguresize{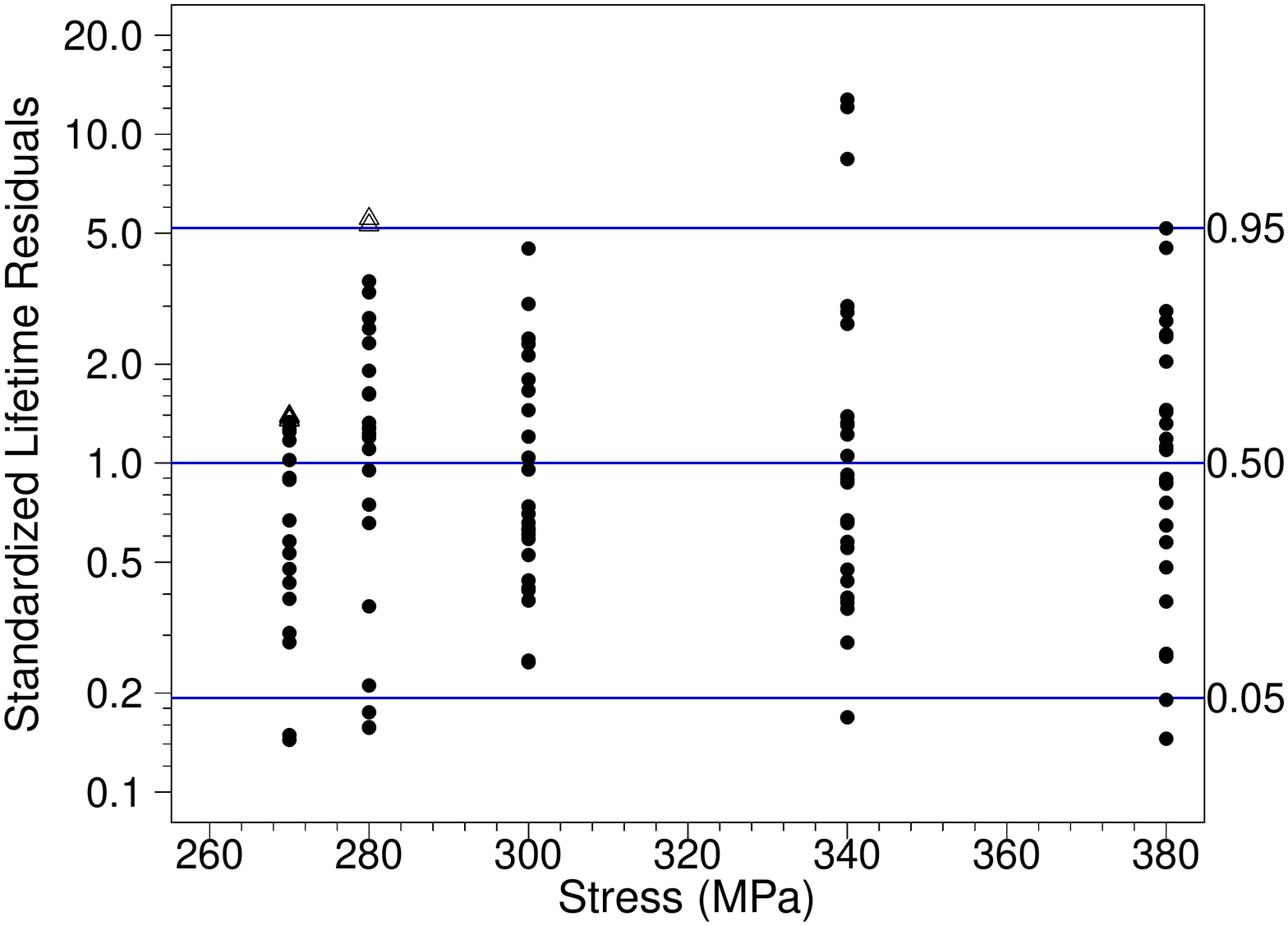}{3.25in} &
\rsplidapdffiguresize{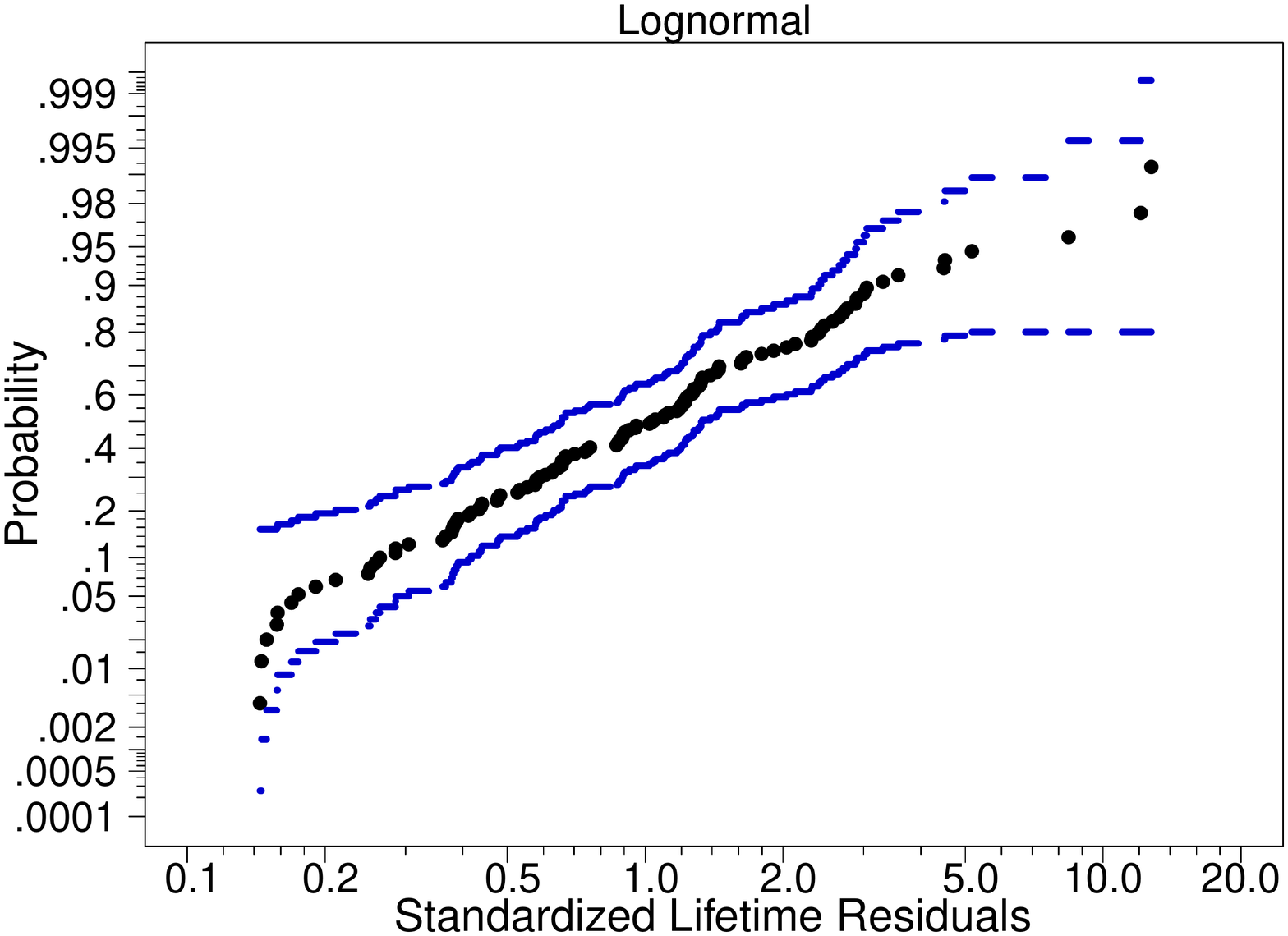}{3.25in}
\end{tabular}
\caption{Fatigue-life residuals from the Box--Cox/Loglinear-$\sigma_{N}$ \SN{}
  model fit to the laminate panel data versus stress~(a) and
  lognormal probability plot~(b).}
\label{figure:LaminatePanelResidual.plots}
\end{figure}
Figure~\ref{figure:LaminatePanelResidual.plots}a is a plot of the
lifetime residuals versus stress, showing one
column of residuals for each of the five stress levels.
The horizontal lines are estimates of the 0.05, 0.50, and 0.95
quantiles of the distribution of the
standardized fatigue-life residuals on the antilog scale.
The residuals for stress levels between 280 and 380
have similar distributions,
indicating that there is no evidence of model inadequacy.
There were eight runouts at $S=270$ MPa
and this is the reason that column is so short.
\end{example}

\subsection{Tolerance Bounds Versus Credible/Confidence Intervals
  for Quantiles} After an \SN{} model has been chosen and fit to the
available data, the results are used by engineers in different
ways. For many applications, estimates and confidence intervals for
lower-tail quantiles of the fatigue-life distribution (at given
$S_{e}$) and/or the fatigue-strength distribution (at given
$N_{e}$) are of particular interest.

In some parts of the \SN{} data modeling literature, there is
discussion of lower one-sided tolerance bounds of the fatigue-life
and/or the fatigue-strength distributions, which are sometimes
called one-sided tolerance intervals
(intervals, by definition, have two endpoints and
having an infinite endpoint does not help
explanation or understanding).
As described in \citet[][Section~2.4.2]{MeekerHahnEscobar2017}, a
one-sided lower $100(1 - \alpha)$\% confidence bound on the $p$
quantile of a distribution is equivalent to a one-sided lower
tolerance bound that one can claim, with $100(1 - \alpha)$\%
confidence, is exceeded by at least a proportion $1 - p$ of that
distribution. In our pedagogical experience, engineers and other
practitioners often confuse the $100(1 - \alpha)$\% confidence level
with the $1 - p$ exceedance probability of a tolerance bound but
that the concept of a small-$p$ lower-tail quantile is easier to
separate from the confidence level.

It is common to report (and most statistical
software packages only provide)
two-sided confidence intervals for specified quantiles.  Note that
the lower endpoint of a two-sided $100(1 - \alpha)$\% confidence
interval can be interpreted as a one-sided lower $100(1 -
\alpha/2)$\% confidence bound (e.g., the endpoints of a two-sided
90\% confidence interval are one-sided 95\% confidence bounds). A
two-sided interval on a quantile, relative to a one-sided bound,
provides more information.  For a quantile in the lower tail
of a fatigue-life or a fatigue-strength distribution, the lower bound
tells how bad things might be; the upper bound tells
\textit{how good things might be}.

The relationship between one-sided confidence (credible) bounds and
two-sided intervals presumes that the procedures provide, at least
approximately, equal error probabilities for each tail and this is
the reason that percentile credible intervals are recommended (as
opposed to highest posterior density) and that simulation-based
confidence intervals \citep[e.g., Chapters 13 and 14 in
][]{MeekerHahnEscobar2017} should calibrate each endpoint separately.

\section{Other \SN{} Regression Relationships and Modeling Examples}
\label{section:other.sn.regression.relationships}

Section~\ref{section:statistical.models.estimate.fatigue.life}
introduced three, relatively simple, \SN{} regression
relationships. Many other such relationships have been
suggested. This section, while not exhaustive, describes several
other \SN{} relationships, illustrates how they fit within our
modular framework, and illustrates the use of two of these
relationships with the Ti64 and nitinol data that were introduced in
Section~\ref{section:motivating.examples}.

\subsection{Physical Explanation of the Curvature and Nonconstant Spread
  in \SN{} Data}
\label{section:physical.explanation.curvature.nonconstant.variance}
Figures~\ref{figure:Ti64350FRm1.plots}a
and~\ref{figure:Nitinol02Model.plots}a are examples of \SN{} data
with strong curvature when plotted on log-log axes. Materials will
exhibit this curvature differently depending on the damage
accumulation mechanisms that are activated by cyclic
loading. Curvature in the \SN{} curve demonstrates that the rate of
damage accumulated per cycle has a strong dependence on the
magnitude of the load amplitude.  Curvature will be greatest at load
levels where the material transitions between
micro-mechanical deformation regimes.  A well documented
example of this phenomena is the transition between elasto-plastic and purely
elastic deformation in high-strength metallic materials.  The \SN{}
relationship tends to be approximately linear (on log-log scales)
when cycling is causing cyclic elasto-plastic deformation, but as
one moves to lower stress levels, the deformation becomes
purely elastic leading to much longer life and this results in
the concave-up curvature.

As noted in
Section~\ref{section:induced.fatigue.life.distribution.neither.vertical.nor.horizontal}
(also see Example~\ref{example:basquin.induced.fatigue.strength} and
Figure~\ref{figure:BasquinFatigueLifeStrength}) when the
\SN{} relationship is linear on log-log axes (i.e., the Basquin
relationship) with $\sigma_{X}$ not depending on $N_{e}$,
the induced fatigue-life distribution will belong to the
same family as the fatigue-strength distribution and will have
constant spread. When, however, the \SN{} relationship has the usual
concave-up curvature described in the previous paragraph, the
induced fatigue-life distribution will have increasing spread as
stress decreases. Technical details for this result
are given in 
Section~\ref{S.section:Curvature.sn.relationships.increased.spread.low.stress}. This
behavior will be illustrated in
Examples~\ref{example:modeling.Ti64.sn.data}
and~\ref{example:modeling.superelastic.nitinol.sn.data}
(and corresponding Figures~\ref{figure:Ti64.model.fit.plots}
and~\ref{figure:Nitinol02.model.fit.plots}).

\subsection{The Modified Bastenaire \SN{} Relationship}
\label{section:modified.bastenaire}
The original \citet{Bastenaire1972} relationship is
\begin{align*}
N=g(S; \betavec) = \frac{A \exp[-C(S-E)]}{S-E}, \quad S>E.
\end{align*}
As illustrated in Figure~\ref{figure:sn.relationships}b, this
model has been modified
\citep[e.g., in][]{ISO12107:2012,HautevilleHermiteLefebvre2022}  to have more flexibility by adding a fourth
parameter giving
\begin{align*}
N=g(S; \betavec) = \dfrac{A \exp \left( -\left[ \dfrac{S-E}{B} \right]^{C} \right)}{S-E}, \quad S>E.
\end{align*}

\subsection{The Nishijima \SN{} Hyperbola Relationship}
\label{section:nishijima.sn.relationships}
The Nishijima \SN{} relationship
\citep{Nishijima1980,Nishijima1985}, illustrated in
Figure~\ref{figure:sn.relationships}c, is
\begin{align*}
  \left [
  \log(S)-E\right] \left[ \log(S)+A\log(N)-B\right]=C.
\end{align*}
where the regression parameters are $\betavec=(A, B, C, E).$
The parameters $A$ and $B$
are, respectively, the negative of the slope and the $\log(N)=0$
intercept of the large-$S$ oblique asymptote (sometimes called the
plastic-zone asymptote); $E$ is the horizontal
asymptote; $\sqrt{C}$ is the vertical distance between
the \SN{} curve and the point where the two asymptotes intersect (all on the
log-log scales of the Figure~\ref{figure:sn.relationships}c plot).
For purposes of specifying a fatigue-strength model
 that can be used to induce a fatigue-life model, the
 relationship can be expressed as
\begin{align}
\label{equation:nishijima.hyperbolic.relationship.hfunction}
    S&=h(N; \betavec) = \exp\left(\frac{
-A \log(N)+B+E + \sqrt{
  \left[A \log(N)-(B-E)\right]^{2} +4C}}{2}\right).
\end{align}

\begin{example}
\label{example:modeling.Ti64.sn.data}
\Exampletitle{Fitting the Nishijima/Lognormal Model to the
  Ti64 \SN{} Data.}
This example is a continuation of Example~\ref{example:Ti64.data}.
A description of the  noninformative/weakly informative
joint prior distribution that was used, additional residual plots,
and other details are in 
Section~\ref{S.section:more.details.Ti64.nishijima}.
\begin{figure}[ht]
\begin{tabular}{cc}
(a) & (b) \\[-3.2ex]
\rsplidapdffiguresize{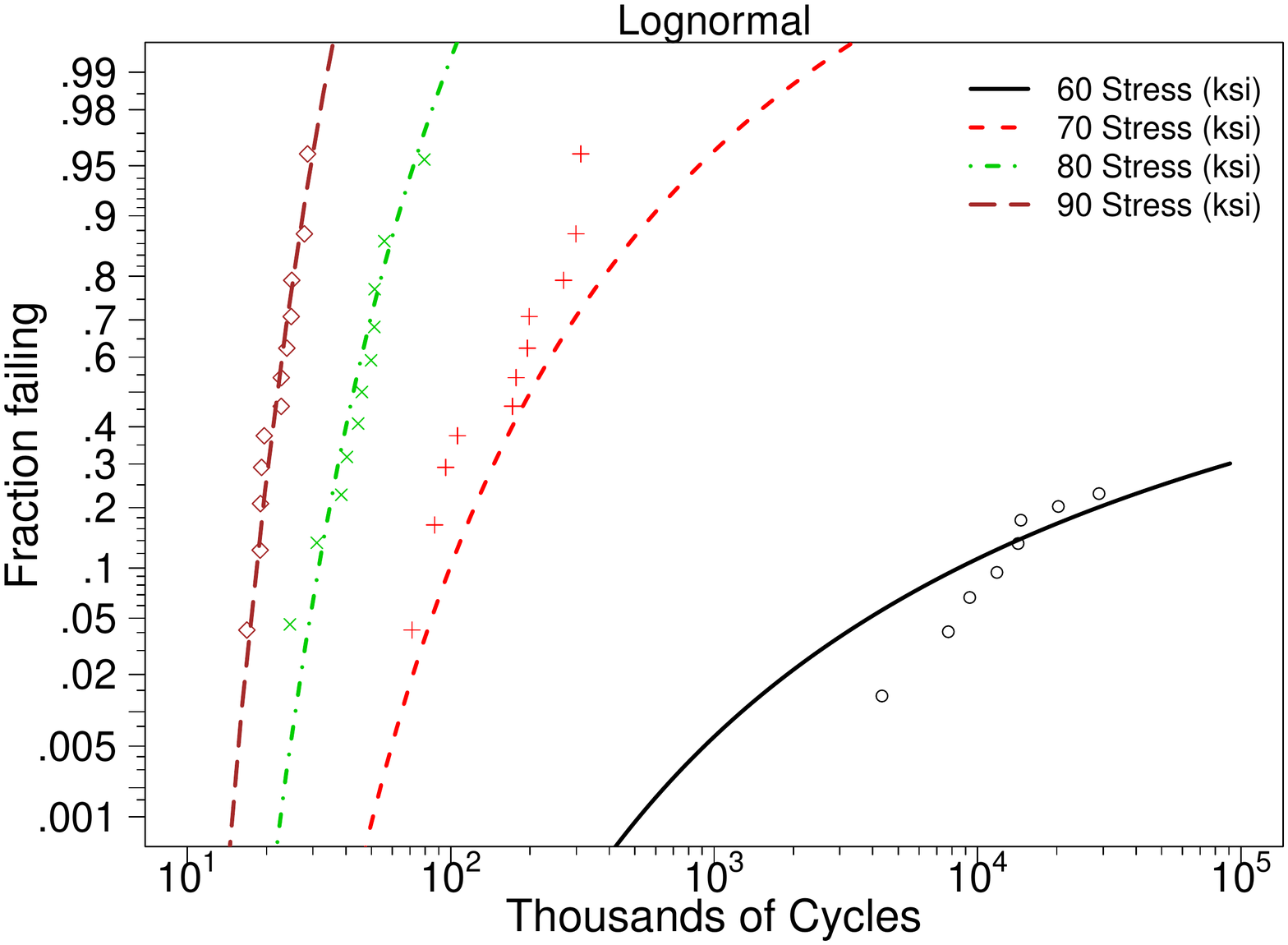}{3.25in} &
\rsplidapdffiguresize{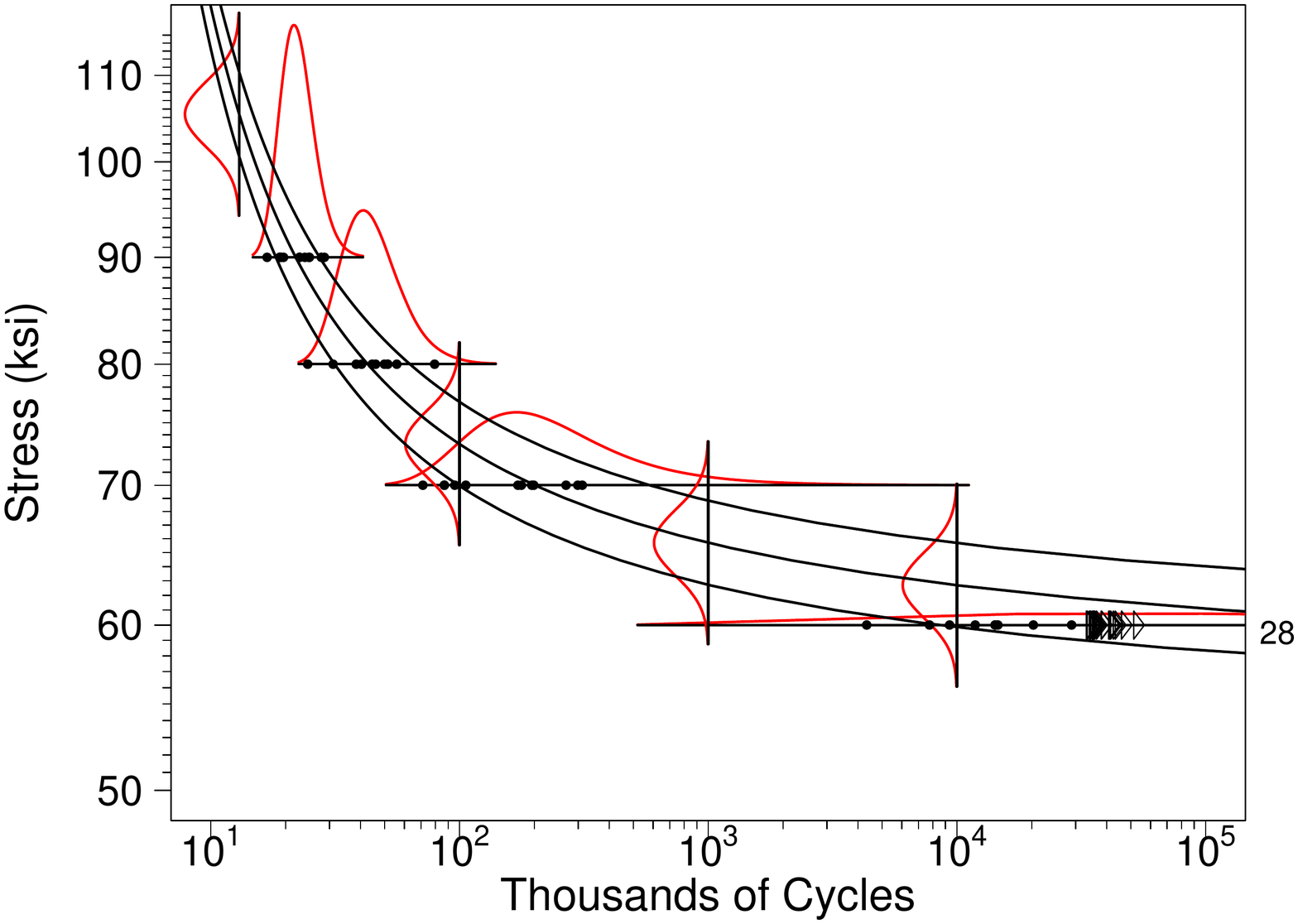}{3.25in}
\end{tabular}
\caption{Lognormal probability plot showing the fatigue-life cdf estimates from
  the Nishijima model fit to the Ti64 \SN{} Data~(a) and the
  corresponding model plot showing 0.10, 0.50, and 0.90 quantile
  curves and densities for fatigue
  strength (vertical) and fatigue life (horizontal)~(b).}
\label{figure:Ti64.model.fit.plots}
\end{figure}
The Nishijima \SN{} relationship
(\ref{equation:nishijima.hyperbolic.relationship.hfunction}) was fit
to the data under the assumption that fatigue strength has a
lognormal distribution with a constant shape parameter $\sigma_{X}$.
The induced fatigue-life model
(Section~\ref{section:induced.fatigue-life.distribution.horizontal.asymptote})
was used to define the log-likelihood in
(\ref{equation:location.scale.likelihood}).
Figure~\ref{figure:Ti64.model.fit.plots}a is a lognormal
probability plot showing, as symbols, the nonparametric estimate of
fraction failing as a function of cycles and the corresponding
regression-model estimates. The agreement is good. The early
failures at 60 ksi deviate from the regression-model estimate but
given the large amount of variability in small order statistics,
this kind of deviation is consistent with the fitted model. As
described in
Section~\ref{section:induced.fatigue-life.distribution.horizontal.asymptote},
the induced fatigue-life cdf
(\ref{equation:quantiles.induced.fatigue.life}) will level off to
$\Phi_{\norm}([\log(S_{e})-\Ehat]/\sigmahat_{X})$ for large values of
$t$. The marginal posterior distribution of this probability at 60
ksi has a median of 0.9754 and results in a 95\% credible interval
 $[0.708,\intervspace 0.9997]$.
Figure~\ref{figure:Ti64.model.fit.plots}b shows the fitted model
superimposed on the same data in
Figure~\ref{figure:Ti64350FRm1.plots}a. Note the vertical
fatigue-strength densities with constant $\sigmahat_{X}=0.0362$ and
the horizontal induced fatigue-life densities with increasing
spread at lower stress levels.
\begin{figure}[ht]
\begin{tabular}{cc}
(a) & (b) \\[-3.2ex]
\rsplidapdffiguresize{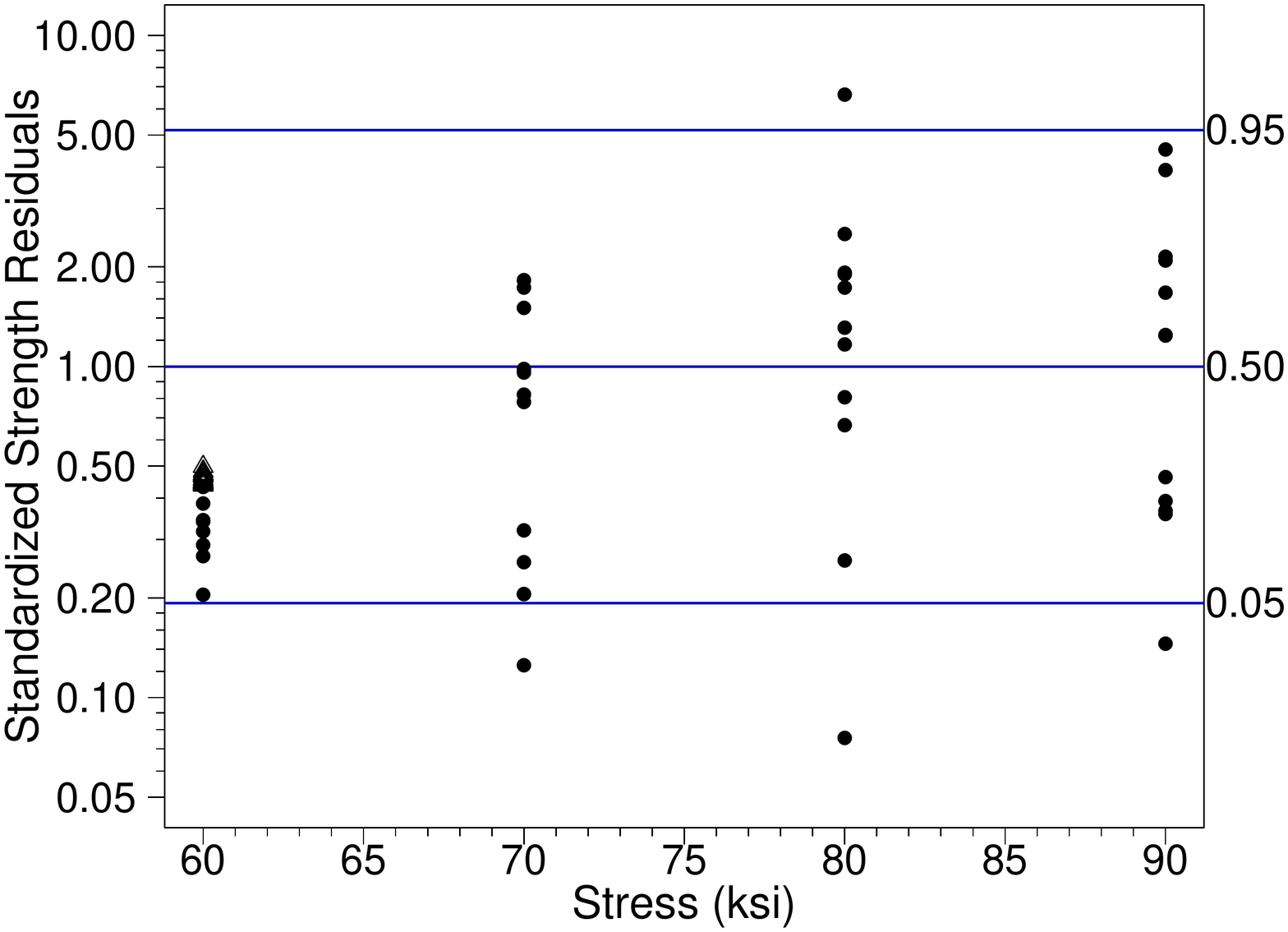}{3.25in} &
\rsplidapdffiguresize{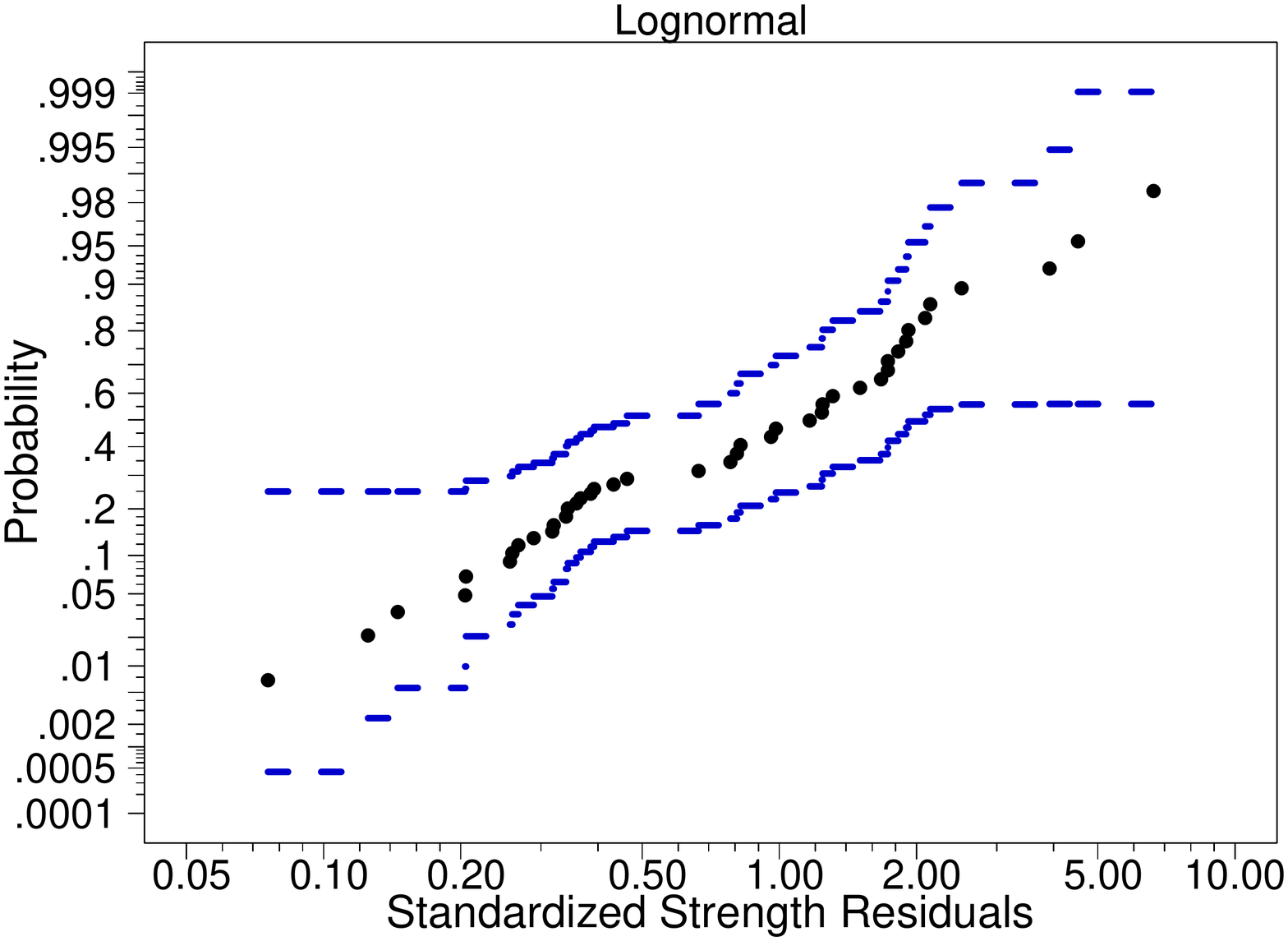}{3.25in}
\end{tabular}
\caption{Fatigue-strength residuals from the Nishijima \SN{}
  model fit to the Ti64 data versus stress~(a) and
  lognormal probability plot~(b).}
\label{figure:Ti64Residual.plots}
\end{figure}
Figure~\ref{figure:Nitinol02Residual.plots}a plots the
standardized residuals of the fatigue-strength distribution computed
from (\ref{equation:standardized.residuals.fatigue.strength}) versus
Stress.
The horizontal lines are estimates of the 0.05, 0.50, and 0.95
quantiles of the distribution of the
standardized fatigue-strength residuals on the antilog scale.
Figure~\ref{figure:Ti64Residual.plots}b is a lognormal probability
plot of the same residuals.
Figure~\ref{figure:Ti64Residual.plots} does not suggest
departures from the assumed model (note that there are 28
right-censored residuals at 60 ksi).
\end{example}

\subsection{The Rectangular Hyperbola \SN{} Relationship}
\label{section:rectangular.hyperbolic.sn.relationships}
The rectangular hyperbola (RH) \SN{} relationship can be written as
\begin{align*}
  \left [\log(N)-B\right]
\left[\log(S)-E\right]=C,
\end{align*}
 where $B$ is a vertical asymptote,
 $E$ is a horizontal asymptote,
 and $C$ controls how fast the \SN{} curve approaches the
 respective asymptotes. All of these parameters are defined on the
 log-log scales that are used in this paper to display \SN{}
 relationships. 
 Figures~\ref{figure:rectangular.hyperbola.regression.model}
 and~\ref{S.figure:sn.models.different.asymptotes.sanity}d
 illustrates this relationship.  For purposes of
 specifying a fatigue-strength model that can be used to induce a
 fatigue-life model and likelihood, the relationship can be
 expressed as
\begin{align*}
  S= h(N; \betavec) = \exp\left[ \frac{C}{\log(N)-B} + E \right].
\end{align*}
The RH model is a limiting case of the Nishijima model that arises
as the plastic-zone slope approaches being vertical, as described in
\citet[][Section~5.6]{LiuMeeker2024}.
Model features that arise from the asymptotes depend on whether the
fatigue-life model is specified (see
Sections~\ref{section:induced.fatigue.strength.distribution.horizontal.asymptote}
and~\ref{section:induced.fatigue.strength.distribution.vertical.asymptote})
or the fatigue-strength model is specified (see
Sections~\ref{section:induced.fatigue-life.distribution.horizontal.asymptote}
and~\ref{section:induced.fatigue-life.distribution.vertical.asymptote}).

\subsection{The Coffin--Manson Relationship}
\label{section:coffin-manson.relationship}
The Coffin--Manson relationship \citep[e.g., pages~748--754
  in][]{Dowling2013}
(also known as the generalized strain-life relationship)
is widely used to model fatigue-life data in
strain-controlled experiments (but can also be used to describe \SN{}
data from stress-controlled experiments).  For this model,
fatigue life $N$ and applied stress $S$ are related
through the relationship
\begin{align}
\label{equation:coffin.manson.basic}
S&=h(N; \betavec) = \Ael (2N)^{b} +  \Apl (2N)^{c}.
\end{align}
This relationship is illustrated in
Figure~\ref{figure:sn.relationships}d.  The terms $\Ael
(2N)^{b}$ and $\Apl (2N)^{c}$ represent separate Basquin relationships for
the elastic and the plastic regimes. Here $\Ael$, $\Apl$,
$b$, and $c$ are material-property parameters to be estimated from
\SN{} data. In particular, $\Ael$ and $\Apl$ are the intercepts of
the lines $\Ael (2N)^{b}$ and $\Apl (2N)^{c}$, respectively, when they are
plotted on log-log axes; $b$ and $c$ are the corresponding
slopes (note that the intercepts are defined as the
value of stress at one half of a cycle when the response is in units
of cycles). The sum of these lines provides a relationship with
concave-up curvature commonly seen in \SN{} data plotted on log-log
scales.

\begin{example}
\label{example:modeling.superelastic.nitinol.sn.data}
\Exampletitle{Fitting the Coffin--Manson/Lognormal Model to the
  Superelastic Nitinol \SN{} Data.}
This example is a continuation of
Example~\ref{example:Nitinol02.data}. A description of the
noninformative joint prior distribution that was
used, additional residual plots, and other details are in the 
Section~\ref{S.section:more.details.nitinol.coffin.manson}.
\begin{figure}[ht]
\begin{tabular}{cc}
(a) & (b) \\[-3.2ex]
\rsplidapdffiguresize{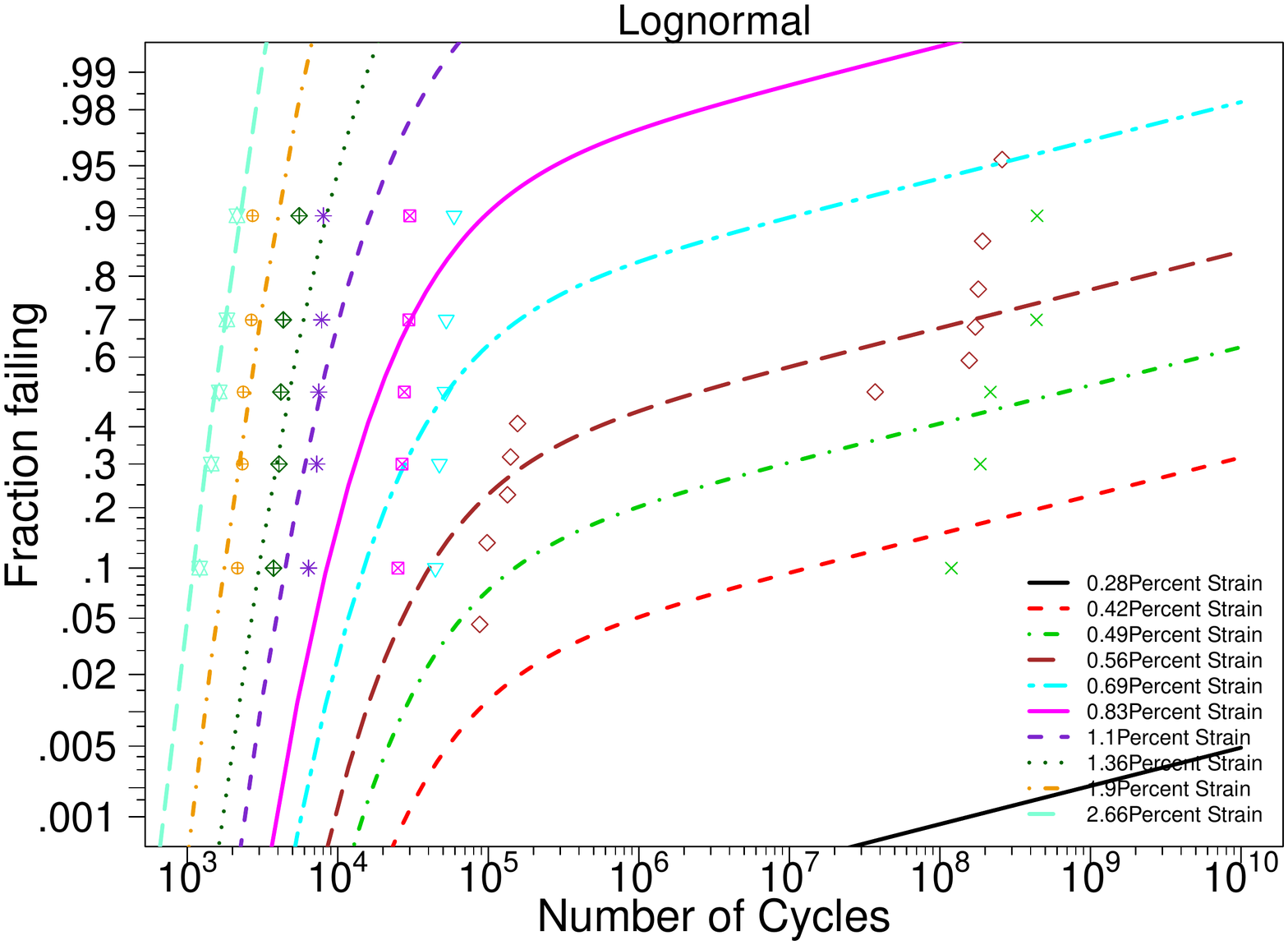}{3.25in} &
\rsplidapdffiguresize{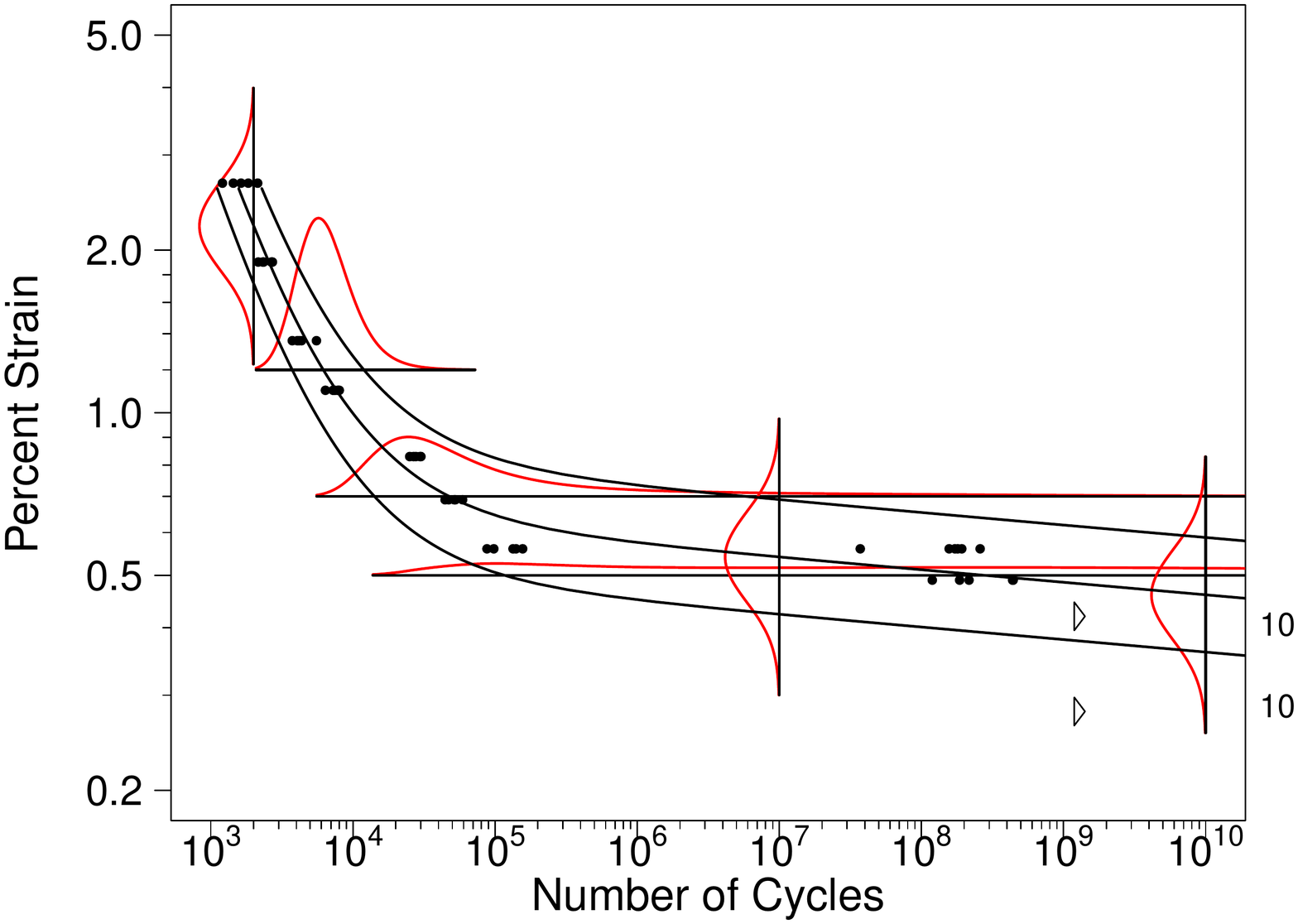}{3.25in}
\end{tabular}
\caption{Lognormal probability plot showing the cdf estimates from
  the Coffin--Manson model fit to the nitinol \SN{} Data~(a) and
  the corresponding model plot showing 0.10, 0.50, and 0.90 quantile curves and
  densities for fatigue life (horizontal) and fatigue strength (vertical)~(b).}
\label{figure:Nitinol02.model.fit.plots}
\end{figure}
The model fitting and likelihood construction and resulting plots
are similar to those described in
Example~\ref{example:modeling.Ti64.sn.data}, except that the
Coffin--Manson \SN{} relationship $S=h(N; \betavec)$ is defined by
(\ref{equation:coffin.manson.basic}) and, because there is neither a
horizontal nor a vertical asymptote, the induced fatigue-life
model is given by
(\ref{equation:cdf.for.induced.fatigue.life}) in
Section~\ref{section:induced.fatigue.life.distribution.neither.vertical.nor.horizontal}.
Figure~\ref{figure:Nitinol02.model.fit.plots}a is a lognormal
probability plot similar to
Figure~\ref{figure:Nitinol02Model.plots}b but with the
Coffin--Manson/lognormal regression model cdf estimates plotted for
the eight levels of strain. Interestingly, the upper tails
of the plotted cdfs are linear, implying that the upper tail of the
distributions behave like a lognormal distribution, in contrast to
the horizontal asymptote in the Nishijima model illustrated in
Figure~\ref{figure:Ti64.model.fit.plots}a.

The bimodality at 0.56\% strain stands out again and is
highly influential, inflating the estimate of spread in the induced
fatigue-life distributions at the lower levels of strain and leading
to lack of fit at the lower levels of strain (i.e., below 0.56\%
strain). \citet{Weaver_etal2022} fit a mixture model to
these nitinol data to accommodate the bimodality. Consideration of
such a model is beyond the scope of this paper but is mentioned as
an area for future research in
Section~\ref{section:concluding.remarks}.

Figure~\ref{figure:Nitinol02.model.fit.plots}b shows the same data
as Figure~\ref{figure:Nitinol02Model.plots}a but now has the 0.10,
0.50, and 0.90 quantile lines and estimated densities superimposed.
Engineers demonstrating the reliability of an artificial heart valve
would typically be interested in the 0.10 quantile of the
fatigue-strength distribution at 600 million cycles (15 years). For
the nitinol data, the marginal posterior draws of the 0.10 quantile of
the lognormal
strength distribution at $N_{e}=600{,}000{,}000$ cycles is
computed using (\ref{equation:quantiles.specified.fatigue.strength})
and these provide the point estimate $0.10223$ and a 95\% credible
interval $[0.0794,\intervspace 0.128]$ in percent strain.

\begin{figure}[ht]
\begin{tabular}{cc}
(a) & (b) \\[-3.2ex]
\rsplidapdffiguresize{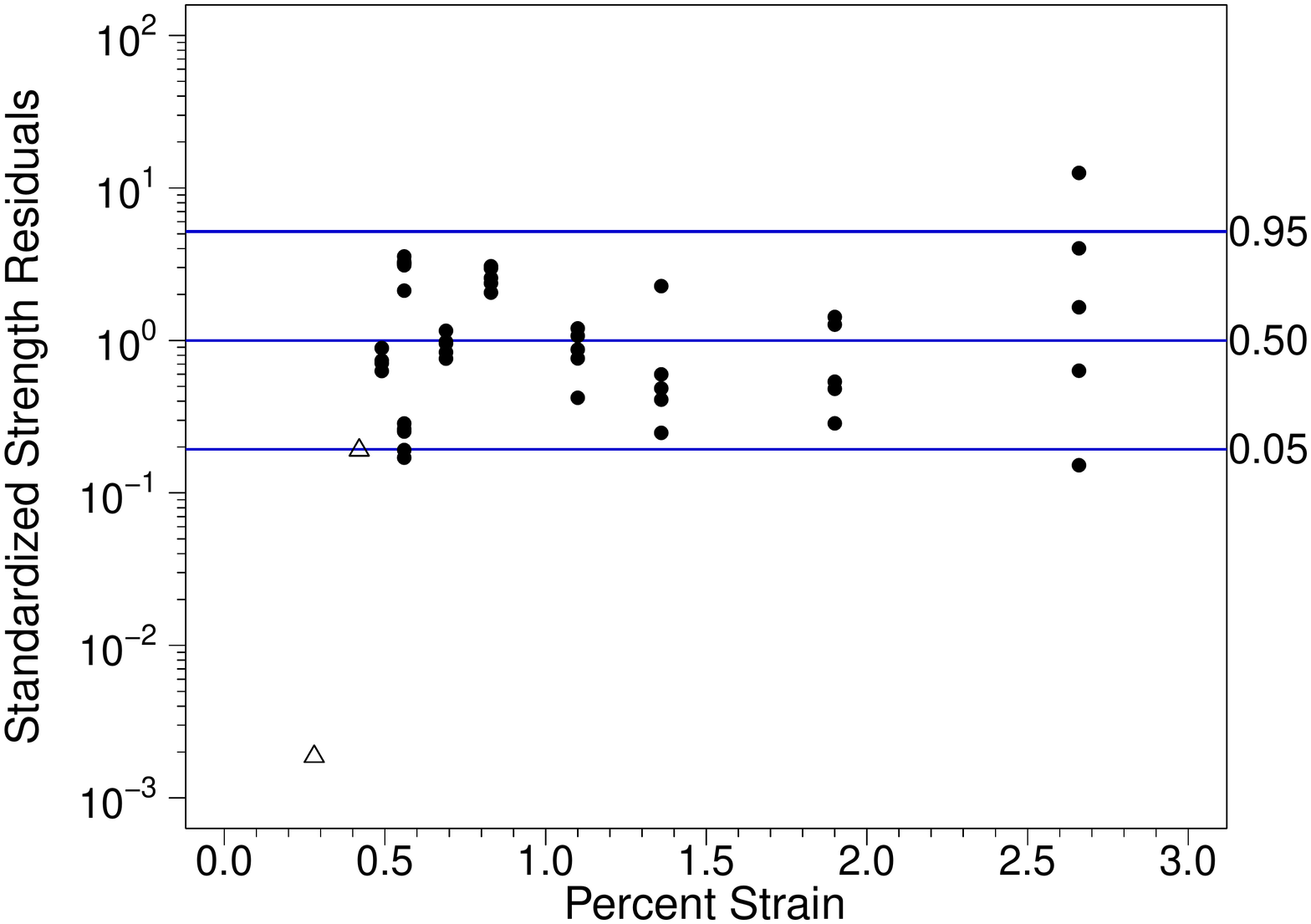}{3.25in} &
\rsplidapdffiguresize{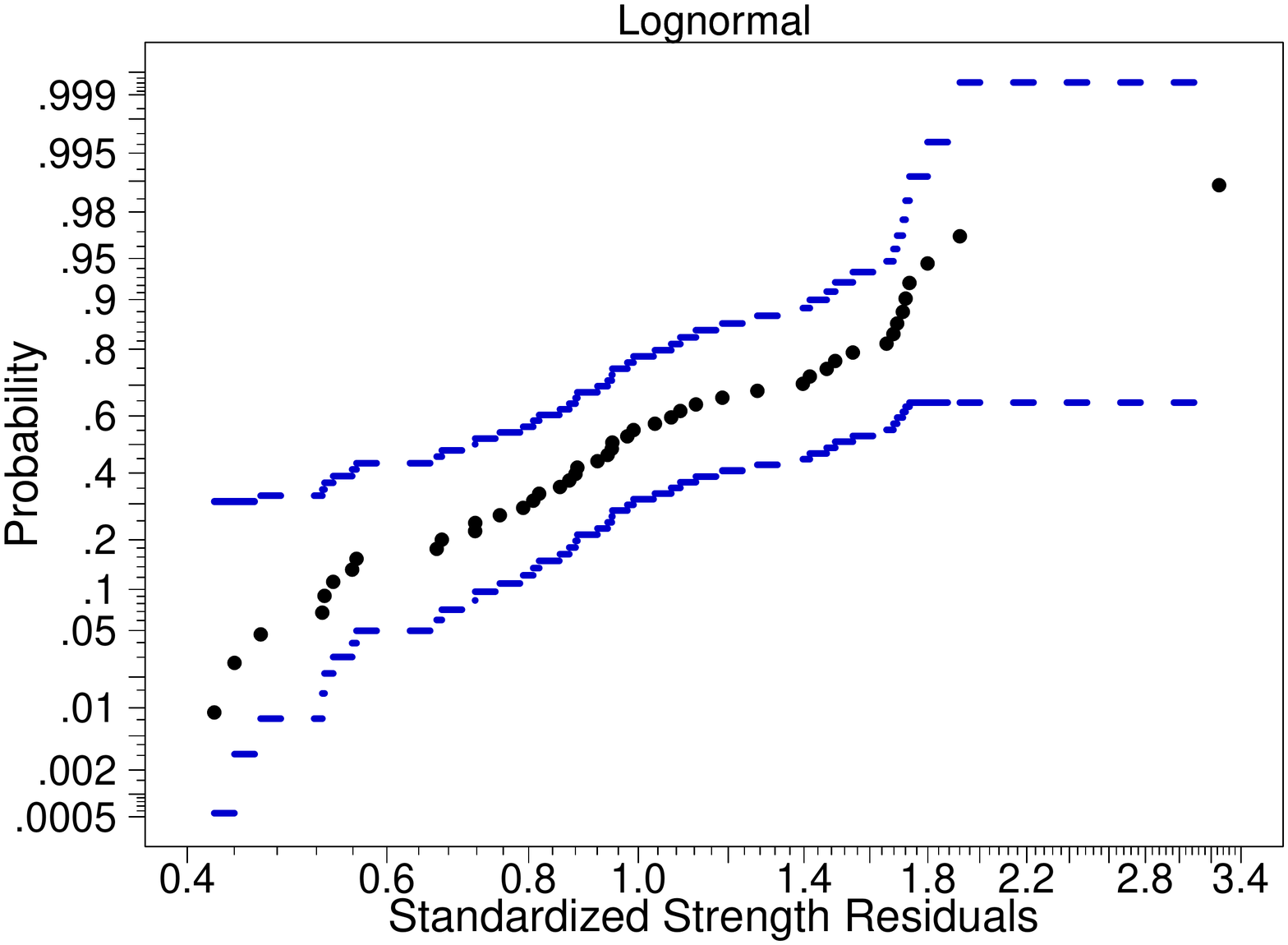}{3.25in}
\end{tabular}
\caption{Fatigue-strength residuals from the Coffin--Manson \SN{}
  model fit to the nitinol data versus strain~(a) and
  lognormal probability plot~(b).}
\label{figure:Nitinol02Residual.plots}
\end{figure}
Figure~\ref{figure:Nitinol02Residual.plots}a plots the
standardized residuals of the fatigue-strength distribution computed
from (\ref{equation:standardized.residuals.fatigue.strength}) versus
\% strain.
The bimodality can be seen in the two clusters of
residuals at 0.56\% strain. Other single clusters can be seen
at 0.49, 0.69, and 0.83\% strain. The small spread within these
clusters suggests, in comparison with the overall spread in the
residuals, that the residuals are not an iid sample. Such
dependence could be due to lack of randomization with respect to factors like
batch, test-machine effects, or the location of specimen wires cut
from the spools. Figure~\ref{figure:Nitinol02Residual.plots}b is a
lognormal probability plot of the same residuals showing that the
lognormal distribution fits well.
\end{example}

\subsection{The Random Fatigue-Limit Model}
\label{section:RFL.model}
\citet{PascualMeeker1999} extended the Stromeyer model
(Section~\ref{section:stromeyer.model}) by allowing the
fatigue-limit~$\gamma$ to vary from unit to unit.
The Random Fatigue-Limit (RFL) model describes both the
curvature and the increased variability at
lower stress levels when plotting \SN{} data on log-log scales.

\subsubsection{The RFL fatigue-life model}
For stress $S_{e}$ conditional on a fixed value of $\gamma>0$,
\begin{align*}
F_{N|\gamma}(t; S_{e}| \gamma)&=\Pr \left(N \le t; S_{e}| \gamma\right)=\Phi
         \left (\frac{\log(t)-\mu(S_{e},\gamma)}{\sigma_{\epsilon}}
         \right), \quad
  t>0, \, S_{e}>0,
\end{align*}
where $\Phi$ is the standard location-scale distribution cdf
corresponding to the conditional log-location-scale distribution for
$N$ (i.e., $N|\gamma$) and $\mu(S_{e}, \gamma)=\beta_{0}+ \beta_{1}\log(S_{e}-\gamma)$.
Then the unconditional distribution of $N$ is obtained by averaging
over the distribution of~$\log(\gamma)$
\begin{align}
\label{equation:rfl.fatigue.life.cdf}
F_{N}(t; S_{e})&=\Pr \left(N \le t; S_{e}\right)=\int_{-\infty}^{\log(S_{e})}\frac{1}{\sigma_{\log(\gamma)}}\Phi
         \left (\frac{\log(t)-\mu(S_{e}, \nu)}{\sigma_{\epsilon}}
         \right) \phi_{\gamma}\left(\frac{\nu-\mu_{\log(\gamma)}}{\sigma_{\log(\gamma)}}\right)d\nu, \quad \quad
  t>0, \,\, S_{e}>0,
\end{align}
where $\phi_{\gamma}$ is the standard location-scale distribution pdf
corresponding to the log-location-scale distribution of $\gamma$,
and the parameters of the model are $\thetavec=(\beta_{0},
\beta_{1}, \sigma_{\epsilon},
\mu_{\log(\gamma)},\sigma_{\log(\gamma)})$. \citet{PascualMeeker1999}
illustrated the fitting of the RFL model for several data sets using
all combinations of Weibull and lognormal distributions for
$N|\gamma$ and $\gamma$.

\subsubsection{The RFL fatigue-strength model}
As with other fatigue-life models for \SN{} data, the RFL model
can be used to define a distribution of fatigue strength $X$ for a
given value of $N_{e}$.
Similar to what was done in
Sections~\ref{section:relationship.fatigue.life.fatigue.strength}
and~\ref{section:alternative.approach.modeling.sn.data}, replacing $t$
with $N_{e}$ and $S_{e}$ with $x$ in the integral of
(\ref{equation:rfl.fatigue.life.cdf})  gives
\begin{align}
\label{equation:rfl.fatigue.strength.cdf}
F_{X}(x; N_{e})&=\Pr \left(X \le x; N_{e}\right)=\int_{-\infty}^{\log(x)}\frac{1}{\sigma_{\log(\gamma)}}\Phi
         \left (\frac{\log(N_{e})-\mu(x, \nu)}{\sigma_{\epsilon}}
         \right) \phi_{\gamma}\left(\frac{\nu-\mu_{\log(\gamma)}}{\sigma_{\log(\gamma)}}\right)d\nu.
\end{align}
Interestingly, as $N_{e} \rightarrow \infty$ in
(\ref{equation:rfl.fatigue.strength.cdf}), the cdf in the integrand
approaches one and the cdf of fatigue strength $X$ approaches the
cdf of the random fatigue-limit~$\gamma$.
There are no closed-form expressions for the quantiles of the RFL
model fatigue-life or fatigue-strength distributions but they can be readily computed by
numerically inverting the cdfs.

\subsection{The Castillo et al. \SN{} Model}
\label{section:Castillo.model}

Castillo et al. (e.g., in \citealp{Castillo_etal1985},
\citealp{CastilloGalambos1987},
\citealp{Castillo_et_al2008},
\citealp{CastilloFernandez-Canteli2009}, and Equation~(2) of
\citealp{Castillo_et_al2019}) suggest an \SN{} model based on the
rectangular hyperbola \SN{} relationship
and a three-parameter Weibull distribution given by
\begin{align}
F(t,x) = 1-\exp \left\{ - \left[ \frac{[\log(t) - B][\log(x) - E] - \gamma}{\eta} \right]^\beta \right\}
\label{equation:logt.logx.castillo.model}
\end{align}
with parameters $\thetavec=(B, E, \gamma, \eta, \beta)$ where $B$ is
a vertical asymptote for log fatigue life (i.e., minimum value for
$\log(N)$), $E$ is a horizontal asymptote for log fatigue strength
(i.e., a fatigue-limit and minimum value for $\log(X)$), and $\gamma$,
$\eta$, and $\beta$ are related to the Weibull distribution
parameters.
Their model derives from a compatibility condition
implying that the fatigue-life and the fatigue-strength
quantile curves coincide, as described for our (different) models in
Section~\ref{section:equivalence.fatigue.life.fatigue.strength.quantile.curves}.

Replacing $x$ with $S_e$,
(\ref{equation:logt.logx.castillo.model}) can be interpreted as the
cdf for fatigue life $N$ at a given level of stress $S_{e}$. That is,
\begin{align*}
F_{N}(t; S_{e}) & = \Pr(N \le t; S_{e}) = F(t,S_{e})\\
& = 1-\exp \left\{ - \left[ \frac{[\log(t) - B][\log(S_{e}) - E] -
    \gamma}{\eta} \right]^\beta \right\},
\end{align*}
where $t > \exp(B + \gamma/[ \log(S_{e}) - E ])$ and $S_{e} > \exp(E)$.
Similarly, replacing $t$ with $N_{e}$,
(\ref{equation:logt.logx.castillo.model}) can be interpreted as the
cdf for fatigue strength $X$ at a given number of  cycles
$N_{e}$. That is,
\begin{align*}
F_{X}(x; N_{e}) & = \Pr(X \le x; N_{e}) = F(N_{e},x) \nonumber \\
& = 1-\exp \left\{ - \left[ \frac{[\log(N_{e}) - B][\log(x) - E] - \gamma}{\eta} \right]^\beta \right\},
\end{align*}
where $x > \exp(E + \gamma/[ \log(N_{e}) - B ])$ and $N_{e} > \exp(B)$.
Expressions for the Weibull parameters, quantile functions for $N$
and $X$, and a plot of the quantile curves are given in 
Section~\ref{S.section:generalization.castillo.sn.model}.

\subsection{A Comparison and Operational Considerations for Choosing
  an \SN{} Model}
\label{section:considerations.choosing.sn.model}
Table~\ref{table:summary.models.sn.data} provides a summary of several
\SN{} models that fit within our modular framework and that we have
either used in our examples or that are commonly used in the fatigue
literature. The table is not meant to be an exhaustive list, but a
sample of the alternative models that are available. Broadly,
there are two categories of models, depending on whether the
fatigue-life or the fatigue-strength model is
specified. Technically, any suitable \SN{} relationship (such as
those given in
Sections~\ref{section:relationship.fatigue.life.fatigue.strength},~\ref{section:simple.sn.relationships},
and~\ref{section:modified.bastenaire}--\ref{section:coffin-manson.relationship})
could be used by specifying either the fatigue-life or the
fatigue-strength model (and having the other be induced). The
specification of Life or Strength given in the third column of
Table~\ref{table:summary.models.sn.data} corresponds to the
specification that we expect would be most useful given the
properties of the resulting \SN{} models and our experience with
analyzing various \SN{} data sets.
\begin{table}
\caption{Summary of Selected Models for \SN{} Data}
\label{table:summary.models.sn.data}
\begin{tabular}{lcllll}
\toprule
&&Model&\multicolumn{1}{c}{Vertical}& \multicolumn{1}{c}{Horizontal}&
Curvature and \\
   &  \multicolumn{1}{c}{\#} &Specified for &Asymptote  &  Asymptote  &  Nonconstant \\
Model            &  Params & Fatigue & for Large $S$  & for Large $N$ & Spread \\
\midrule
Basquin  (inverse-power) & 2+1=3          &  Life & No         &  No            & No  \\[1ex]
Box--Cox/loglinear-$\sigma_{N}$ & 3+2=5     & Life & Yes        &  No          &Yes\\[1ex]
Stromeyer/loglinear-$\sigma_{N}$ & 3+2=5     &  Life & No        &  Yes     &Yes\\[1ex]
Box--Cox & 3+1=4     & Strength & Yes        &  No          &Yes\\[1ex]
Stromeyer & 3+1=4     &  Strength & No        &  Yes     &Yes\\[1ex]
Nishijima  & 4+1=5 &  Strength & No   &  Yes     & Yes \\[1ex]
Coffin--Manson & 4+1=5 & Strength &       No       &  No            & Yes \\[1ex]
Bastenaire & 3+1=4      &  Strength &No       &  Yes  & Yes \\[1ex]
Modified Bastenaire & 4+1=5      &  Strength &No       &  Yes  & Yes \\[1ex]
Rectangular hyperbola & 3+1=4 &  Strength & Yes   &  Yes     & Yes \\[1ex]
Castillo et al. & 3+2=5 &  Both & Yes  &  Yes     & Yes \\[1ex]
Random fatigue-limit & 3+2=5  & Life &  No  &  Yes          & Yes \\[1ex]
\bottomrule
\end{tabular}
\begin{minipage}[t]{0.9\linewidth}
The number of parameters in the \# Params column is the sum of the number of
parameters in the \SN{} relationship and those that describe variability
 in the statistical model.
\end{minipage}
\end{table}

As part of our research,  beyond the three examples presented in the
paper, we fit the Basquin,
Box--Cox/loglinear-$\sigma_{N}$, Nishijima, Coffin--Manson, and RFL
models to 18 different \SN{} data sets (and other models to a smaller
number of data sets) covering a wide range of materials,
specimen types, and sample sizes.
Based on those experiences and our knowledge of the nature of
the different models, the remainder of this section provides
recommendations on how to
choose which model or models to use in a particular situation
(as in most statistical modeling applications,
it is generally important to fit and compare different models).

Key features of the various models are the existence (or not) of coordinate
(i.e., vertical or horizontal) asymptotes in the \SN{} relationship
and the way that variability (including changes in variability as a
function of stress) is described. 
Figure~\ref{S.figure:sn.models.different.asymptotes.sanity}
illustrates the fitting of four \SN{} models,
with different combinations of the existence of
asymptotes or
not, to a version of the nitinol data. Figure~\ref{S.figure:sn.models.different.asymptotes.residuals}
provides corresponding plots of the residuals versus strain.
Plots like these, for the many data sets,
helped inform the following discussion.

\subsubsection{Models with no coordinate asymptotes }
Because of its simplicity, the Basquin (inverse-power rule)
statistical model (Section~\ref{section:Basquin.model} and
Figure~\ref{figure:BasquinFatigueLifeStrength}) is the most
common model fit to \SN{} data and it is appropriate when testing is done
at relatively high stress levels (plastic range) where the
relationship between log-life and log-stress is approximately linear
with constant spread.  Such tests are frequently conducted to
compare fatigue-life distributions for factors such as different treatments, test
conditions like temperature or frequency, formulations of
product materials, or different mechanical designs. When units are
tested at high stress levels to estimate fatigue-life at lower
stress levels (accelerated testing), the Basquin model will
provide conservative estimates of low-stress fatigue-life quantiles,
relative to models that describe the concave-up curvature typically
seen at low stress levels
\citep[e.g., Examples~19.11--19.14 in][]{MeekerEscobarPascual2021}.

The Coffin--Manson relationship
(Section~\ref{section:coffin-manson.relationship} and
Figure~\ref{figure:sn.relationships}c) is appropriate
when there is curvature in the \SN{} data plotted on log-log axes
but no evidence for the
existence of a fatigue limit. When  used with a
specified fatigue-strength distribution with constant $\sigma_{X}$
(as suggested in \citet{Falk2019} and as we recommend),
the model describes the increase in
spread at lower stress levels.

\subsubsection{Models with a horizontal asymptote}
The Stromeyer
(Section~\ref{section:stromeyer.model}), Bastenaire
(Section~\ref{section:modified.bastenaire} and
Figure~\ref{figure:sn.relationships}b), Nishijima
(Section~\ref{section:nishijima.sn.relationships} and
Figure~\ref{figure:sn.relationships}d), and the Random fatigue-limit
(Section~\ref{section:RFL.model}) \SN{} models all have a horizontal
asymptote that suggests the possible existence of a fatigue limit. A
fatigue limit does not have to exist to use these models, as long as
the model fits well and there is no extrapolation in stress. In
such cases, the model provides valid inferences for lower-tail
quantiles of the fatigue-life and fatigue-strength distributions.
For \SN{} data with the common concave-up shape when plotted on
log-log axes, we found the properties
of the induced fatigue-life model (for a specified
fatigue-strength model with constant $\sigma_{X}$) have
better agreement with physical reality (also see
Section~\ref{section:induced.fatigue-life.distribution.horizontal.asymptote})
when compared to a specified fatigue-life model.

\subsubsection{Models with a vertical asymptote}
The Box--Cox model
(Section~\ref{section:box.cox.relationship} and
Figure~\ref{figure:sn.relationships}a) has a vertical asymptote.
The rectangular hyperbola model
(Section~\ref{section:rectangular.hyperbolic.sn.relationships} and 
Figure~\ref{S.figure:sn.models.different.asymptotes.sanity}d) and the
Castillo et al. model (Section~\ref{section:Castillo.model} and 
Figure~\ref{S.figure:CastilloFatigueLifeQuantiles}) have both
vertical and horizontal asymptotes. The vertical asymptote
is related to some interesting features of these models.
First, the asymptote is related to the smallest number of cycles where a
failure could occur, even as stress amplitude approaches
infinity. Second \citep[as noted, for example, in Section~4
  of][]{ToasaUmmenhofer2018},
the shape of the \SN{} relationship does not agree
with the most commonly seen behavior of \SN{} data at higher stress levels.
Finally, as shown in 
Figure~\ref{S.figure:CastilloFatigueLifeQuantiles}, for small values
of $N_{e}$, the spread in the induced distribution of fatigue strength $X$
can increase dramatically. As described in
Section~\ref{section:box.cox.relationship}, these issues are not of
concern if this asymptotic behavior occurs outside the range where
the model would be used (e.g., Figure~\ref{figure:LaminatePanelModel.plots}b).

\section{Concluding Remarks and Areas for Future Research}
\label{section:concluding.remarks}
This paper outlines a modular framework for specifying,
fitting, checking statistical models for \SN{} fatigue data. The
framework includes most of the \SN{}
relationships previously suggested in the fatigue
literature.  We illustrated the use of flexible Bayesian methods with
noninformative or weakly informative prior distributions to estimate
fatigue-life and fatigue-strength models. We
illustrated the methods using \SN{} data from three different
materials and specimen types and described our
experiences with many other data sets
and types of materials.

When modeling \SN{} data, how should one choose whether to specify
the fatigue-life model (resulting in an induced fatigue-strength
model) or specify a fatigue-strength model (resulting in an induced
fatigue-life model)? When there is curvature in the \SN{}
relationship (which is common in HCF applications),
given the manner in which it naturally
describes increasing spread at lower stress levels (as explained in
Section~\ref{section:physical.explanation.curvature.nonconstant.variance}),
and other reasons given in
Section~\ref{section:important.advantages.of.specifying.fatigue.strength.distribution},
we strongly favor the approach that specifies the fatigue-strength
model (Examples~\ref{example:modeling.Ti64.sn.data}
and~\ref{example:modeling.superelastic.nitinol.sn.data}). What
reasons are there to continue to use the approach that specifies the
fatigue-life model
(Example~\ref{example:box-cox.loglin.laminate.panel.data})? It is
traditional, widely known, and software is readily available. We see
no other advantages.

The following are areas where further research is needed.
\begin{itemize}
\item
There is a need to develop practical methods for
designing statistically efficient experiments to obtain \SN{}/\eN{} data
(how many and which levels of stress, number of specimens, and how
to allocate them to stress levels). Although existing results for
planning accelerated life tests \citep[e.g., Chapter 6
  in][]{Nelson1990a} may provide insight, there are important
differences. Often there is no need to extrapolate in
stress (although there may be extrapolation into the lower
tails of both the fatigue-life and the fatigue-strength
distributions). Depending on the application, inferences are
generally needed for fatigue life over a range of stress values or
fatigue-strength quantiles at particular points in time. Tools to
quantify estimation precision for these quantities for proposed
experimental designs are needed. \citet{King_etal2016} describe such
work for estimating fatigue-life distributions for a particular
fatigue-life model. Their methods could be extended to focus on
fatigue-strength distributions and other models.
\item
Our modeling has focused on experiments in which stress
\textit{amplitude} is the experimental variable. Mean stress (or
equivalently, the min/max stress ratio), temperature, cycling
frequency, and surface condition/treatments are
additional factors that are often studied in fatigue
experiments. For example, \citet{Pascual2003} and \citet{King_etal2016}
illustrate the use of such multiple explanatory variable fatigue
modeling.  The models and methods presented in this paper can be
readily extended to allow for such additional explanatory variables.
\item
In experimental studies, it is important to understand and take
account of important sources of variability. In fatigue testing,
batch-to-batch (also called heat-to-heat or blend-to-blend)
variability can be
importantly large. Traditionally, careful experimenters would test
the same number of specimens from each heat at each stress level.
This equally represents the heats across the
stress levels. \citet{Nelson1984} provides an example and shows
how to assess, graphically, whether there
is heat-to-heat variability. A more quantitative approach would be
needed to assess statistical significance of suspected
batch-to-batch variability and assess whether efforts to reduce
variability are successful. The methods presented in this paper
could be readily extended to model batch-to-batch variability, in a
manner similar to that used in
\citet[][Section~23.4]{MeekerEscobarPascual2021} to describe
batch-to-batch variability in an accelerated life test.
\item
We have seen numerous examples of \SN/\eN{} data where there is a
bimodal distribution of lifetimes (e.g., in the nitinol example
presented here), usually at an intermediate level of stress.
Various explanations have been suggested for this
phenomenon. These include material defects (similar to multiple
failure modes) and batch-to-batch variability. Appropriate models to
describe such behavior need to be developed. \citet{Weaver_etal2022}
give an example of such a model.
\item
There is extensive existing knowledge of material properties. For
example, \citet[][page 751]{Dowling2013} provides a table containing
nominal values for the parameters of the Coffin--Manson model for
different materials.  \citet{MMPDS2021} contains a large amount of
information about materials properties for different alloys that
used in aerospace applications. This kind of information, combined
with general engineering principles, could be
used to help inform prior distributions for estimating the
parameters of \SN{} models.
\item
  Statisticians \citep[e.g.,][]{Koenker2005} have developed quantile
  regression methods that might be useful for the analysis of
  fatigue data. These methods do not require specification of a
  particular failure-time distribution and have been used
  effectively to model regression data with nonconstant
  spread. However, such methods require much larger sample sizes
  than the fully
  parametric modes that are traditionally used in fatigue data modeling.
\end{itemize}

\section*{Acknowledgments}
We would like to thank
Charles Annis,
Necip Doganaksoy,
Woong Kim,
Larry Leemis,
Lu Lu,
Wayne B. Nelson,
Peter Parker, and
an anonymous referee
for providing helpful comments on an earlier version of our paper.
We would also like to thank Professors
Enrique Castillo and Alfonso Fern{\'a}ndez-Canteli
for helping us to understand some aspects of their approach to modeling
fatigue data.
Elena Garc\'{i}a-S\'{a}nchez provided much useful
information and references to us explaining how manufacturers and
regulators interact to assure aircraft safety.

\appendix

\section{Overview of the Materials in Appendices}
The purpose of the appendix is to provide additional technical
details including derivations, additional examples, simulation results, and
other technical details. This appendix is organized as follows.
Section~\ref{S.section:technical.details.fatigue.life.distribution.specified}
provides additional technical details (i.e., beyond what is in the
main paper) for \SN{} regression models where the fatigue-life model
is specified and the fatigue-strength model is induced. Similarly,
Section~\ref{S.section:technical.details.fatigue.strength.distribution.specified}
provides additional technical details for \SN{} regression models
where the fatigue-strength model is specified and the fatigue-life
model is induced.
Section~\ref{S.proofs.technical.results.stated.in.paper} provides
proofs of some of the technical results stated in the main paper.
Section~\ref{S.section:comparison.sn.model.shapes} compares the
different basic \SN{} model shapes and illustrates the importance of
using residual analysis to help compare and choose such an \SN{}
relationship.
Section~\ref{S.section:examples.comparing.lognormal.and.weibull.distributions.fit.sn.data}
compares lognormal and Weibull probability plots for nine \SN{} data
sets based on fatigue tests for nine different materials and
specimen types.
Section~\ref{S.section:generalization.castillo.sn.model} provides,
for the Castillo et al. \SN{} model (described in
Section~\ref{section:Castillo.model}), additional technical details
and
characteristics. Section~\ref{S.section:more.details.numerical.examples}
gives the prior distributions, numerical results and other details
for the three data analysis/modeling examples in the paper.

\section{Technical Details for \SN{} Regression Models Where the
  Fatigue-Life Model is Specified and the Fatigue-Strength
  Model is Induced}
\label{S.section:technical.details.fatigue.life.distribution.specified}
This section outlines additional technical details of
specifying a fatigue-life model  and using it to induce
a fatigue-strength model.

\subsection{Basic  \SN{} Relationships for
  \texorpdfstring{$N=\gfun(S;\betavec)$}{N=G(S)}
  and General Assumptions}
Here we consider  \SN{} relationships of the type
\begin{align}
\label{S.equation:N.generic.life.lifetime.model}
  N&=\gfun(S;\betavec),
\end{align}
where $\betavec$ is a vector of regression model parameters and
$\gfun(x;\betavec)$ satisfies the following general conditions:
\begin{itemize}
\item
$\gfun(x;\betavec)$ is positive; that is $\gfun(x;\betavec)>0$ for
  $0<x<\infty.$
\item
$\gfun(x;\betavec)$ is strictly decreasing in $x.$
\item
$\gfun(x;\betavec)$ is differentiable for all $x.$
\end{itemize}
There are potentially two asymptotes for $\log[\gfun(S; \betavec)]$:
A horizontal asymptote at $E=\log(S)$ and a vertical
asymptote at $B=\log(N)$, as illustrated in
Figure~\ref{S.figure:CastilloFatigueLifeQuantiles}
for the case $\exp(B)=\exp(E)=1$.

\subsection{The Specified Fatigue-Life Model}
\label{S.section:specified.fatigue.life.distribution}
The random variable $N$ is the observed number of cycles for a unit
at stress amplitude $S_{e}$. Based on the $\SN$
relationship~(\ref{S.equation:N.generic.life.lifetime.model}),
taking logs, replacing $S$ with
$S_{e}$ and adding an error term $\epsilon$ gives
\begin{align}
\label{S.equation:general.model.for.N}
  \log(N)&=\log[\gfun(S_{e};\betavec)]+\sigma_{N} \epsilon,
\end{align}
where $\epsilon$ has a location-scale distribution with $\mu=0$ and
$\sigma=1,$ and $\sigma_{N}$ is constant.  Thus the
log-location-scale cdf $F_{N}(t; S_{e})$ for fatigue life $N$ is
\begin{align}
  \label{S.equation:FL.cdf.model.FL.to.FS}
  \begin{aligned}
F_{N}(t;S_{e})&=\Pr\left(N \le t\right)=\Pr \left[\log(N) \le \log(t) \right]\\
&=\Phi\left[\frac{\log(t)-\log[\gfun(S_{e};\betavec)]}{\sigma_{N}}\right],\,\,\, t>0, \,\, S_{e}>0.
  \end{aligned}
  \end{align}
  This cdf has the standard properties of a cdf for a positive
  random variable. In particular, $\lim_{t \downarrow 0}
  F_{N}(t;S_{e})=0$ and $\lim_{t \to \infty} F_{N}(t;S_{e})=1$.

Then the pdf of $N$ is
\begin{align}
\nonumber f_{N}(t; S_{e})&=\frac{d}{d t} F_{N}(t; S_{e})\\
\label{S.equation:FL.pdf.model.FL.to.FS}
 &=\frac{1}{ t \sigma_{N}} \phi\left[
  \frac{\log(t)-\log[\gfun(S_{e};\betavec)]}{\sigma_{N}}\right],\,\,\,
t>0.
\end{align}
The quantiles of $F_{N}(t; S_{e})$ are the solution to $F_{N}(t_{p})=p.$ Using
(\ref{S.equation:FL.cdf.model.FL.to.FS}),
\begin{align}
    \label{S.equation:FL.quantiles.FL.to.FS}
  t_{p}&=\exp \left [\log[\gfun(S_{e};\betavec)]+\Phi^{-1}(p)\sigma_{N}
\right ],\,\,\, 0<p<1, \,\, S_{e}>0.
\end{align}

\subsection{Additional Results for Induced Fatigue-Strength Models}
\label{S.section:induced.fatigue.strength.distribution}
The induced fatigue-strength cdfs (and corresponding quantile
functions) are described, in general terms, depending on
whether the \SN{} relationship has asymptotes or not, in
Sections~\ref{section:induced.fatigue.strength.distribution.no.asymptotes},~\ref{section:induced.fatigue.strength.distribution.horizontal.asymptote},
and~\ref{section:induced.fatigue.strength.distribution.vertical.asymptote}
of the main paper. This section provides some additional,
potentially useful,
results not given there.

\subsubsection{The induced fatigue-strength cdf for the Basquin model}
\label{S.section:technical.details.distributions.basquin}
Example~\ref{example:basquin.induced.fatigue.strength} in
Section~\ref{section:induced.fatigue.strength.distribution.no.asymptotes}
of the main paper provides details on the induced fatigue-strength cdf
for the Basquin model.

\subsubsection{The induced fatigue-strength cdf for the Stromeyer model}
\label{S.section:technical.details.distributions.stromeyer}

For the Stromeyer model (Section~\ref{section:stromeyer.model} of
the main paper),
the induced fatigue-strength cdf $F_{X}(x; N_{e})$ is obtained
from (\ref{equation:general.fatigue.strength.cdf.ls}) by using
\begin{align*}
\log[\gfun(x;\betavec)]&=\beta_{0} + \beta_{1}\log(x-\gamma), \,\,\, x>\gamma.
\end{align*}
Note that (with $\beta_{1}<0$) there is a horizontal asymptote at
$E=\log(\gamma)$ and thus $\lim_{x \downarrow \gamma}
[\beta_{0}+\beta_{1} \log(x-\gamma)]=\infty$.  This implies $ \lim_{x
  \downarrow \gamma} F_{X}(x; N_{e})=0$, and thus $\gamma$ is a
threshold parameter for the fatigue-strength distribution, implying that
fatigue strength $X$ will never be less than $\gamma$.

To obtain the quantile function $x_{p}(N_{e})$ for $N_{e}$, use
(\ref{equation:fatigue.strength.quantile}) in the main paper with
\begin{align}
\label{equation:ginverse.for.stromeyer}
  \gfun^{-1}(w;\betavec)&=\gamma+\exp \left[
    \frac{\log(w)-\beta_{0}}{\beta_{1}}
    \right].
\end{align}
Because of the horizontal asymptote, as $p\to 0,$
$x_{p}(N_{e}) \to \exp(E)=\gamma$ is a lower bound on the
fatigue-life quantile.

\subsubsection{The induced fatigue-strength cdf for the
  Box--Cox model}
\label{S.section:technical.details.distributions.box.cox}

For the Box--Cox model (Section~\ref{section:box.cox.relationship}),
the induced fatigue-strength cdf $F_{X}(x; N_{e})$ is obtained
from (\ref{equation:general.fatigue.strength.cdf.ls}) by using
\begin{align}
\label{S.equation:log.gfun.boxcox}
\log[\gfun(x;\betavec)]&=\beta_{0} + \beta_{1}\nu(x;\lambda)=
\begin{cases}
\beta_{0} + \beta_{1}\left(\dfrac{x^{\lambda} -1}{\lambda}\right)& \text{if $\lambda \ne  0$}\\[2ex]
\beta_{0} + \beta_{1}\log(x)& \text{if $\lambda=0$}.
\end{cases}
\end{align}
Because $\lambda<0$, $-1/\lambda$ is an upper bound for
$\nu(X;\lambda)$ and thus
\begin{align*}
B=\lim_{x \to \infty}\log[g(x;\betavec)]&=\lim_{x \to \infty}\log[\beta_{0}+\beta_{1}\nu(x;\lambda)]=\beta_{0}-\frac{\beta_{1}}{\lambda}
\end{align*}
is a vertical asymptote. As described in
Section~\ref{section:induced.fatigue.strength.distribution.vertical.asymptote},
this vertical asymptote results in a discrete atom of probability of size
\begin{align*}
1-\Phi\left(\frac{\log(N_{e}) - (\beta_{0}-\beta_{1}/\lambda)}{\sigma_{N}}   \right)
\end{align*}
 at $\infty$. This discrete atom corresponds to the limiting proportion of
 units for which $N>N_{e}$ as $x \rightarrow \infty$.  This can be
 interpreted as the (physically questionable) proportion of units
 that would survive $N_{e}$ cycles, even as stress approaches
 $\infty$.

To obtain the quantile function $x_{p}(N_{e})$ at $N_{e}$, use
(\ref{equation:fatigue.strength.quantile}) in the main paper with
\begin{align*}
  \gfun^{-1}(t;\betavec)&=\left\{
  1+\lambda\left[
    \frac{\log(t)-\beta_{0}}{\beta_{1}}
    \right]
  \right \}^{1/\lambda}.
\end{align*}
The quantiles are finite for
$0<p<\Phi([\log(N_{e}) - (\beta_{0}-\beta_{1}/\lambda)]/\sigma_{N})$.

\subsection{Expressions for the Induced Fatigue-Strength
  Model pdfs}
\label{S.section:pdfs.for.induces.distributions}
The pdf corresponding to the fatigue-strength cdf
(\ref{equation:general.fatigue.strength.cdf.ls}) in the main paper for the random
variable $X$ (induced from a specified fatigue-life distribution in
Section~\ref{section:induced.fatigue.strength.distribution.no.asymptotes})
is
\begin{align}
  \label{S.equation:induced.fatigue.strength.pdf.ls}
  f_{X}(x; N_{e}) &=\frac{dF_{X}(x; N_{e})}{dx}=\frac{1}{\sigma_{N}}
  \left |\frac{d}{d x}\log[\gfun(x;\betavec)]\right |\,
  \phi\left[\frac{\log(N_{e})-\log[\gfun(x;\betavec)]}{\sigma_{N}}
    \right].
\end{align}
The expression for $d\log[\gfun(x;\betavec)]/dt$ depends on the
  particular model being used. The following sections give
  $\gfun(x;\betavec)$ for some models.

\subsubsection{Expression for the induced fatigue-strength pdf
for the Basquin model}
For the Basquin model, the fatigue-strength pdf can be obtained by substituting
\begin{align*}
\frac{d}{d x} \log[\gfun(x;\betavec)]&=\frac{\beta_{1}}{x},\quad \quad x>0,
\end{align*}
into (\ref{S.equation:induced.fatigue.strength.pdf.ls}).

\subsubsection{Expression for the induced fatigue-strength pdf
for  the Stromeyer model}
For the Stromeyer model (Section~\ref{S.section:technical.details.distributions.stromeyer}),
the fatigue-strength pdf can be obtained by substituting
\begin{align*}
\frac{d}{d x} \log[\gfun(x;\betavec)]&=\frac{\beta_{1}}{x
  -\gamma},\quad \quad x>\gamma,
\end{align*}
into (\ref{S.equation:induced.fatigue.strength.pdf.ls}).
Because $\gamma$ is a threshold parameter, the pdf is positive only when~$x>\gamma.$

\subsubsection{Expression for the induced fatigue-strength pdf
for the Box--Cox model}
For the Box--Cox model (Section~\ref{S.section:technical.details.distributions.box.cox}),
the fatigue-strength pdf can be obtained by substituting
\begin{align*}
\frac{d}{d x}\log[\gfun(x;\betavec)]&=\frac{\gfun'(x;\betavec)}{\gfun(x;\betavec)}= \beta_{1} x^{\lambda-1},\quad \quad x>0,
\end{align*}
into (\ref{S.equation:induced.fatigue.strength.pdf.ls}). Because of
the discrete atom of probability at  $\infty$ (see
Section~\ref{section:box.cox.relationship} in the main paper),
(\ref{S.equation:induced.fatigue.strength.pdf.ls}) will, in this
case, integrate to
\begin{align*}
\Phi\left(\frac{\log(N_{e})-(\beta_{0}-\beta_{1}/\lambda)}{\sigma_{N}}   \right).
\end{align*}

\section{Technical Details for \SN{} Regression Models Where the
  Fatigue-Strength Model is Specified and the Fatigue-Life
  Model is Induced}
\label{S.section:technical.details.fatigue.strength.distribution.specified}

This section outlines additional technical details of
specifying the fatigue-strength model and using it to induce a
fatigue-life model.

\subsection{Basic  \SN{} Relationships for
  \texorpdfstring{$S=\hfun(N;\betavec)$}{S=h(N)}
  and General Assumptions}
Here we consider \SN{} relationships of the type
\begin{align}
\label{S.equation:S.generic.life.lifetime.model}
  S&=\hfun(N;\betavec),
\end{align}
where $\betavec$ is a vector of regression model parameters and  $\hfun(t;\betavec)$
satisfies  the following general conditions:
\begin{itemize}
\item
$\hfun(t;\betavec)$ is positive; that is $\hfun(t;\betavec)>0$ for
  $0<t<\infty.$
\item
$\hfun(t;\betavec)$ is strictly decreasing in $t.$
\item
$\hfun(t;\betavec)$ is differentiable for all $t.$
\end{itemize}
There are potentially two asymptotes for $\log[\hfun(N; \betavec)]$:
A horizontal asymptote at $E=\log(S)$ and a vertical
asymptote at $B=\log(N)$, as illustrated in
Figure~\ref{S.figure:CastilloFatigueLifeQuantiles}.

\subsection{Additional Motivation for Specifying the
  Fatigue-Strength Distribution}
\label{S.section:additional.motivation.specify.fatigue.strength}
Section~\ref{section:alternative.approach.modeling.sn.data} of the
main paper outlined
the important advantages of specifying the fatigue-strength
distribution in \SN{} modeling.
\begin{figure}
\begin{tabular}{cc}
\phantom{MM}Annealed aluminum wire (a) & \phantom{MMMM}C35 Steel (b) \\[-3.2ex]
\rsplidapdffiguresize{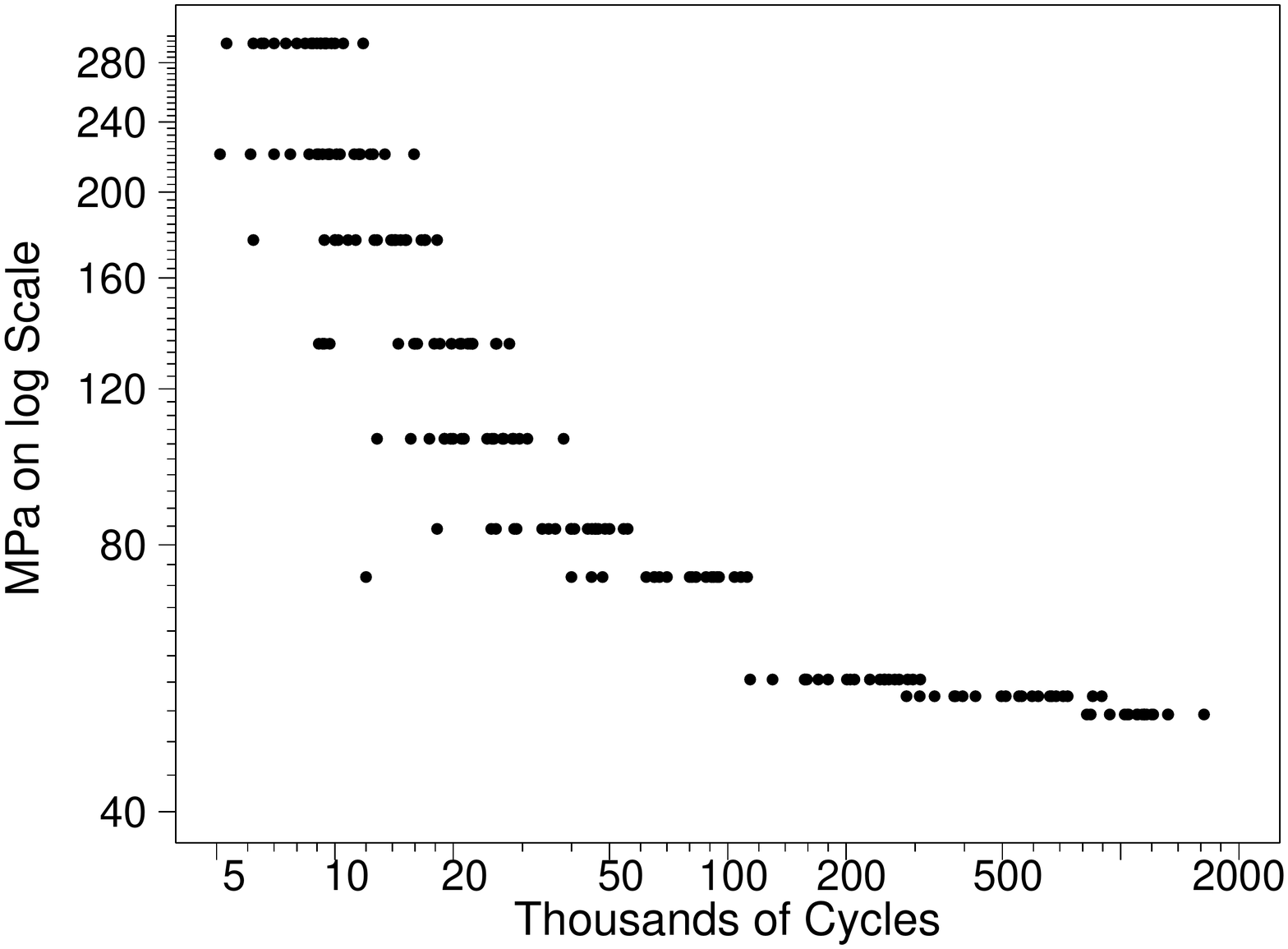}{3.25in}&
\rsplidapdffiguresize{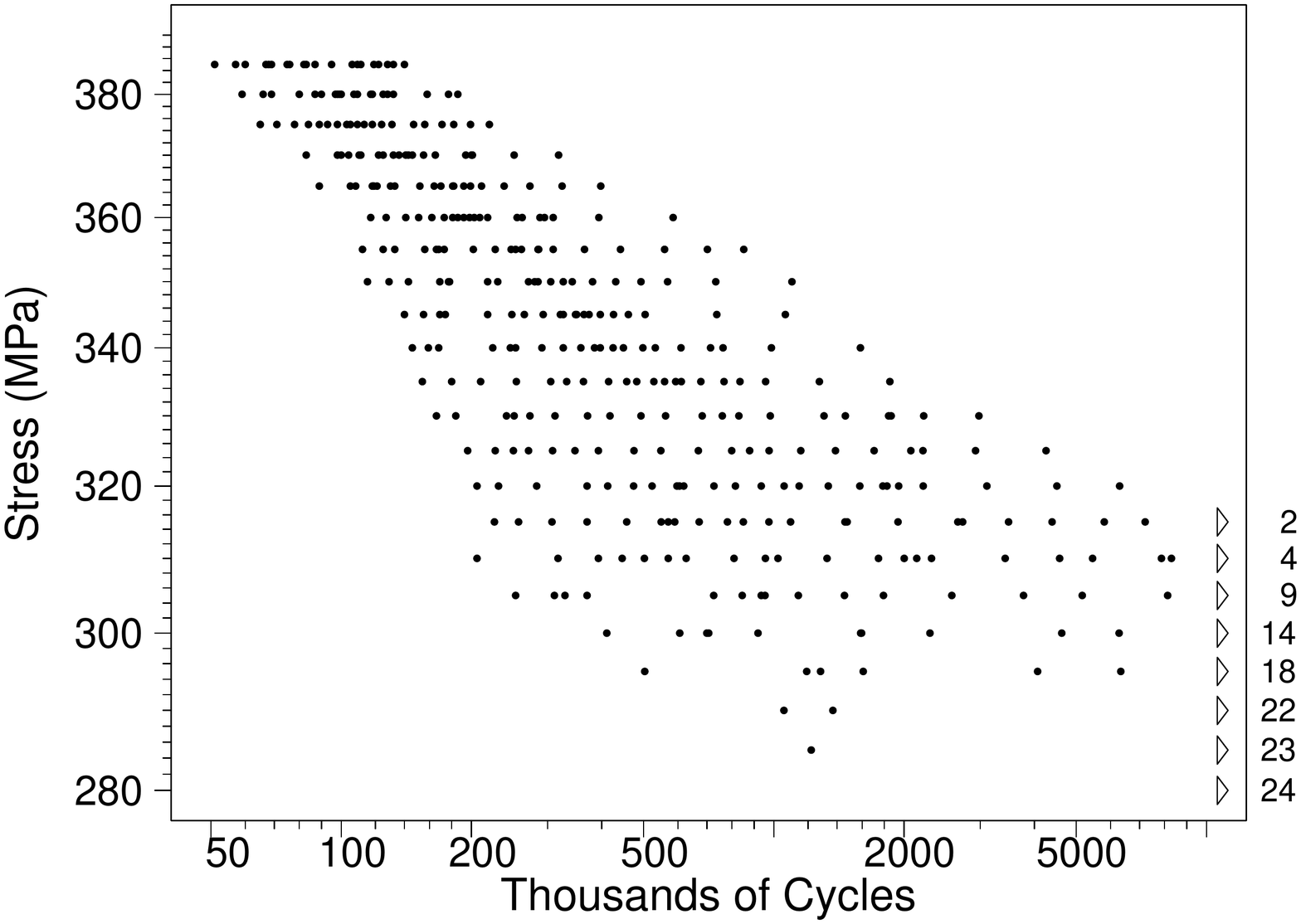}{3.25in}\\
\phantom{MM}Inconel 718 (c) & \phantom{MM}Titanium02 (d) \\[-3.2ex]
\rsplidapdffiguresize{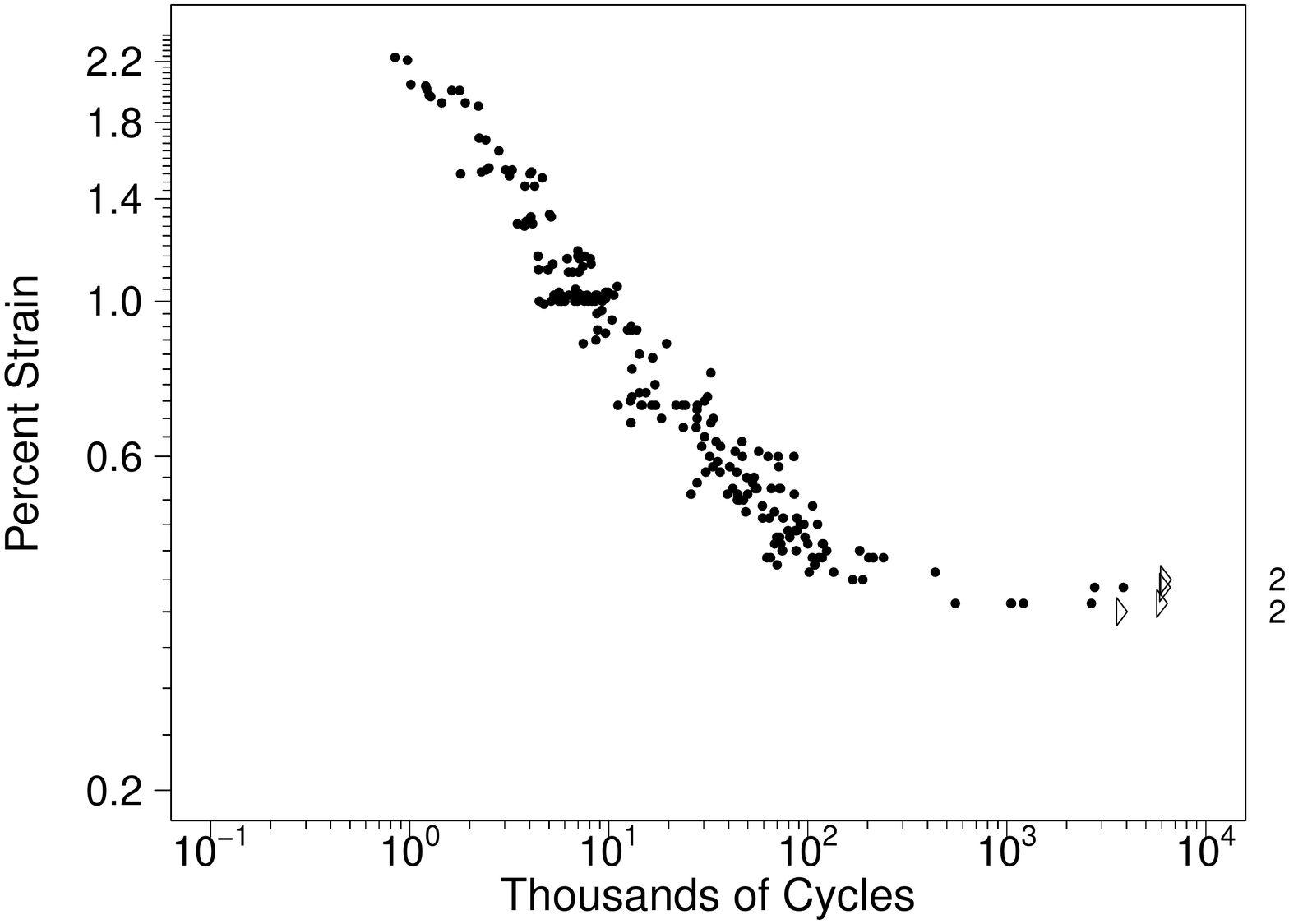}{3.25in}&
\rsplidapdffiguresize{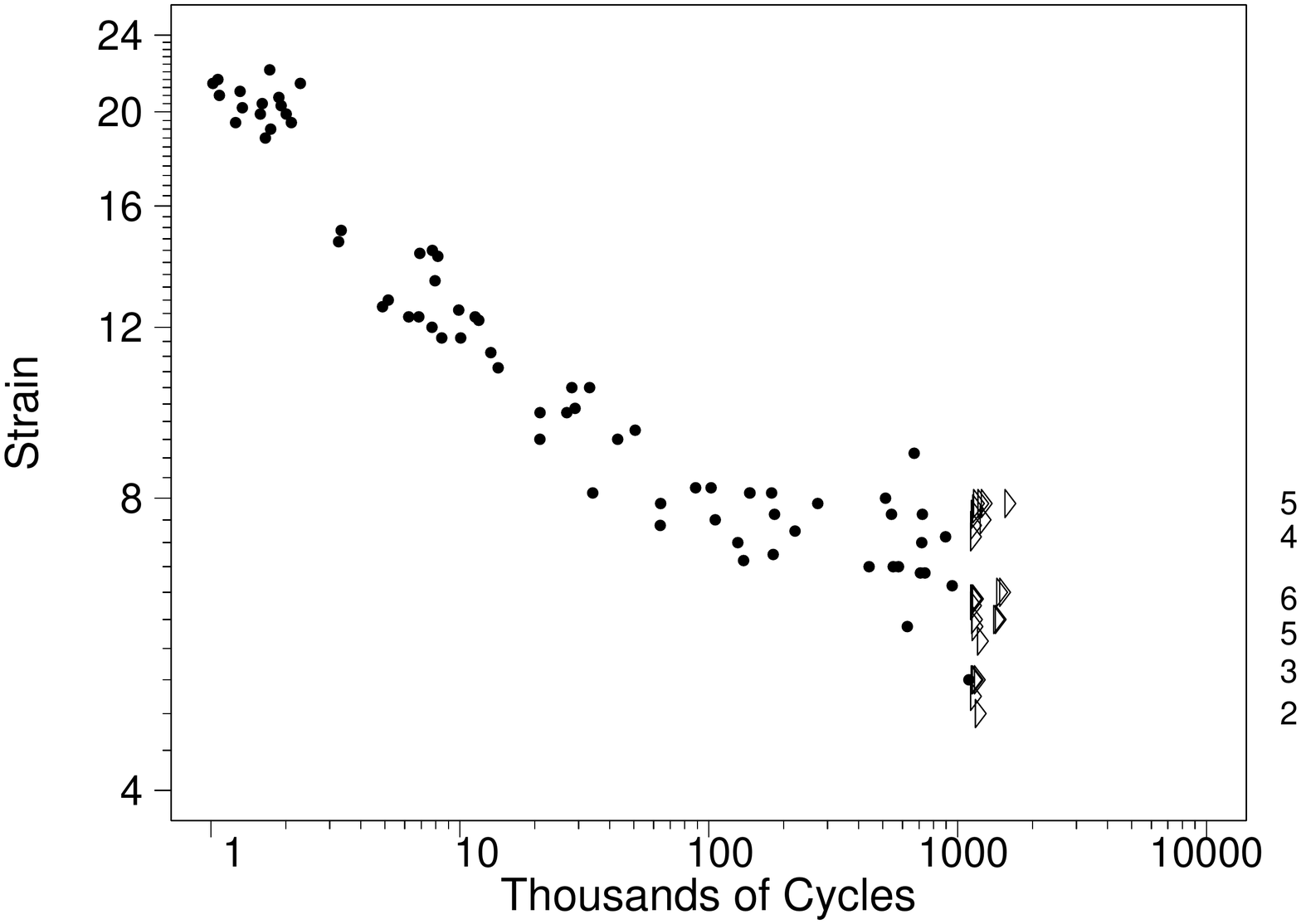}{3.25in}\\
\phantom{MM}Steel wire (e) & \phantom{MM}Holman concrete (f) \\[-3.2ex]
\rsplidapdffiguresize{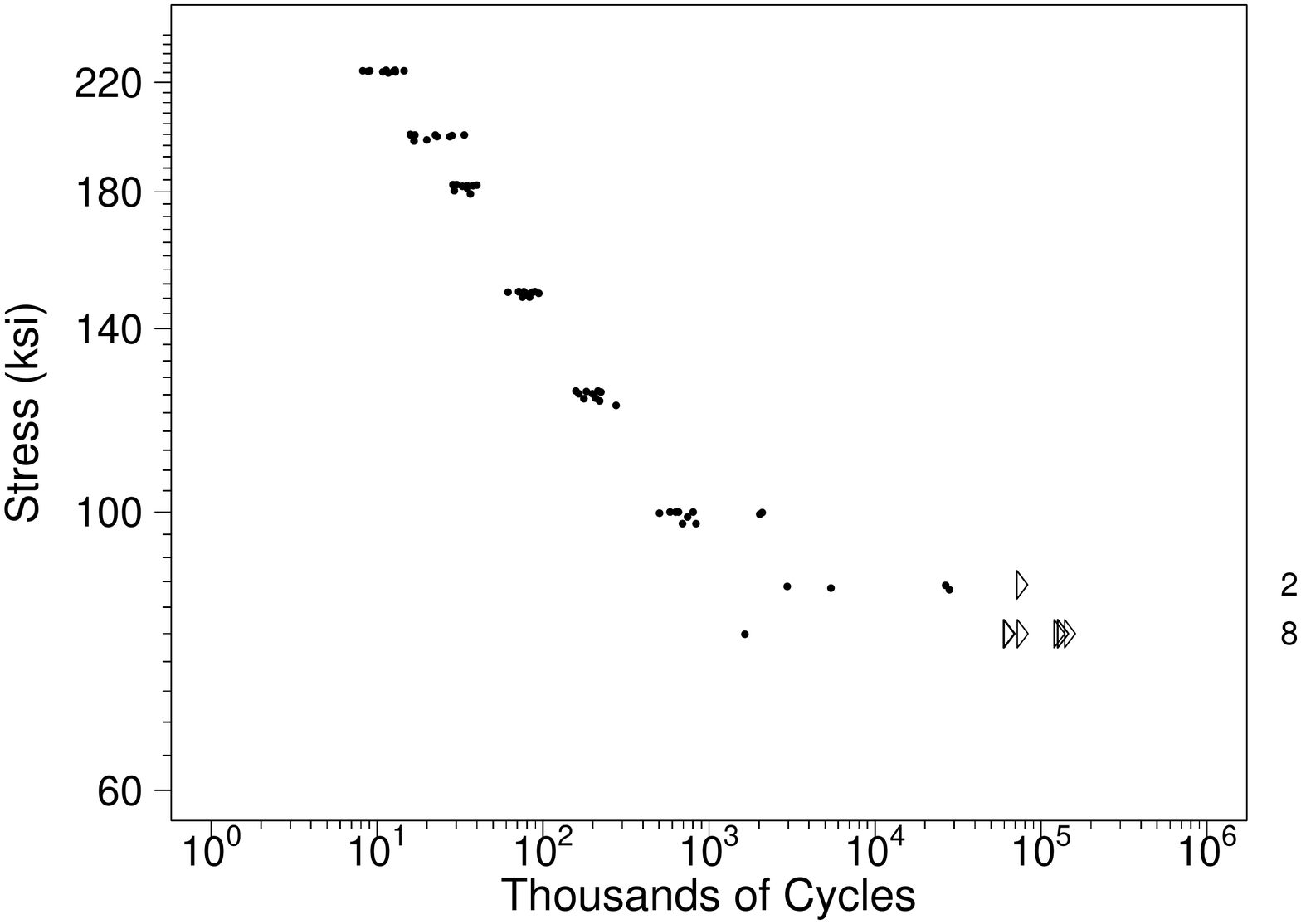}{3.25in}&
\rsplidapdffiguresize{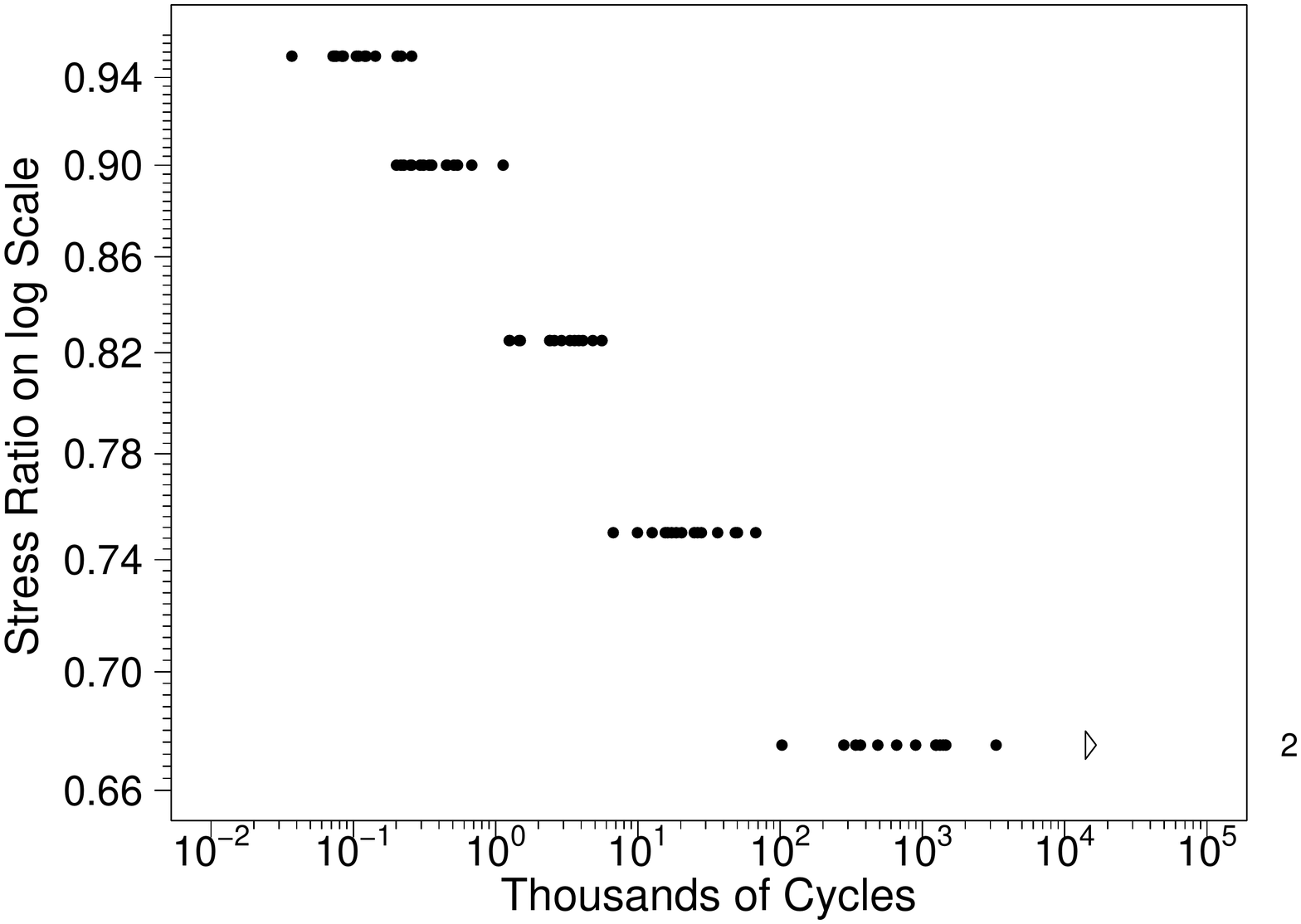}{3.25in}
\end{tabular}
\caption{Scatterplots for the annealed aluminum wire~(a), C35 steel~(b),
Inconel 718~(c),
Titanium02~(d),  steel wire~(e), and Holman concrete~(f) \SN{} data.}
\label{S.figure:datasets.contrasting.spread.fl.fs}
\end{figure}
Because the variability in
fatigue-strength $X$ tends not to depend strongly on the number of
cycles $N_{e}$ one can, if there is curvature in the \SN{}
relationship, avoid having to include an additional model component to
describe nonconstant $\sigma_{N}$ as was done in
Section~\ref{section:fitting.fatigue.life.model} of the main paper.
Figure~\ref{S.figure:datasets.contrasting.spread.fl.fs} shows
scatter plots for six high-cycle-fatigue (HCF) data sets. The data
sets in Figure~\ref{S.figure:datasets.contrasting.spread.fl.fs}
were chosen because cycling was done at a large number of levels of
stress/strain (in contrast to the three examples in the
main paper). Having so many stress/strain levels allows us to see,
empirically, the relatively constant variability in fatigue-strength
as a function of cycles. This approximate constancy in the
fatigue-strength distribution spread is in contrast
to the often sizable increase in spread in fatigue-life distribution at lower
stress/strain levels. Thus, model specification is simplified.
Also, for the reasons given
Section~\ref{section:physical.explanation.curvature.nonconstant.variance}
of the main paper, we have observed, in many of the data sets that we
have analyzed, that the \textit{induced} fatigue-life
model tends to nicely describe fatigue-life data.

An important advantage of specifying a fatigue-strength model in
which the fatigue-strength distributions have
\begin{itemize}
  \item
a scale parameter
  that depends on the number of cycles $N_{e}$ but
  \item
a constant
  shape  (i.e., shape and spread do not depend on the number of cycles $N_{e}$)
\end{itemize}
is that the fatigue-strength distributions and the induced fatigue
life distributions meet compatibility condition (cdfs are non-decreasing and
quantile lines for different levels of $p$ are monotone decreasing
and do not cross). Note that these compatibility conditions do not
hold in all cases when one used a model component that allows the
distribution shape to depend on stress, as in the
loglinear-$\sigma_{N}$ model component described in
Section~\ref{section:loglinear.sigma.component} of the main paper
and used in Example~\ref{example:box-cox.loglin.laminate.panel.data}
with the laminate panel data.

\subsection{The Specified Fatigue-Strength Distribution}
\label{S.section:specified.fatigue.strength.distribution}
The fatigue-strength random variable $X$ is the (unobservable)
lowest level of applied stress that would result in a
failure at a given number of cycles $N_{e}$ cycles. Based on the $\SN$
relationship~(\ref{S.equation:S.generic.life.lifetime.model}), taking logs,
replacing $S$ with $X$ and replacing $N$ with $N_{e},$ and adding an
error term $\epsilon$ gives
\begin{align}
\label{S.equation:general.model.for.S}
  \log(X)&=\log[\hfun(N_{e};\betavec)]+\sigma_{X} \epsilon,
\end{align}
where $\epsilon$ has a location-scale distribution with $\mu=0$ and
$\sigma=1,$ and $\sigma_{X}$ is constant.  Thus the
log-location-scale cdf $F_{X}(x; N_{e})$ for strength $X$ is
\begin{align}
  \label{S.equation:FS.cdf.model.FS.to.FL}
  \begin{aligned}
F_{X}(x;N_{e})&=\Pr\left(X \le x\right)=\Pr \left[\log(X) \le \log(x) \right]\\
&=\Phi\left[\frac{\log(x)-\log[\hfun(N_{e};\betavec)]}{\sigma_{X}}\right],\,\,\, x>0,\,\,N_{e}>0.
  \end{aligned}
  \end{align}
  This cdf has the standard properties of a cdf for a non-negative
  random variable. In particular, $\lim_{x \downarrow 0}
  F_{X}(x;S_{e})=0$ and $\lim_{x \to \infty} F_{X}(x;S_{e})=1$.

\subsection{Additional Results for Induced Fatigue-Life Models}
\label{S.section:induced.fatigue.life.distribution}
The induced fatigue-life cdfs (and corresponding quantile
functions) are described, depending on
whether the \SN{} relationship has asymptotes or not, in
Sections~\ref{section:induced.fatigue.life.distribution.neither.vertical.nor.horizontal},~\ref{section:induced.fatigue-life.distribution.horizontal.asymptote},
and~\ref{section:induced.fatigue-life.distribution.vertical.asymptote} of the main paper. In this section we provide some additional
results not given there.

\subsubsection{The induced fatigue-life distribution for the Stromeyer model}
\label{S.section:induced.fatigue.life.stromeyer}
The Stromeyer (Section~\ref{section:stromeyer.model}) \SN{}
relationship is
\begin{align}
N & =\exp\left[\beta_{0} + \beta_{1}\log(S-\exp(\horiasym))\right ].
\end{align}
Solving for $S$ gives
\begin{align}
\label{S.equation:Stromeyer.relationship.forS}
 S&=\hfun(N;\betavec)=\exp(\gamma)+\exp \left ( \frac{\log(N)-\beta_{0}}{\beta_{1}} \right),
\end{align}
 where $N>0$ and $\beta_{1}<0.$
Then the
 induced fatigue-life cdf $F_{N}(t; S_{e})$ is given by
 (\ref{equation:cdf.for.induced.fatigue.life}) in the main paper.
 The Stromeyer model has a horizontal
 asymptote at $E=\log(\gamma)=\log(S)$ because with $\beta_{1}<0$,
\begin{align*}
\lim_{t \to \infty}\log[ \hfun(t;\betavec)]&=\log\left[\gamma+\exp \left ( \frac{t-\beta_{0}}{\beta_{1}} \right)\right]=\log(\gamma).
\end{align*}
 Thus, as noted in
 Section~\ref{section:induced.fatigue-life.distribution.horizontal.asymptote}
 of the main paper,
 $F_{N}(t;S_{e})$ has a discrete atom of probability at $\infty$ equal to
 the value in (\ref{equation:atom.of.prob.due.hor.asym}).

To obtain the quantile function $t_{p}(S_{e})$ for $N$, use
\begin{align}
\label{S.equation:hfun.inverse.Stromeyer}
\hfun^{-1}(S)&=\exp \left[\beta_{0}+\beta_{1}\log(S-\gamma) \right]
\end{align}
in (\ref{equation:quantiles.induced.fatigue.life}) of the main paper
but, as noted in
Section~\ref{section:induced.fatigue-life.distribution.horizontal.asymptote},
because of the discrete atom of probability at $\infty$, $t_{p}(S_{e})$ is
only finite for $0 \le p <
\Phi\left[(\log(S_{e})-\log(\gamma))/\sigma_{X}\right]$ and thus there is a
limit, depending on $S_{e}$, for the largest finite quantile.

\subsubsection{The induced fatigue-life cdf for the
  Box--Cox model}
\label{S.section:induced.fatigue.life.box.cox}
The Box--Cox  \SN{} relationship is $N =\exp\left[\beta_{0} +
  \beta_{1}\nu(S;\lambda)\right]$, where
\begin{align}
\label{S.equation:box.cox.relationship}
\begin{aligned}
\nu(S;\lambda)&=
\begin{cases}
\dfrac{S^{\lambda} -1}{\lambda} & \text{if $\lambda \ne  0$}\\[2ex]
\log(S) & \text{if $\lambda=0$}.
\end{cases}
\end{aligned}
\end{align}
In fatigue applications, $\beta_{1}<0$ and $\lambda \leq 0.$
For $\lambda < 0$, solving for $S$ in (\ref{S.equation:box.cox.relationship}),
\begin{align}
\label{S.equation:BoxCox.relationship.forS}
S&=\hfun(N;\betavec)=
\left[
  1+\frac{\lambda}{\beta_{1}} \left[ \log(N)-\beta_{0} \right]\right ]^{1/\lambda}.
\end{align}
As described in
Section~\ref{section:induced.fatigue-life.distribution.vertical.asymptote}
of the main paper, the induced fatigue-life cdf is given by
(\ref{equation:induced.fatigue.life.cdf.vertical.asymptote}). The
Box--Cox model has a vertical asymptote at
$B=\beta_{0}-\beta_{1}/\lambda$, and this implies that $N >
\exp(\beta_{0}-\beta_{1}/\lambda)$ and thus the cdf has a threshold
parameter.

To obtain the quantile function $t_{p}(S_{e})$ for $N$, use
\begin{align}
\nonumber 
\hfun^{-1}(S)&=N=\exp\left[\beta_{0} + \beta_{1}\nu(S;\lambda)\right]
\end{align}
in (\ref{equation:quantiles.induced.fatigue.life}) of the main paper.
As noted in
Section~\ref{section:induced.fatigue-life.distribution.vertical.asymptote},
because of the threshold parameter, the quantiles $t_{p}(S_{e})$ of
$F_{N}(t;S_{e})$ are the same as
(\ref{equation:quantiles.induced.fatigue.life}) but as $p\to 0,$
$t_{p}(S_{e}) \to \exp(\beta_{0}-\beta_{1}/\lambda).$ Thus the
lower bound on the fatigue-life quantile is
$\exp(\beta_{0}-\beta_{1}/\lambda)$.

\subsection{Expressions for the Induced Fatigue-Life Model pdfs}
\label{S.section:pdfs.for.induced.distributions}

The pdf of an induced fatigue-life random variable $N$ is
\begin{align}
  \label{S.equation:induced.fatigue.life.pdf.ls}
  f_{N}(t; S_{e})
  &=\frac{dF_{N}(t; S_{e})}{dt}=\frac{1}{\sigma_{X}} \left |\frac{d}{d t}\log[\hfun(t;\betavec)]\right |\, \phi\left[\frac{\log(S_{e})-\log[\hfun(t;\betavec)]}{\sigma_{X}}   \right].
\end{align}
Expressions for $d\log[\hfun(t;\betavec)]/dt$ depend on the
  particular $\hfun(t;\betavec)$ relationship and are given for some models in the
  following subsections.

\subsubsection{Expressions for the induced  fatigue-life pdf for
  the Coffin--Manson model}
For the Coffin--Manson model, following
from~(\ref{equation:coffin.manson.basic})
in the main paper and replacing $N$ with $t$ gives
\begin{align*}
S=h(t;\betavec)=A_{el} (2t)^{b} +  A_{pl} (2t)^{c}
\end{align*}
and thus
\begin{align*}
 \hfun'(t;\betavec)=d \, \hfun(t;\betavec)/dt&=
2bA_{el} (2t)^{b-1} +  2cA_{pl} (2t)^{c-1}.
\end{align*}
Then
\begin{align*}
 \frac{d}{d t}\log[\hfun(t;\betavec)]
 &=\frac{\hfun'(t;\betavec)}{\hfun(t;\betavec)}= \frac{2b A_{el} (2t)^{b-1} +  2c A_{pl} (2t)^{c-1}}{A_{el} (2t)^{b} +  A_{pl} (2t)^{c}}.
\end{align*}
This can be substituted into
(\ref{S.equation:induced.fatigue.life.pdf.ls}) to give the
needed expression for the Coffin--Manson model pdf.

\subsubsection{Expressions for the induced  fatigue-life pdf for
the Nishijima relationship}
For the Nishijima relationship, following
from~(\ref{equation:nishijima.hyperbolic.relationship.hfunction})
replacing $N$ with $t,$ gives
\begin{align*}
    S &=h(t; \betavec)=\exp\left(\frac{
-A \log(t)+B+E + \sqrt{
  \left[A \log(t)-(B-E)\right]^{2} +4C}}{2}\right).
\end{align*}
and thus
\begin{align*}
  \frac{d \log[\hfun(t;\betavec)]}{dt}&=
  \frac{1}{2}\,\frac{d}{d t}\left[-A\, \log(t)+B+E+\sqrt{\left[A\,\log(t)-(B-E)\right]^{2}+4C}\right]\\
  &=\frac{A}{2 t}
  \left[
    -1+
\frac{A\,\log(t)-(B-E)}{\sqrt{\left[A\,\log(t)-(B-E)\right]^{2}+4C}}
\right].
\end{align*}
This can be substituted into
(\ref{S.equation:induced.fatigue.life.pdf.ls}) to give the needed
expression for the Nishijima model pdf.
Note that this pdf does not integrate to 1
due to the discrete atom of probability at $t=\infty$ (see
Section~\ref{section:induced.fatigue-life.distribution.horizontal.asymptote})
given in~(\ref{equation:atom.of.prob.due.hor.asym}).

\subsubsection{Expression for the induced fatigue-life pdf
  for the Stromeyer model}
For the Stromeyer model,
\begin{align*}
  \frac{d \log[\hfun(t;\betavec)]}{dt}&=\left(\frac{1}{t\, \beta_{1} }\right)\,
  \dfrac{\exp \left ( \dfrac{\log(t)-\beta_{0}}{\beta_{1}}\right)}
         {\gamma+\exp \left (\dfrac{\log(t)-\beta_{0}}{\beta_{1}}\right)
          }.\end{align*}
This can be substituted into
(\ref{S.equation:induced.fatigue.life.pdf.ls}) to give the needed
expression for the Stromeyer fatigue-life pdf.
Due to the discrete atom of probability at $\infty$
(see
Sections~\ref{section:induced.fatigue-life.distribution.horizontal.asymptote}
and~\ref{S.section:induced.fatigue.life.stromeyer}), this
pdf integrates to $\Phi\left[ \left( \log(S_{e}) - E\right)/\sigma_{X}  \right]< 1$.

\subsubsection{Expression for the induced fatigue-life pdf
for the Box--Cox model}
For the Box--Cox model,
\begin{align*}
  \frac{d \log[\hfun(t;\betavec)]}{dt}&=\left(\frac{1}{t\, \beta_{1} }\right)\,
  \dfrac{1}
         {1+\dfrac{\lambda}{\beta_{1}} \left( \log(t)-\beta_{0} \right)}.
\end{align*}
This can be substituted into
(\ref{S.equation:induced.fatigue.life.pdf.ls}) to give the needed
expression for the pdf.  Because of the threshold parameter, caused
by the vertical asymptote in the \SN{} relationship,  noted in
Sections~\ref{section:induced.fatigue-life.distribution.vertical.asymptote}
and~\ref{S.section:technical.details.distributions.box.cox},
this pdf is positive only when $t>\exp(B)=\exp(\beta_{0}-\beta_{1}/\lambda).$

\section{Technical Details of Results Stated in the Main Paper}
\label{S.proofs.technical.results.stated.in.paper}

\subsection{Equivalence of Fatigue-Life and Fatigue-Strength
  Quantile Curves}
\label{S.section:quantile.curve.equivalence}
Section~\ref{section:equivalence.fatigue.life.fatigue.strength.quantile.curves}
stated that, for \SN{} relationships that have neither a horizontal
nor a vertical asymptote, the fatigue-life and fatigue-strength
models have the same quantile curves.
Section~\ref{S.section:quantile.curve.equivalence.proof} proves that
result and
Section~\ref{S.section:quantile.curve.equivalence.exceptions}
describes the behavior of the exceptional cases, for extreme
values of $p$, when the \SN{} relationship has one or two coordinate
asymptotes.
\subsubsection{Proof of the equivalence of fatigue-life and fatigue-strength
  quantile curves when there is neither a horizontal
nor a vertical asymptote}
\label{S.section:quantile.curve.equivalence.proof}
Consider the quantile function for the specified fatigue-strength
distribution in
(\ref{equation:quantiles.specified.fatigue.strength}).  Changing
variable names $x_{p}(N_{e})$ to $S_{e}$ and $N_{e}$ to
$t_{p}(S_{e})$ gives
\begin{align}
\label{S.equation:quantiles.specified.fatigue.strength.name.change}
S_{e} = \exp(\log[\hfun(t_{p}(S_{e});\betavec)] + \Phi^{-1}(p) \sigma_{X}).
\end{align}
Solving (\ref{S.equation:quantiles.specified.fatigue.strength.name.change}) for
$t_{p}(S_{e})$ gives
\begin{align*}
t_{p}(S_{e})&=\hfun^{-1}\left(\exp \left[\log(S_{e})-
  \Phi^{-1}(p)\sigma_{X} \right];\betavec \right),
\end{align*}
which agrees with the quantile function for the induced fatigue-life
model in (\ref{equation:quantiles.induced.fatigue.life}),
giving the needed result.
There is a similar and parallel result showing the equivalence of
the quantiles curves for a specified fatigue-life model and an
induced fatigue-strength model.

\subsubsection{The effect of coordinate
  asymptotes on the equivalence of
fatigue-life and fatigue-strength quantile curves}
\label{S.section:quantile.curve.equivalence.exceptions}

This section describes the effect that coordinate asymptotes, when
they exist,
have on the behavior of quantile
curves and the equivalence of
fatigue-life and fatigue-strength quantile curves.  For the
\textit{specified} fatigue-life and fatigue-strength distributions,
the quantiles functions are defined for all values of $0 < p <
1$. This can be seen for a \textit{specified} fatigue-life
distribution by comparing (\ref{equation:general.model.for.N})
and~(\ref{equation:fatigue.life.failure.time.model.quantile}) in
Section~\ref{section:relationship.fatigue.life.fatigue.strength} and
for a \textit{specified} fatigue-strength distribution by comparing
(\ref{equation:logx.general.strength.cdf})
and~(\ref{equation:quantiles.specified.fatigue.strength}) in
Section~\ref{section:specifying.fatigue.strength.distribution} and
noting that there are no restrictions on the error $\epsilon$.

If the \SN{} relationship has a coordinate asymptote (horizontal or
vertical or both), the behavior of the \textit{induced} fatigue-strength and
fatigue-life models have special characteristics.
For situation where the fatigue-life model is specified,
the top row of Table~\ref{S.table:quantile.function.behavior.with.asymptotes}
summarizes detailed information given in
Sections~\ref{section:induced.fatigue.strength.distribution.horizontal.asymptote},
\ref{section:induced.fatigue.strength.distribution.vertical.asymptote},
about the effect that coordinate asymptotes have on the \textit{induced}
fatigue-strength cdfs $F_{X}(x;N_{e} )$ and the quantile functions $x_{p}(N_{e})$.
Similarly, for situation where the fatigue-strength
model is specified, the bottom row of
Table~\ref{S.table:quantile.function.behavior.with.asymptotes}
summarizes detailed information given in
\ref{section:induced.fatigue-life.distribution.horizontal.asymptote},
and
\ref{section:induced.fatigue-life.distribution.vertical.asymptote}
about the effect that coordinate asymptotes have on the \textit{induced}
fatigue-life cdfs $F_{N}(t;S_{e} )$ and corresponding
quantile functions $t_{p}(S_{e})$.
\begin{table}
\caption{Description of Quantile Function Behavior for Extreme
  Values of $p$ for the
  Induced Fatigue Distributions when the \SN{} Relationship
  Has One or Two Coordinate
  Asymptotes
\label{S.table:quantile.function.behavior.with.asymptotes}}
 \begin{tabular}{|p{2.0in}|p{2.1in}|p{2.1in}|}
\hline
\centering Induced Distribution         & \centering Horizontal Asymptote $E$ & \centering Vertical Asymptote $B$\arraybackslash\\
\hline
Fatigue Strength $X$ \newline
cdf $F_{X}(x;N_{e})$ \newline
Quantile Function $x_{p}(N_{e})$
&Section~\ref{section:induced.fatigue.strength.distribution.horizontal.asymptote}\newline
$\exp(E)$ is a threshold parameter for the induced fatigue-strength cdf
$F_{X}(x;N_{e})$. Fatigue strength $X$ cannot be less than
$\exp(E)>0$. As $p\to 0,$ $x_{p}(N_{e}) \to \exp(E)$.
&Section~\ref{section:induced.fatigue.strength.distribution.vertical.asymptote}\newline
The induced fatigue-strength cdf $F_{X}(x;N_{e})$ has a discrete atom of probability of size
$1-\Phi\left(\dfrac{\log(N_{e})-B}{\sigma_{N}}\right)$ at $\infty$
corresponding to the limiting
probability that fatigue life $N$ is greater than $N_{e}$ when $x$
is large. Finite fatigue-strength
quantiles exist only for $0 < p < \Phi\left(\dfrac{\log(N_{e})-B}{\sigma_{N}}\right)$.\\[10.0ex]
\hline
        Fatigue Life $N$ \newline
        cdf  $F_{N}(t;S_{e})$ \newline
 Quantile Function $t_{p}(S_{e})$
&
Section~\ref{section:induced.fatigue-life.distribution.horizontal.asymptote}\newline
The induced fatigue-life cdf $F_{N}(x;S_{e})$ has a discrete atom of probability of size
$1-\Phi\left(\dfrac{\log(S_{e})-E}{\sigma_{X}}   \right)$ at $\infty$
corresponding to the limiting
probability that fatigue strength $X$ is greater than applied stress
$S_{e}$ when $t$ is large.  Finite fatigue-life
quantiles exist only for $0 < p < \Phi\left(\dfrac{\log(S_{e})-E}{\sigma_{X}}\right)$.
&
Section~\ref{section:induced.fatigue-life.distribution.vertical.asymptote}\newline
$\exp(B)$
is a threshold parameter for the induced fatigue-life cdf $F_{N}(t;S_{e})$. Fatigue life $N$  cannot be less than
$\exp(B)>0$. As $p\to 0,$
$t_{p}(S_{e}) \to \exp(B)$.\\
        \hline
    \end{tabular}
\end{table}

The table illustrates an interesting duality.
The following describes each of the four special cases outlined in
Table~\ref{S.table:quantile.function.behavior.with.asymptotes}.
\begin{itemize}
\item
Following the development in
Section~\ref{section:induced.fatigue.strength.distribution.vertical.asymptote}
for an induced fatigue-strength model, because of the discrete atom of probability at infinity, for a given $p$, the
quantile will be at infinity until cycles level
$N_{e}=\exp[B+\Phi^{-1}(p)\sigma_{N}]$ after which it will follow
(\ref{equation:fatigue.strength.quantile}) and the fatigue-life and
fatigue-strength quantile curves will agree.
\item
Following the development in
Section~\ref{section:induced.fatigue-life.distribution.horizontal.asymptote},
for an induced fatigue-life model, because of the discrete atom of probability at infinity, for a given $p$, the
quantile will be at infinity until stress level
$S_{e}=\exp[E+\Phi^{-1}(p)\sigma_{X}]$ after which it will follow
(\ref{equation:fatigue.life.failure.time.model.quantile}) and
the fatigue-life and fatigue-strength quantile curves will agree.
\item
Following the development in
Sections~\ref{section:induced.fatigue.strength.distribution.horizontal.asymptote}
and~\ref{section:induced.fatigue-life.distribution.vertical.asymptote}
the induced cdfs have a threshold parameter. In either case,
however, the quantile functions $t_{p}(S_{e})$ and $x_{p}(N_{e})$ map
out the same curve (because one is the inverse function of the
other, as shown by the proof in
Section~\ref{S.section:quantile.curve.equivalence.proof}) even
though one of the quantile functions approaches the threshold parameter as $p
\rightarrow 0$.
\end{itemize}

\subsection{Proof that Concave-up Curvature in the \SN{} Relationship Induces Fatigue-Life
  Distributions with Increasing Spread at Lower Stress Levels}
\label{S.section:Curvature.sn.relationships.increased.spread.low.stress}

Suppose that a fatigue-stress model with constant $\sigma_{X}$ is
specified and the induced fatigue-life model is as given in
Section~\ref{section:induced.fatigue.life.distribution.neither.vertical.nor.horizontal}
of the main paper, resulting in a quantile function
\begin{align}
\nonumber
t_{p}(S_{e}) &=\hfun^{-1}\left(\exp
\left[\log(S_{e})-\Phi^{-1}(p)\sigma_{X}\right]\right)\\
\label{S.equation:quantiles.induced.fatigue.life}
              &=\gfun\left(\exp
\left[\log(S_{e})-\Phi^{-1}(p)\sigma_{X}\right]\right),
\end{align}
which is the same as (\ref{equation:quantiles.induced.fatigue.life})
in the main paper except here we suppress the dependency on the
parameter vector $\betavec$ and introduce $g(x)$ to simplify
notation. Suppose $t_{p}(S_{e})$ is differentiable, decreasing, and
strictly concave-up in $\log(S_{e})$, the latter implying
\begin{align}
\label{S.equation:due.to.concave.up}
  \frac{\partial^{2}  \log[t_{p}(S_{e})] }{\partial [\log(S_{e})]^{2}}>0.
\end{align}
Consider $t_{p}(S_{e}) =\gfun\left(w_{p}\right)$ where $w_{p}=\exp
\left[\log(S_{e})-\Phi^{-1}(p)\sigma_{X}\right]$ can be interpreted
as the \textit{pseudo reverse} fatigue-strength $p$ quantile (pseudo
reverse because of the minus sign and that the center of the
fatigue-strength distribution is taken to be $S_{e}$), which then,
according to (\ref{S.equation:quantiles.induced.fatigue.life}), gets
mapped, through the \SN{} relationship $g(x)$, to the fatigue-life
quantile $t_{p}(S_{e})$ at stress $S_{e}$.

First note that $\partial w_{p}/\partial \log(S_{e})=w_{p}$.  Using
the chain rule, the first partial derivative with respect to
$\log(S_{e}),$ is
\begin{align}
\label{S.equation:first.partial.wrt.log.Se}
  \frac {\partial \log \left[ t_{p}(S_{e})\right]} {\partial
    \log(S_{e})}&=\frac{\partial
    \log\left[\gfun\left(w_{p}\right)\right]}{\partial
    \log(S_{e})}=\frac{\partial
    \log\left[\gfun\left(w_{p}\right)\right]}{\partial w_{p}}\,
  \frac{\partial w_{p}}{\partial \log(S_{e})}=\frac{ \gfun'(w_{p})\,
    w_{p}}{\gfun\left(w_{p}\right)},
\end{align}
where $\gfun'(w_{p})= d\gfun(w_{p})/dw_{p}$.  Using the result
in (\ref{S.equation:first.partial.wrt.log.Se}) and the chain rule
again, the second partial derivative with respect to $\log(S_{e})$
is
\begin{align}
  \nonumber \frac {\partial^{2} \log \left[ t_{p}(S_{e})\right]}
            {\partial [\log(S_{e})]^{2}}&= \frac{\partial }{\partial
              \log(S_{e})} \left[ \frac{ \gfun'(w_{p})\,
                w_{p}}{\gfun \left(w_{p}\right)} \right]=
            \frac{\partial }{\partial w_{p}} \left [ \frac{
                \gfun'(w_{p})\, w_{p}}{\gfun\left(w_{p}\right)}
              \right]\, \frac{\partial w_{p}}{ \partial
              \log(S_{e})}\\
\label{S.equation:second.partial.logtp.function.of.w}
                &=\frac{\partial }{\partial w_{p}} \left [ \frac{
    \gfun'(w_{p})\, w_{p}}{\gfun\left(w_{p}\right)} \right]\, w_{p}=
w_{p}\, \frac{\partial }{\partial w_{p}} \left [ \frac{
    \gfun'(w_{p})\, w_{p}}{\gfun\left(w_{p}\right)} \right]
\end{align}
which, from the concave-up property in
(\ref{S.equation:due.to.concave.up}), is positive.  Then, because
$w_{p}$ is positive
\begin{align}
\label{S.equation:concave.up.property.of.logtp}
\frac{\partial }{\partial w_{p}} \left [ \frac{ \gfun'(w_{p})\,
    w_{p}}{\gfun\left(w_{p}\right)} \right]>0.
\end{align}
Using the result in
(\ref{S.equation:first.partial.wrt.log.Se}) and the chain rule
again, the second mixed partial derivative with respect to $p$ is
\begin{align}
\nonumber \frac {\partial^{2} \log \left[ t_{p}(S_{e})\right]}
          {\partial p\, \partial \log(S_{e})}&= \frac {\partial
          }{\partial p}\left[\frac{ \gfun'(w_{p})\,
              w_{p}}{\gfun\left(w_{p}\right)}\right]= \frac
          {\partial }{\partial w_{p}}\left[\frac{ \gfun'(w_{p})\,
              w_{p}}{\gfun\left(w_{p}\right)}\right] \frac{\partial
            w_{p}}{\partial p}\\
\label{S.equation:second.partial.logtp.p.and.logse}
&=\frac{\partial }{\partial w_{p}} \left [ \frac{ \gfun'(w_{p})\,
    w_{p}}{\gfun\left(w_{p}\right)} \right]\, \left[
  -\frac{\sigma_{X}}{\phi\left[ \Phi^{-1}(p)\right]} \right] w_{p},
\end{align}
where
\begin{align*}
\frac{\partial w_{p}}{\partial p} &= \left[
  -\frac{\sigma_{X}}{\phi\left[ \Phi^{-1}(p)\right]} \right]w_{p}
\end{align*}
and $\phi(z)=d\Phi(z)/dz.$ Then, because $\partial w_{p}/\partial p$
is negative and (\ref{S.equation:concave.up.property.of.logtp}),
\begin{align}
\label{S.equation:first.partial.logtp.decreasing}
\frac {\partial^{2} \log \left[ t_{p}(S_{e})\right]}
         {\partial p\, \partial \log(S_{e})}& < 0.
\end{align}

Now we show that the difference between two fatigue-life quantiles
decreases when $\log(S_{e})$ increases. Consider $0<p_{L}<p_{U}<1.$
The sign of the derivative in
(\ref{S.equation:first.partial.logtp.decreasing}) implies that
\begin{align*}
  \frac{\partial  \log[t_{p_{U}}(S_{e})] }{\partial \log(S_{e})}
  <  \frac{\partial  \log[t_{p_{L}}(S_{e})] }{\partial \log(S_{e})}.
\end{align*}
Equivalently,
\begin{align}
\label{equation:decreasing.derivatives}
  \frac{\partial  \log[t_{p_{U}}(S_{e})] }{\partial \log(S_{e})}
  -  \frac{\partial  \log[t_{p_{L}}(S_{e})] }{\partial \log(S_{e})} <0.
\end{align}
Let $\Delta [\log(t_{p})] = \log[t_{p_{U}}(S_{e})]-
\log[t_{p_{L}}(S_{e})]$ be the difference between the two
fatigue-life quantiles on the log scale.  Using
(\ref{equation:decreasing.derivatives}), the derivative of the
difference of the log-quantiles is
\begin{align*}
  \frac{\partial \Delta [\log(t_{p})]}{\partial \log(S_{e})
  }&=\frac{\partial}{\partial \log(S_{e})}\left(
  \log[t_{p_{U}}(S_{e})]- \log[t_{p_{L}}(S_{e})]\right)\\ &=
  \frac{\partial \log[t_{p_{U}}(S_{e})]}{\partial \log(S_{e})} -
  \frac{\partial \log[t_{p_{L}}(S_{e})]}{\partial \log(S_{e})}<0.
\end{align*}
Consequently, the difference between the fatigue-life quantiles
decreases when the stress increases which implies that the spread in
the fatigue-life distribution increases when the stress decreases.

\subsection{A Generalization of Relationship between the
  Specified Fatigue-Strength Model and the Induced Fatigue-Life Model}
\label{S.section:generalization.specifiedfsm.induced.flm}
This section generalizes the specify-fatigue-strength-model approach
that we introduce in
Section~\ref{section:alternative.approach.modeling.sn.data} of the
main paper. Here we use a scale-shape parameter distribution as a
basis for the specified fatigue-strength model. The scale parameter
is controlled by the specified monotone decreasing \SN{}
relationship and the shape parameter(s) are constant. This model has
the log-location-scale-based model
Section~\ref{section:alternative.approach.modeling.sn.data} as a
special case but still satisfies compatibility conditions. For
example, fatigue-strength and fatigue-life quantile
curves are decreasing and do not cross or bend
back; fatigue-strength and fatigue-life cdfs are non-decreasing.

\subsubsection{A statistical model for fatigue-strength}
\label{S.section:specifying.fatigue.strength.distribution}
Suppose that the logarithm of the fatigue-strength random variable
$X$ at a \textit{given} number of cycles $N_{e}$ is
\begin{align}
\label{S.equation:logx.general.strength.model}
\log(X) = \log[\hfun(N_{e};\betavec)] +   \epsilon,
\end{align}
where $S=\hfun(N;\betavec)$ is a positive monotonically
decreasing \SN{} regression relationship of known form,
$\betavec$ is a vector of regression parameters, and $\epsilon$
is a random error such that the distribution of $Z=\exp(\epsilon)$
is a standard cdf (i.e., a distribution with a scale parameter equal to 1)
denoted by $\Psi(z; \kappavec)$ having
one or more constant shape parameters in $\kappavec.$

Then for any given number of cycles
$N_{e}$, fatigue strength $X$ has the cdf
\begin{align}
  \label{S.equation:FS.cdf.scale.shape}
  F_{X}(x; N_{e})&= \Pr(X \le x; N_{e})= \Pr\left[ \log(X) \le \log(x)\right ]\\
  &= \Pr\left[ \left(\log[\hfun(N_{e};\betavec)] +   \epsilon \right)\le \log(x) \right]\\
 &= \Pr\left(   \epsilon \le \log(x)-\log[\hfun(N_{e};\betavec)]  \right)\\
 &= \Pr\left[    \exp(\epsilon) \le \frac{x }{\hfun(N_{e};\betavec)}\right]\\
 &= \Psi\left(\frac{x}{h(N_{e};\betavec)}; \kappavec\right), \,\,\,
  x>0, \,  N_{e}>0,
\end{align}
where $\hfun(N_{e};\betavec)$ is a scale parameter and $\kappavec$
contains one or more shape parameters of the distribution of $X$.
Besides
distributions in the log-location-scale family, this formulation
allows distributions such as the Birnbaum-Saunders, gamma, and
generalized gamma distributions \citep[described, e.g., in Chapter 4
  of][]{MeekerEscobarPascual2021}.

The $p$ quantile of the fatigue-strength random variable $X$
is obtained by solving $p=F_{X}[x_{p}(N_{e});N_{e}]$ for $x_{p}(N_{e})$, giving
\begin{align}
\label{S.equation:quantiles.specified.fatigue.strength}
x_{p}(N_{e}) = \hfun(N_{e};\betavec)\,   \Psi^{-1}\left(p;  \kappavec\right), \,\,\,
  0<p<1, \,\, N_{e}>0.
\end{align}

\subsubsection{The induced fatigue-life model when
  \texorpdfstring{$\hSN$}{log[h(N)]}
  has neither a vertical nor a horizontal
  asymptote}
\label{S.section:induced.fatigue.life.distribution.neither.vertical.nor.horizontal}
For the moment, suppose that the positive monotonically decreasing \SN{}
relationship $S=\hfun(N;\betavec)$ has neither a vertical nor a
horizontal asymptote (e.g., the Basquin or Coffin--Manson \SN{}
relationship).
In our model, based on the definitions of fatigue strength and
fatigue life (described in
Section~\ref{section:relationship.fatigue.life.and.fatigue.strength}),
the random variables $X$ and $N$ share the same random-error term.
Replacing $N_{e}$ with $N$ and $X$ with $S_{e}$ in
(\ref{S.equation:logx.general.strength.model}) gives
\begin{align}
\label{S.equation:distribution.general.h.sn.relationship}
\log(S_{e}) -\log[\hfun(N;\betavec)]&=   \epsilon.
\end{align}
In this role switching, fatigue life $N$ at given $S_{e}$
replaces fatigue strength $X$ at
given $N_{e}$, but the random variables $X$ and $N$
have the same $ \epsilon$ error
term.  Equation
(\ref{S.equation:distribution.general.h.sn.relationship}) implies
that $(\log(S_{e})-\log[\hfun(N;\betavec)])$ has the same
distribution as~$\epsilon.$ Thus the induced cdf of fatigue life $N$ is
\begin{align}
\nonumber
F_{N}(t; S_{e})&=\Pr(N \le t; S_{e})=\Pr\left [\hfun(N;\betavec)>\hfun(t;\betavec) \right]\\
\nonumber
&=\Pr\left(\log [\hfun(N;\betavec)] > \log [\hfun(t;\betavec)]\right)=\Pr\left(-\log [\hfun(N;\betavec)] < -\log [\hfun(t;\betavec)]\right)\\
\nonumber
  &=\Pr\left(\left(\log(S_{e})-\log [\hfun(N;\betavec)]\right) < \log(S_{e})-\log [\hfun(t;\betavec)]\right)\\
  &=
\label{S.equation:cdf.for.induced.fatigue.life}
  \Psi\left[\frac{S_{e}}{\hfun(t;\betavec)}; \kappavec \right], \,\,\,
  t>0, \, S_{e}>0.
\end{align}

The $p$ quantile of the fatigue-life random variable $N$
is obtained by solving $p=F_{N}[t_{p}(S_{e});S_{e}]$ for $t_{p}(S_{e})$, giving
\begin{align}
\label{S.equation:quantiles.induced.fatigue.life.psi}
t_{p}(S_{e}) = \hfun^{-1} \left (
\frac{S_{e}}{\Psi^{-1}(p; \kappavec)};\betavec \right), \,\,\,
  0<p<1, \,\, S_{e}>0.
\end{align}

The fatigue-life density is $f_{N}(t; S_{e})=dF_{N}(t; S_{e})/dt$
which can be computed from
\begin{align*}
  f_{N}(t; S_{e})&= \frac{dF_{N}(t; S_{e})}{dt} =  \frac{\partial}{d
    t} \Psi\left[\frac{S_{e}}{\hfun(t;\betavec)}   ; \kappavec \right]\\
  &=
  \psi \left[\frac{S_{e}}{\hfun(t;\betavec)}; \kappavec
  \right ] \frac{\partial}{d t}\left[\frac{S_{e}}{\hfun(t;\betavec)}   \right]\\
    &=
  \psi \left[\frac{S_{e}}{\hfun(t;\betavec)}; \kappavec
    \right ] \left[\frac{S_{e}}{\hfun^{2}(t;\betavec)}   \right]
  \left |\frac{\partial \hfun(t;\betavec)}{d t}\right |,
\end{align*}
where $\psi(z; \kappavec ) = d\Psi(z; \kappavec) /dz$.

\subsubsection{The induced fatigue-life model when
  \texorpdfstring{$\hSN$}{log[h(N)]}
has a horizontal asymptote}

When $\hfun(N;\betavec)$ has a horizontal asymptote,
$\lim_{t \to \infty} \hfun(N;\betavec)=\exp(E)>0$. Then
  \begin{align*}
  \lim_{t \to \infty}
F_{N}(t; S_{e})&=
   \lim_{t \to \infty}
   \Psi\left[\frac{S_{e}}{\hfun(t;\betavec)}; \kappavec     \right]\\
  &= \Psi\left[\frac{S_{e}}{\exp(E)};\kappavec \right]<1,
  \end{align*}
  which implies that the cdf $F_{N}(t; S_{e})$ has a discrete atom of probability at  $\infty.$ The size of the discrete
atom is
\begin{align*}
1-\Psi\left(\frac{S_{e}}{\exp(E)};\kappavec \right).
\end{align*}

\subsubsection{The induced fatigue-life model when
  \texorpdfstring{$\hSN$}{log[h(N)]}
 has a vertical asymptote}
When there is a vertical asymptote, as $t$ decreases toward
$\exp(B),$ $\hfun(t;\betavec)$ is unbounded.
Thus
\begin{align*}
 \lim_{t \downarrow \exp(B)}
 F_{N}(t; S_{e})&=0.
\end{align*}
Consequently, $\exp(B)$ is a threshold parameter.

\subsection{Some History and an Alternative Way to View the
  Relationship among Fatigue-Strength  cdfs, Fatigue-Life cdfs, and
  their Respective Quantile Functions}
\label{S.section:history.alternative.view}
Early researchers studying and developing statistical modeling and
methods for experimental fatigue data recognized the existence of
and relationships among fatigue-strength cdfs, fatigue-life cdfs,
and their respective quantile functions. As mentioned in
Section~\ref{section:contributions} of the main paper, examples
include
\citet{FreudenthalGumbel1956,Weibull1956,Armitage1961,BastenaireBastienPomey1961},
and \citet{CastilloGalambos1987}. It appears, from the examples that
they present in their publications, that much of the knowledge that
early researchers obtained about proper methods for modeling fatigue
data came from relatively large fatigue-life data sets that were
obtained across numerous research projects (some of which are
presented in Figure~\ref{S.figure:datasets.contrasting.spread.fl.fs}
and Figures~\ref{S.figure:LognormalWeibullCompareSet01}--\ref{S.figure:LognormalWeibullCompareSet03}).
They made extensive use
of N versus S scatter plots and probability plots (both Weibull and
lognormal) based on nonparametric estimates of the fatigue-life
distribution cdf (sometimes called plotting positions) at each level of stress used
in the experiments. Observations about special features like
curvature in \SN{} plots and changes in distribution spread as a function of stress
in probability plots
were noted. However, the early work used notation and terminology
somewhat different from what is in common use today.

The purpose of this section is to review some of those early ideas
and link them to the more modern presentation in our paper.
\citet{Armitage1961} and \citet{BastenaireBastienPomey1961} provide
a nice summary of the work done in the 1950s and we will introduce
some notation from the latter to help establish the links.
\citet{BastenaireBastienPomey1961}, for example, say
\begin{align}
  \textrm{Probability of fracture} &= F(S, N)
\end{align}
and in other works, the function is written as $P(S, N)$ and has
been called the ``PSN function'' or the ``PSN field.''
One difficulty with the $P(S, N)$ and similar notation is that it can lead to
confusion about what is random and what is fixed in the statistical
model \citep[although many of the early papers, e.g,][did explicitly define
  notation for fatigue-life and fatigue-strength cdfs or corresponding
  survival functions]{Weibull1956}.

To link this older notation to the more precisely specified
parametric probability model and technical material in our paper
and this appendix, suppose that $P(u, v)$ is a function determined
by the \SN{} regression relationship and specified probability model
for fatigue life (when the fatigue-life model is specified as in
Section~\ref{section:statistical.models.estimate.fatigue.life}) or
fatigue strength (when the fatigue-strength model is specified as in
Section~\ref{section:alternative.approach.modeling.sn.data}).  The
dummy argument $u$ is associated with strain, stress or strength, depending
on the context, and $v$ is associated time (number of cycles).

For concreteness, we define $P(u, v)$ according to the
particular rather general probability model introduced in
Section~\ref{S.section:specifying.fatigue.strength.distribution}
where $S=\hfun(N;\betavec)$ is, for fixed $\betavec$, a positive
monotonically decreasing function of $N$ that serves as an \SN{}
regression relationship and a scale parameter for the
fatigue-strength random variable $X$
while $\Psi(z; \kappavec)$ is the cdf of a
random variable with a scale parameter equal to 1 and one or more
constant shape parameters in $\kappavec.$ The idea behind this model
is that the explanatory variable $N_{e}$ changes the scale but not
the shape of the strength distributions. One reason for this
formulation is that all of the sometimes-desirable compatibility
conditions (e.g., that quantile lines are decreasing and do not
cross)
will always hold. To simplify the presentation in the
rest of this section we suppress the dependency of
$\hfun(N;\betavec)$ on $\betavec$ and the dependency of $\Psi(z;
\kappavec)$ on $\kappavec$.

Then the \textit{specified}
fatigue-strength cdf (\ref{S.equation:FS.cdf.scale.shape})
can be written as
\begin{align*}
  F_{X}(x; N_{e})&= \Pr(X \le x; N_{e})= P(x, N_{e})= \Psi\left(\frac{x}{h(N_{e})}\right), \,\,\,
  x>0, \,  N_{e}>0.
\end{align*}
As in (\ref{S.equation:quantiles.specified.fatigue.strength}), the
$p$ quantile of the fatigue-strength random variable $X$ is
\begin{align}
\label{S.equation:quantiles.specified.fatigue.strength.P}
x_{p}(N_{e}) &= P_{X}^{-1}(p, N_{e})= \hfun(N_{e})\,
\Psi^{-1}\left(p \right), \,\,\,
  0<p<1, \,\, N_{e}>0.
\end{align}
where $P_{X}^{-1}(p, N_{e})$ is the inverse of $P(u, v)$ for the $u$
argument.

 By changing the names of the arguments in $P(x, N_{e})$, as in
 Section
 \ref{S.section:induced.fatigue.life.distribution.neither.vertical.nor.horizontal},
 the \textit{induced} cdf of fatigue life $N$ is
\begin{align*}
F_{N}(t; S_{e})&=\Pr(N \le t; S_{e})=P(S_{e}, t) =\Psi\left(\frac{S_{e}}{\hfun(t)} \right), \,\,\,
  t>0, \, S_{e}>0.
\end{align*}
It is important to note that $\hfun(t)$ in this cdf is \textit{not}
a scale parameter for $N$; instead it controls the shape and spread
of the distribution of $N$.
As in (\ref{S.equation:quantiles.induced.fatigue.life.psi}), the $p$
quantile of the fatigue-life random variable $N$ is
\begin{align}
\label{S.equation:quantiles.induced.fatigue.life.P}
t_{p}(S_{e}) = P_{N}^{-1}(p, S_{e}) =\hfun^{-1} \left (
\frac{S_{e}}{\Psi^{-1}(p)} \right), \,\,\,
  0<p<1, \,\, S_{e}>0.
\end{align}
where $P_{N}^{-1}(p, S_{e})$ is the inverse of $P(u, v)$ for the $v$
argument.

Now if one equates $S_{e}$ to
(\ref{S.equation:quantiles.specified.fatigue.strength.P}) and solves
for $N_{e}$, the result is the same as
(\ref{S.equation:quantiles.induced.fatigue.life.P}). This establishes
that the quantile curves for the fatigue-strength distribution
are the same as those for the fatigue-life distribution.

Besides being technically interesting, the results in this section
have practical value for the implementation of statistical methods
for the analysis of fatigue data. In particular
\begin{itemize}
  \item
  Once one has an algorithm to compute the fatigue-strength cdf
  (fatigue-life cdf), that  same algorithm can be used to compute the
  fatigue-life cdf (fatigue-strength cdf) by just changing the
  definition of the arguments of the function.
\item
  Corresponding pdfs will not have the same form, but can be
  obtained by differentiating the cdf by the appropriate argument
  (depending on which pdf is needed).
\item
  The existence of closed-form expressions for the quantile
  functions depends whether closed-form expressions for the needed
  inverse functions are available or not. To make a plot of quantile
  curves, only one algorithm is required and if quantile estimates (and
  confidence/credible intervals) are needed and no close-for
  expression is available, simple root-finding methods can be used.
\end{itemize}

\section{A Comparison of \SN{} Relationship Shapes and How to Choose
the Best One}
\label{S.section:comparison.sn.model.shapes}

\begin{figure}
\begin{tabular}{cc}
\phantom{MM}Coffin Manson (a) & \phantom{MM}Nishijima~(b) \\[-2.2ex]
\rsplidapdffiguresize{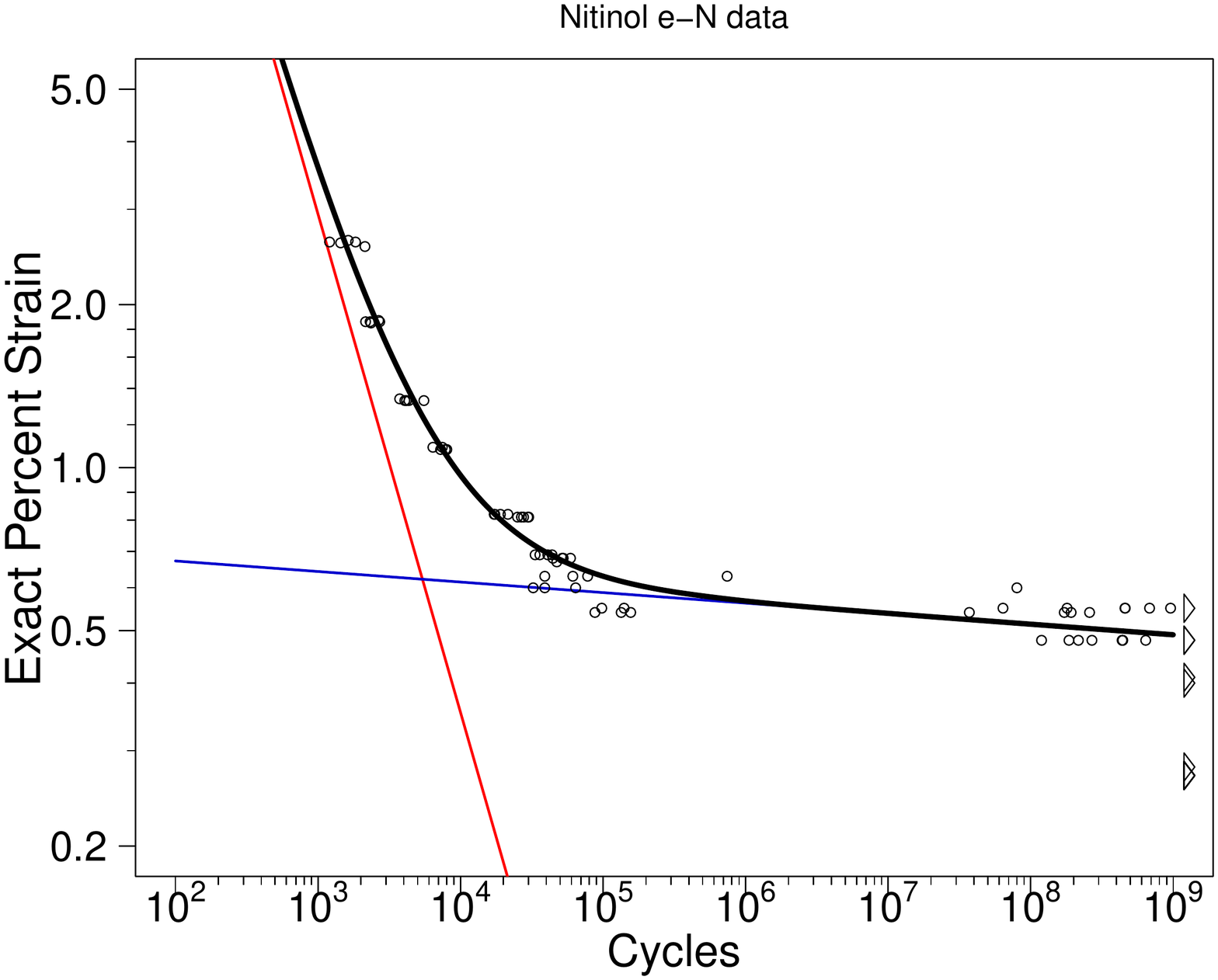}{3.25in}&
\rsplidapdffiguresize{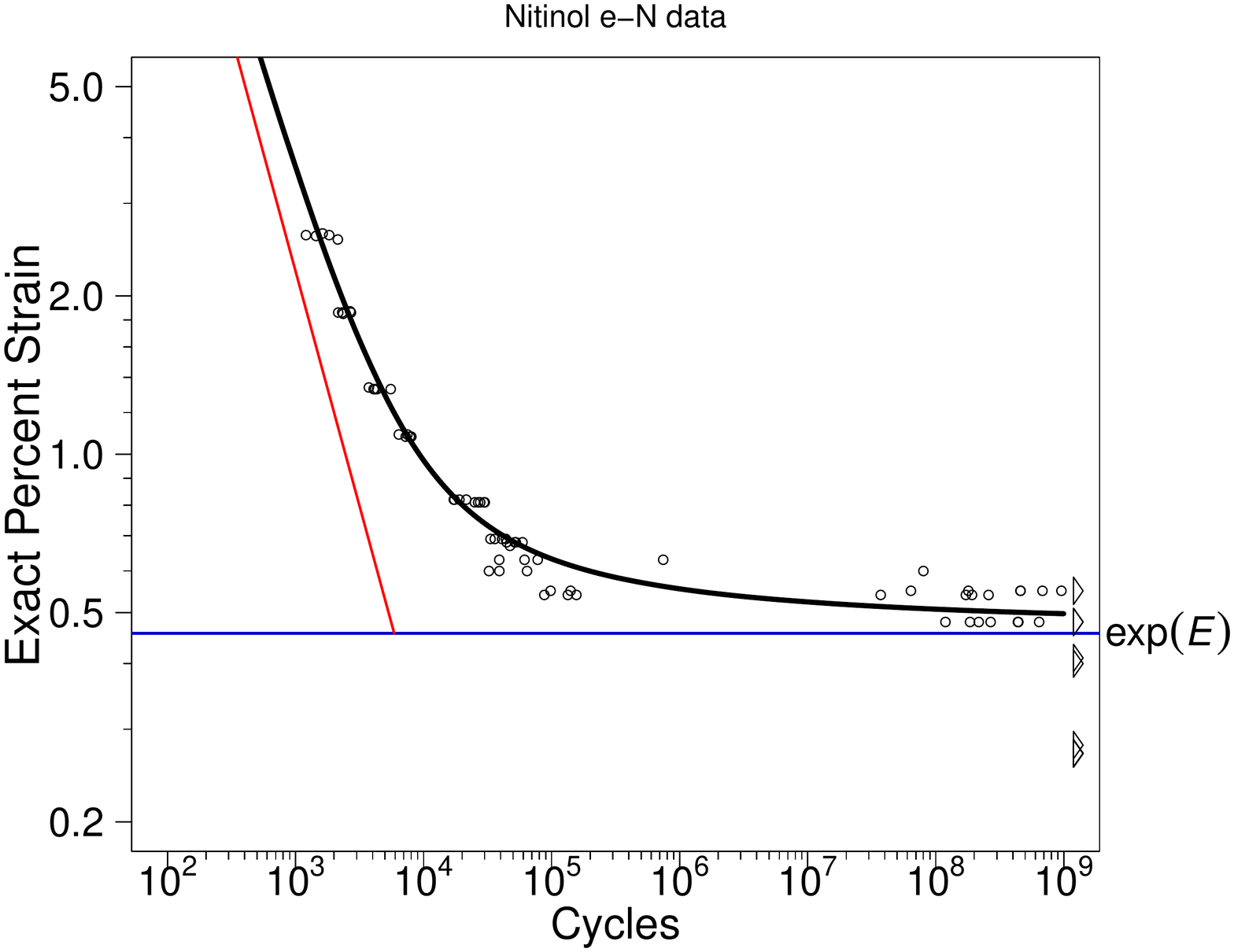}{3.25in}\\
\phantom{MM}Box--Cox (c) & \phantom{MM}Rectangular Hyperbola (d) \\[-2.2ex]
\rsplidapdffiguresize{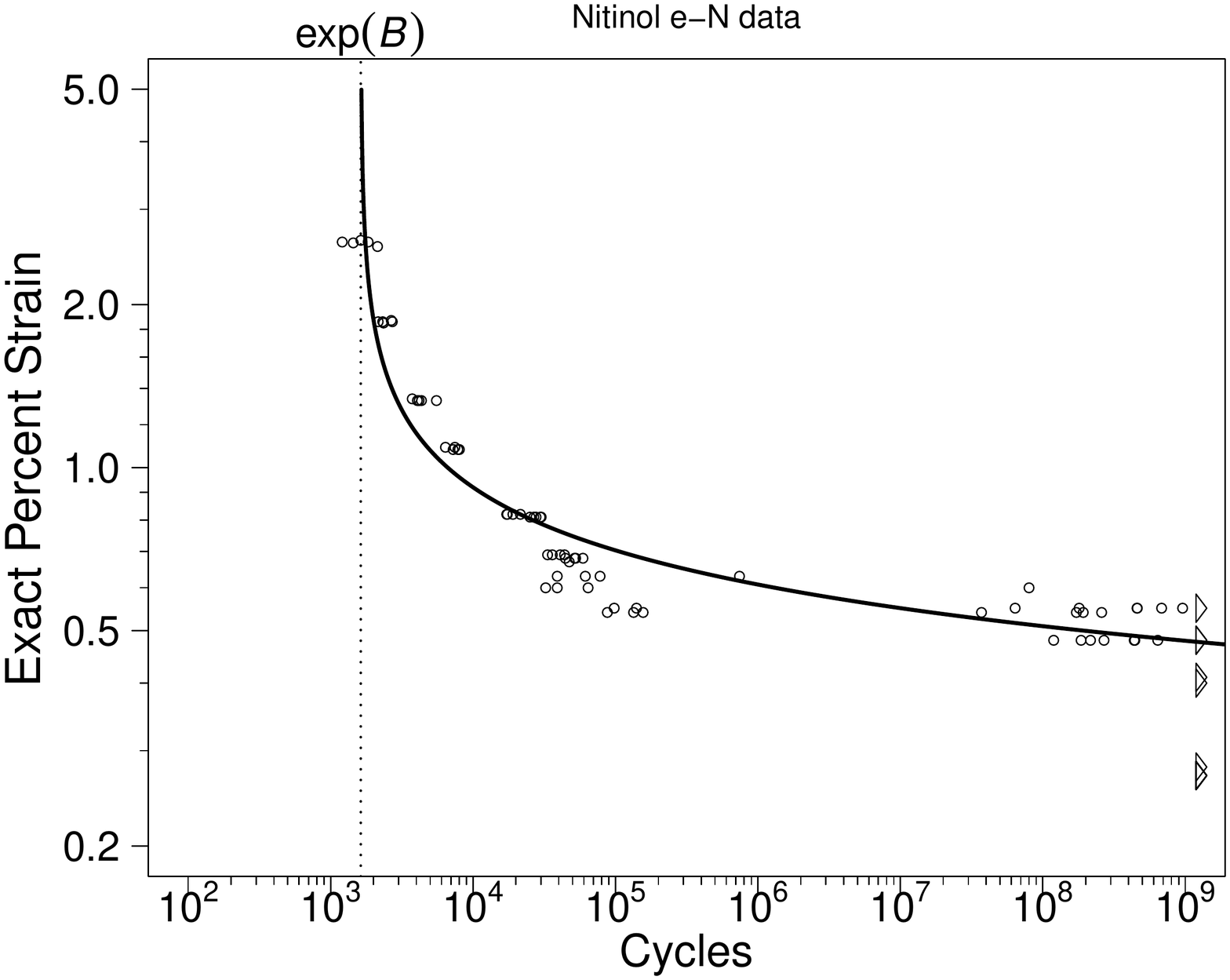}{3.25in}&
\rsplidapdffiguresize{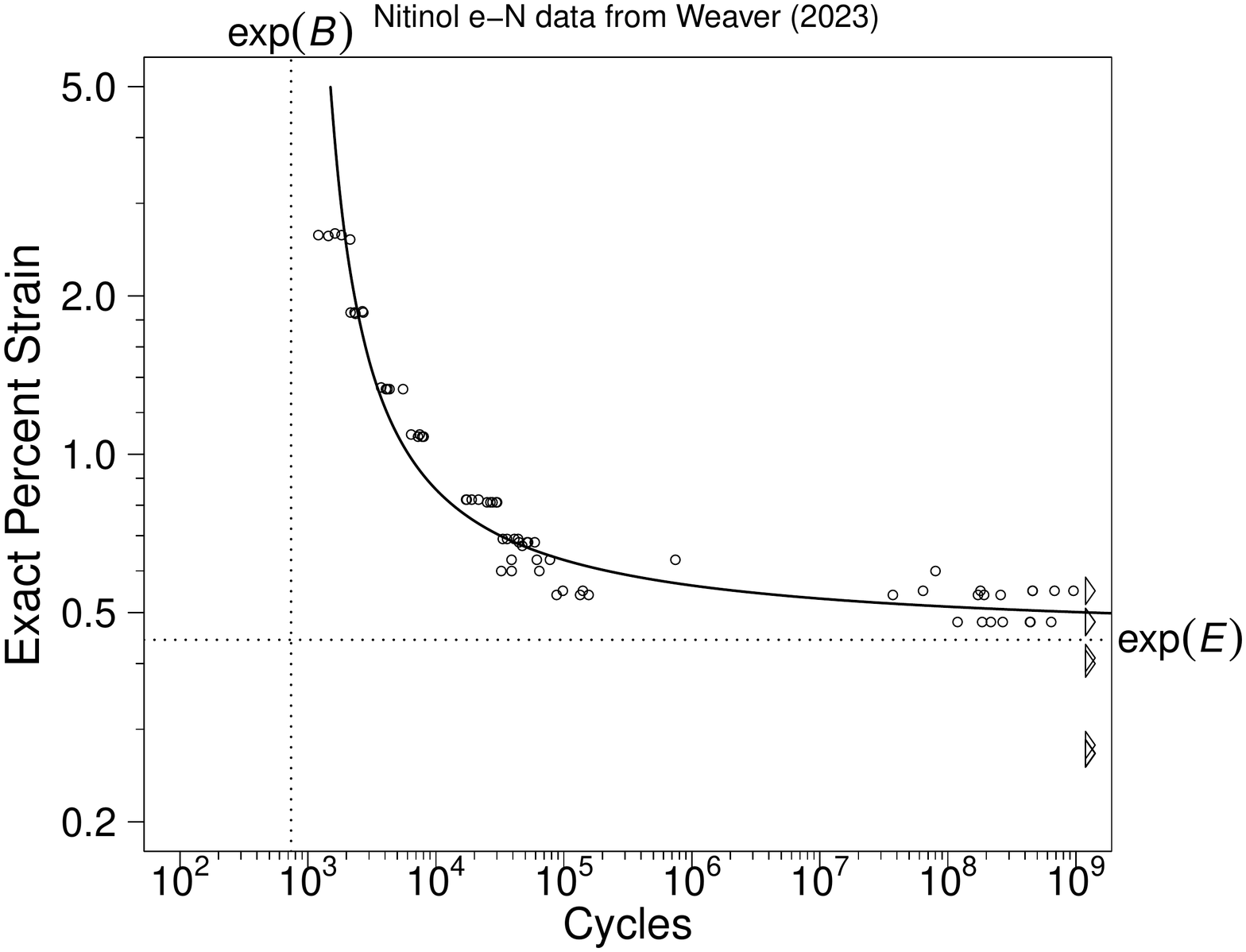}{3.25in}
\end{tabular}
\caption{Median \SN{} estimates for models fit to the complete nitinol
  data for the Coffin--Manson relationship with \textit{no coordinate
  asymptotes}~(a); Nishijima relationship with
  a \textit{horizontal asymptote}~(b); Box--Cox relationship with a
  \textit{vertical asymptote}~(c); and a rectangular hyperbola
  relationship with \textit{both horizontal and vertical asymptotes}~(d).}
\label{S.figure:sn.models.different.asymptotes.sanity}
\end{figure}

The existence of coordinate asymptotes is an important
distinguishing characteristic for the \SN{} relationships.
Figure~\ref{S.figure:sn.models.different.asymptotes.sanity}
shows the \SN{} relationships for four of the models considered in
the main paper for all four combinations of the existence of coordinate
asymptotes.

As described and illustrated in
Section~\ref{section:using.residuals.as.model.checking.diagnostics}
and Examples~\ref{example:modeling.Ti64.sn.data}
and~\ref{example:modeling.superelastic.nitinol.sn.data} of the
main paper, residual analysis is an important tool for statistical model
building. Providing further compelling support for the use of
residual analysis,
Figure~\ref{S.figure:sn.models.different.asymptotes.sanity}
shows the fitted median curve for four different \SN{} models, fit
to nitinol data used in \citet{Weaver_etal2022} (including additional data taken at a different
laboratory).
\begin{figure}
\begin{tabular}{cc}
\phantom{MM}Coffin Manson (a) & \phantom{MM}Nishijima~(b) \\[-3.2ex]
\rsplidapdffiguresize{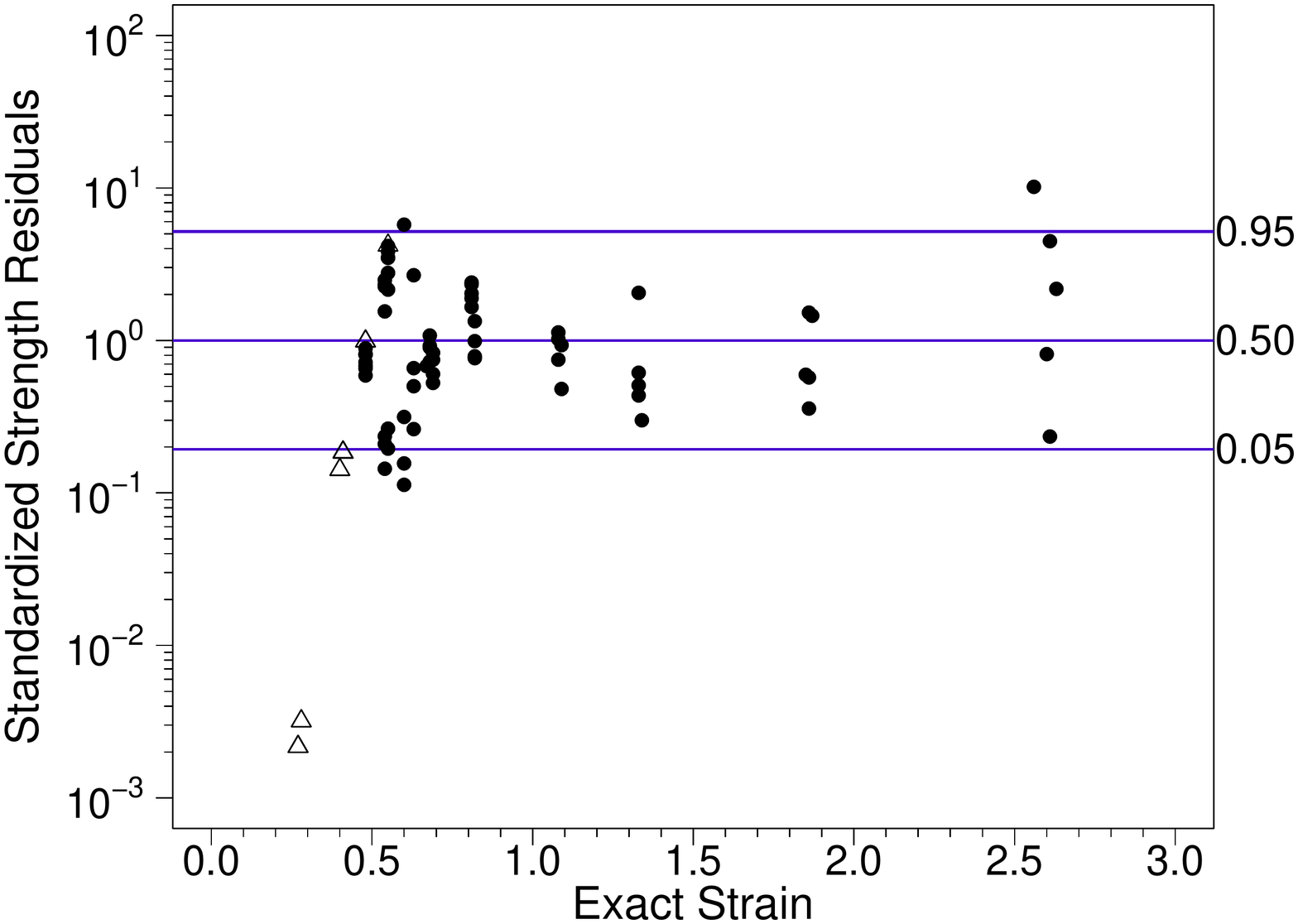}{3.25in}&
\rsplidapdffiguresize{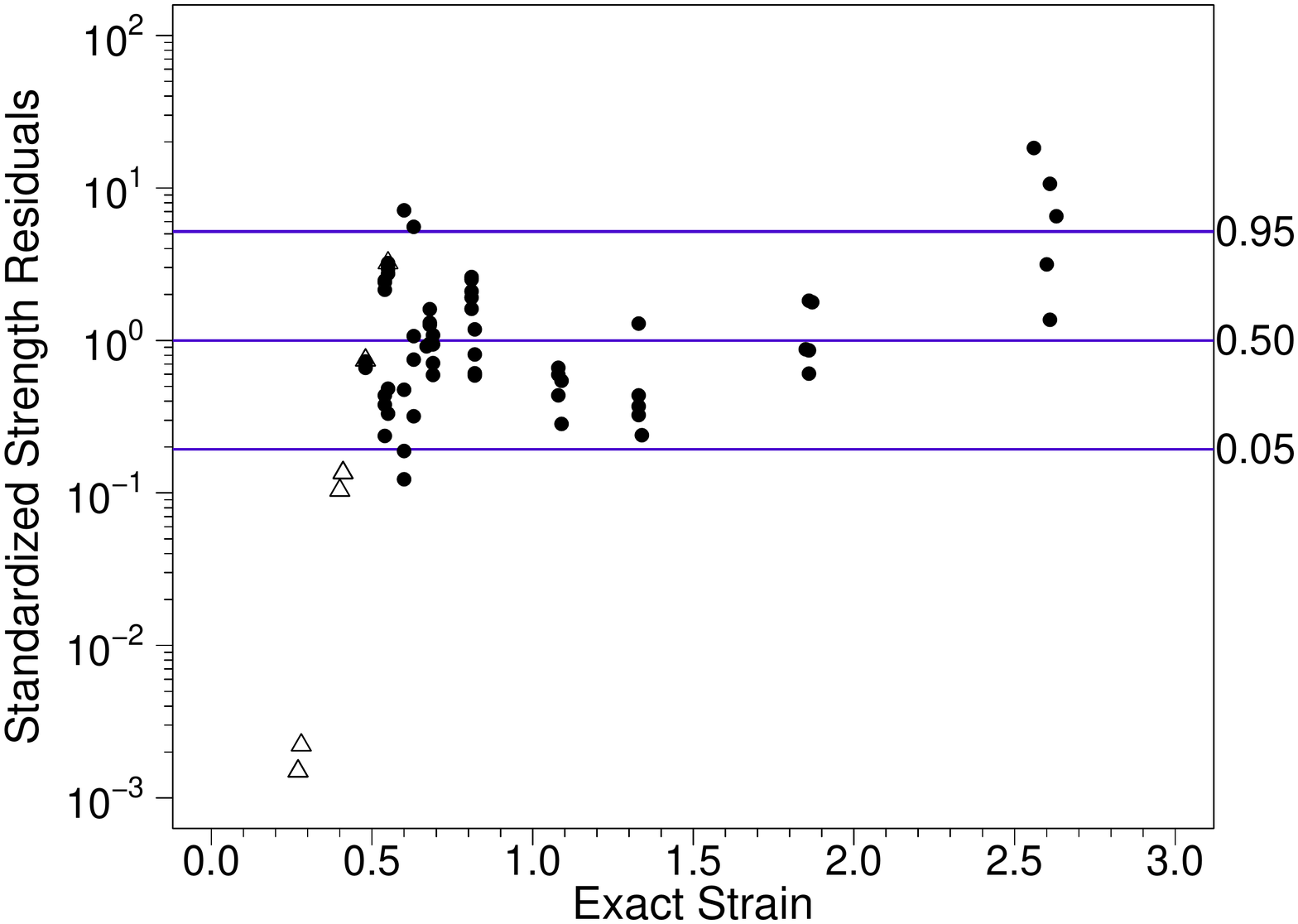}{3.25in}\\
\phantom{MM}BoxCox (c) & \phantom{MM}Rectangular Hyperbola (d) \\[-3.2ex]
\rsplidapdffiguresize{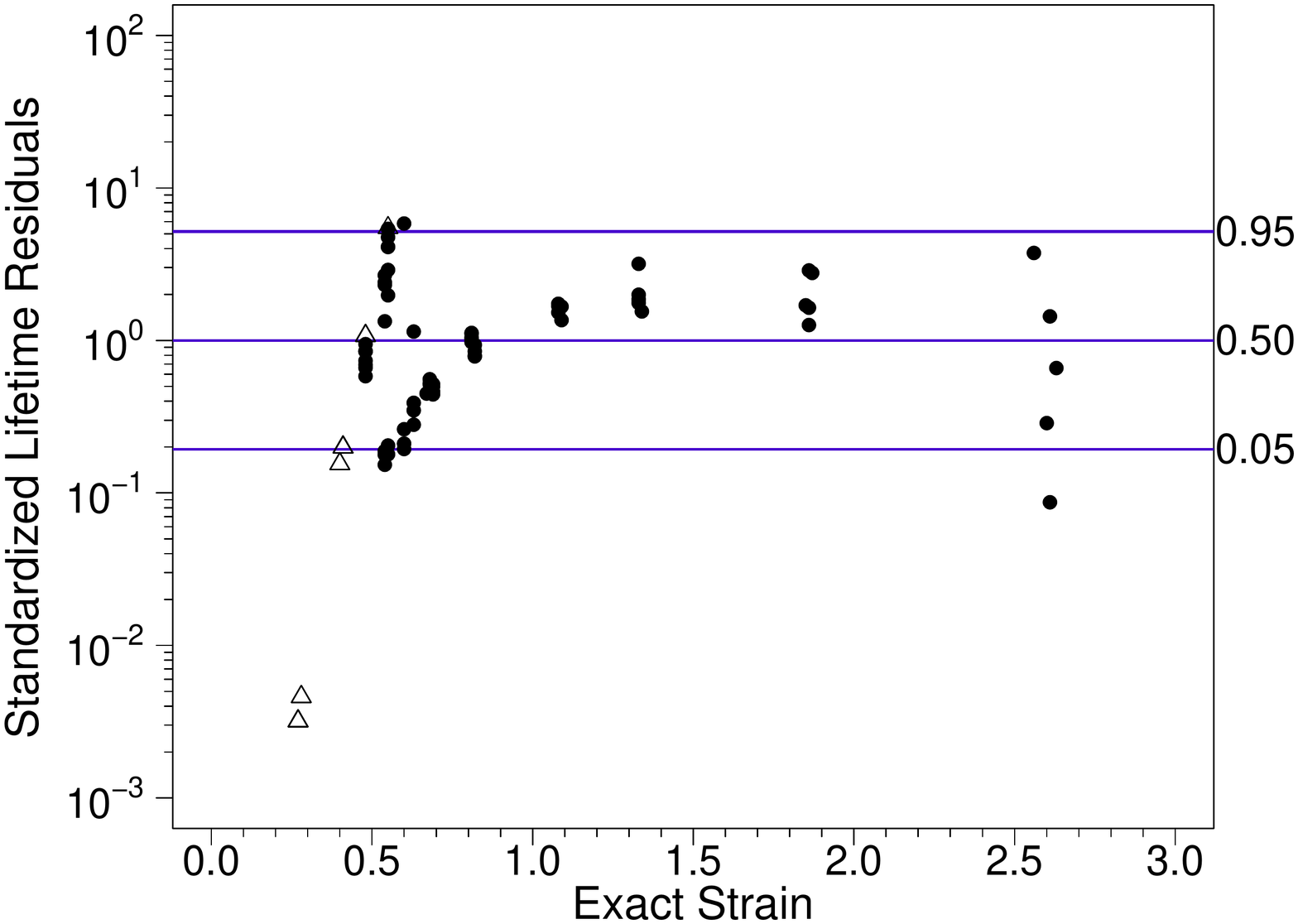}{3.25in}&
\rsplidapdffiguresize{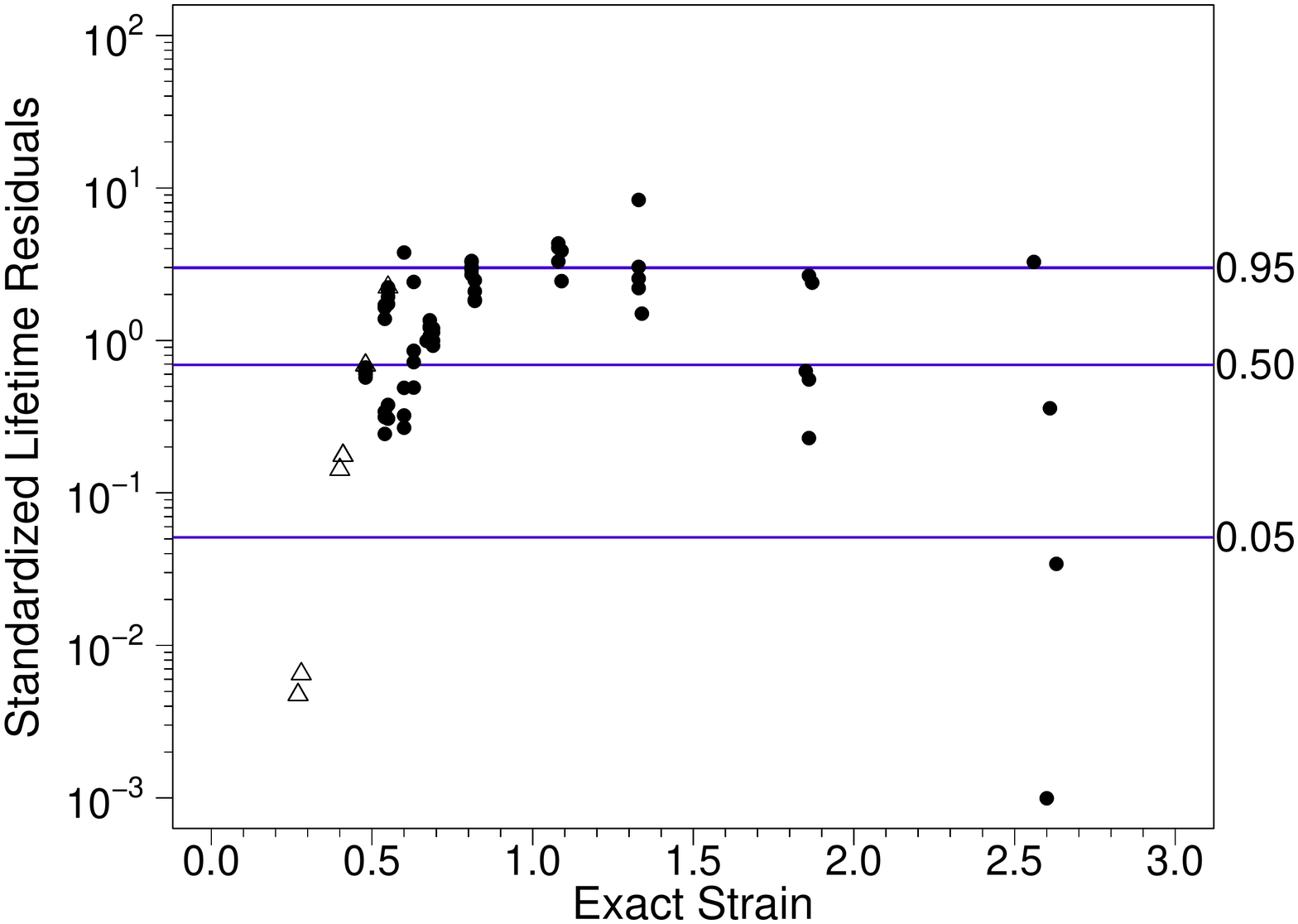}{3.25in}
\end{tabular}
\caption{Residuals versus strain plots for models fit to the complete
  nitinol data: Coffin--Manson~(a); Nishijima~(b); Box--Cox~(c); and rectangular
  hyperbola~(d).}
\label{S.figure:sn.models.different.asymptotes.residuals}
\end{figure}
There is some evidence of lack of fit for the Box--Cox model and to
a lesser degree for the rectangular hyperbola model. The residuals
versus strain plots in
Figure~\ref{S.figure:sn.models.different.asymptotes.residuals}
provide more complete and accurate information about the adequacy of
the different models. In particular, there appears to be an
upside-down-U pattern for the Box--Cox and to a lesser degree for
the rectangular hyperbola models, perhaps caused by the steeply
increasing (due to the vertical asymptote) \SN{} curve in the
high-strain region. The residual plots highlight
the more subtle differences between the Coffin--Manson and Nishijima
\SN{} relationships. Taking into account the runouts,
across different values of stress/strain the distribution
of the residuals for the Coffin--Manson have empirical distributions
that are more constant than those for the Nishijima model.

\section{Examples Comparing Lognormal and Weibull Distributions Fit
  to \SN{} Data from Different Materials and Specimen Types}
\label{S.section:examples.comparing.lognormal.and.weibull.distributions.fit.sn.data}

\begin{figure}
\begin{tabular}{cc}
\multicolumn{2}{c}{Annealed Aluminum Wire \SN{} Data}\\[-3.2ex]
\rsplidapdffiguresize{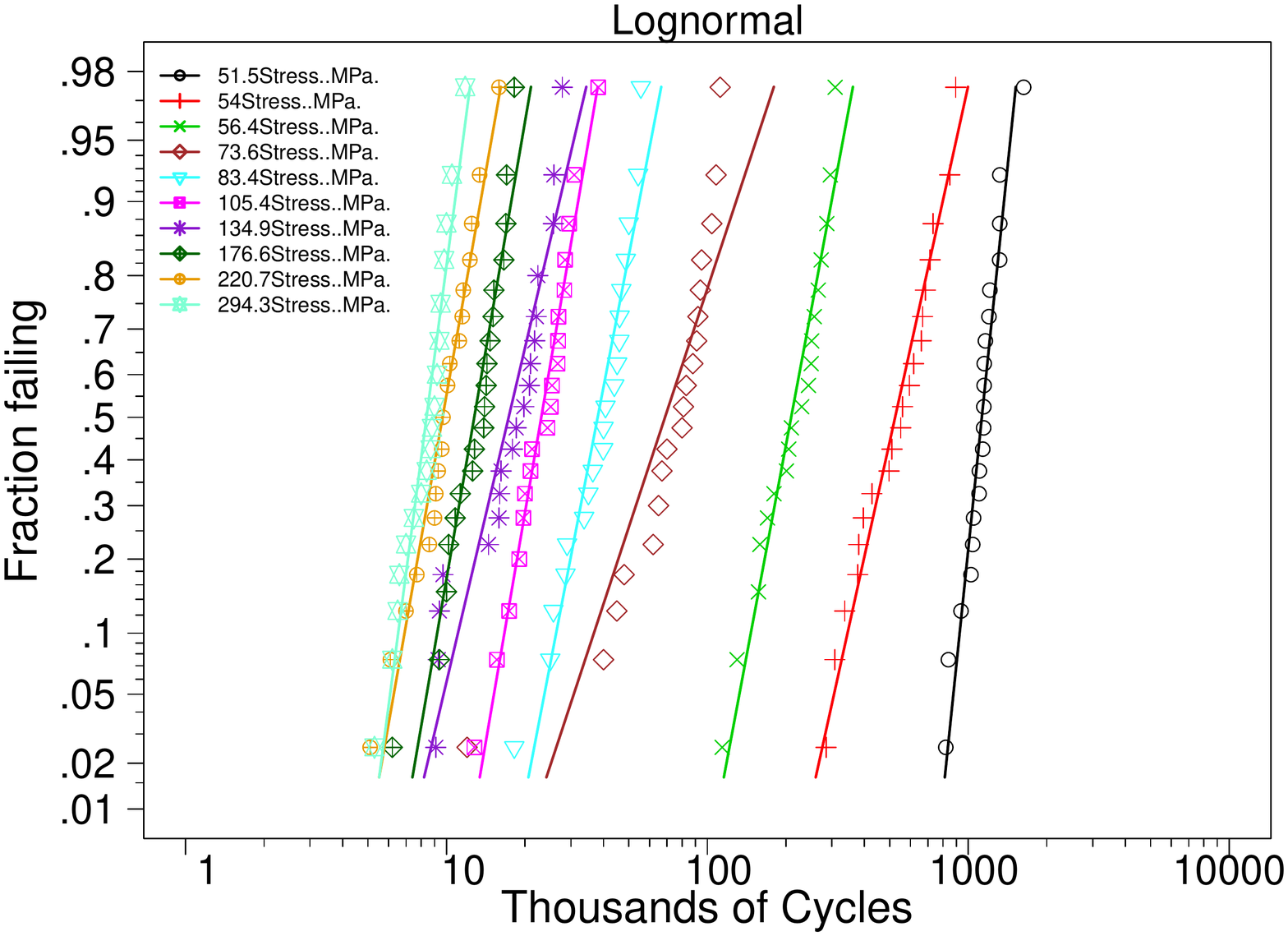}{3.25in}&
\rsplidapdffiguresize{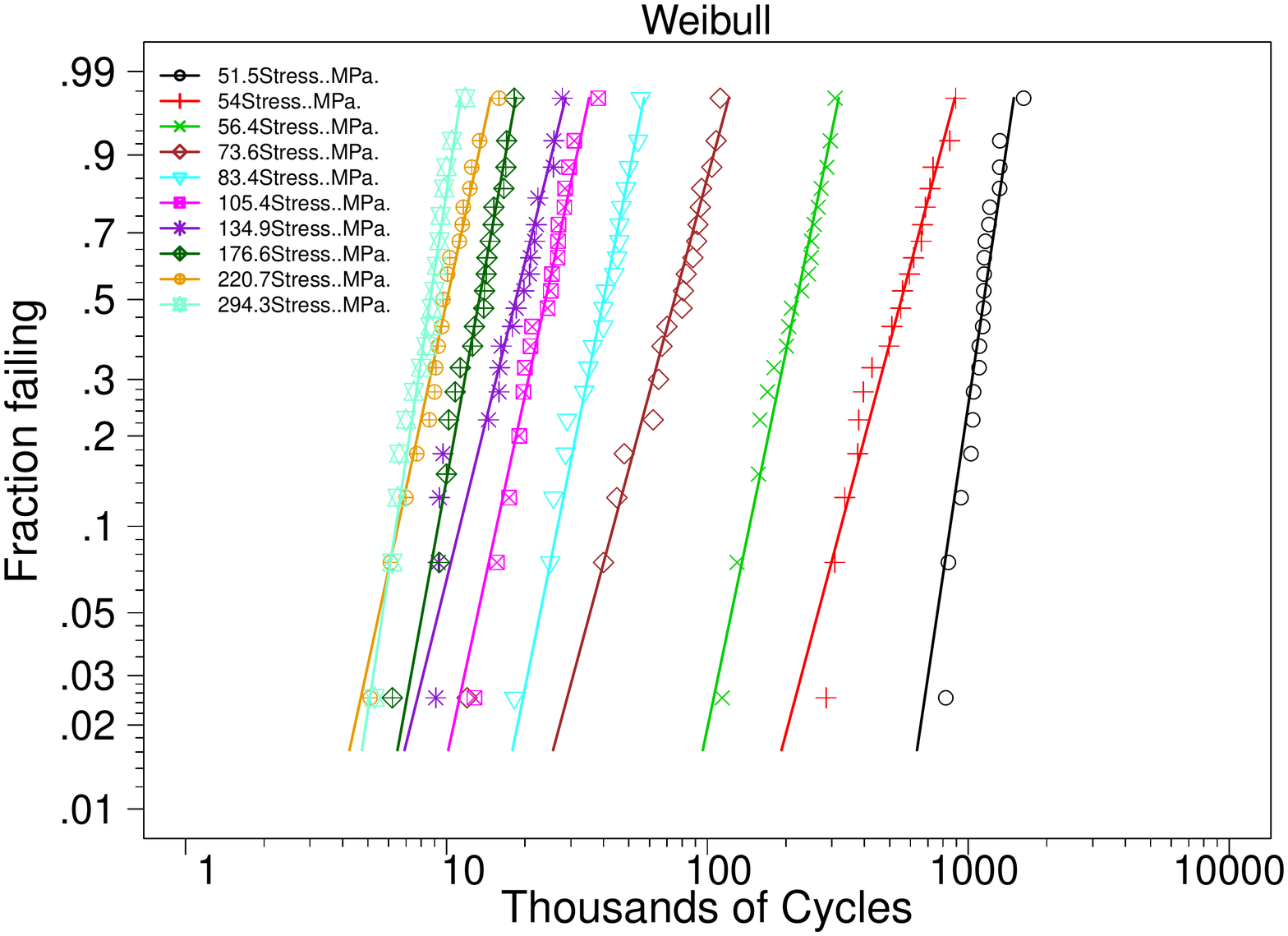}{3.25in}\\
\multicolumn{2}{c}{Annealed Electrolytic Copper Wire \SN{} Data}\\[-3.2ex]
\rsplidapdffiguresize{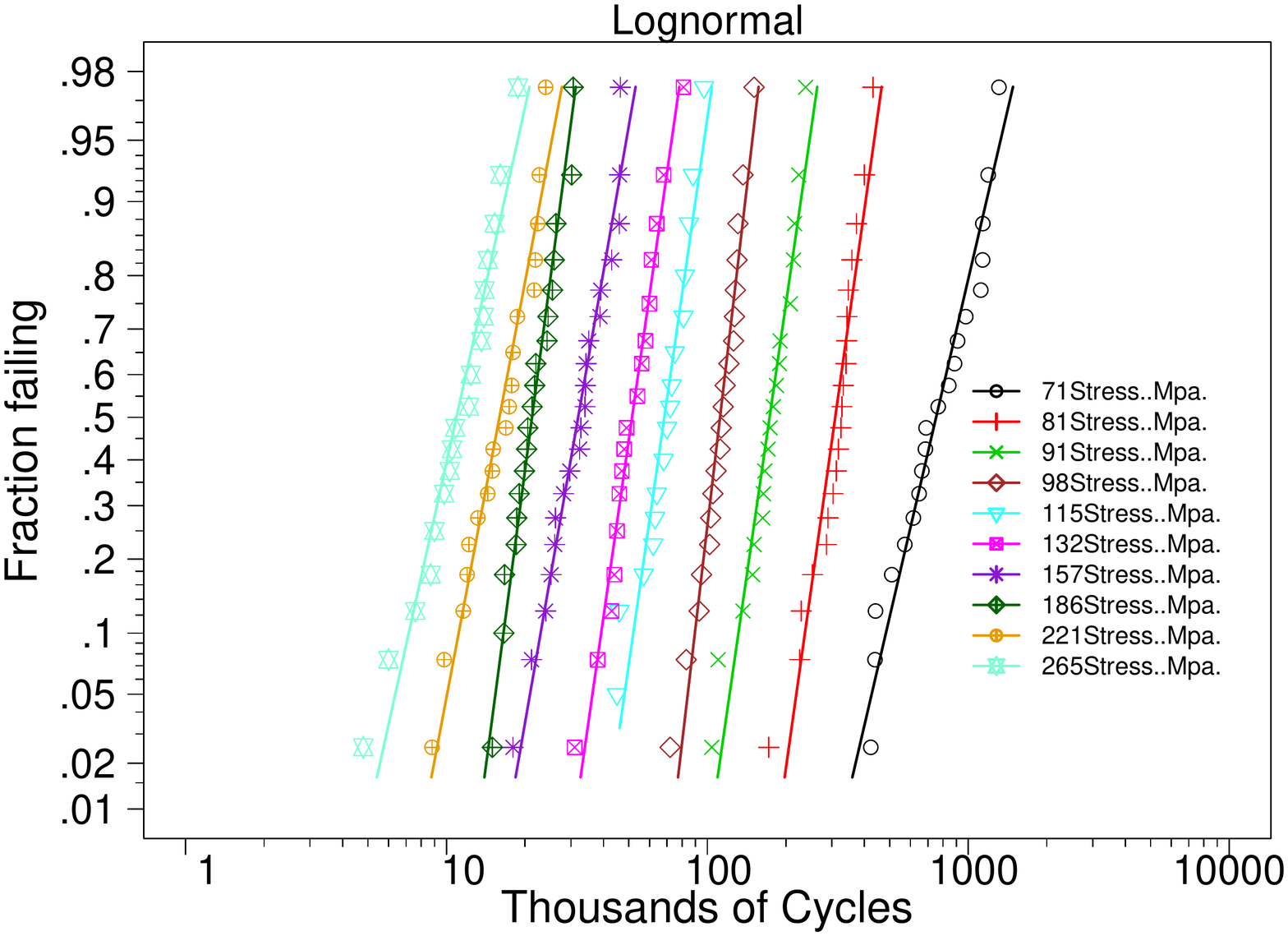}{3.25in}&
\rsplidapdffiguresize{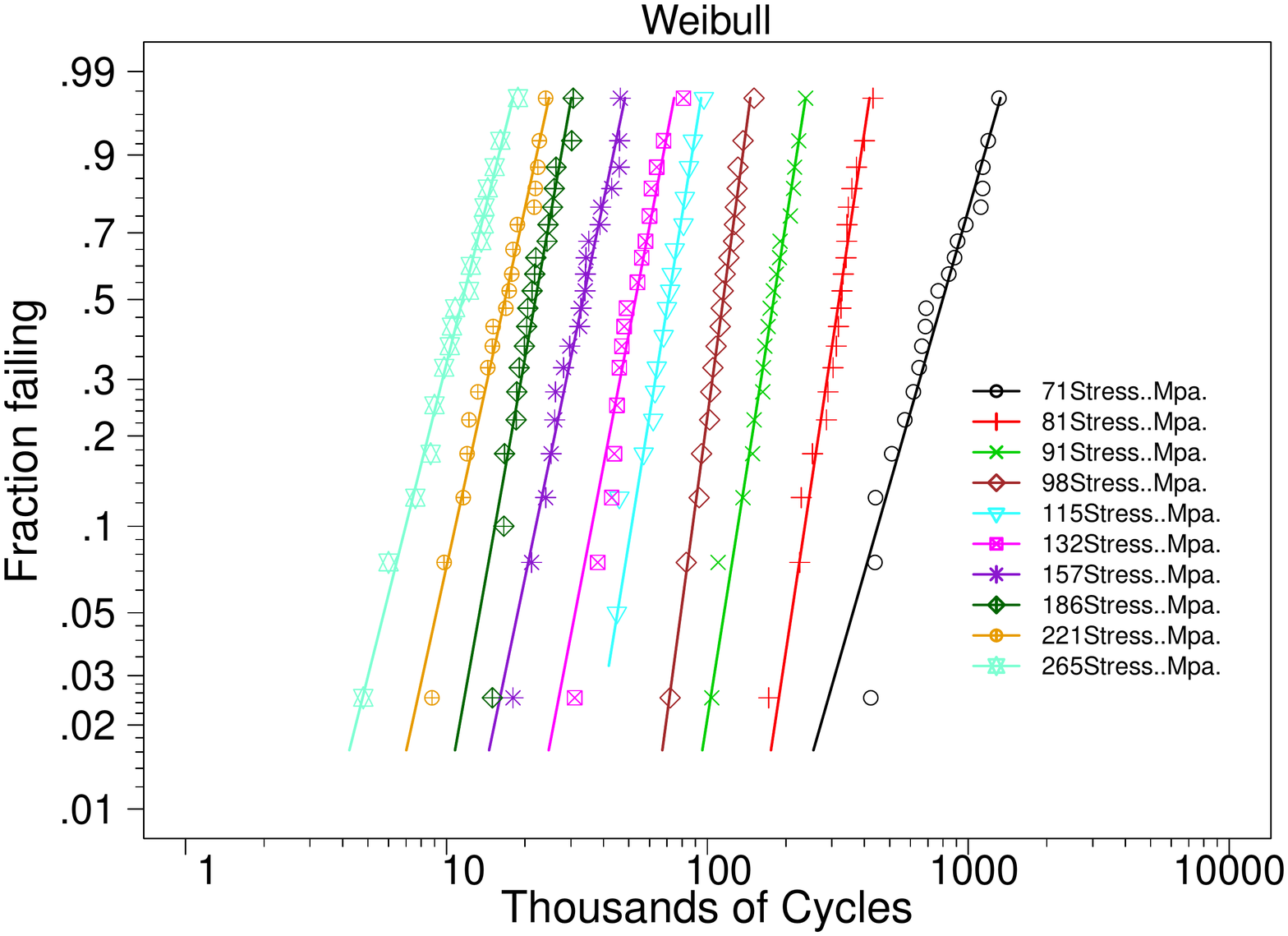}{3.25in}\\
\multicolumn{2}{c}{Annealed ARMCO Iron Wire \SN{} Data}\\[-3.2ex]
\rsplidapdffiguresize{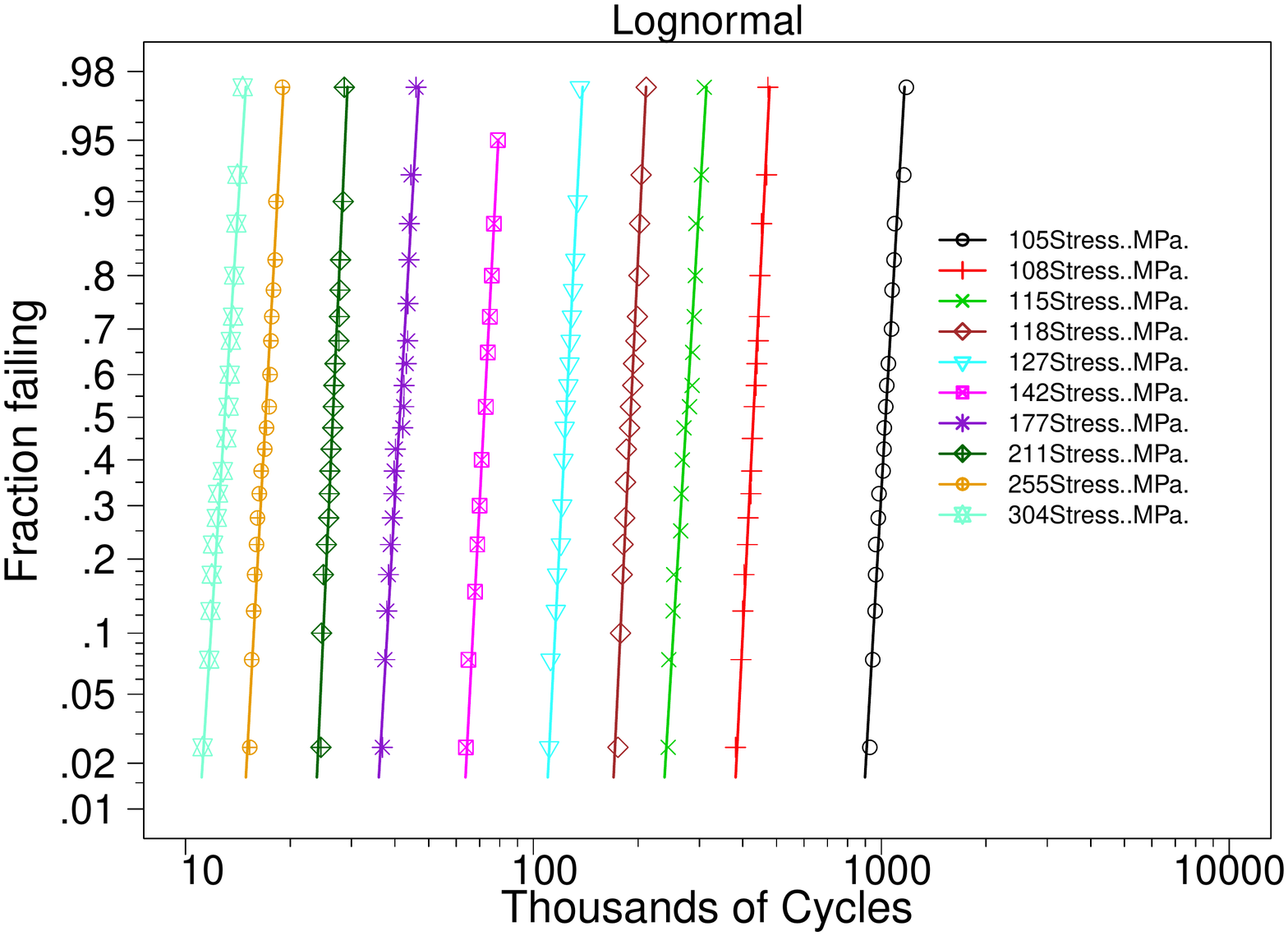}{3.25in}&
\rsplidapdffiguresize{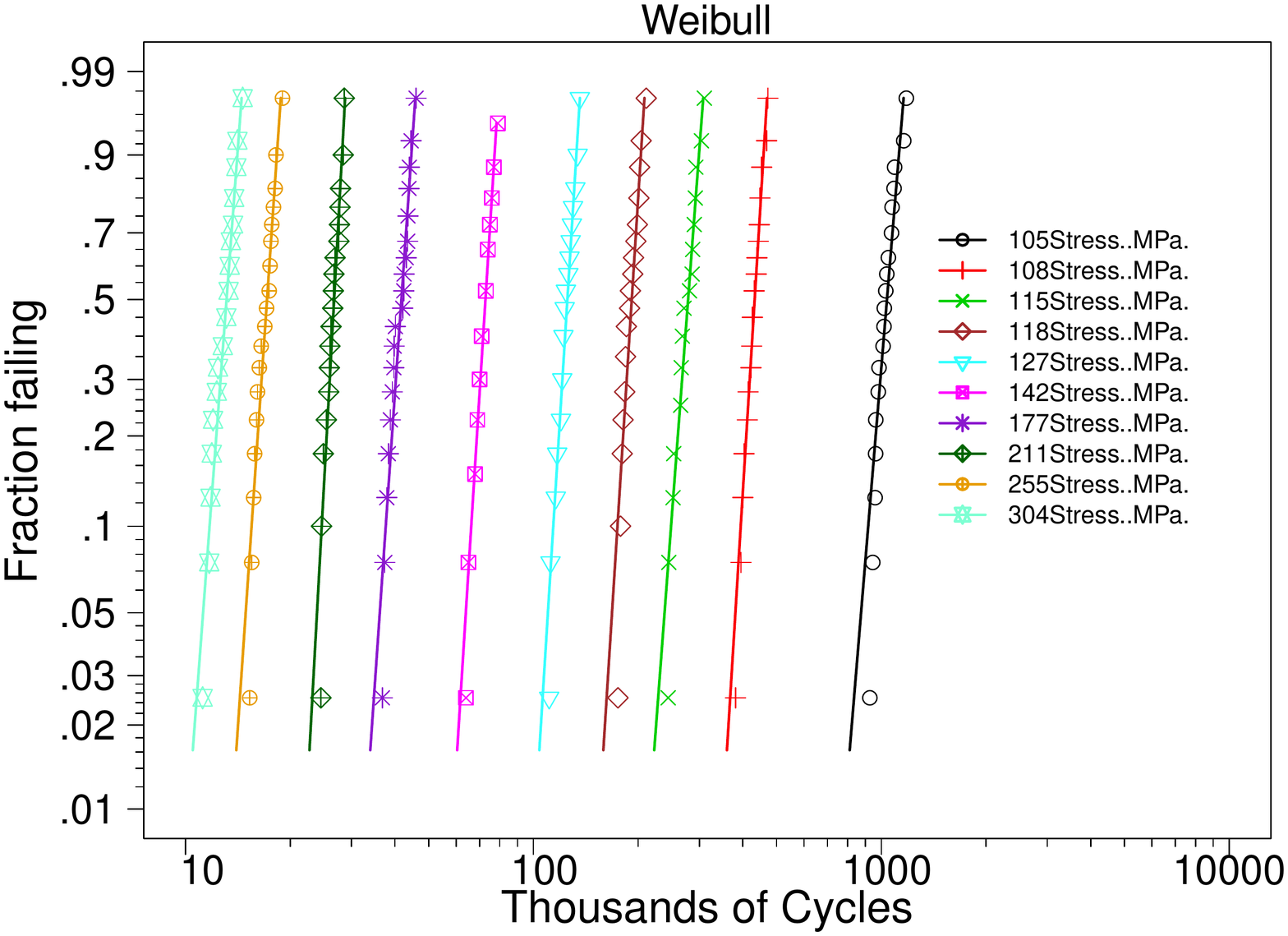}{3.25in}
\end{tabular}
\caption{Comparison of lognormal (left) and Weibull (right)
  distribution probability plots for the
  annealed aluminum wire,
annealed electrolytic copper wire, and annealed ARMCO iron wire \SN{} data.}
\label{S.figure:LognormalWeibullCompareSet01}
\end{figure}

\begin{figure}
\begin{tabular}{cc}
\multicolumn{2}{c}{Aluminum 6061-T6 Coupons \SN{} Data}\\[-3.2ex]
\rsplidapdffiguresize{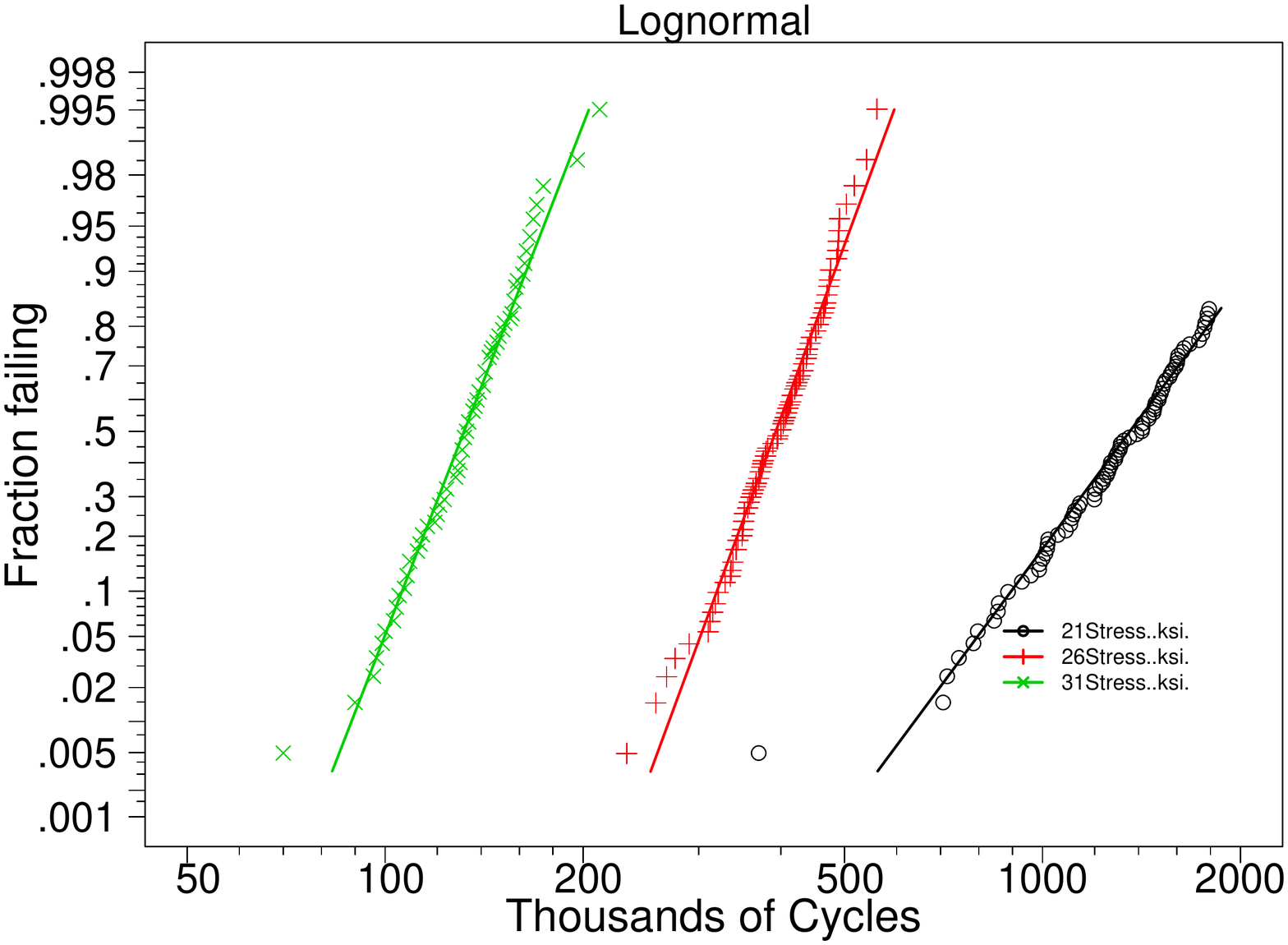}{3.25in}&
\rsplidapdffiguresize{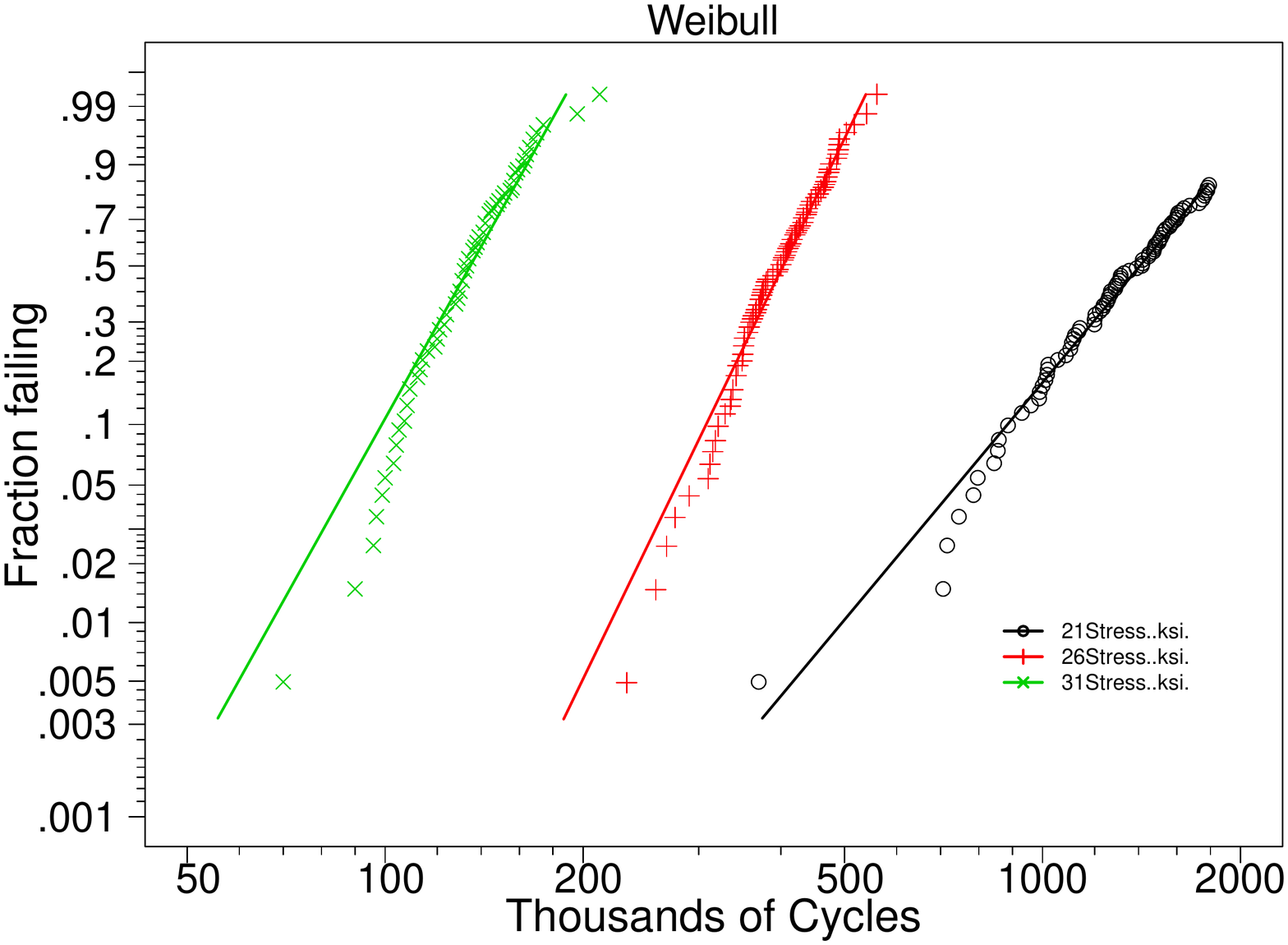}{3.25in}\\
\multicolumn{2}{c}{C35 Steel Slightly Notched Specimens  \SN{} Data (systematic
  subset of stress levels)}\\[-3.2ex]
\rsplidapdffiguresize{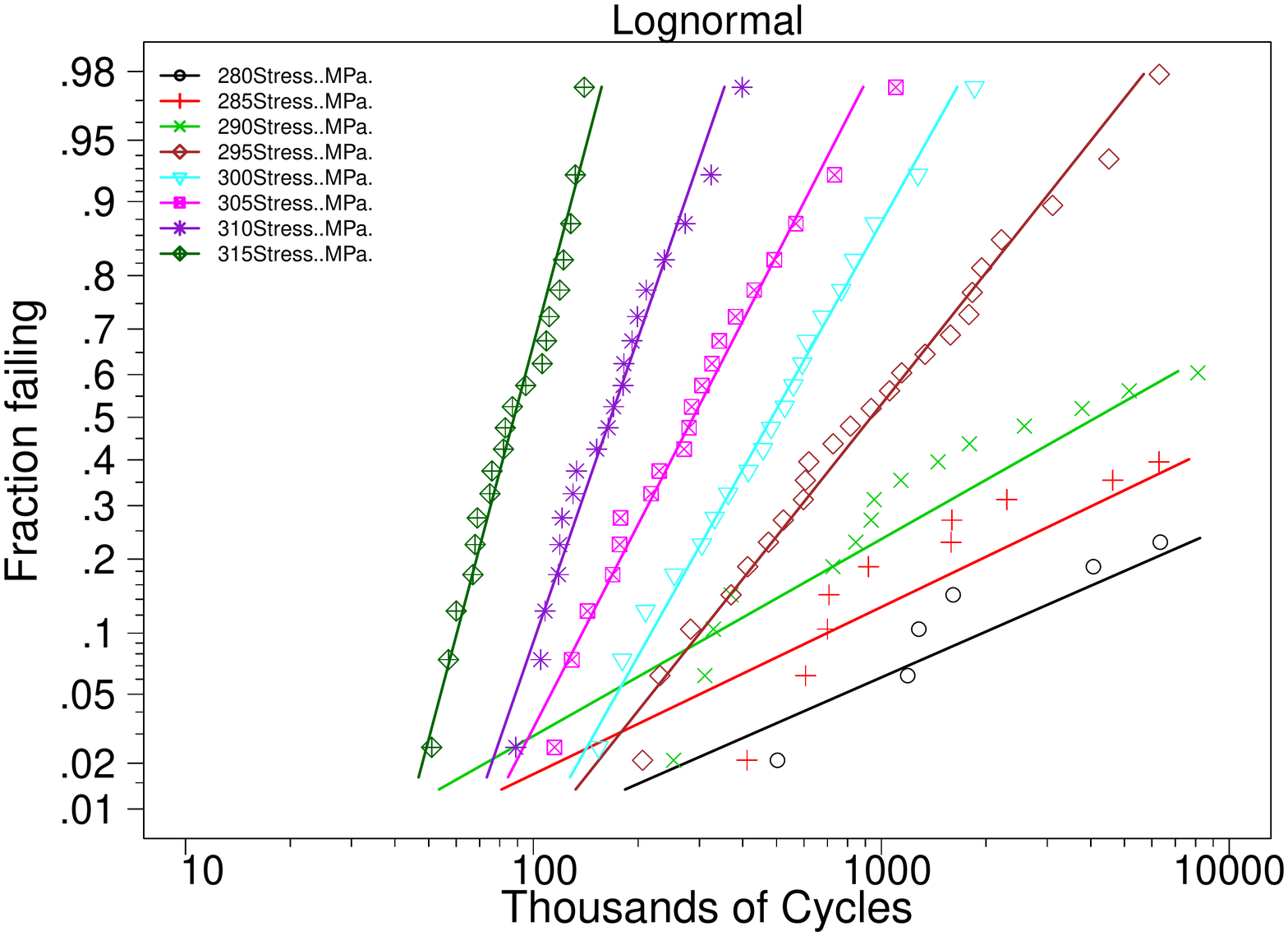}{3.25in}&
\rsplidapdffiguresize{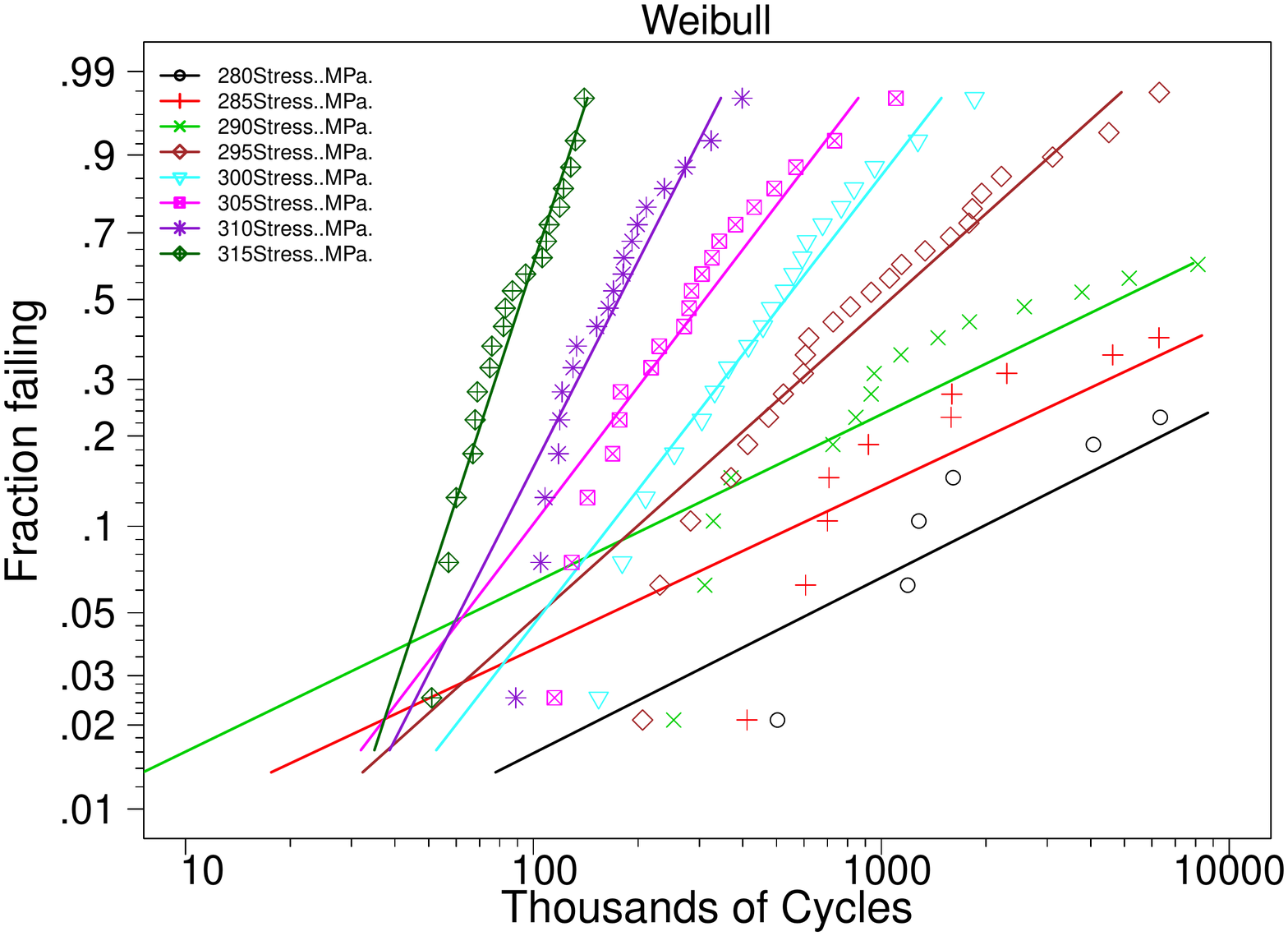}{3.25in}\\
\multicolumn{2}{c}{Holman Concrete \SN{} Data}\\[-3.2ex]
\rsplidapdffiguresize{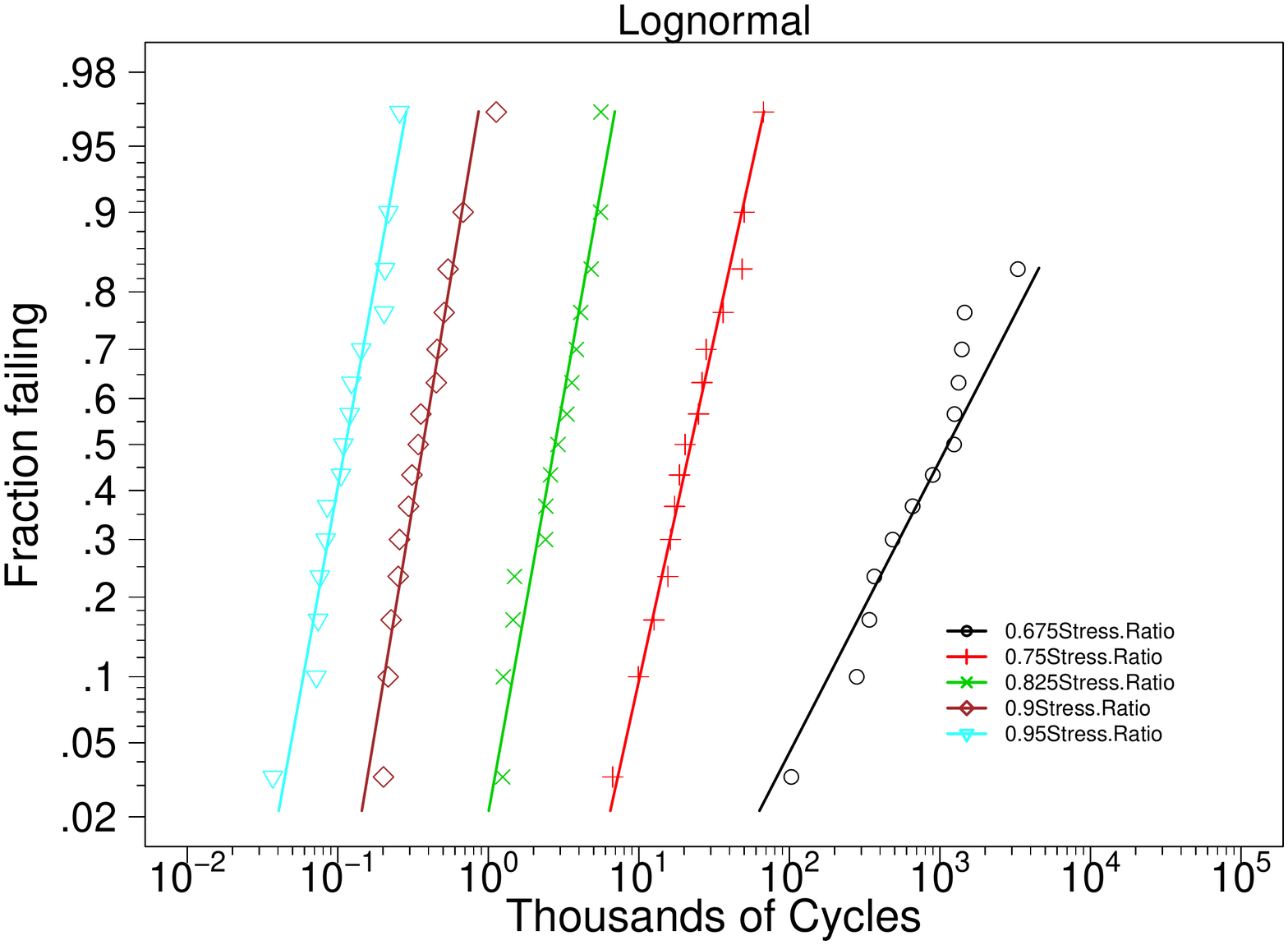}{3.25in}&
\rsplidapdffiguresize{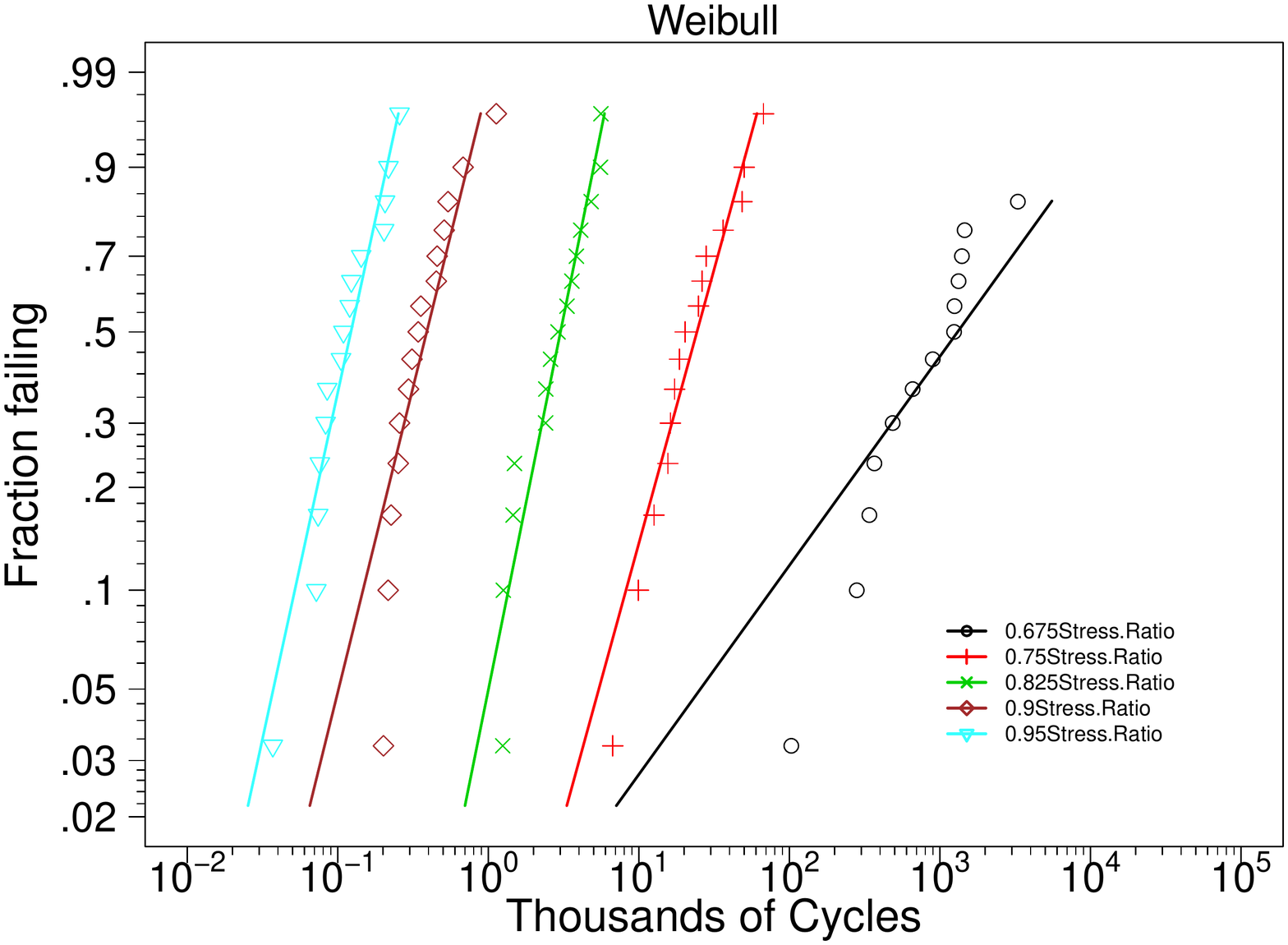}{3.25in}
\end{tabular}
\caption{Comparison of lognormal (left) and Weibull (right)
  distribution probability plots for the
aluminum 6061-T6 coupons,
  slightly notched C35 steel specimens, and the Holman concrete \SN{} data.}
\label{S.figure:LognormalWeibullCompareSet02}
\end{figure}

\begin{figure}
\begin{tabular}{cc}
\multicolumn{2}{c}{Nitinol Wire Rotating Bend \SN{} Data with 0.56
  Strain Omitted}\\[-3.2ex]
\rsplidapdffiguresize{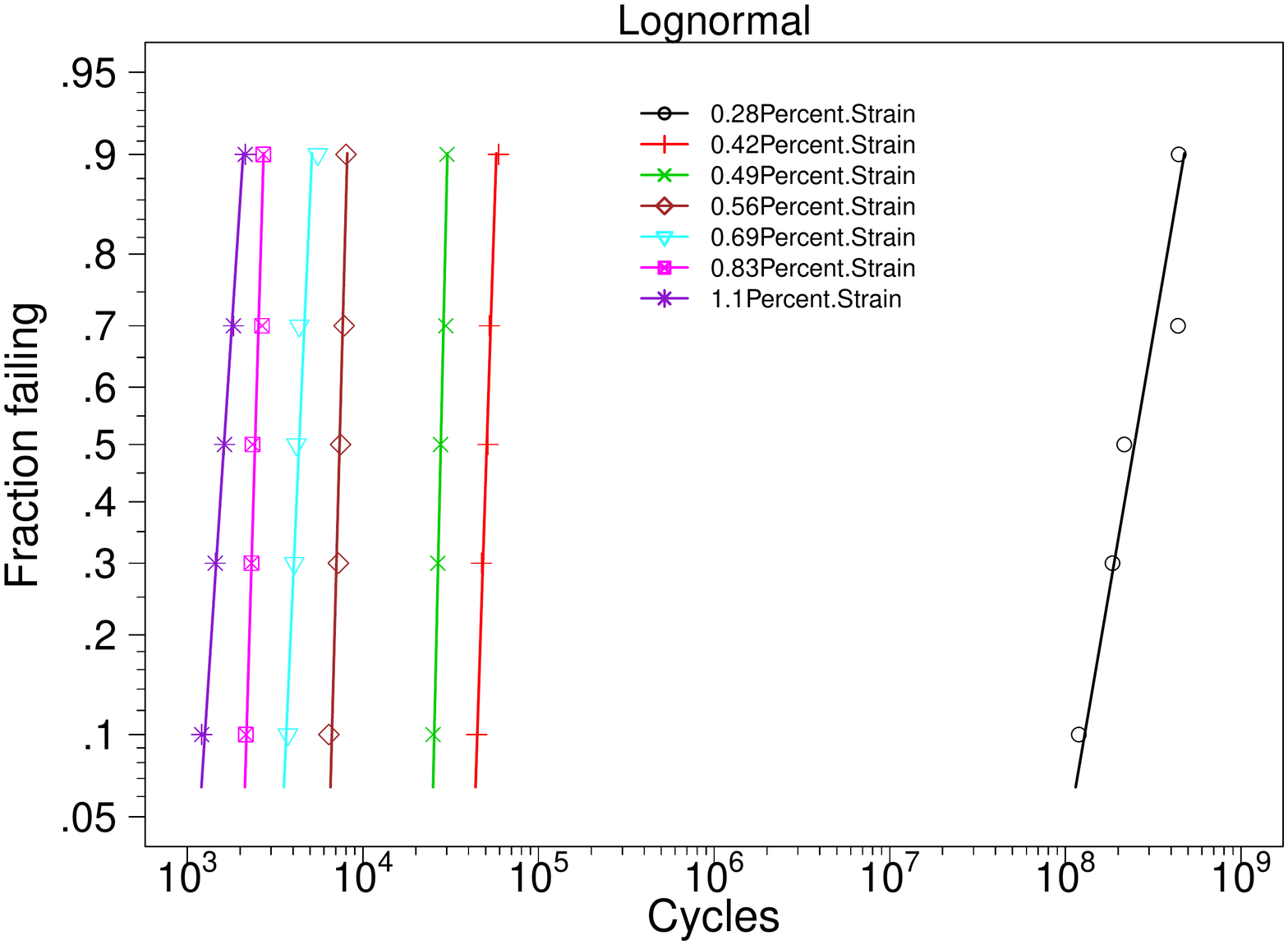}{3.25in}&
\rsplidapdffiguresize{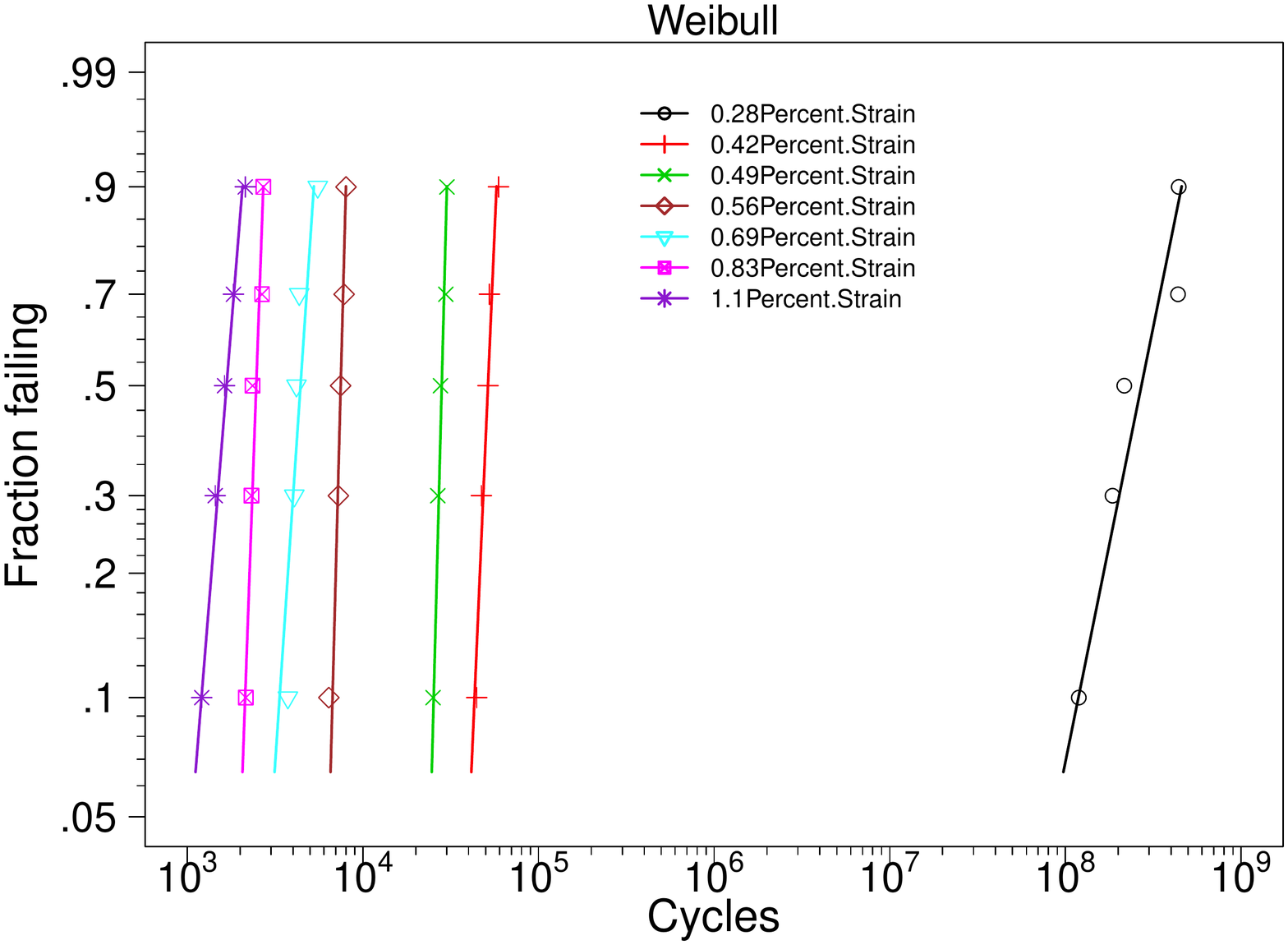}{3.25in}\\
\multicolumn{2}{c}{Sharply Notched Specimens of Aluminum 2024-T4 \SN{} Data} \\[-3.2ex]
\rsplidapdffiguresize{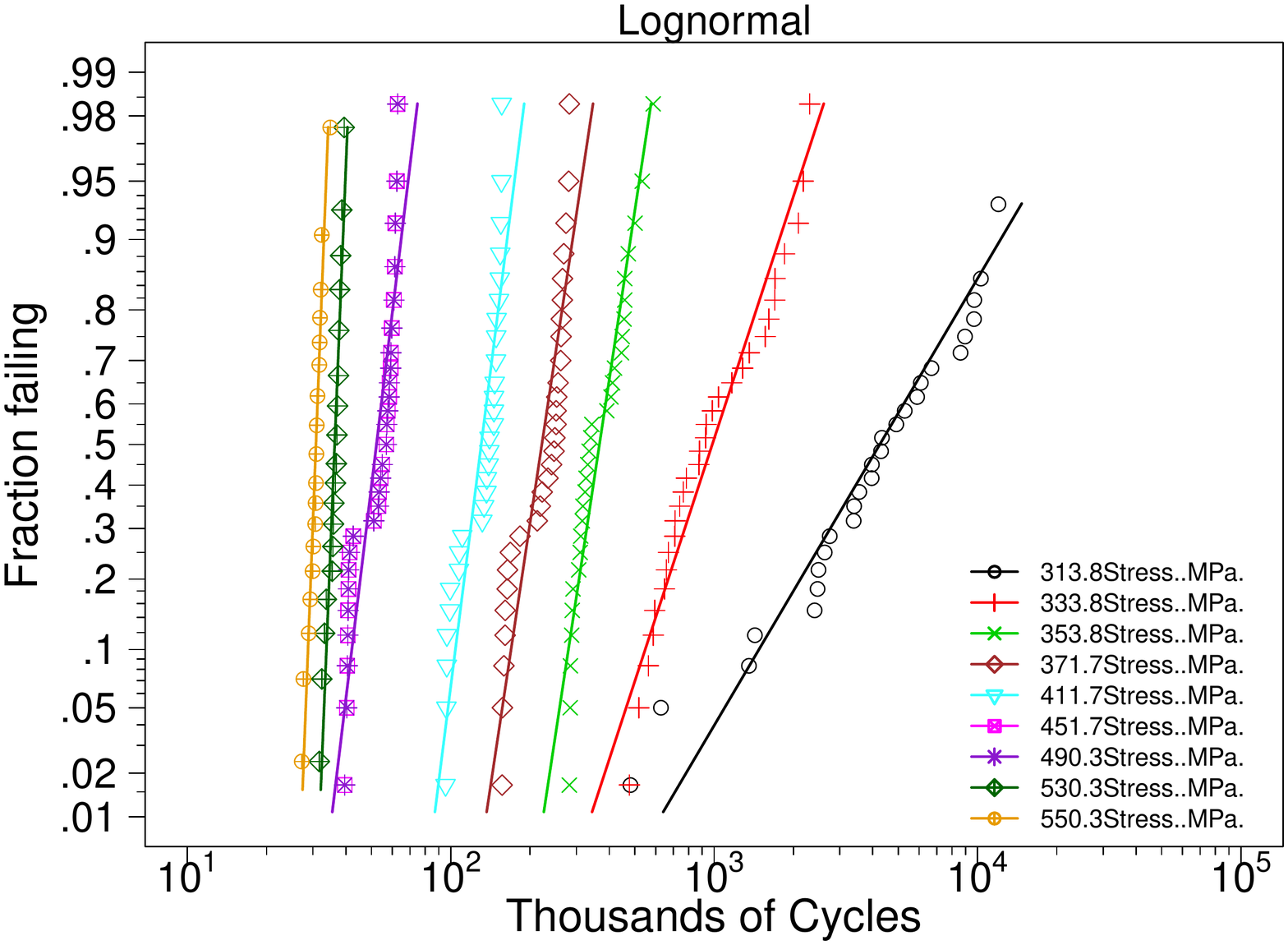}{3.25in}&
\rsplidapdffiguresize{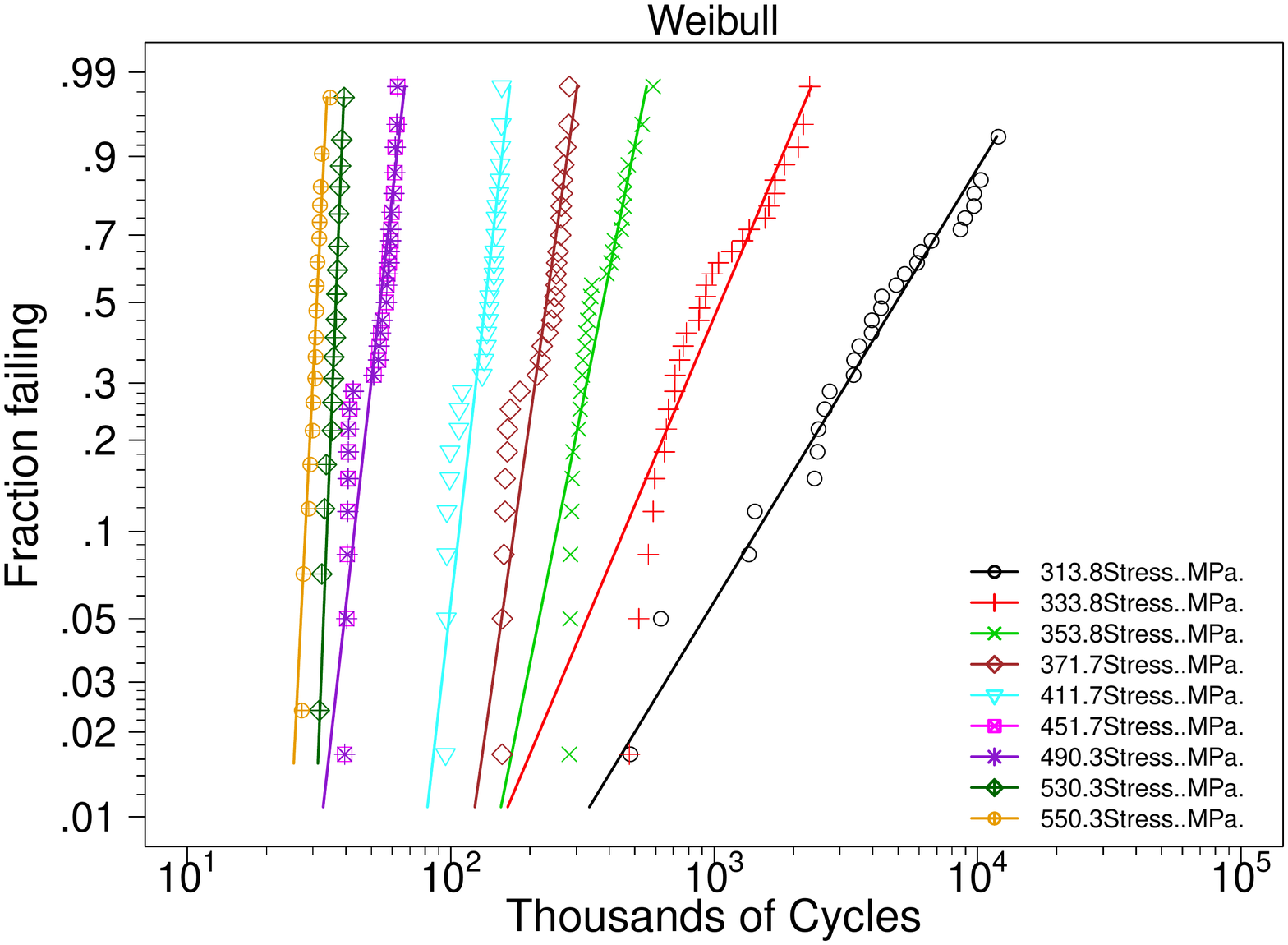}{3.25in}\\
\multicolumn{2}{c}{Ti64 350F $R=-1$ \SN{} Data} \\[-3.2ex]
\rsplidapdffiguresize{./pdfFigures/Ti64_350F_Rm1_groupi_lognormal.pdf}{3.25in}&
\rsplidapdffiguresize{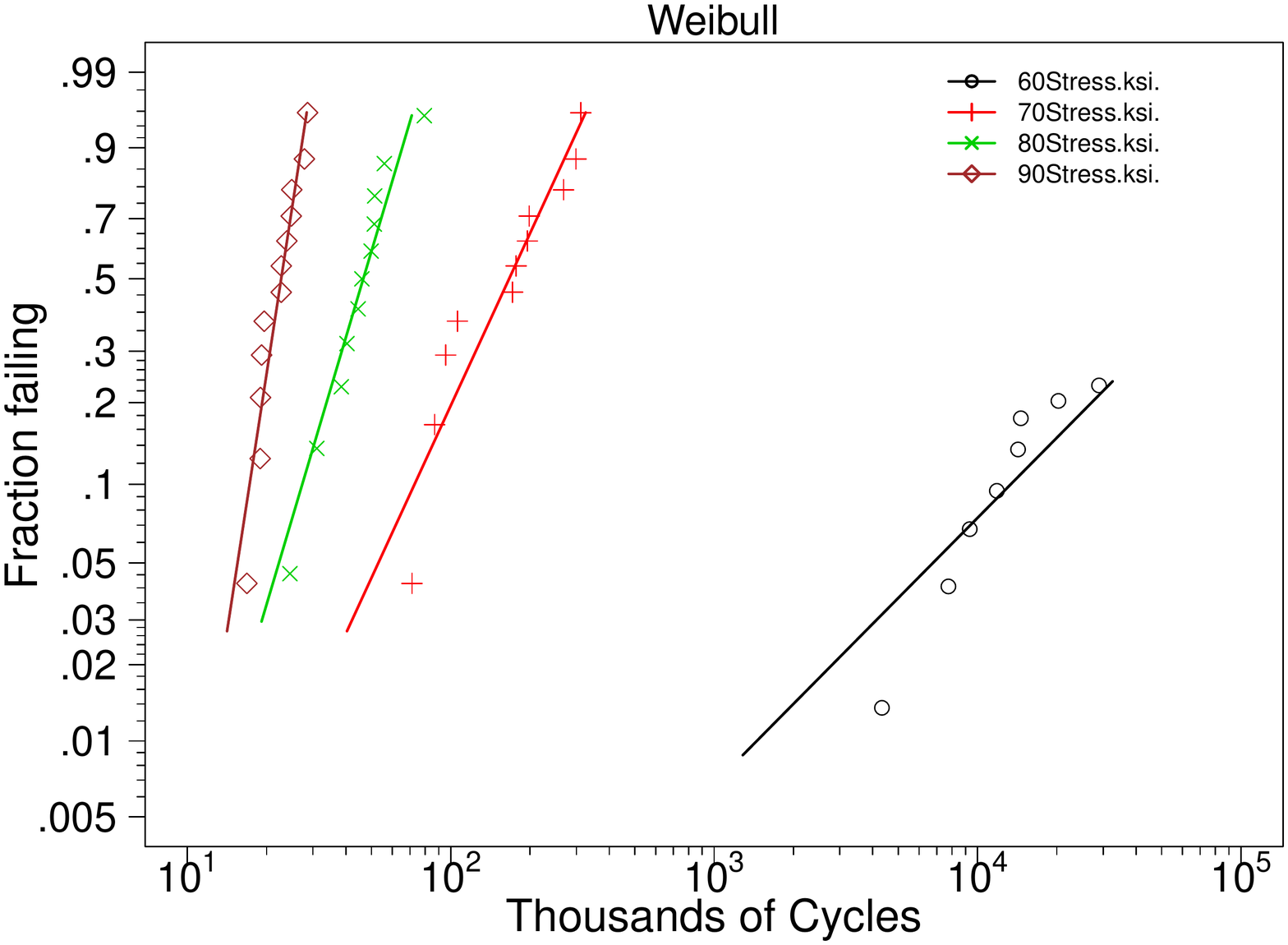}{3.25in}
\end{tabular}
\caption{Comparison of lognormal (left) and Weibull (right)
  distribution probability plots for the nitinol wire rotating bend,
  sharply notched aluminum 2024-T4 Specimens, and the Ti64 350F
  $R=-1$ \SN{} data.}
\label{S.figure:LognormalWeibullCompareSet03}
\end{figure}

Figure~\ref{figure:LaminatePanelProbability.plots} in the main paper
compared the
lognormal and Weibull multiple probability plots for the laminate
panel \SN{} data, showing that the lognormal distribution fit much
better, as would be suggested by the cumulative damage failure
mechanism (Section~\ref{section:choosing.fatigue.life.distribution}).
Figures~\ref{S.figure:LognormalWeibullCompareSet01},~\ref{S.figure:LognormalWeibullCompareSet02},
and~\ref{S.figure:LognormalWeibullCompareSet03} provide side-by-side
comparisons of lognormal and Weibull multiple probability plots for
nine additional \SN{} data sets of various different materials and
specimen types.

The three wire data sets in
Figure~\ref{S.figure:LognormalWeibullCompareSet01} came from
\citet{Freudenthal1952}.
In Figure~\ref{S.figure:LognormalWeibullCompareSet02}, the aluminum
6061-T6 data were used in \citet{BirnbaumSaunders1969b}.  For this
data set, the
failure times larger than 1800 thousand cycles were
converted to right censored observations at that
point because the fit in the upper tail was bad (interest is focused
on the lower tail and those upper-tail observation could bias
lower-tail estimates).  The C35 steel data came from tests of
slightly notched specimens and were given in
\citet{Maennig1968}. These data were subsequently analyzed in
\citet{CastilloGalambos1987} and \citet{Castillo_et_al2019}. The
concrete \SN{} data came from \citet{Holmen1979,Holmen1982} and were
subsequently analyzed in \citet{CastilloHadi1995}, the rejoinder
of \citet{PascualMeeker1999}, and \citet{Castillo_etal2007}.

In Figure~\ref{S.figure:LognormalWeibullCompareSet03}, the 0.02
inch diameter nitinol wire rotating bend \SN{} data are a subset of
the data presented in \citet{Weaver_etal2022} that were generated in the FDA
laboratories. The \SN{} data based on sharply notched specimens of
2024-T4 aluminum alloy specimens were given in
\citet{ShimokawaHamaguchi1979} but also analyzed in
\citet{Shen1994}. The Ti64 data (same as used in
Examples~\ref{example:Ti64.data}
and~\ref{example:modeling.Ti64.sn.data}) have not appeared in any
previous publication.

Among all nine comparisons, the Weibull distribution fit better or
just as well as the lognormal only for the four tests on wire
specimens. This is not surprising given the weakest-link nature of
the fatigue-to-fracture mechanism in a wire. For the other five
examples, the lognormal distribution generally provided a better fit
across the different levels of stress or strain.

Note that these kinds of comparisons are convenient when fatigue
tests are (as they often are) conducted by allocating a substantial
number of test units to particular levels of stress so that separate
nonparametric (and parametric) estimates can be computed at those
levels of stress. In experiments where there is a large number of
unique level of stress with few repeats such probability plots are
not possible and distributional assessment has to be made on the
basis of residual probability plots after finding a regression
relationship that fits the data well, as described in
Section~\ref{section:using.residuals.as.model.checking.diagnostics}
of the main paper and illustrated in
Figures~\ref{figure:LaminatePanelResidual.plots}b,
\ref{figure:Ti64Residual.plots}b, and \ref{figure:Nitinol02Residual.plots}b.

In
Section~\ref{section:important.advantages.of.specifying.fatigue.strength.distribution}
of the main paper and
Section~\ref{S.section:additional.motivation.specify.fatigue.strength}
of this appendix, we provided strong motivation for and
recommendations to specify the fatigue-strength model instead of the
fatigue-life model. One might then question the value of probability
plots of the fatigue-life distributions at the different test
levels of stress, especially because the form of the induced life
distributions will differ from the distributions of the specified
fatigue-strength model when the \SN{} relationship is nonlinear when
plotted on log-log axes. As illustrated in
Figures~\ref{figure:Ti64.model.fit.plots}a and
Figure~\ref{figure:Nitinol02Model.plots}a, especially at the
higher levels of stress or strain (because the  \SN{} relationship
is approximately linear when
plotted on log-log axes), the shapes/spreads of the
fatigue-strength and the fatigue-life distributions are similar.

\section{Further Explanation of the Castillo et al. \SN{} Model}
\label{S.section:generalization.castillo.sn.model}
This section provides additional technical details for the Castillo et al.
model given in (\ref{equation:logt.logx.castillo.model}) and
described in Section~\ref{section:Castillo.model} of the main paper. In
particular, we provide explicit expressions for the Weibull
parameters and quantile functions for the fatigue life $N$ and
fatigue strength $X$ random variables.

\subsection{The Distribution of Fatigue Life}
In the main paper, we stated that by starting with
(\ref{equation:logt.logx.castillo.model}) one can replace
$x$ with $S_e$ and the result can be interpreted as the cdf
for fatigue life $N$ at a given level of stress $S_{e}$:
\begin{align}
F_{N}(t; S_{e}) & = \Pr(N \le t; S_{e}) = F(t,S_{e}) \nonumber \\ &
= 1 - \exp \left\{ - \left[ \frac{\log(t) - \gamma_{N}}{\eta_{N}}
  \right]^\beta \right\}, \quad t > \exp(\gamma_{N}), \, S_{e} >
\exp(E)
\label{S.equation:probN.castillo}
\end{align}
where $\gamma_{N} = B + \gamma/[ \log(S_{e}) - E ]$ and $\eta_{N} =
\eta/[ \log(S_{e}) - E ]$.  Also, it can be shown that
$Y_{N}=\log(N)$ given $S_{e}$ has a three-parameter Weibull
distribution with scale parameter $\eta_{N}$, threshold parameter
$\gamma_{N}$, and shape parameter $\beta$. Also,
\begin{align*}
W = \left( \frac{Y_{N} -\gamma_{N}}{\eta_{N}} \right)^{\beta} =
\left( \frac{\log(N) -\gamma_{N}}{\eta_{N}} \right)^{\beta}
\end{align*}
at a given $S_{e}$ has an exponential distribution with scale
parameter 1 or, equivalently, a two-parameter Weibull distribution
with scale and shape parameters both equal to 1. Then
\begin{align}
\epsilon = \log(W) = \log\left[ \left(
  \frac{Y_{N}-\gamma_{N}}{\eta_{N}} \right)^{\beta} \right] =
\log\left[ \left( \frac{\log(N)-\gamma_{N}}{\eta_{N}}
  \right)^{\beta} \right]
\label{S.equation:LS.N.error.castillo}
\end{align}
has a smallest extreme value (Gumbel) distribution with location
parameter $\mu=0$ and scale parameter $\sigma=1$.  Solving for $N$
in (\ref{S.equation:LS.N.error.castillo}) gives
\begin{align}
N = \exp\left[ \gamma_{N} + \eta_{N} [\exp(\epsilon)]^{1/\beta}
  \right] = \exp \left[ \left( B + \frac{\gamma}{\log(S_{e}) - E}
  \right) + \left( \frac{\eta}{\log(S_{e}) - E} \right)
        [\exp(\epsilon)]^{1/\beta} \right].
\label{S.equation:N.weibull.error.castillo}
\end{align}

\subsection{Distribution of Fatigue Strength}
In the main paper, we stated that by starting with
(\ref{equation:logt.logx.castillo.model}) one can replace
$t$ with $N_{e}$ and the result can be
interpreted as the cdf for fatigue strength
$X$ at a given number of  cycles
$N_{e}$:
\begin{align}
F_{X}(x; N_{e}) & = \Pr(X \le x; N_{e}) = F(N_{e},x) \nonumber \\ &
= 1 - \exp \left\{ - \left[ \frac{\log(x) - \gamma_{X}}{\eta_{X}}
  \right]^\beta \right\}, \quad x > \exp(\gamma_{X}), \, N_{e} >
\exp(B)
\label{S.equation:probx.castillo}
\end{align}
where $\gamma_{X} = E + \gamma/[ \log(N_{e}) - B ]$ and $\eta_{X} =
\eta/[ \log(N_{e}) - B ]$.  Also, it can be shown that
$Y_{X}=\log(X)$, the logarithm of fatigue strength
given $N_{e}$ has a three-parameter Weibull
distribution with scale parameter $\eta_{X}$, threshold parameter
$\gamma_{X}$, and shape parameter $\beta$. Then, parallel to
(\ref{S.equation:LS.N.error.castillo}),
\begin{align}
\epsilon = \log\left[ \left( \frac{Y_{X}-\gamma_{X}}{\eta_{X}}
  \right)^{\beta} \right] = \log\left[ \left(
  \frac{\log(X)-\gamma_{X}}{\eta_{X}} \right)^{\beta} \right]
\label{S.equation:LS.X.error.castillo}
\end{align}
has a standard smallest extreme value (Gumbel) distribution.
Solving for $X$ in (\ref{S.equation:LS.X.error.castillo}) gives
\begin{align}
X = \exp\left[ \gamma_{X} + \eta_{X} [\exp(\epsilon)]^{1/\beta}
  \right] = \exp \left[ \left( E + \frac{\gamma}{\log(N_{e}) - B}
  \right) + \left( \frac{\eta}{\log(N_{e}) - B} \right)
        [\exp(\epsilon)]^{1/\beta} \right].
\label{S.equation:X.weibull.error.castillo}
\end{align}

\subsection{Quantiles of the Fatigue-life and the Fatigue-Strength Distributions}
The fatigue-life $p$ quantile curve is obtained by setting
$p=F_{N}[t_{p}(S_{e}); S_{e}]$ in (\ref{S.equation:probN.castillo}).
Solving for $t_{p}(S_{e})$ gives
\begin{align*}
t_{p}(S_{e}) & = \exp\left[ \gamma_{N} + \eta_{N}
  [-\log(1-p)]^{1/\beta} \right]\\ &= \exp \left[ \left( B +
  \frac{\gamma}{\log(S_{e}) - E} \right) + \left(
  \frac{\eta}{\log(S_{e}) - E} \right) [-\log(1-p)]^{1/\beta}
  \right], \quad 0 < p < 1, \, S_{e} > \exp(E).
\end{align*}
Similarly, the fatigue-strength $p$ quantile is obtained by setting
$p=F_{X}[x_{p}(N_{e}); N_{e}]$ in (\ref{S.equation:probx.castillo}),
from which $x_{p}(N_{e})$ is derived as
\begin{align*}
x_{p}(N_{e}) &= \exp\left[ \gamma_{X} + \eta_{X}
  [-\log(1-p)]^{1/\beta} \right]\\ &= \exp \left[ \left( E +
  \frac{\gamma}{\log(N_{e}) - B} \right) + \left(
  \frac{\eta}{\log(N_{e}) - B} \right) [-\log(1-p)]^{1/\beta}
  \right], \quad 0 < p < 1, \, N_{e} > \exp(B).
\end{align*}

\subsection{Comments}
\begin{figure}
\centering
\rsplidapdffiguresize{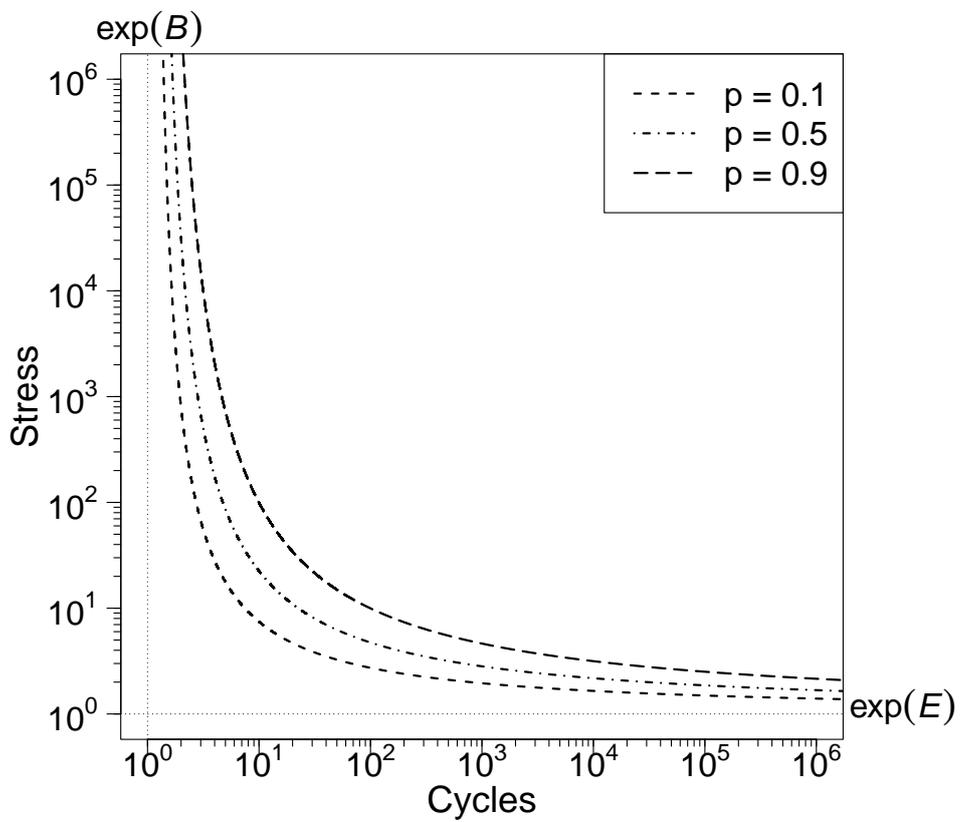}{6.0in}
\caption{The 0.10, 0.50, and 0.90 quantile curves for the Castillo
  et al. rectangular-hyperbola Weibull \SN{} model with $\exp(E)=\exp(B)=1$,
  $\gamma=3$, $\beta=2$, and $\eta=5$.}
\label{S.figure:CastilloFatigueLifeQuantiles}
\end{figure}

Observe that
\begin{itemize}
\item The error terms $\epsilon$ for the random variables $N$ in
  (\ref{S.equation:N.weibull.error.castillo}) and $X$ in
  (\ref{S.equation:X.weibull.error.castillo}) have the same
  distribution and that either of these equations leads back to the
  cdfs derived from (\ref{equation:logt.logx.castillo.model}) in the
  main paper.

\item
Setting $p=F[t_{p}(S_{e}),S_{e}]$ in
(\ref{equation:logt.logx.castillo.model})
in the main paper leads to
\begin{align}
[\log(t_{p}(S_{e})) - B][\log(S_{e}) - E] = \gamma + \eta [-\log(1-p)]^{1/\beta}.
\label{S.equation:life.quantile.curve.castillo}
\end{align}
Analogously, setting $p=F[N_{e}, x_{p}(N_{e})]$ in
(\ref{equation:logt.logx.castillo.model}) in the main paper leads to
\begin{align}
[\log(N_{e}) - B][\log(x_{p}(N_{e})) - E] = \gamma + \eta [-\log(1-p)]^{1/\beta}.
\label{S.equation:strength.quantile.curve.castillo}
\end{align}
Equations (\ref{S.equation:life.quantile.curve.castillo}) and
(\ref{S.equation:strength.quantile.curve.castillo}) correspond to
Equation~(18) of \cite{Castillo_etal1985} and the Castillo et
al.~(1985) model in Table~A.21 on page~195 and Equation~(A.26) on
page~204 of \cite{CastilloFernandez-Canteli2009}.

\item The respective $p$ quantiles of $N$ and $S$ coincide as
  illustrated in Figure~\ref{S.figure:CastilloFatigueLifeQuantiles}.
  The right-hand side of
  (\ref{S.equation:life.quantile.curve.castillo}) and
  (\ref{S.equation:strength.quantile.curve.castillo}) decreases as
  $\gamma \rightarrow 0$ which implies that the $p$ quantile curve
  moves closer to the vertical asymptote $\exp(B)$ and horizontal
  asymptote $\exp(E)$.
\end{itemize}

\section{\mbox{  }More Details from the Data Analysis/Modeling Examples}
\label{S.section:more.details.numerical.examples}
The data analysis/modeling examples in the main paper focus on the
main features of the fitted models. To save space, less interesting,
but potentially useful details are presented in this section.

\subsection{More Details for the
  \texorpdfstring{Box--Cox/Loglinear-$\sigma_{N}$}{Box--Cox/Loglinear-sigma}
  \SN{} Model
  Fit to the Laminate Panel Data}
\label{S.section:more.details.laminate.panel.boxcox.loglinear}
Example~\ref{example:box-cox.loglin.laminate.panel.data} in the main
paper gives a summary of the results of fitting the
Box--Cox/loglinear-$\sigma_{N}$ \SN{} model to the laminate panel
data. Note that in this example, the fatigue-life model is specified
and the fatigue-strength model is induced.

\subsubsection{Model and prior distributions}
The lognormal distribution regression model fit to the laminate
panel data was
\begin{align*}
\log(N/N_{\textrm{max}})&= \mu(S) + \sigma(S) \epsilon,
\end{align*}
where
\begin{align*}
\mu(S) &= \beta_{0}+ \beta_{1}\nu(S/S_{\textrm{max}};\lambda)\\
\sigma(S) &= \beta^{[\sigma]}_{0}+ \beta^{[\sigma]}_{1} \log(S/S_{\textrm{max}}).
\end{align*}
\begin{align*}
\nu(S;\lambda) &=
\begin{cases}
\dfrac{S^{\lambda} -1}{\lambda} & \text{if $\lambda \ne  0$}\\[2ex]
\log(S) & \text{if $\lambda=0$},
\end{cases}
\end{align*}
is the Box--Cox power transformation, and $\epsilon  \sim \NORM(0,1).$
Here stress $S$ and number of cycles $N$
were scaled for numerical reasons
\citep[as described in][]{LiuMeeker2024}.
The unknown traditional parameters (TPs) to be estimated from the data are $\beta_{0}$,
$\beta_{1}$, $\lambda$, $\beta^{[\sigma]}_{0}$, and $\beta^{[\sigma]}_{1}$.

As suggested in \citet[][Section~6.2]{LiuMeeker2024}, the MCMC
algorithm was run to generate posterior draws of the unrestricted
stable parameters (USPs). The USPs are $\lambda$ and
\begin{align*}
	\log(\sigma_{\Lowx}) & =  \beta^{[\sigma]}_{0}+\beta^{[\sigma]}_{1}\log(S_{\Highx})\\
	\log(\sigma_{\Highx}) & =  \beta^{[\sigma]}_{0}+\beta^{[\sigma]}_{1}\log(S_{\Lowx})\\
	\log(t_{\Lowx}) & =  \beta_{0}+\beta_{1}\nu(S_{\Highx};\lambda) + \Phi^{-1}(0.50)\sigma_{\Lowx}\\
	\log(t_{\Highx}) & =  \beta_{0}+\beta_{1}\nu(S_{\Lowx};\lambda) + \Phi^{-1}(0.50)\sigma_{\Highx}
\end{align*}
where $S_{\Lowx}$ and $S_{\Highx}$ are, respectively, the smallest
and the largest levels of stress in the data where failures were observed.

The marginal prior distributions for the USPs
$\log(\sigma_{\Lowx})$,
$\log(\sigma_{\Highx})$, $\lambda$, $\log(t_{\Lowx})$, and
$\log(t_{\Highx})$ were chosen to be flat. The resulting
noninformative joint prior for the USPs results in Bayesian
posterior estimates that are close to the ML estimates.

\subsubsection{Posterior draws and parameter estimates}
\begin{figure}[ht]
\begin{tabular}{cc}
(a) & (b) \\[-1.0ex]
\rsplidapdffiguresize{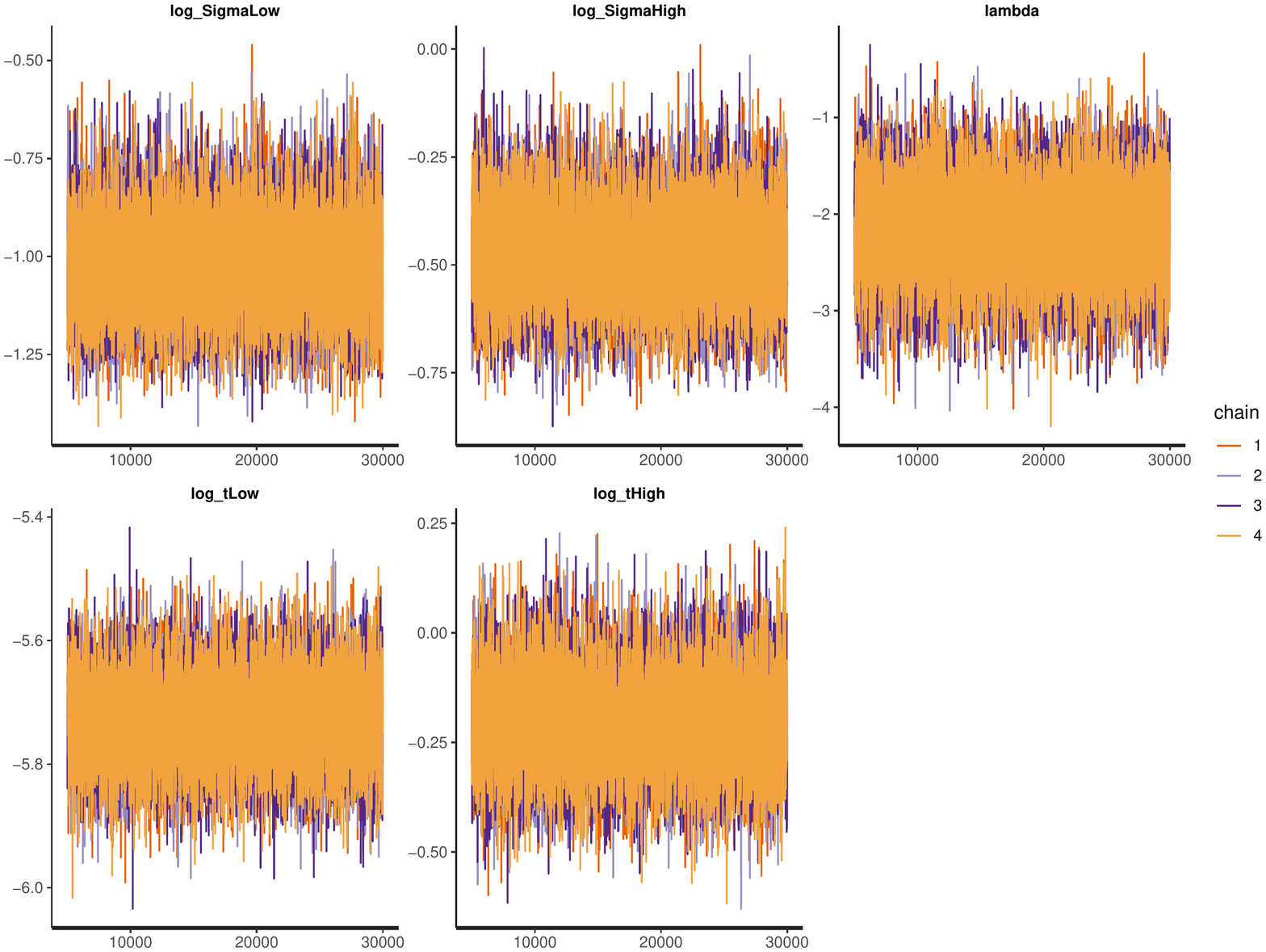}{3.25in} &
\rsplidapdffiguresize{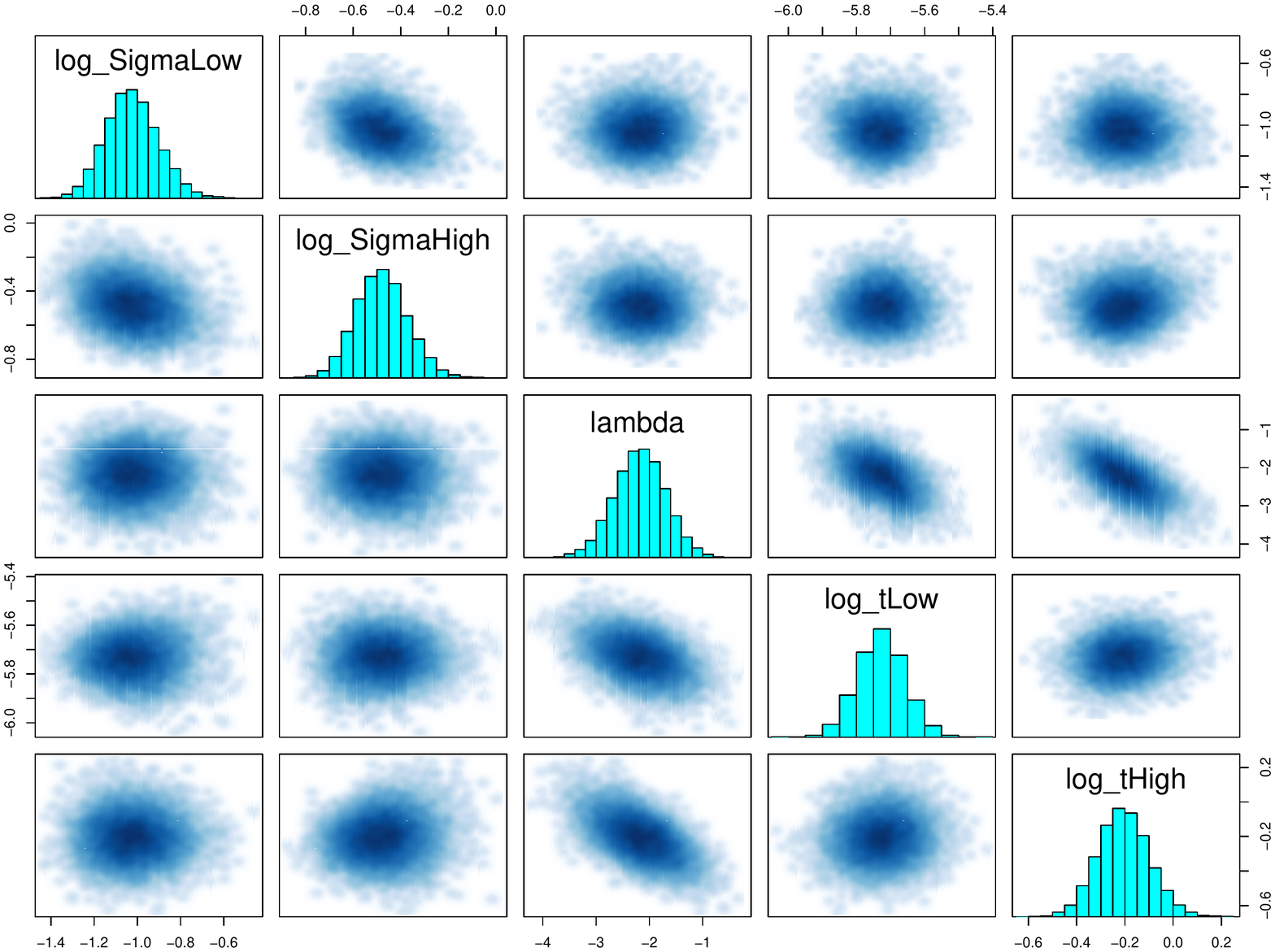}{3.25in}
\end{tabular}
\caption{Trace plots~(a) and a pairs plot~(b) of the
  posterior draws for the Box--Cox/loglinear-$\sigma_{N}$ model fit
  to the laminate panel data.}
\label{S.figure:LaminatePanel.estimation.plots}
\end{figure}

The following table is a summary of the posterior draws for the USPs.
\begin{verbatim}
Inference for Stan model: BoxCoxLogLinSigma.
4 chains, each with iter=30000; warmup=5000; thin=5;
post-warmup draws per chain=5000, total post-warmup draws=20000.

               mean se_mean   sd  2.5%   25%   50%   75% 97.5% n_eff Rhat
log_SigmaLow  -1.02       0 0.12 -1.25 -1.11 -1.03 -0.94 -0.77 19655    1
log_SigmaHigh -0.48       0 0.11 -0.68 -0.55 -0.48 -0.41 -0.26 19668    1
lambda        -2.18       0 0.48 -3.14 -2.50 -2.17 -1.85 -1.24 19766    1
log_tLow      -5.73       0 0.07 -5.86 -5.77 -5.73 -5.68 -5.59 19385    1
log_tHigh     -0.20       0 0.11 -0.41 -0.28 -0.21 -0.13  0.01 19507    1
\end{verbatim}
Figure~\ref{S.figure:LaminatePanel.estimation.plots}a shows
trace plots for the parameters (indicating good mixing) and
Figure~\ref{S.figure:LaminatePanel.estimation.plots}b the
corresponding pairs plot.

The following table is a summary of the marginal posterior draws for
the Traditional Parameters based on UnScaled data (TPUSs),
computed as described in
\citet[][Sections~4.4 and~4.5]{LiuMeeker2024}.
\begin{verbatim}
Laminate Panel Test Data Lognormal distribution BoxCoxLogLinSigma model
 Summary of the marginal posteriors for the
   Traditional Parameters based on UnScaled data (TPUSs)
      based on 20000 Stan draws
                     2.5%       25%       50%       75%     97.5%
Beta0            7.09e+03  1.86e+05  1.12e+06  6.61e+06  2.12e+08
Beta1           -6.62e+08 -1.66e+07 -2.44e+06 -3.45e+05 -8.73e+03
lambda          -3.12e+00 -2.50e+00 -2.18e+00 -1.85e+00 -1.23e+00
Beta0.log.Sigma  2.18e+00  6.36e+00  8.48e+00  1.05e+01  1.46e+01
Beta1.log.Sigma -2.67e+00 -1.96e+00 -1.60e+00 -1.23e+00 -5.03e-01
\end{verbatim}

\subsection{More Details for the  Nishijima/Lognormal  \SN{} Model
  Fit to the Ti64 Data}
\label{S.section:more.details.Ti64.nishijima}

\subsubsection{Comparison of lognormal and Weibull distributions}
\begin{figure}[ht]
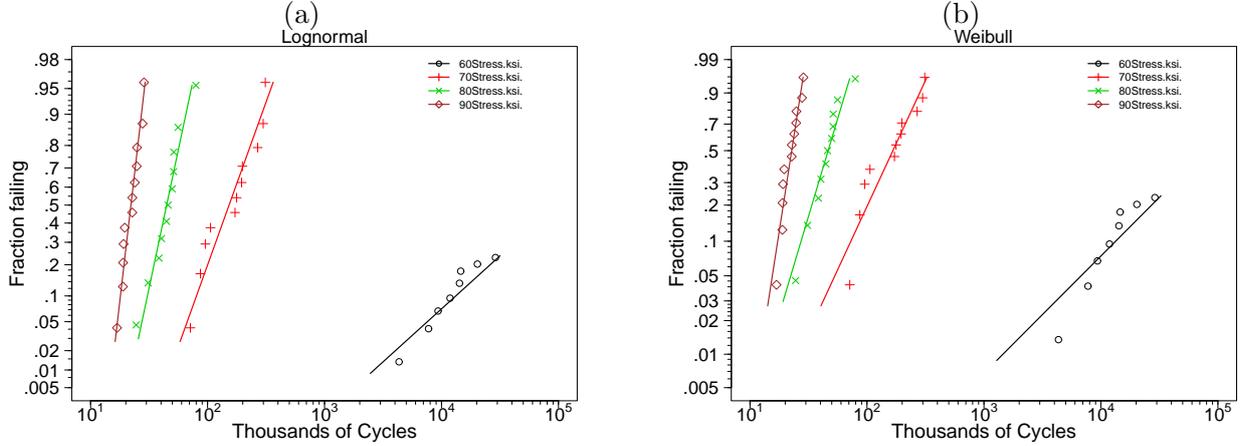

\begin{tabular}{cc}
(a) & (b) \\[-3.2ex]
\rsplidapdffiguresize{./pdfFigures/Ti64_350F_Rm1_groupi_lognormal.pdf}{3.25in} &
\rsplidapdffiguresize{./pdfFigures/Ti64_350F_Rm1_groupi_weibull.pdf}{3.25in}
\end{tabular}
\caption{A comparison of lognormal~(a) and Weibull~(b) probability
  plots for the Ti64 data.}
\label{S.figure:Ti64.lognormal.Weibull.probability.plots}
\end{figure}

Figure~\ref{S.figure:Ti64.lognormal.Weibull.probability.plots}
uses lognormal and Weibull probability plots to compare the two
distributions fit to the Ti64 data. The lognormal distribution fits better.
Although this example uses a specified fatigue-strength model,
probability plots of the fatigue-life distributions are useful
because the fatigue-life and fatigue-strength distributions have
similar shapes along parts of the \SN{} relationship that are
approximately linear. The \SN{} relationship is often approximately
linear where stress levels are higher.

\subsubsection{Model and prior distributions}
Example~\ref{example:modeling.Ti64.sn.data} in the main
paper gives a summary of the results of
fitting the Nishijima \SN{} model to the Ti64 data.
 Note that in this example, the fatigue-strength model is specified
and the fatigue-life model (providing the basis for the likelihood
function) is induced.
The lognormal distribution regression model for fatigue-strength $X$
was
\begin{align*}
  X= \log[h(N; \betavec)] + \sigma_{X} \epsilon
\end{align*}
where stress $S$ and number of cycles $N$
were scaled for numerical reasons (as described in
\citet{LiuMeeker2024}), and $\epsilon  \sim \NORM(0,1).$
Specifically, $S/S_{\textrm{max}}$ was substituted for $S$ and  $N/N_{\textrm{max}}$ was substituted for $N$.
The Nishijima \SN{} relationship is
\begin{align*}
S&= h(N; \betavec) = \exp\left(\frac{
	-A \log(N)+B+E + \sqrt{
		\left[A \log(N)-(B-E)\right]^{2} +4C}}{2}\right),
\end{align*}
The unknown traditional parameters (TPs) to be estimated from the data are $A$,
$B$, $C$, $E$, and $\sigma_{X}$.

As suggested in \citet[][Section~5.2]{LiuMeeker2024}, the MCMC
algorithm was run to generate posterior draws of the unrestricted
stable parameters (USPs).
First we define three points on the \SN{} curve
\begin{align*}
  S_{\Lowx} & =  h(N_{\Highx};\beta) \\
  S_{\Midx} & = h(N_{\Midx};\beta) \\
  S_{\Highx} & =  h(N_{\Lowx};\beta),
\end{align*}
where $N_{\Lowx}$ is the smallest value of $N$ in the data,
$N_{\Highx}$ is the largest value of $N$ in the data that is a failure, and $N_{\Midx} =
\exp[(\log(N_{\Lowx}) + \log(N_{\Highx}))/2]$. Also,
\begin{align*}
  \log(S_{\MidU}) & =  [\log(S_{\Highx})+\log(S_{\Lowx})]/2 \\
  \log(S_{\MidL}) & =  2\frac{[\log(S_{\Lowx})-E][\log(S_{\Highx})-E]}{\log(S_{\Lowx})+\log(S_{\Highx})-2E}+E.
\end{align*}
Then, following \citet{LiuMeeker2024}, the USPs are defined as
\begin{align*}
& \logSLow  = \log(S_{\Lowx}) \\
& \logDeltaHighLow = \log[\log(S_{\Highx}) - \log(S_{\Lowx})] \\
& \qlogisp = \qlogis(p) \\
& \logDeltaSLowE = \log[\log(S_{\Lowx}) - E] \\
& \logSigmaX=\log(\sigma_{X}),
\end{align*}
where $\qlogis=\log[p/(1-p)]$ is the logit transformation and
\begin{align*}
p &=
    \frac{\log(S_{\MidU})  -\log(S_{\Midx})}{\log(S_{\MidU})-\log(S_{\MidL})}.
\end{align*}

\begin{figure}
\rsplidapdffiguresize{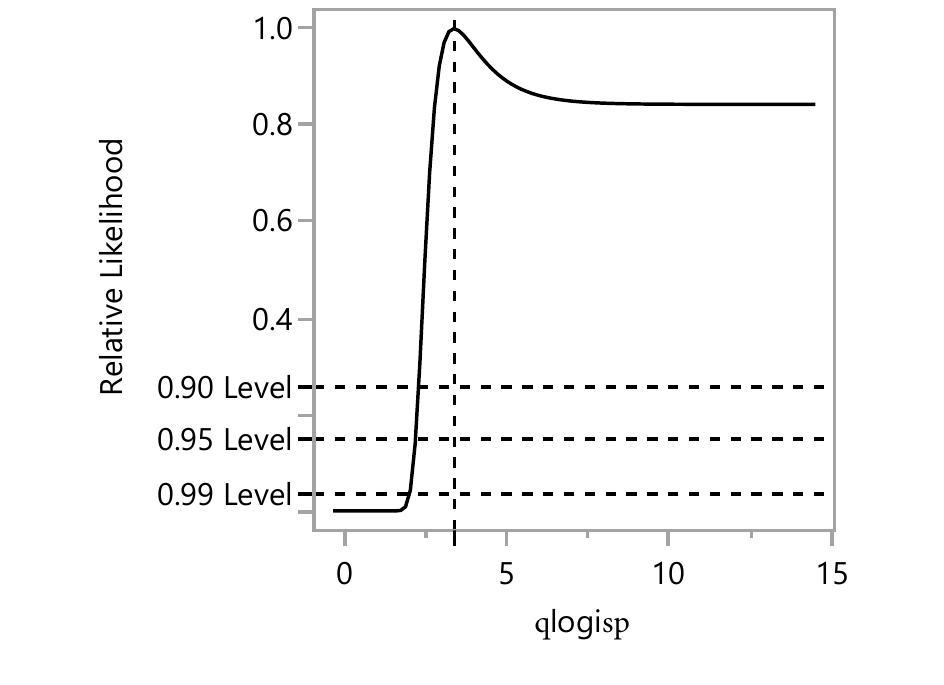}{6.0in}
\caption{Relative profile likelihood for $\qlogisp$ for the
  Nishijima/Lognormal model fit to the Ti64 data.}
\label{S.figure:Ti64.qlogis.Profile}
\end{figure}

Marginal priors for the USPs $\logSLow$, $\logDeltaHighLow$,
$\logDeltaSLowE$, and $\logSigmaX$ were specified as
flat. Attempting to use a flat prior on $\qlogisp$, in addition,
resulted in an improper posterior distribution.
Figure~\ref{S.figure:Ti64.qlogis.Profile} is a relative profile
likelihood for $\qlogisp$. The improper posterior results from the
flatness of the relative profile
likelihood for large values of $\qlogisp$.
As shown in
\citet[][Section~5.6]{LiuMeeker2024} when $\qlogisp \rightarrow
\infty$, the Nishijima model approaches a rectangular hyperbola
model and the relative profile
likelihood indicates that the  rectangular hyperbola
model is also consistent with the
data. This and the relative profile
likelihood for $\qlogisp$ in Figure~\ref{S.figure:Ti64.qlogis.Profile}
suggest two alternative priors for $\qlogisp$ that will result in
a proper posterior distribution:
\begin{itemize}
\item
A weakly informative prior that does not allow $\qlogisp$ to be too
large (large values of $\qlogisp$ cause numerical instability),
effectively fitting the Nishijima model. The exact
specification of the prior location
is not critical (because inferences about distribution
probabilities and quantiles will not be affected) as long as the
lower tail covers small values of $\qlogisp$ where the profile
relative likelihood is
very close to zero and upper tail of the prior distribution is
on the flat part of the profile relative likelihood for $\qlogisp$
so that $\qlogisp$ cannot be too large.
We used a normal distribution with a 0.005
quantile of -10 and a 0.995 quantile of 10.
\item
  An informative prior that constrains $\qlogisp$ to be a somewhat
  large value so that the fitted model is, essentially, a
  rectangular hyperbola model. Again, the exact values are not critical as
  long as the prior distribution is almost entirely on the flat part
  of the profile relative likelihood for $\qlogisp$. We used a normal
  distribution with a 0.005 quantile of 10 and a 0.995 quantile of
  11.
\end{itemize}
We experimented with both alternatives and present summaries of both
in the following two subsections. For estimation of lower-tail
quantiles of the fatigue-life or the fatigue-strength distributions
at lower levels of stress, the Nishijima model is conservative,
relative to the Rectangular Hyperbola model.
For this reason, the results we present in the main paper
use the weakly informative prior that constrains $\qlogisp$ to
effectively fit the Nishijima model.

\subsubsection{Posterior draws and parameter estimates for the
  weakly informative prior for \texorpdfstring{\qlogisp}{qlogisp}
  (Nishijima model)}
\begin{figure}[ht]
\begin{tabular}{cc}
(a) & (b) \\[-1.0ex]
\rsplidapdffiguresize{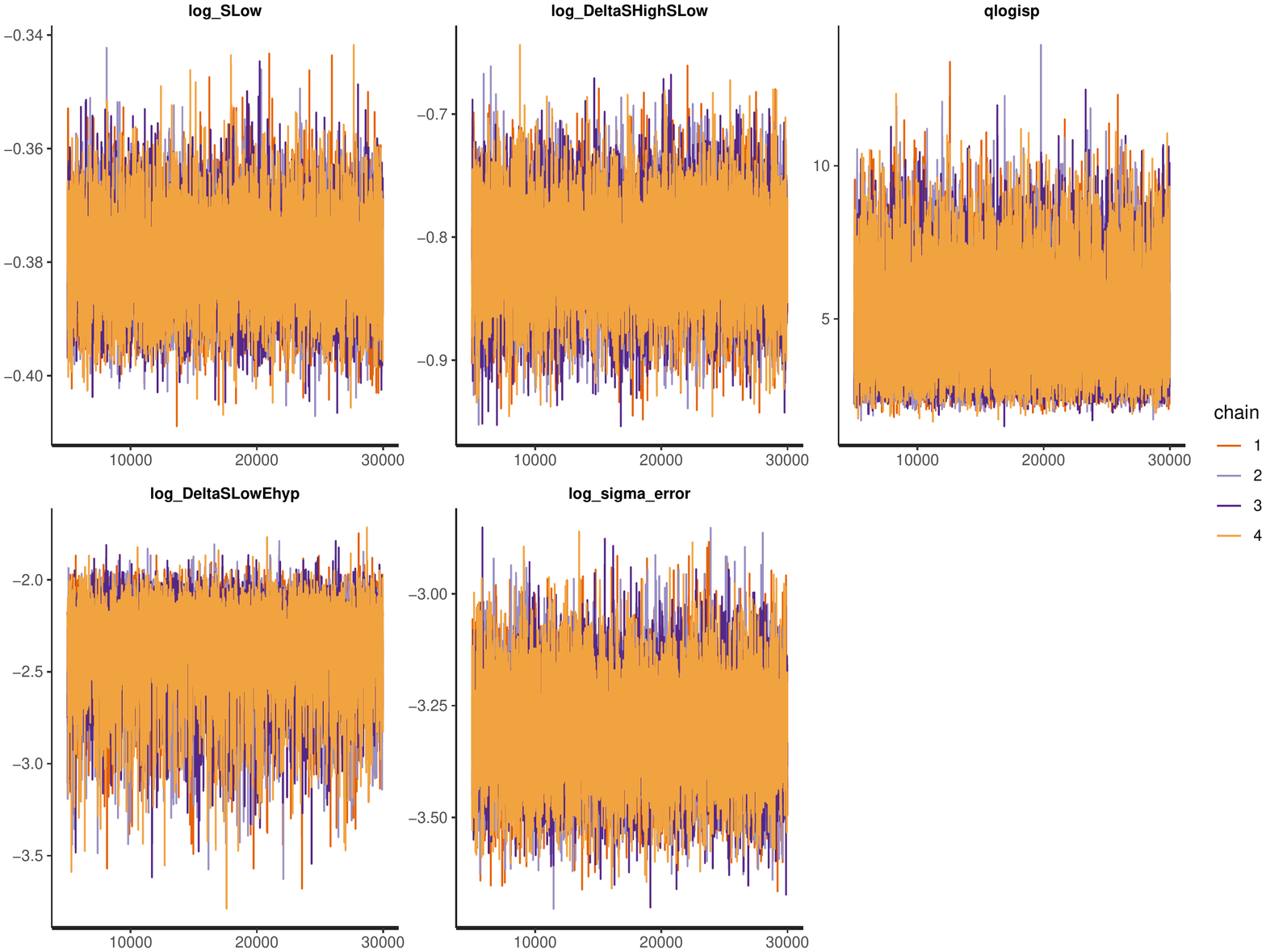}{3.25in} &
\rsplidapdffiguresize{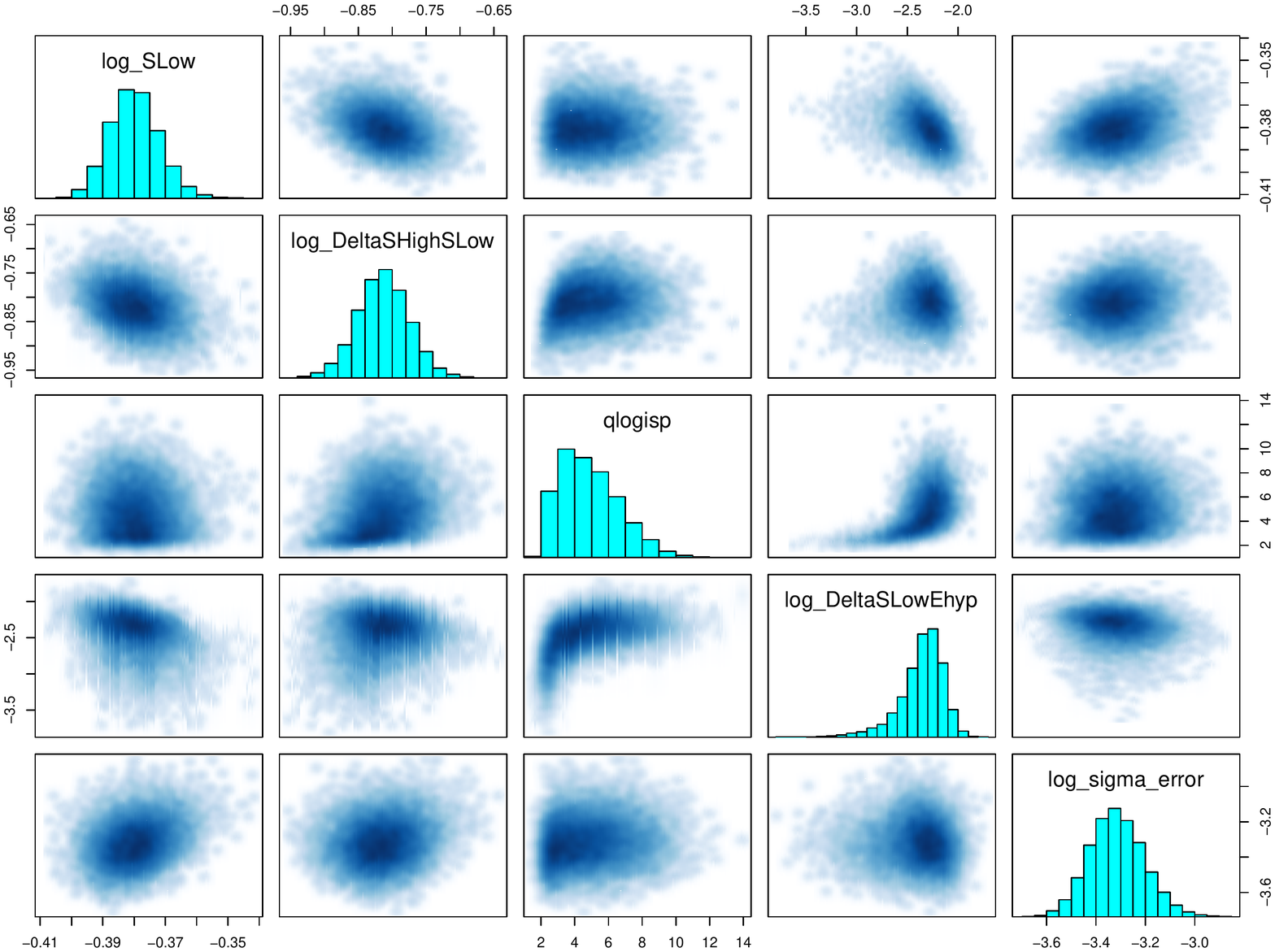}{3.25in}
\end{tabular}
\caption{Trace plots~(a) and a pairs plot~(b) of the
  posterior draws for the Nishijima model fit
  to the Ti64 data.}
\label{S.figure:Ti64.estimation.plots}
\end{figure}

The following table is a summary of the posterior draws for the USPs.
\begin{verbatim}
Inference for Stan model: Nishijima.
4 chains, each with iter=30000; warmup=5000; thin=5;
post-warmup draws per chain=5000, total post-warmup draws=20000.

                     mean se_mean   sd   2.5%    25%    50%    75%  97.5% n_eff Rhat
log_SLow            -0.38    0.00 0.01  -0.39  -0.39  -0.38  -0.37  -0.36 19737    1
log_DeltaSHighSLow  -0.81    0.00 0.04  -0.89  -0.84  -0.81  -0.79  -0.74 19593    1
qlogisp              4.89    0.01 1.74   2.35   3.50   4.64   6.02   8.76 18714    1
log_DeltaSLowEhyp   -2.36    0.00 0.22  -2.93  -2.46  -2.32  -2.21  -2.03 17796    1
log_sigma_error     -3.31    0.00 0.11  -3.52  -3.39  -3.32  -3.24  -3.08 19478    1
sigma_error          0.04    0.00 0.00   0.03   0.03   0.04   0.04   0.05 19456    1
lp__               215.28    0.01 1.61 211.28 214.47 215.62 216.47 217.37 19426    1
\end{verbatim}
Figure~\ref{S.figure:Ti64.estimation.plots}a shows
trace plots for the parameters (indicating good mixing) and
Figure~\ref{S.figure:Ti64.estimation.plots}b the
corresponding pairs plot.

The following table is a summary of the marginal posterior draws for
the TPUSs, computed as described in
\citet[][Sections~4.4 and~4.5]{LiuMeeker2024}.
\begin{verbatim}
Ti64 350F R=-1 Lognormal distribution Nishijima model
 Summary of the marginal posteriors for the
   Traditional Parameters based on UnScaled data (TPUSs)
      based on 20000 Stan draws
          2.5%    25%    50%     75%    97.5%
Ahyp    0.3200 1.0600 3.3800 13.4000 208.0000
Bhyp    5.1600 5.9700 8.2800 18.3000 215.0000
Chyp    0.1230 0.8030 3.1000 13.0000 211.0000
Ehyp    3.9800 4.0100 4.0200  4.0400   4.0700
sigma.X 0.0295 0.0337 0.0362  0.0391   0.0457
\end{verbatim}

\subsubsection{Posterior draws and parameter estimates for the
  informative prior for \texorpdfstring{\qlogisp}{qlogisp} (rectangular hyperbola model)}
\begin{figure}[ht]
\begin{tabular}{cc}
(a) & (b) \\[-1.0ex]
\rsplidapdffiguresize{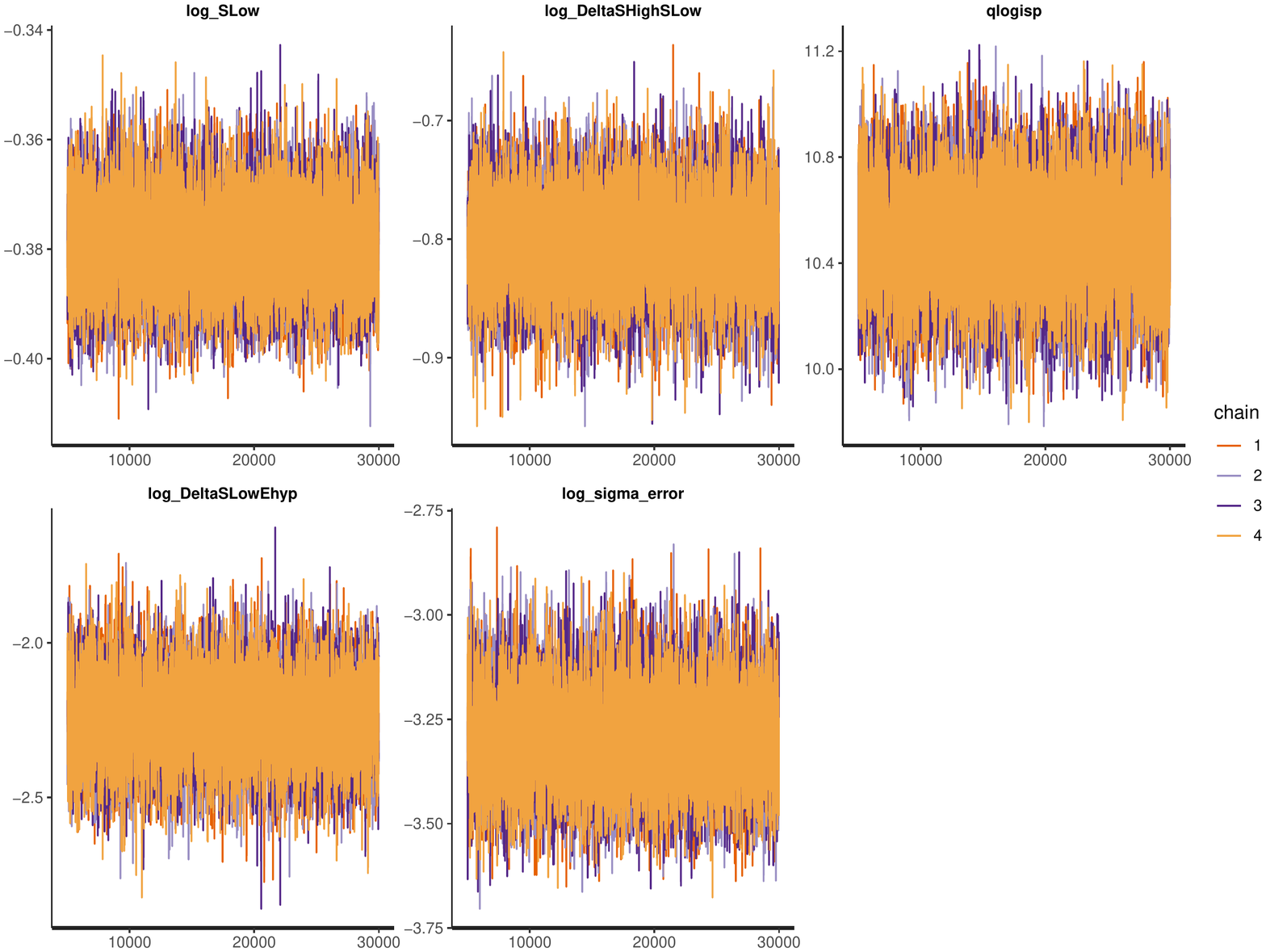}{3.25in} &
\rsplidapdffiguresize{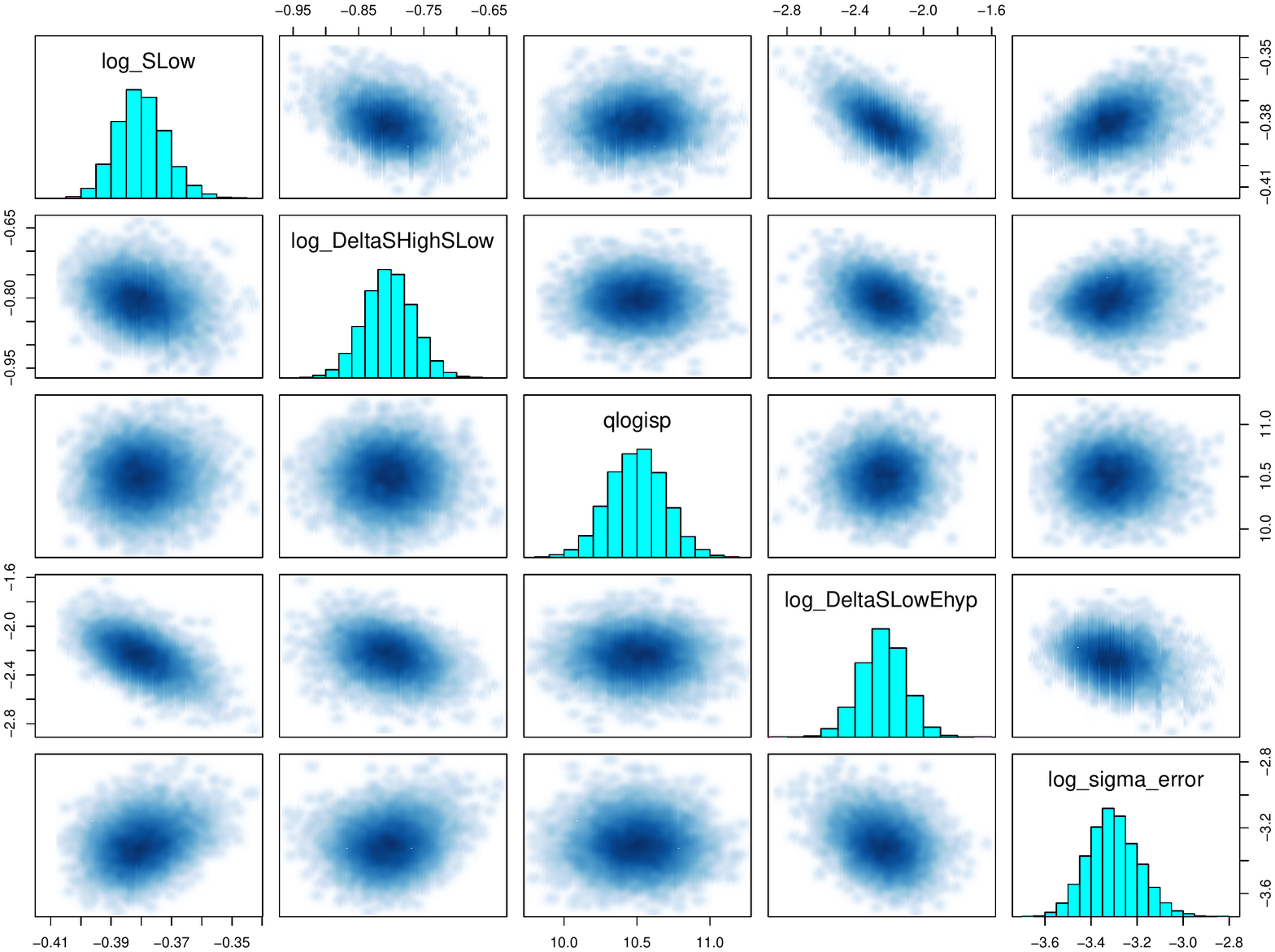}{3.25in}
\end{tabular}
\caption{Trace plots~(a) and a pairs plot~(b) of the
  posterior draws for the Rectangular Hyperbola model fit
  to the Ti64 data.}
\label{S.figure:Ti64.estimation.plots.RHLimit}
\end{figure}

The following table is a summary of the posterior draws for the USPs.
\begin{verbatim}
Inference for Stan model: Nishijima.
4 chains, each with iter=30000; warmup=5000; thin=5;
post-warmup draws per chain=5000, total post-warmup draws=20000.

                    mean se_mean   sd  2.5%   25%   50%   75% 97.5% n_eff Rhat
log_SLow           -0.38       0 0.01 -0.40 -0.39 -0.38 -0.37 -0.36 19994    1
log_DeltaSHighSLow -0.80       0 0.04 -0.88 -0.83 -0.80 -0.78 -0.73 19513    1
qlogisp            10.50       0 0.19 10.12 10.37 10.50 10.63 10.89 19534    1
log_DeltaSLowEhyp  -2.24       0 0.13 -2.50 -2.32 -2.24 -2.15 -1.99 19599    1
log_sigma_error    -3.30       0 0.11 -3.51 -3.38 -3.31 -3.23 -3.08 19175    1
\end{verbatim}
Figure~\ref{S.figure:Ti64.estimation.plots.RHLimit}a shows
trace plots for the parameters (indicating good mixing) and
Figure~\ref{S.figure:Ti64.estimation.plots.RHLimit}b the
corresponding pairs plot.
Note that the marginal posterior distribution of $\qlogis$ matches
the informative prior described above. This is because the relative
profile likelihood for $\qlogis$ is flat over the range of the
specified informative prior.

The following table is a summary of the marginal posterior draws for
the TPUSs, computed as described in
\citet[][Sections~4.4 and~4.5]{LiuMeeker2024}.
\begin{verbatim}
Ti64 350F R=-1 Lognormal distribution Nishijima model
 Summary of the marginal posteriors for the
   Traditional Parameters based on UnScaled data (TPUSs)
      based on 20000 Stan draws
            2.5%      25%      50%      75%    97.5%
Ahyp    781.0000 1.02e+03 1.18e+03 1.35e+03 1.75e+03
Bhyp    564.0000 9.91e+02 1.21e+03 1.44e+03 1.95e+03
Chyp    640.0000 9.43e+02 1.17e+03 1.42e+03 2.06e+03
Ehyp      3.9700 4.00e+00 4.01e+00 4.03e+00 4.05e+00
sigma.X   0.0299 3.41e-02 3.66e-02 3.95e-02 4.65e-02
\end{verbatim}

\subsubsection{Additional residual analyses}
\label{S.section:Ti64.additional.residual.analyses}
\begin{figure}[ht]
\begin{tabular}{cc}
(a) & (b) \\[-3.2ex]
\rsplidapdffiguresize{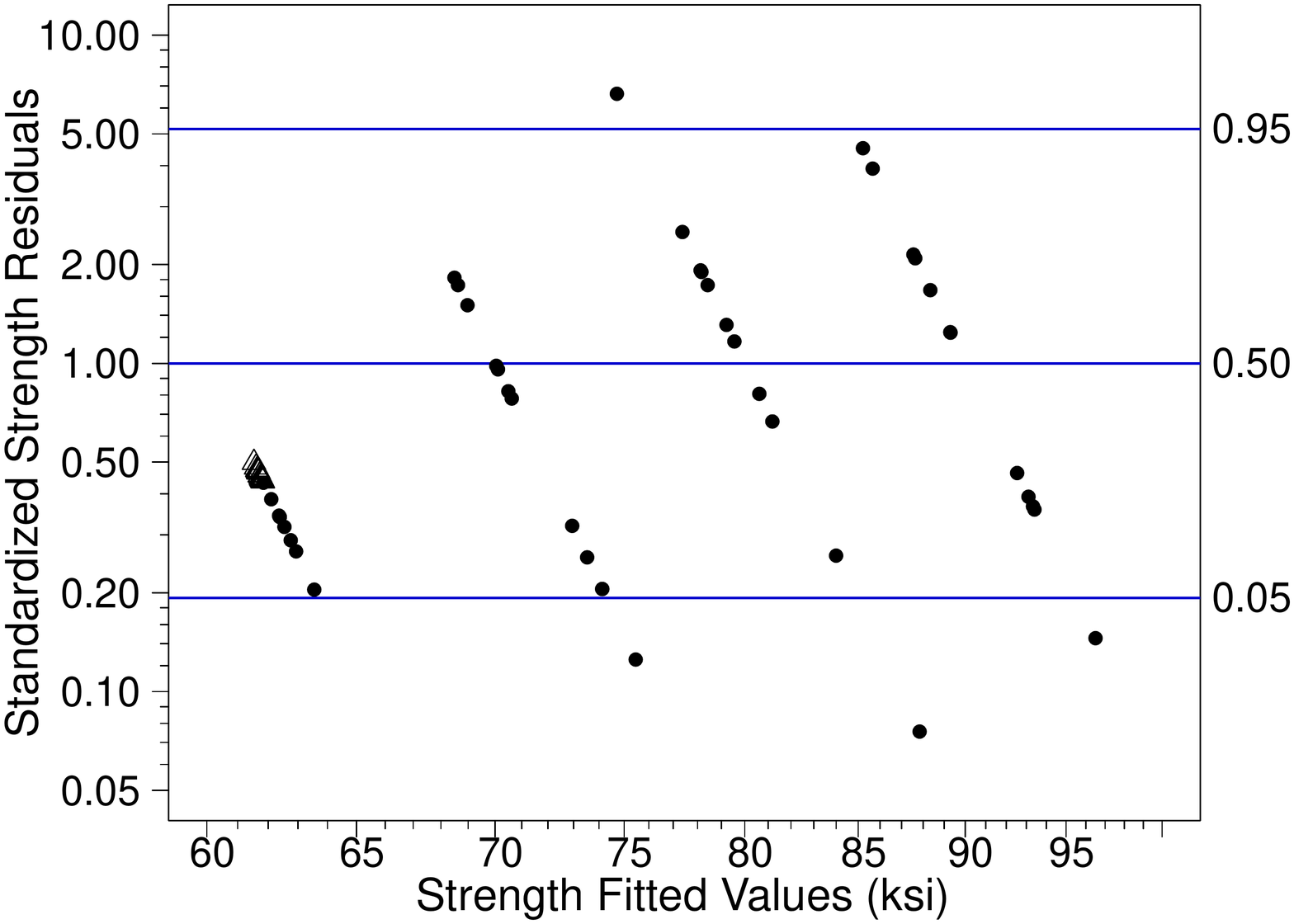}{3.25in} &
\rsplidapdffiguresize{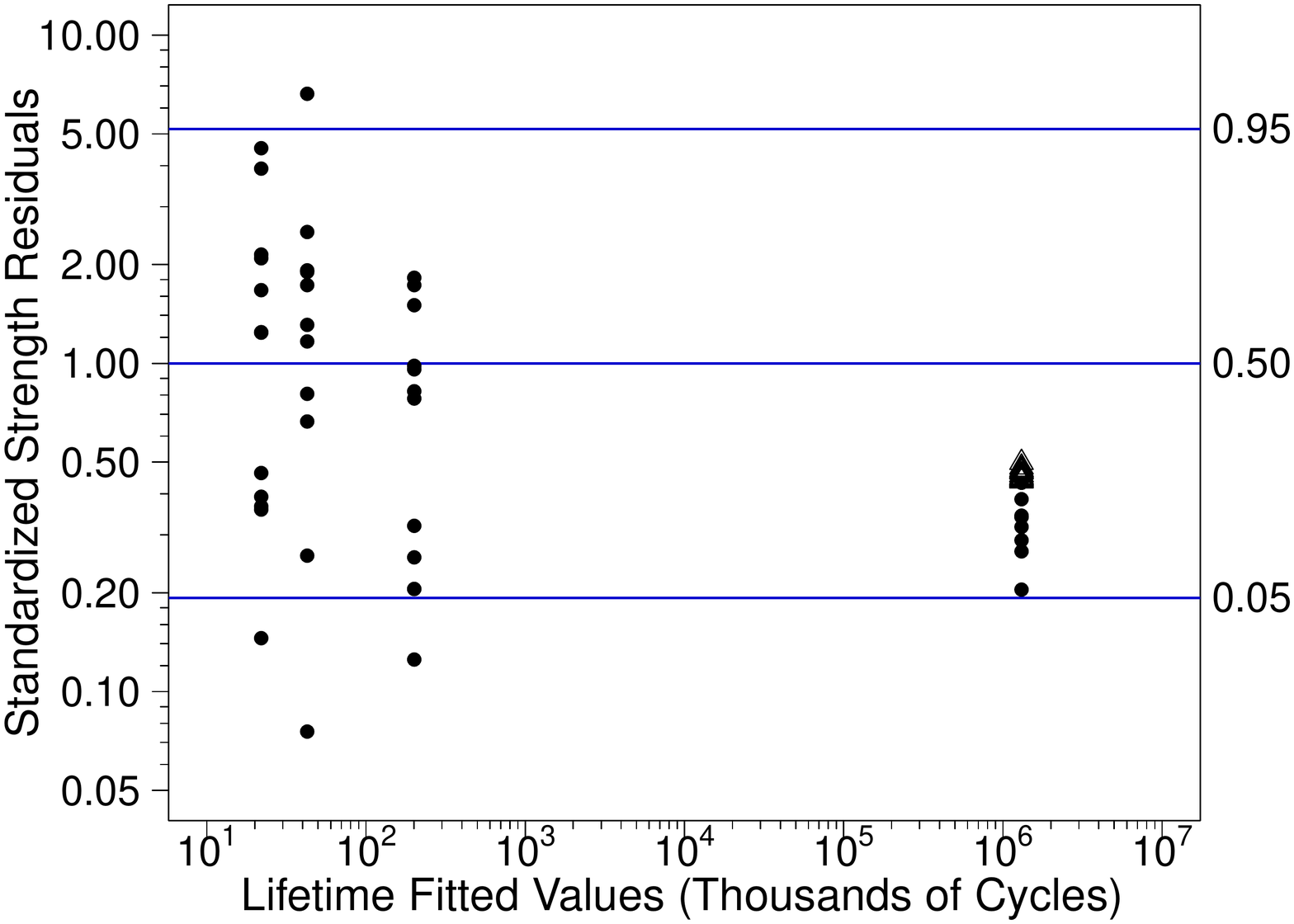}{3.25in}
\end{tabular}
\caption{Residuals from the Nishijima model fit to the Ti64 data versus lifetime fitted values~(a) and
  strength fitted values~(b).}
\label{S.figure:Ti64Residual.fitted.values.plots}
\end{figure}

Figure~\ref{S.figure:Ti64Residual.fitted.values.plots} shows plots
of the standardized strength residuals based on the fit of the
Nishijima model fit to the Ti64 data versus lifetime~(a) and
strength~(b) fitted values.
In the residuals versus lifetime fitted values plot, each column of
residuals corresponds to one of the four stress levels that were
used in the experiment. Taking into account the 28 out of 37 runouts
at 60 ksi, there is no evidence of any departure from the assumed
model.

In the residuals versus strength fitted values plot, again, each
group corresponds to a level of stress used in the experiment. The
plot of residuals versus strength fitted values is, however, harder
to interpret because the fitted values also depend on the
number of cycles for each data point.

\subsection{More Details for the  Coffin--Manson/Lognormal \SN{} Model
  Fit to the Superelastic Nitinol Data}
\label{S.section:more.details.nitinol.coffin.manson}

\subsubsection{Comparison of lognormal and Weibull distributions}
\begin{figure}[ht]
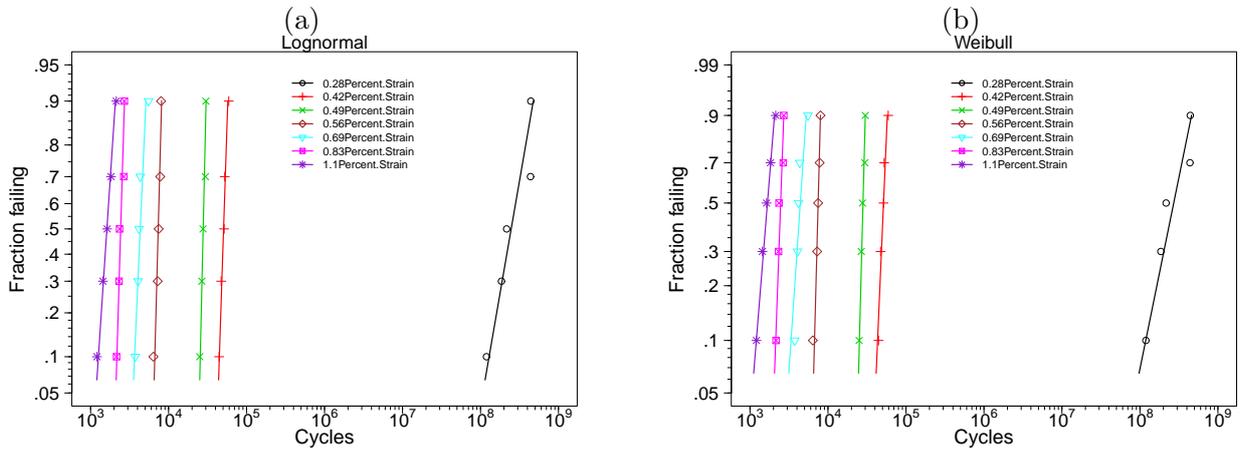

\begin{tabular}{cc}
(a) & (b) \\[-3.2ex]
\rsplidapdffiguresize{./pdfFigures/Nitinol02x_groupi_lognormal.hide.pdf}{3.25in} &
\rsplidapdffiguresize{./pdfFigures/Nitinol02x_groupi_weibull.hide.pdf}{3.25in}
\end{tabular}
\caption{A comparison of lognormal~(a) and Weibull~(b) probability
  plots for the nitinol data with the bimodal data at 0.56\% strain removed.}
\label{S.figure:Nitinol.lognormal.Weibull.probability.plots}
\end{figure}

Figure~\ref{S.figure:Nitinol.lognormal.Weibull.probability.plots}
uses lognormal and Weibull probability plots to compare the two
distributions fit to the nitinol data. Both distributions fit well.

\subsubsection{Model and prior distributions}
Example~\ref{example:modeling.superelastic.nitinol.sn.data} in the main
paper gives a summary of the results of fitting the Coffin--Manson
model to the superelastic nitinol \SN{} data.

 Note that in this example, the fatigue-strength model is specified
and the fatigue-life model (providing the basis for the likelihood
function) is induced.
The lognormal distribution regression model for fatigue-strength $X$
was
\begin{align*}
  X= \log[h(N; \betavec)] + \sigma_{X} \epsilon
\end{align*}
where stress $S$ and number of cycles $N$
were scaled for numerical reasons (as described in
\citet{LiuMeeker2024}), and $\epsilon  \sim \NORM(0,1).$
Specifically, $S/S_{\textrm{max}}$ was substituted for $S$ and  $N/N_{\textrm{max}}$ was substituted for $N$.
The Coffin--Manson \SN{} relationship is
\begin{align*}
S = h(N; \betavec)= \Ael (2N)^{b} +  \Apl (2N)^{c}  ,
\end{align*}
The unknown parameters to be estimated from the data are
$\Ael>0$, $\Apl \geq 0$, $b \leq 0$, $c<0$, with the constraint $|c|>|b|$.

As suggested in \citet[][Section~4.2]{LiuMeeker2024}, the MCMC
algorithm was run to generate posterior draws of the unrestricted
stable parameters (USPs).
First we define two points on the \SN{} curve
\begin{align*}
S_{\Lowx} & =  \Ael(2N_{\Highx})^{b}+\Apl(2N_{\Highx})^{c}\\
S_{\Highx} & =  \Ael(2N_{\Lowx})^{b}+\Apl(2N_{\Lowx})^{c},
\end{align*}
where $N_{\Lowx}$ is the smallest value of $N$ in the data and
$N_{\Highx}$ is the largest value of $N$ in the data that is a failure.
Then, following \citet{LiuMeeker2024}, the USPs are defined as
\begin{align*}
& \logSLow  = \log(S_{\Lowx}) \\
& \logDeltaHighLow = \log[\log(S_{\Highx}) - \log(S_{\Lowx})] \\
  &  \qlogisp = \qlogis(b/\limitSlope)\\
  & \logDeltaSlopes  =  \log\left(\limitSlope-c\right), \textrm{ and}\\
& \logSigmaX=\log(\sigma_{X}),
\end{align*}
where $\qlogis=\log[p/(1-p)]$ is the logit transformation and
\begin{align*}
\limitSlope & = \frac{\log(S_{\Lowx})-\log(S_{\Highx})}{\log(N_{\Highx})-\log(N_{\Lowx})}.
\end{align*}

The marginal prior distributions for the USPs $\logSLow$,
$\logDeltaHighLow$, $\qlogisp$, $\logDeltaSlopes$, and $\logSigmaX$,
were chosen to be flat. The resulting noninformative joint prior for
the USPs results in Bayesian posterior estimates that are close to
the ML estimates.

\subsubsection{Posterior draws and parameter estimates}
\begin{figure}[ht]
\begin{tabular}{cc}
(a) & (b) \\[-1.0ex]
\rsplidapdffiguresize{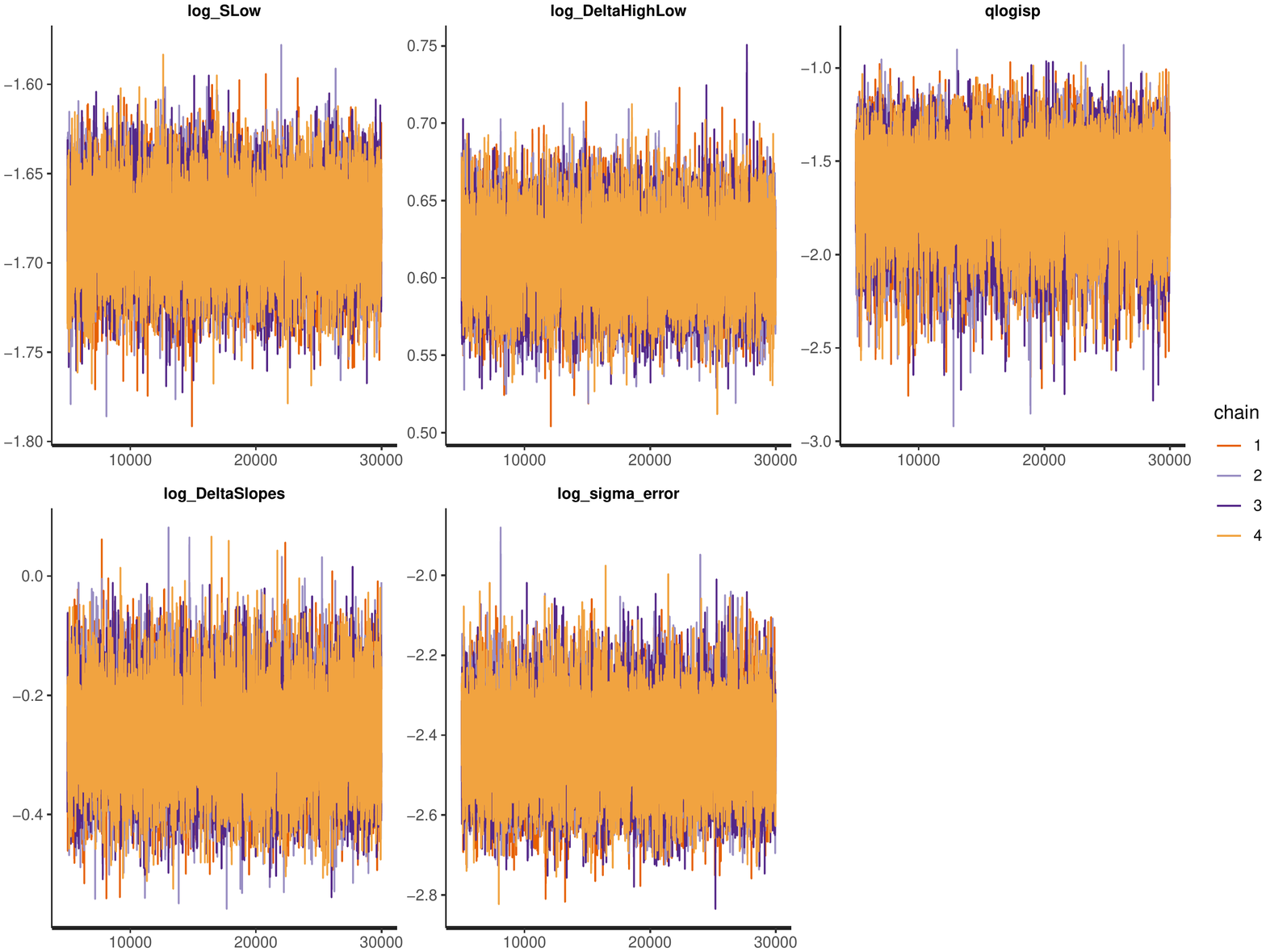}{3.25in} &
\rsplidapdffiguresize{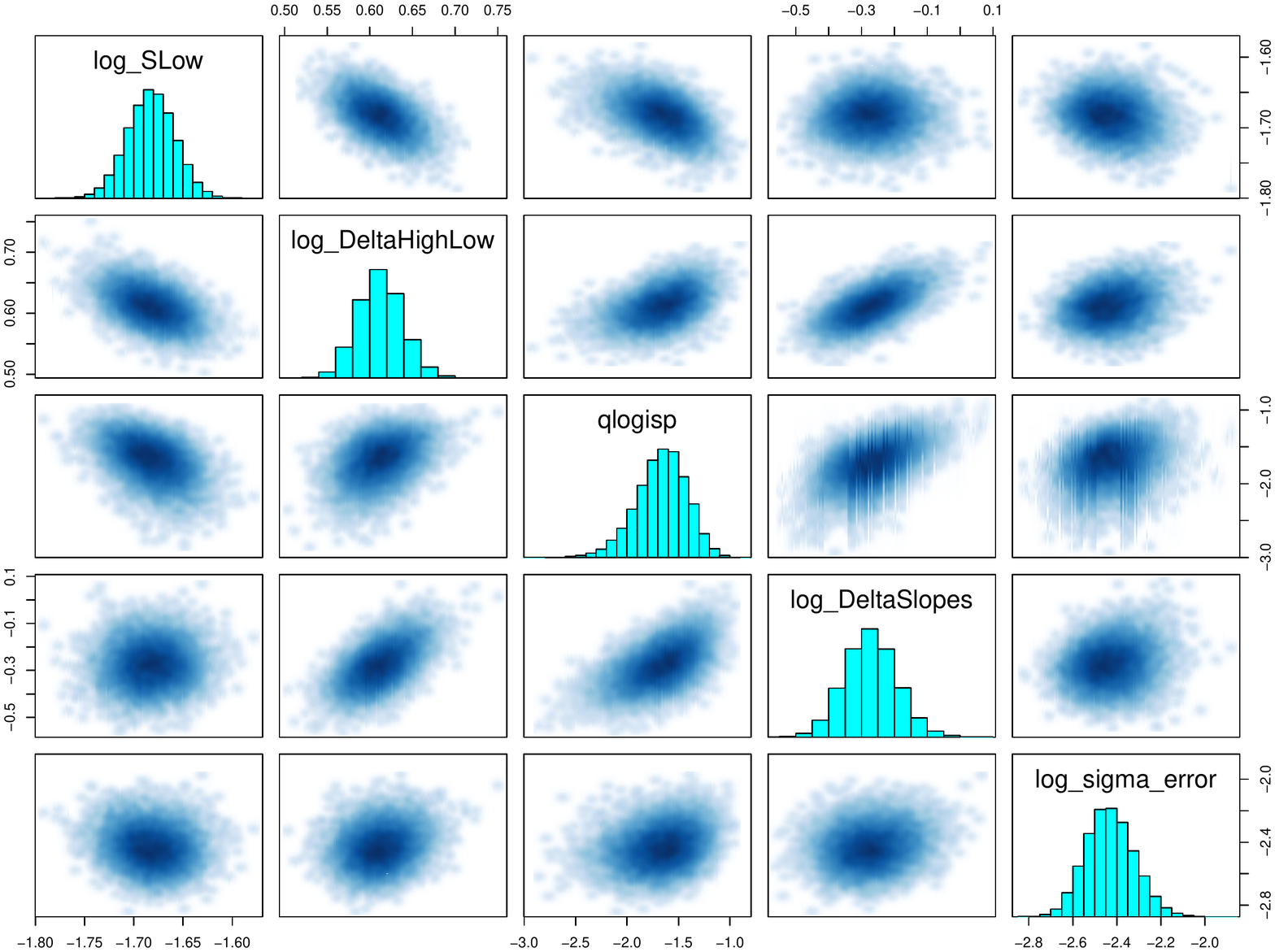}{3.25in}
\end{tabular}
\caption{Trace plots~(a) and a pairs plot~(b) of the
  posterior draws for the Coffin--Manson model fit
  to the nitinol data.}
\label{S.figure:nitinol.estimation.plots}
\end{figure}

The following table is a summary of the posterior draws for the USPs.
\begin{verbatim}
Inference for Stan model: CoffinManson.
4 chains, each with iter=30000; warmup=5000; thin=5;
post-warmup draws per chain=5000, total post-warmup draws=20000.

                  mean se_mean   sd  2.5%   25%   50%   75% 97.5% n_eff Rhat
log_SLow         -1.68       0 0.02 -1.73 -1.70 -1.68 -1.67 -1.63 19137    1
log_DeltaHighLow  0.61       0 0.03  0.56  0.59  0.61  0.63  0.66 17993    1
qlogisp          -1.67       0 0.25 -2.20 -1.82 -1.65 -1.50 -1.23 17819    1
log_DeltaSlopes  -0.27       0 0.08 -0.42 -0.33 -0.27 -0.22 -0.11 18460    1
log_sigma_error  -2.43       0 0.11 -2.64 -2.51 -2.43 -2.36 -2.20 19169    1
\end{verbatim}
Figure~\ref{S.figure:nitinol.estimation.plots}a shows
trace plots for the parameters (indicating good mixing) and
Figure~\ref{S.figure:nitinol.estimation.plots}b the
corresponding pairs plot.

The following table is a summary of the marginal posterior draws for
the TPUSs, computed as described in
\citet[][Sections~4.4 and~4.5]{LiuMeeker2024}.
\begin{verbatim}
Nitinol e-N data from Falk (2019) Lognormal distribution CoffinManson model
 Summary of the marginal posteriors for the
   Traditional Parameters based on UnScaled data (TPUSs)
      based on 20000 Stan draws
            2.5%       25%       50%       75%    97.5%
Ael       0.6700    0.7500    0.7970    0.8500    0.962
Apl    1130.0000 2020.0000 2850.0000 4060.0000 8750.000
bel      -0.0332   -0.0265   -0.0231   -0.0198   -0.014
cpl      -1.0400   -0.9480   -0.9050   -0.8640   -0.794
sigmaX    0.0718    0.0814    0.0877    0.0944    0.110
\end{verbatim}

\subsubsection{Additional residual analyses}
\label{S.section:nitinol.additional.residual.analyses}
\begin{figure}[ht]
\begin{tabular}{cc}
(a) & (b) \\[-3.2ex]
\rsplidapdffiguresize{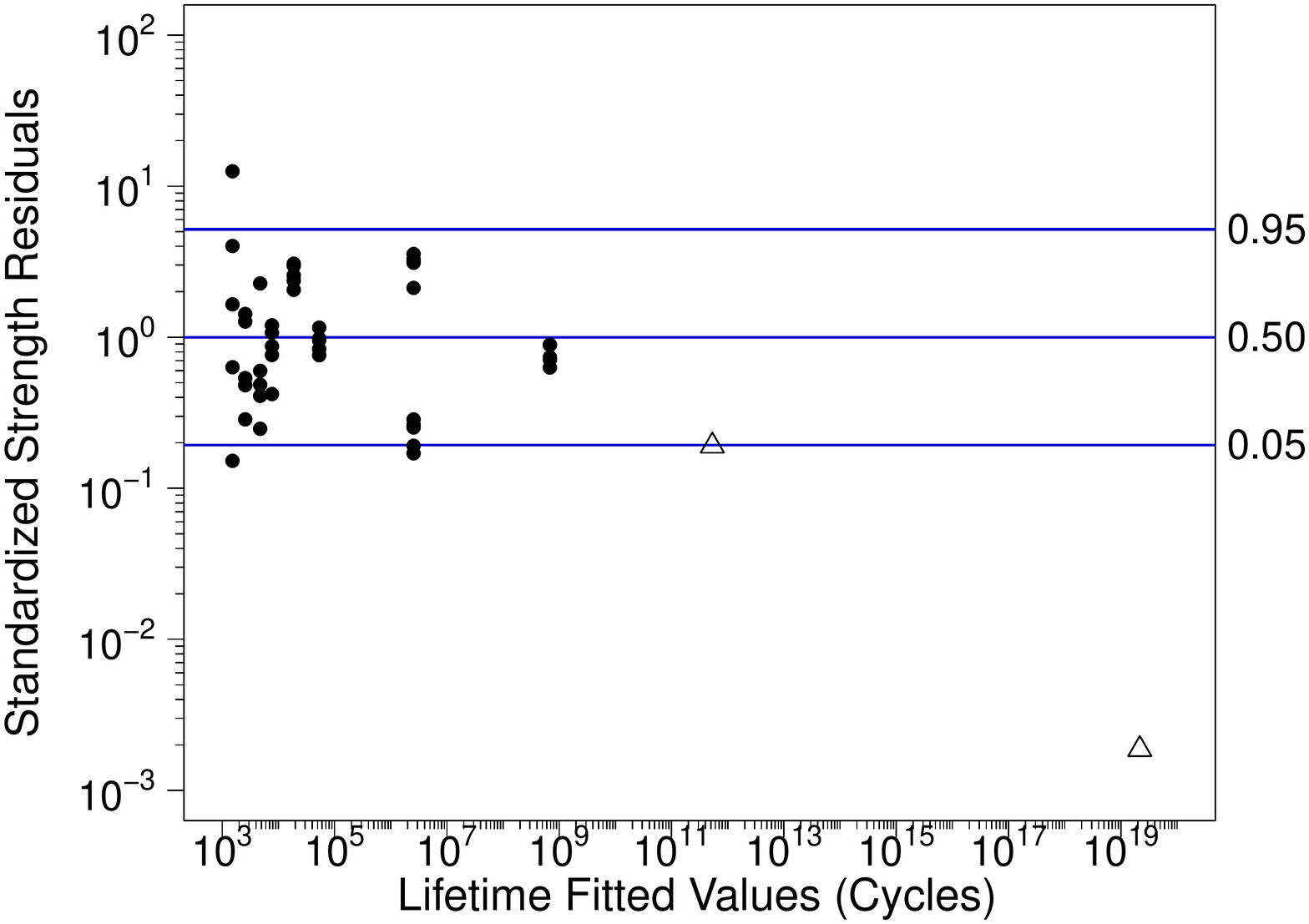}{3.25in} &
\rsplidapdffiguresize{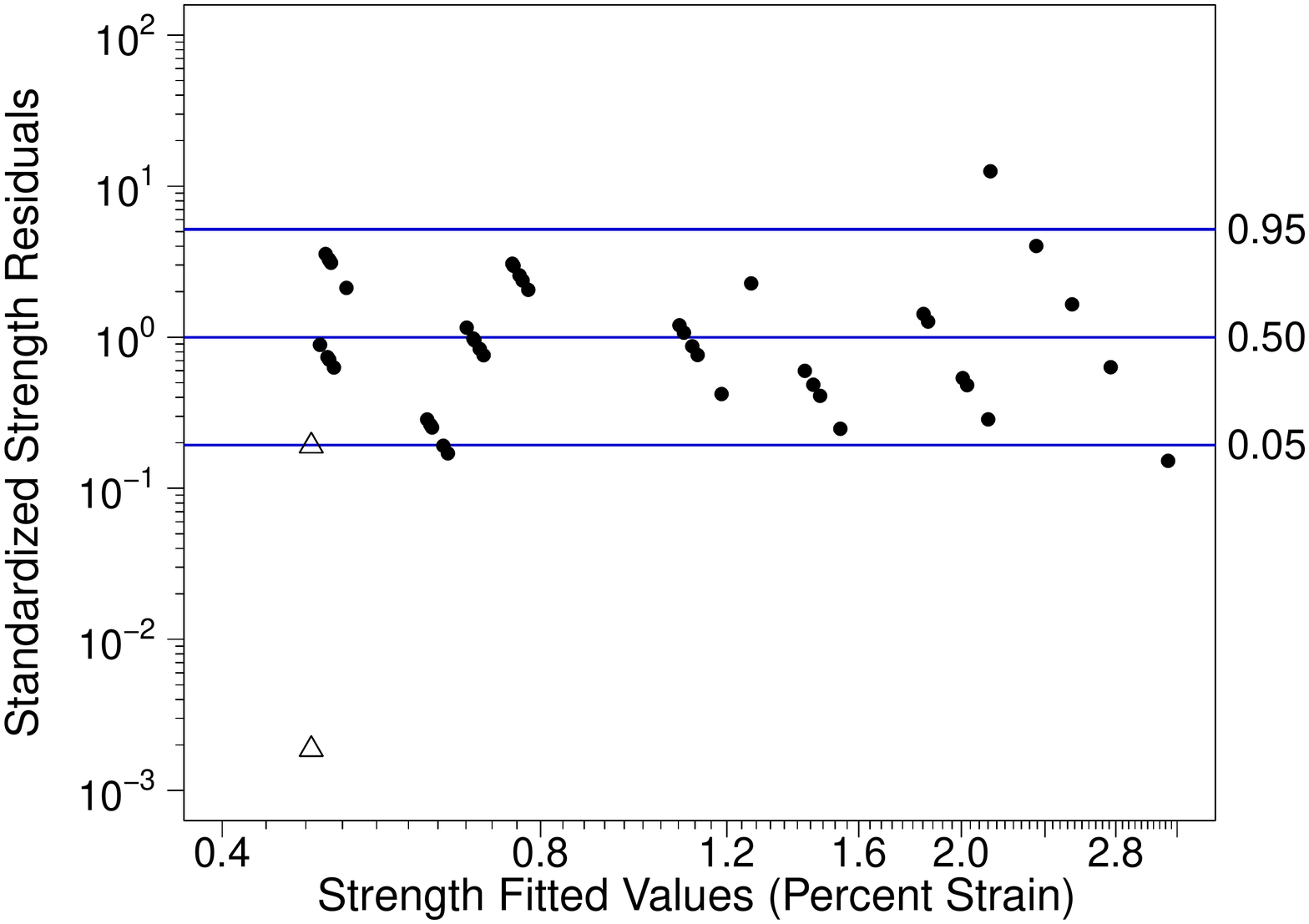}{3.25in}
\end{tabular}
\caption{Residuals from the Coffin--Manson model fit to the nitinol
  data versus lifetime fitted values~(a) and
  strength fitted values~(b).}
\label{S.figure:NitinolResidual.fitted.values.plots}
\end{figure}

Figure~\ref{S.figure:NitinolResidual.fitted.values.plots} shows plots
of the standardized strength residuals based on the fit of the
Coffin Manson model to the nitinol data versus lifetime~(a) and
strength~(b) fitted values.

\addcontentsline{toc}{section}{\protect\numberline{}References}

\end{document}